\documentclass[%
aip,
amsmath,amssymb,
reprint,%
]{revtex4-1}

\usepackage{graphicx}% Include figure files
\usepackage{dcolumn}% Align table columns on decimal point
\usepackage{bm}% bold math
\usepackage[utf8]{inputenc}
\usepackage[T1]{fontenc}
\usepackage{mathptmx}

\usepackage{graphicx}% Include figure files
\usepackage{caption}
\usepackage{subcaption}

\usepackage{xcolor}

\begin{document}
	
	\preprint{AIP/123-QED}
	
	\title[On the dynamics of the jet wiping process]{On the Dynamics of the jet Wiping Process: Numerical Simulations and Modal Analysis}
	% Force line breaks with \\
	
	\author{David Barreiro-Villaverde}
	\email{david.barreiro1@udc.es}
	\affiliation{ 
		CITIC Research, Universidade da Coruña, Campus de Elviña, 15071 A Coruña, Spain 
	}%
	\author{Anne Gosset}%
	\affiliation{ 
		Technological Research Center (CIT), Universidade da Coruña, Campus de Esteiro, 15403 Ferrol, Spain 
	}%
	\author{Miguel A. Mendez}
	\affiliation{%
		von Karman Institute for Fluid Dynamics, Waterloosesteenweg 72, Sint-Genesius-Rode, Belgium
	}%
	
	\date{\today}% It is always \today, today,
	
	\begin{abstract}

		We analyze the flow of a planar gas jet impinging on a thin film, dragged by a vertical moving wall. In the coating industry, this configuration is known as jet wiping, a process in which impinging jets control the thickness of liquid coatings on flat plates withdrawn vertically from a coating bath. We present three-dimensional (3D) two-phase flow simulations combining Large Eddy Simulation (LES) and Volume of Fluid (VOF). Three wiping configurations are simulated and the results are validated with experimental data from previous works. Multiscale modal analysis is used to analyze the dynamic interaction between the gas flow and the liquid film. In particular, we present a combination of Multiscale Proper Orthogonal decomposition (mPOD)  and correlation analysis. The mPOD is used to identify the dominant travelling wave pattern in the liquid film flow, and the temporal structures are used to determine the most correlated flow features in the gas jet. This allows for revealing a two-dimensional (2D) mechanism for wave formation in the liquid coat. Finally, we use the numerical results to analyze the validity of some of the critical assumptions underpinning the derivation of integral film models of jet wiping.
		
	\end{abstract}
	
	\maketitle

	\section{\label{sec:intro}Introduction}

\begin{figure}
		\includegraphics[width=\linewidth]{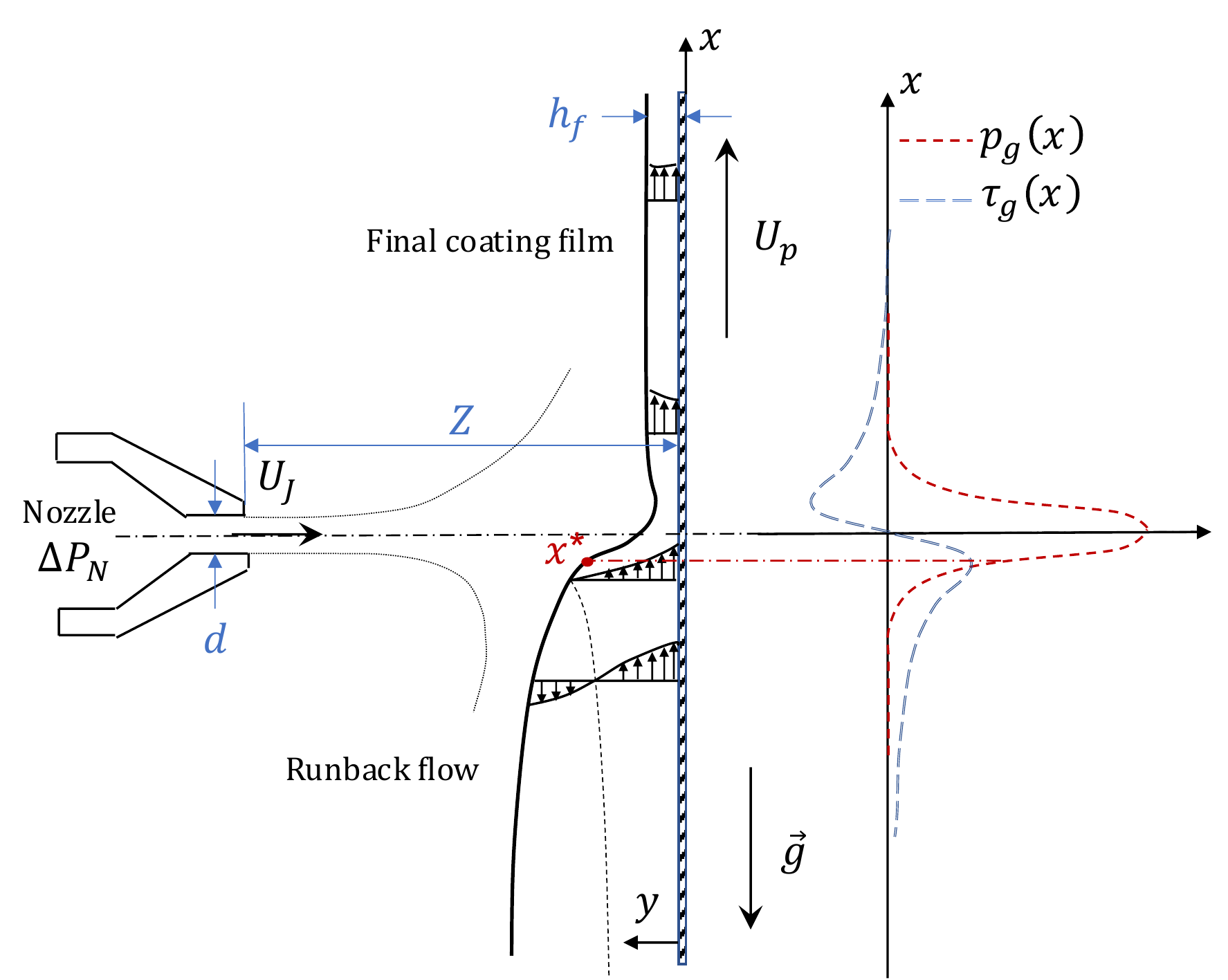}% Here is how to import EPS art
		\caption{\label{fig:sketch} Schematic of the jet wiping process, recalling the main operating parameters. The substrate moves upwards (positive $x$ direction) at a speed $U_p$. The average final film thickness is denoted $h_f$. The pressure and shear stress distributions produced by the impinging jet at the film interface are denoted $P_g(x)$ and $\tau_g(x)$, respectively, and their typical shape is represented on the right hand side. The location at which the pressure gradient is maximum is the wiping point, $x^*$.}
	\end{figure}

	The jet wiping, also known as jet stripping, is a continuous coating process in which impinging jets control the thickness of a liquid layer on an upward-moving substrate. This liquid is dragged by viscosity as the substrate emerges from a bath. In the case of planar jet wiping, the process involves two slot jets impinging on each side of the strip, acting literally as \emph{air-knives}. Fig.~\ref{fig:sketch} shows a schematic of the process. An impinging jet allows reducing the coating layer thickness by introducing a pressure gradient $\partial_x p_g(x)$ and an interface shear stress distributions $\tau_g (x)$, qualitatively sketched in Fig.~\ref{fig:sketch}.

	These two quantities are known as \emph{wiping actuators}, which force part of the liquid to reverse direction and form a \emph{run-back} flow. The process is characterized by the nozzle slot opening ($d$), its distance to the substrate ($Z$), stagnation pressure ($\Delta P_N$, which leads to a jet exit velocity $U_J$) and the substrate speed ($U_p$). The relevant fluid properties are the gas and the liquid densities ($\rho_g,\rho_l$) and viscosities ($\mu_g$,$\mu_l$) as well as the gas-liquid surface tension ($\sigma$).
	
	Jet wiping is very popular in the galvanization industry because of its comparatively low maintenance cost, high productivity and energy efficiency. This process has also been widely used for photographic films manufacturing and in the paint industry \cite{Cohen1992,TAPPI}. Simple models to predict the average film thickness $h_f$ as a function of all the operating parameters have been developed \cite{Thornton1976,Tuck1983,Ellen1984,Buc1997} and successfully implemented for design and control of industrial lines. 
	
	The window of operating conditions of these lines, however, is significantly limited by the occurrence of two unsteady phenomena. The first is known as \emph{splashing}, and consist in the violent break-up of the run-back flow. In galvanization lines, this phenomenon endangers the plant operators due to the high temperatures of the liquid zinc and must therefore be always avoided. Gosset and Buchlin\cite{Gosset2007b} provide empirical correlation to predict the onset of splashing while Myrillas \emph{et al.}\cite{Myrillas2011a} analyzed different techniques to delay its occurrence.
	
	The second limiting phenomenon is known as \emph{undulation}, and consists in the appearance of large amplitude and long-wavelength waves in the final coating. After solidification, these waves affect the quality of the final product. Although the long-wavelength nature of the undulation is reminiscent of interface instabilities in gravity-driven falling films\cite{Kalliadasis2012,Alekseen1994,Demekhin2007}, several works\cite{Ellen1984,Hocking2011,Gosset2007} have shown that these do not result from an intrinsic instability of the liquid coat. On the other hand, it has been recently shown\cite{Johnstone2019,Mendez2020} that the liquid film is sensitive to jet disturbances such as pulsation and oscillations, if these occur at sufficiently low frequencies. The present work analyzes the origin of jet disturbances and their impact one the coating film.
	
	Jet oscillations were first observed by Myrillas et al.\cite{Myrillas2011,Myrillas2013}, who simulated the process using high fidelity 2D and 3D simulations, combining Large Eddy Simulations (LES) for the gas flow and Volume of Fluid (VOF) treatment of the two phases. The dynamics of the interaction between the gas jet and the liquid film was found to be qualitatively in agreement with high-speed flow visualizations in a laboratory scale experiment: the impinging jet featured large scale oscillations, coupled with the pulsation of the liquid film in the run-back flow. Similar behaviour was reported by Pfeiler et al.\cite{Pfeiler2018}, who analyzed a more challenging industrial configuration albeit limiting the analysis to 2D simulations.
	
	%Jet oscillations were first observed by Myrillas \emph{et al.}\cite{Myrillas2011,Myrillas2013} using 2D and 3D Computational Fluid Dynamics (CFD) simulations of the process. The high fidelity model combines Large Eddy Simulations (LES) for the gas flow and Volume of Fluid (VOF) treatment of the two phases. The dynamics of the interaction between the gas jet and the liquid film was found to be qualitatively in agreement with high-speed flow visualizations in a laboratory scale experiment: the impinging jet featured large scale oscillations, coupled with the pulsation of the liquid film in the run-back flow. Similar behaviour was reported by Pfeiler \emph{et al.}\cite{Pfeiler2018}, who analyzed a more challenging industrial configuration albeit limiting the analysis to 2D simulations.
	
	The dynamics of the jet wiping was investigated experimentally by Gosset \emph{et al.}\cite{Gosset2019} and Mendez \emph{et al.}\cite{Mendez2019} combining Time-Resolved Particle Image Velocimetry (TR-PIV) for the gas flow, Laser-Induced Fluorescence (LIF)- based interface tracking of the impinged liquid film, and Light Absorption (LABS) thickness measurement for the final coating flow. These works highlighted the link between the frequency content of the jet oscillation and the frequency content of the coating undulation over a range of operating conditions. Moreover, the analysis of the jet flow field revealed flow structures and patterns that are encountered in the confinement-driven oscillation of impinging jets, identified via TR-PIV and multiscale modal decompositions by Mendez \emph{et al.}\cite{Mendez2018a}.
	These experimental campaign focus nevertheless on operating conditions which are far from industrial conditions in galvanizing lines, as a complete similarity cannot be reached at a laboratory scale.
	
	Similar studies in galvanizing conditions are still unfeasible from an experimental side, and computationally prohibitive from the numerical side, as discussed by Aniszewski \emph{et al.} \cite{Aniszewski2019}. These authors were the first to present a 3D, two-phase flow simulation of the process in galvanizing conditions and managed to resolve the wiping meniscus region accurately using an adaptive grid approach. The computational cost, however, prevented the analysis beyond the initial stage of the wiping, when the gas jet first impinges on the liquid coating. 
	
	As of today, the only tools available to analyze \emph{some} of the mechanisms at the origin of the undulation in galvanizing conditions are simplified integral models, recently extended by Mendez \emph{et al.}\cite{Mendez2020}. These models were used to derive the response of the liquid film to a set of perturbations and identify, for various operating conditions, the range of frequencies yielding the largest disturbance amplification. However, these integral models are based on two critical simplifying assumptions. The first assumption is that the gas-liquid interaction obeys a one-way coupling, with the wiping actuators assumed to be independent of the liquid film dynamics and modelled using experimental \cite{Beltaos1973,Tu1996} and numerical \cite{Naphade2005,Elsaadawy2007} correlations derived for jets impinging on dry walls. The second assumption is that the velocity profile within the liquid film is approximately parabolic, with the simplest model also assuming self-similarity.

	This work analyzes the dynamics of the jet wiping process using high-fidelity 3D LES-VOF simulations with three main objectives. First, by considering some of the experimental test cases presented in Mendez \emph{et al.}\cite{Mendez2019}, we validate this numerical approach using experimental data. Second, we extend the modal analysis presented by Mendez \emph{et al.}\cite{Mendez2019} using Multiscale Proper Orthogonal Decomposition (mPOD) to both the gas jet field and the final coating thickness distribution. The mPOD\cite{Mendez2019a,Mendez2018b,Mendez2020_MST} is a data-driven decomposition which allows identifying coherent patterns in data according to both energy contribution \emph{and} frequency range. In this work, the mPOD is performed on the liquid film thickness contours to retrieve the dominant travelling wave patterns. Then, the temporal evolution of these patterns is correlated with the gas flow, to identify the dominant flow structures evolving within the range of frequencies characterizing the undulation. Finally, we analyze the validity of the simplifying assumptions supporting integral models, namely the long-wave formulation and the self-similarity of the velocity profiles within the liquid film.
		
	Section \ref{sec:numerical_methods} presents the investigated test cases and the numerical methodology employed for the LES-VOF simulations. Section \ref{SecIII} presents the data processing and briefly reviews the fundamental of mPOD. Section \ref{SecIV} reviews the fundamentals of integral modelling and the simplifying assumptions tested in this work. Section \ref{sec:results} presents and discusses the results. Conclusions and perspectives are drawn in section \ref{sec:conclusions}.

	\section{\label{sec:numerical_methods}Test Cases and Numerical Methods}

	\subsection{\label{sec:test_cases} Test cases}

	The test cases analyzed in this work are three of the experiments described in previous studies\cite{Mendez2019,Mendez2018}. The substrate is assumed to be flat, and withdrawn from a bath of isothermal coating.
	The working liquid is dipropylene-glycol, with density $\rho_l=1023$ kg/m$^3$, dynamic viscosity $\mu_l= 0.075$ Pa$\cdot$s and surface tension $\sigma = 0.032$ N/m. The gas is air, with density $\rho_g=1.2$ kg/m$^3$, and dynamic viscosity $\mu_g= 1.776 \times 10^{-5}$ Pa$\cdot$s. The coating liquid dragged by the substrate is impinged by a slot air jet, with a nozzle opening $d=1.3$ mm. The internal and external geometry of the nozzle reproduces the one of the experiments. The substrate speed is constant and equal to $U_p=0.34$ m/s. The selected wiping conditions include two standoff distances ($Z= 25.2$ and $18.5$ mm, corresponding to normalized distances $\hat{Z}=Z/d=14.2$ and $17.2$), and two nozzle pressures for $\hat{Z}=14.2$ ($\Delta P_N=425$ and $875$ Pa). The three configurations are summarized in Table \ref{tab:config} in which the dimensionless wiping number $\Pi_g=\Delta P_N d/\left(\rho_l g Z^2 \right)$, known to be representative of the process \cite{Gosset2019}, is indicated. 
	
	\begin{table}
		\caption{\label{tab:config} Conditions of the three analyzed test cases, in terms of dimensionless stand-off distance $\hat{Z}=Z/d$, nozzle pressure $\Delta P_N$ and wiping number $\Pi_g=\Delta P_N d/ (\rho_l g Z^2)$. The remaining set of parameters is recalled in the text.}
		\begin{ruledtabular}
			\begin{tabular}{cccc}
				Case $\#$ & $\hat{Z}$ [-] & $\Delta P_N$ [Pa] & $\Pi_g$ [-] \\
				\hline
				\\
				1 & $14.2$ & 425 & $0.16$ \\
				%2 & $19.4$ & 425 & $0.09$ \\
				2 & $14.2$ & 875 & $0.33$ \\
				3 & $19.4$ & 875 & $0.18$ \\
			\end{tabular}
		\end{ruledtabular}
	\end{table}
	
	\begin{figure*}
		\includegraphics[width=\linewidth]{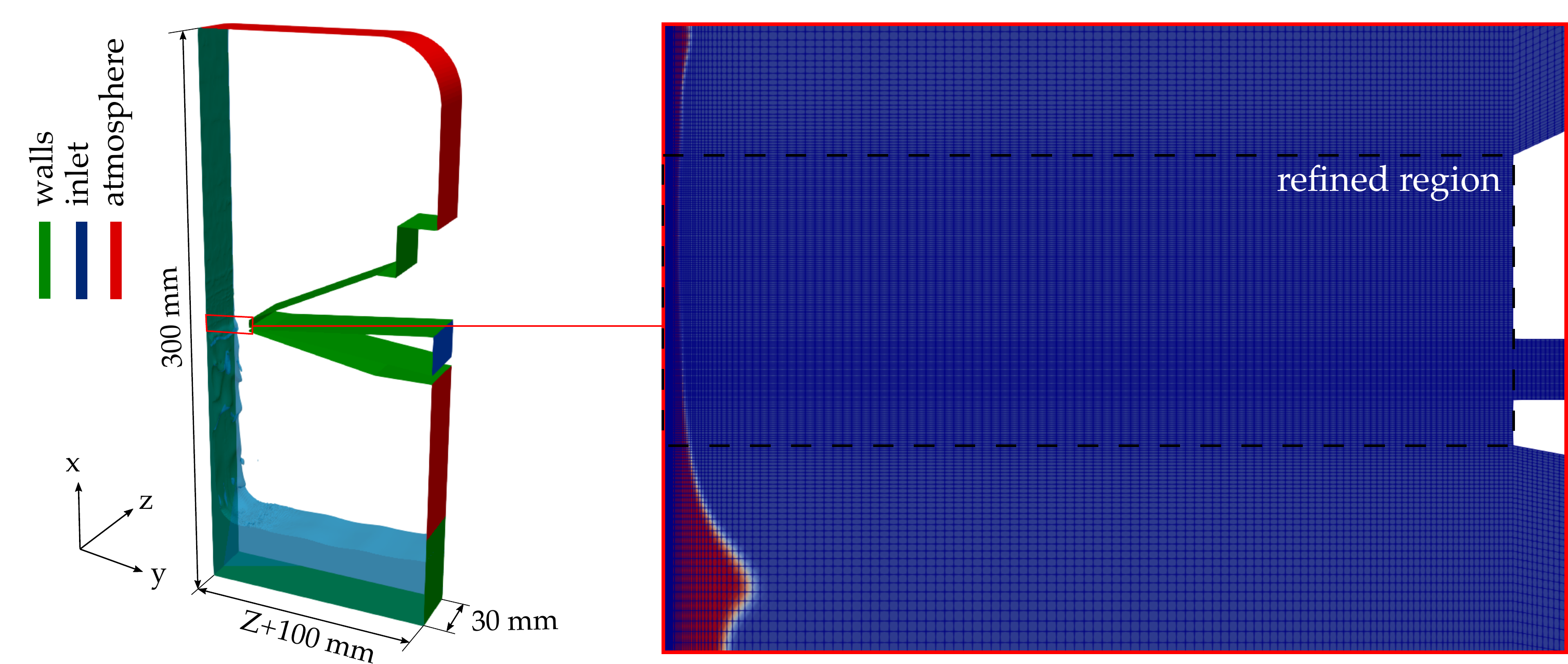} %{meshSketch_v2.jpg}% Here is how to import EPS art
		\caption{\label{fig:boundaries} Left: 3D view of the computational domain with the prescribed boundary conditions and relevant units 2D representation of the computational domain, mesh details and boundary conditions. The front and back patches are defined as cyclic and the mesh is extruded 30 mm in the direction perpendicular to the present plane.}
	\end{figure*}
	
	The domain used for computations is represented in Fig. \ref{fig:boundaries} (left). It includes the coating bath, the moving flat substrate emerging from it, and a portion of the slot nozzle. The domain spans $300$ ${mm}$ in the direction of the moving substrate ($x$), $100$ $\mbox{mm}$ in addition to the stand-off distance $Z$ in the jet axis direction ($y$), and $30$ $mm$ in the span-wise direction ($z$).
	
	The domain is discretized with a structured mesh built in \textit{blockMesh}. The grid features 10 million cells for cases 1 and 2, and 12 million for case 3 at larger $Z$ (cf. Table \ref{tab:config}).
	The cell size in the refined region, shown in Fig. \ref{fig:boundaries} (right) is $\Delta x \approx 50$ $\mu m$,  while $\Delta y$ ranges from $2$ $\mu m$ close to the substrate to $200$ $\mu m$ at the exit of the nozzle. The cell width around the interface is of the order of $50$ $\mu m$. The cell size in $z$ is set constant an equal to $\Delta z=500$ $\mu m$. 
	
	The domain boundaries open to atmosphere are modelled as outlet conditions in which the reference pressure is set to 0. The solid boundaries are defined as walls, and the gas jet flow is established through an inlet flow condition in which the stagnation pressure $\Delta P_N$ is prescribed. The computations are initialized with a $40$ $mm$ high coating bath at the bottom of the domain and a $1$ mm thick liquid film on the moving strip, in order to speed up the development of the liquid film flow. 
	This value is loosely based on the expected thickness of the film dragged out by the substrate in absence of wiping, following the optimal conditions from Derjaguin \cite{Derjaguin1993} (see Rio \emph{et al.}\cite{Rio2017} for a recent review) and is computed from the viscous-gravity balance as $h_0=\sqrt{\nu_l U_p / g}=1.67 mm$. Once the gas jet flow is activated and the wiping process starts, the computation reaches a fully developed state after approximately two flow-through, equivalent to twice the time needed for the liquid film to leave the domain after wiping $(\approx 1$ s).

	\subsection{\label{sec:numerical_method}  Numerical methodology}
	
	The numerical simulations were carried out using a combination of Large Eddy Simulation (LES) to model turbulence in the gas flow and Volume of Fluid (VOF) to account for the two-phase nature of the problem.
	The VOF method, developed by Hirt and Nichols \cite{Hirt1981}, is based on the transport of a color function  which represents the liquid volume fraction $\alpha$ in each computational cell. This function is equal to 1 when it is filled with liquid, and 0 when it is filled with gas. The gas-liquid interface is located in the cells where $0< \alpha <1$. The local fluid properties $\phi$ are computed as a weighted average of the properties of the liquid ($\phi_l$) and the gas ($\phi_g$), depending on $\alpha$:
	
	\begin{eqnarray}
	\phi = \phi_l\alpha + \phi_g(1-\alpha) 
	\end{eqnarray}
	
The weighted properties are introduced in the incompressible Navier-Stokes equations, leading to:

\begin{subequations}
\begin{eqnarray}
\label{eq:N_S_1}
\nabla \cdot \textbf{U} = 0
\end{eqnarray}
\begin{multline}
\label{eq:N_S_2}
  \frac{\partial \left( \rho \mathbf{U} \right)}{\partial t} + \nabla \cdot \left( \rho \mathbf{U} \otimes \mathbf{U} \right)   =  - \nabla p + \rho \mathbf{g} + \nabla \cdot [ 2(\mu + \mu_t) \mathbf{S}]  + F_\sigma
%\rho \frac{D \mathbf{U}}{Dt} =  - \nabla p + \rho g + \nabla \cdot [ 2(\mu + \mu_t) \mathbf{S}]  + S_\sigma
\end{multline}
\end{subequations}

where $\textbf{U}=(U,V,W)$ is the velocity, $p$ the pressure, $\mu$ the dynamic viscosity, $\mathbf{S}=(\nabla \textbf{U} + \nabla^{T}\textbf{U})/2$ the strain rate tensor and $F_\sigma$ the term accounting for the Laplace pressure due to surface tension at the interface. These equations are combined with a transport equation for the volume fraction $\alpha$:

\begin{equation}
\label{eq:advection_alpha_org}
\frac{\partial \alpha}{\partial t} +  \nabla \cdot (\alpha \textbf{U})  = 0
\end{equation}

The computations were performed using the \textit{interFoam} solver from OpenFoam\cite{MarquezDamian2013}, which implements an algebraic VOF formulation. In contrast to geometric VOF methods using the Simplified Line Interface Calculation (SLIC) by Noh and Woodward \cite{Noh1976} or the Piecewise Linear Interface Calculation (PLIC) by Rider and Kothe \cite{Rider1998}, this approach does not include a geometrical reconstruction of the interface. To limit the numerical diffusion of the interface, an artificial relative velocity term $U_r = U_l - U_g$ is defined in order to compress the gas-liquid boundary, where $U_g$ is the velocity of the gas and $U_l$ the velocity of the liquid in the interface. Thus, the advection equation of the volume fraction $\alpha$ is modified according to the definition of the velocity $U = U_l\alpha + U_g(1-\alpha)$ to include a compressive term:

\begin{equation}
\label{eq:advection_alpha}
\frac{\partial \alpha}{\partial t} +  \nabla \cdot (\textbf{U} \alpha) +  \nabla \cdot \left(\textbf{U}_r \alpha (1-\alpha) \right)
 = 0
\end{equation}

The major advantage of this method relies is its low computational cost and simple implementation \cite{Deshpande2012, Larsen2019}. For the flux corrected transport scheme, the algorithm implemented in \textit{interFoam} is known as the Multidimensional Universal Limiter with Explicit Solution solver (MULES \cite{MarquezDamian2013}), which adjusts the compression by means of the face fluxes and a user-defined parameter $C_\alpha$ to guarantee boundedness and stability (validated by Deshpande \emph{et al.}\cite{Deshpande2012}). 

The surface tension force $F_{\sigma}$ is computed from the $\alpha$ field using the continuum surface force (CSF) model by Brackbill \emph{et al.}\cite{Brackbill1992}, which allows converting the surface force into a volumetric force acting upon the smeared interface. It reads:

\begin{equation}
\label{eq:Ssigma}
F_{\sigma}=\sigma \kappa \textbf{n} \delta
\end{equation} where $\sigma$ is the liquid surface tension, $\textbf{n}=\nabla \alpha /\mid \nabla \alpha \mid$ is the unit normal vector to the interface, $\kappa=-\nabla \cdot \textbf{n}$ is the local interface curvature, and $\delta$ the Dirac function that equals unity at the interface.

The two-way coupling between interfaces and turbulence in two phase flows is a complex subject, and its numerical treatment is not yet mature \cite{FAN2020106876}. In the present study, the Large Eddy Simulation (LES) method is considered an adequate compromise solution for the modelling of the turbulent behaviour of the gas jet. The coupling of LES with multiphase flows modeling is commonly referred to as Large Eddy Interface Simulations (LEIS); the state of the art and the perspectives of this method are presented by Lakehal \emph{et al.} \cite{Lakehal2018}. The LES consists in the filtering of the turbulence scales based on the grid scale \cite{Sagaut}. The scales smaller than the grid (sub-grid scales) are modeled as they are supposed to have a universal and isotropic behaviour. The larger scales are more case-dependent and crucial in the global mechanism we are interested in, so their accurate resolution is of capital importance. LES implies therefore a spatial low-pass filtering of every flow field variable by convolution with a spatial filter. When applied to the two-phase conservation Eqs.~\ref{eq:N_S_1}-\ref{eq:N_S_2}-\ref{eq:advection_alpha}, several subgrid scale (SGS) terms arise in their filtered version (see Eqs.~45 to 48 in Lakehal \emph{et al.} \cite{Lakehal2018}).

In this work, the advective SGS term (also present in single phase flows) is accounted for using the Smagorinsky model with an eddy viscosity approximation, in which the turbulent eddy viscosity $\mu_t$ reads:

\begin{eqnarray}
\label{eq:turb_visc}
\mu_t = \rho (C_s \Delta x_{cell})^2 |\mathbf{S}|\,,
\end{eqnarray} where $C_s$ is the Smagorinsky model constant, set to 0.158 (usual values are in the range 0.1-0.2), $\Delta x_{cell}$ is the cell size and $|\mathbf{S}|=\sqrt{2 \mathbf{S} : \mathbf{S}}$ is Frobenious norm of the the strain rate tensor. 

Due to the presence of an interface, two additional subgrid terms arise (Eqs.~47 and 48 in Lakehal \emph{et al.} \cite{Lakehal2018}). These represent the influence of unresolved surface tension and subgrid interface deformations on the filtered flow. Different modeling strategies are proposed for these SGS contributions in litterature \cite{LABOURASSE20071,LIOVIC2007504,toutant_jump_2009,liovic_subgrid-scale_2012,ketterl_-priori_2018,jofre_near-interface_2020}, but the relative importance of each term is very case dependent. Any modeling attempt requires therefore an a priori analysis with DNS to quantify the impact of each subgrid term on the under-resolved case. This process is often led on relatively academic test cases\cite{fulgosi_direct_2003,LABOURASSE20071,LIOVIC2007504,toutant_dns_2008,VINCENT2008898,ketterl_-priori_2018,jofre_near-interface_2020}. In more industrial configurations, such an approach is not possible, and due to a lack of generality in the modeling strategies, the interface subgrid terms are often ignored \cite{lacanette_macroscopic_2006,bianchi_3d_2007,rek_numerical_2017}. In the present case, we know beforehand (from experiments) the typical scales of interface deformations related to undulation. The later are much larger than the grid, so the interface is expected to be well resolved. On the other hand, we are dealing with an interface that is continuous and smooth, so the unresolved mass transfer due to filtering is expected to be low, unlike in flows with a dispersed phase (e.g. jet atomization \cite{bianchi_3d_2007,herrmann_sub-grid_2013,ketterl_-priori_2018}). 
	
	Regarding numerical schemes, an implicit first order Euler scheme is used to compute the time derivatives and a second order Gauss linear for convective, diffusive and pressure terms. The pressure-velocity coupling is ensured by the PISO (Pressure-Implicit Split-Operator) algorithm with two correctors at every time step. 
	The MULES algorithm was used with 4 correctors, and with the interface compression parameter $C_\alpha$ set to 1, meaning that there is no additional relaxation of the interface compression. The time step is adjustable to ensure either a maximum CFL criteria of 0.95 on the overall flow, or a maximum interface CFL of 0.2, yielding to time steps between $1.5 \times 10^{-6}$ and $7 \times 10^{-7}$ s. 
	
	The computations were run in parallel in 288 Intel E5-2680v3 CPUs at the Centro de Supercomputación de Galicia (CESGA). The time required for 1 second of real flow computation ranges between 400 and 800 hours.
	
	The total simulated flow time $t_s$ is 2 seconds, starting from a fully developed state, and the variable fields are sampled at a frequency $f_s$ of 1 KHz, yielding $n_t=2000$ time snapshots of the flow.
	
	\section{\label{SecIII}Multiscale Modal Analysis of the Wiping}
	
	The multiscale analysis of the jet wiping presented in Section \ref{sec:results} consists in two steps. The first step is a multiscale Proper Orthogonal Decomposition (mPOD) of the film thickness contour maps, to identify the dominant traveling wave patterns. The second step is a correlation analysis, to identify the coherent structures in the jet flow that are most correlated with the wave formation. Section \ref{SecIIIA} recalls the fundamental of mPOD while section \ref{SecIIIB} and \ref{SecIIIC} report on the data processing for the liquid film and the gas flow side respectively.
	
	\subsection{\label{SecIIIA}The Multiscale Proper Orthogonal Decomposition}
	
	The mPOD is a linear technique for dimensionality reduction, combining the advantages of the two most popular alternatives, namely the Proper Orthogonal Decomposition (POD\cite{Siro3,Holmes1996}) and the Dynamic Mode Decomposition (DMD\cite{Schmid2010,Rowley2}).
	
	Every linear decomposition can be seen as a modal expansion in which a dataset $d(\mathbf{x},t)$, sampled over a spatial domain $\mathbf{x}:=(x,y)$ and evolving in time $t\in [0,T]$, is written as a summation of modes with variable separated form:
	
	\begin{equation}
	\label{MODAL}
	d(\mathbf{x},t)=\sum^{R}_{r=1} \sigma_r \phi_r(\mathbf{x}) \psi_r(t)
	\end{equation} where $\sigma_r$ is the mode amplitude, $\phi_r(\mathbf{x})$ is the mode's spacial structure and $\psi_r (t)$ is the mode temporal evolution. The structures $\phi_r$ and $\psi_r$ form, respectively, a basis for spatial distribution and temporal evolution of the data.

	The main limitation of POD and DMD are recalled elsewhere \cite{Mendez2019a,MODULO,Mendez2018a,Mendez2019a}. Briefly, POD modes are characterized by temporal structures $\psi_r(t)$ with unconstrained frequency content, derived under an energy optimality criteria. In datasets featuring coherent patterns of comparable energy contribution but largely different frequency content, the POD leads to modes that are linked to different scales. On the contrary, DMD modes are characterized by harmonic temporal structures $\psi_r (t)=e^{\mathrm{j}\omega_r t}$, with $\omega_r \in \mathbb{C}$ and $\mathrm{j}=\sqrt{-1}$. This generally leads to poor convergence and no time-localization of the modes.
	
	The mPOD modes are derived using a combination of energy optimality of the POD and the spectral purity of the DMD. In particular, mPOD modes are optimal within a certain frequency bandwidth. This combination is achieved by combining the POD with classic Multi-resolution Analysis (MRA) which is performed on the temporal correlation matrix 
	
	\begin{equation}
	\mathbf{K}(t_l,t_n)=\langle d(\mathbf{x},t_l), d(\mathbf{x},t_n) \rangle_X =\int_{x}  d(\mathbf{x},t_l) d(\mathbf{x},t_n) d \mathbf{x}_i\,,
	\label{K_DECO}\end{equation} having introduced the inner product in space $\langle \rangle_X$.
	
	The temporal structures of the POD modes are computed as eigenfunctions of the correlation matrix, i.e. $\mathbf{K} \psi_r =\lambda_r \psi_r$. The mPOD first partitions $\mathbf{K}$ into the contribution of $M$ scales using a set of filters with impulse response $\mathbf{h}_1, \mathbf{h}_2\dots \mathbf{h}_M$:  
	
	\begin{equation}
	\label{K_Deco}
	\mathbf{K}= \sum^M_{m=1} \mathbf{K} \circledast \mathbf{h}_m = \mathbf{K}^{(1)}+\mathbf{K}^{(2)}+\dots \mathbf{K}^{(M)}\,,
	\end{equation} where $\circledast$ denotes the 2D convolution operator. The mPOD temporal structures are eigenfunctions of the different contributions, i.e. $\mathbf{K}^{(m)} \psi^{(m)}_r =\lambda^{(m)}_r \psi^{(m)}_r$. 
	The filters in Eq.~\ref{K_Deco} have separable impulse response and have non-overlapping band-pass regions in the frequency domain. This means that given the filter transfer function
	
	\begin{equation}
	\mathbf{H}_m(f_i,f_j)=\frac{1}{n^2_t}\sum_{l} \sum_{n} \mathbf{h}_m(t_l,t_n) e^{-2\pi (f_i t_l+f_n t_k)}\,,
	\end{equation} the product $\mathbf{H}_m \mathbf{H}_n$ is identically zero for all $m\neq n$. 
	
	The mPOD limits Eq.~\ref{K_DECO} to an approximation of the correlation matrix. In particular, only the contributions defined by filters with unitary transfer function along the diagonal, i.e. $\mathbf{H}_m (f_n,f_n)\neq0$ are kept.
	Then, it is possible to show\cite{Mendez2019a} that the eigenvectors of all the correlation matrices $\mathbf{K}^{(m)}$ are mutually orthogonal. Moreover, their frequency content is defined by the transfer function $\mathbf{H}_m$ of the filter identifying the corresponding scale; hence structures from different scales have no common frequencies. That is, given $\psi^{(m)}_r$ the r-th eigenvectors of the correlation $\mathbf{K}^{(m)}$, and given
	
	\begin{equation}
	\widehat{\psi}^{(m)}_r(f)=\frac{1}{\sqrt{2\pi}}\int_{-\infty}^{\infty} \psi^{(m)}_r(t) e^{-\mathrm{j} 2\pi f t}
	\end{equation} its frequency spectra, the product $\psi^{(m)}_r \psi^{(n)}_j$ is identically zero for all $m\neq n$ and $\widehat{\psi}^{(m)}_r(f_n)=0$ if $\mathbf{H}_m(f_n,f_n)=0$.
	
	The resulting temporal basis $\psi=[\psi^{(1)},\psi^{(2)}\dots \psi^{(M)}]$, collecting the eigenvectors of all scales is the mPOD basis. This is orthogonal by construction and spans $\mathbb{R}^{n_t\times1}$, hence allows for a lossless decomposition while having modes that are optimal within the prescribed range of frequencies.
	
	The associated spatial structures are then computed via projection and normalization:

	\begin{equation}
	\label{Proj}
	\sigma_r \phi_r (\mathbf{x})=\langle d(\mathbf{x},t),\psi_r(t) \rangle_T=\int_T d(\mathbf{x},t) \psi_r(t) dt\,,
	\end{equation} with $\sigma_r=||\langle d(\mathbf{x},t),\psi_r(t)\rangle_T||_2$. 
	The mPOD decomposition has been performed using the opensource software package MODULO\cite{MODULO}. The decomposed data is the thickness contourmaps of the liquid film, extracted from the CFD analysis and pre-processed as described in the following subsection.
	
	\subsection{\label{SecIIIB}Liquid Film Analysis}
	
	The detection of the interface is carried out from the $\alpha$ field along $z$-constant planes $z=[5:0.5:25]$ $\mbox{mm}$. 
	
	For each of these planes, the CFD data is interpolated on two Cartesian meshes, due to the different resolution requirements in the final film and the run-back flow. The interpolation mesh for the final film region ($x>0$) is set to $x\in[-5:0.1:150]$ $mm$ $y\in[0:0.01:0.001]$ $mm$, while the interpolation mesh for the run-back flow ($x<0$) is set to $x\in[-20:0.1:-5]$ $mm$ $y\in[0:0.025:0.006]$ $mm$. After the interpolation of the volume fraction, an image processing routine based on morphological erosion and dilation was used to detect and filter out droplets produced by the interface splashing and gas bubbles entrained within the liquid film. Then, the interface detection is performed using gradient-based edge detection techniques on the pre-processed contours of $\alpha$, in each of the $z$ planes.
	
\begin{figure}
		\includegraphics[width=\linewidth]{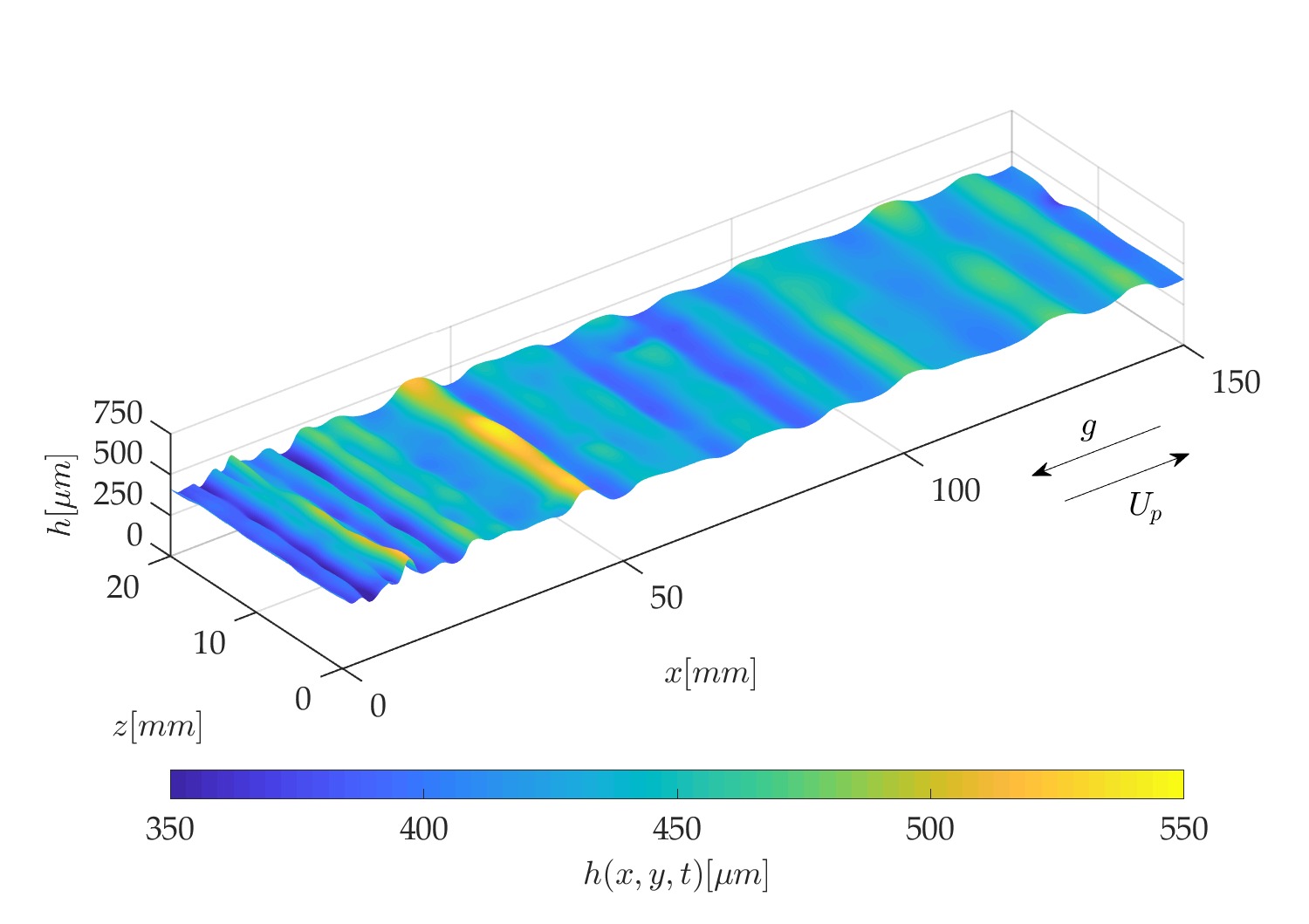}% Here is how to import EPS art
		\caption{\label{fig:liquidFilm_3D} Snapshot of the final film thickness distribution after pre-processing for Case 1. The interface is shown horizontally for plotting convenience: the direction of the strip motion $\vec{U}_p$ and gravity $\vec{g}$ is indicated in the figure.}
	\end{figure}

	The resulting thickness contour maps $h(\mathbf{x},t)$ are resampled over a Cartesian mesh. A snapshot of the final film thickness interface reconstruction (in $x>0$) for Case 1, after interpolation and morphological pre-processing, is shown in Fig.~\ref{fig:liquidFilm_3D}. 
	
	These thickness contourmaps are the main focus of the wave pattern analysis via modal decomposition. The decomposition is nevertheless not performed directly on the extracted thickness maps, as the waves in the run-back flow have much larger amplitudes than those of the final coating film. This difference biases the decomposition towards the dynamics of the run-back flow.
	
	To give comparable weight to both final coat and run-back flow, a normalization step is performed prior to the modal decomposition. This step, which allows for avoiding the need for weighted inner products, reads
	
	\begin{eqnarray}
	\label{eq:data_normalization}
	\check{h}(\mathbf{x},t)= \frac{h(\mathbf{x},t)-\bar{h}(\mathbf{x})}{\sigma_h(\mathbf{x})}\,,
	\end{eqnarray} where $\overline{h}(\mathbf{x})$ is the time average thickness $\sigma_h(\mathbf{x})$ is the distribution of the thickness standard deviation. The normalized thickness field obtained with Eq.~\ref{eq:data_normalization} is decomposed using the mPOD described in section \ref{SecIII}.

\begin{figure}
		\includegraphics[width=\linewidth]{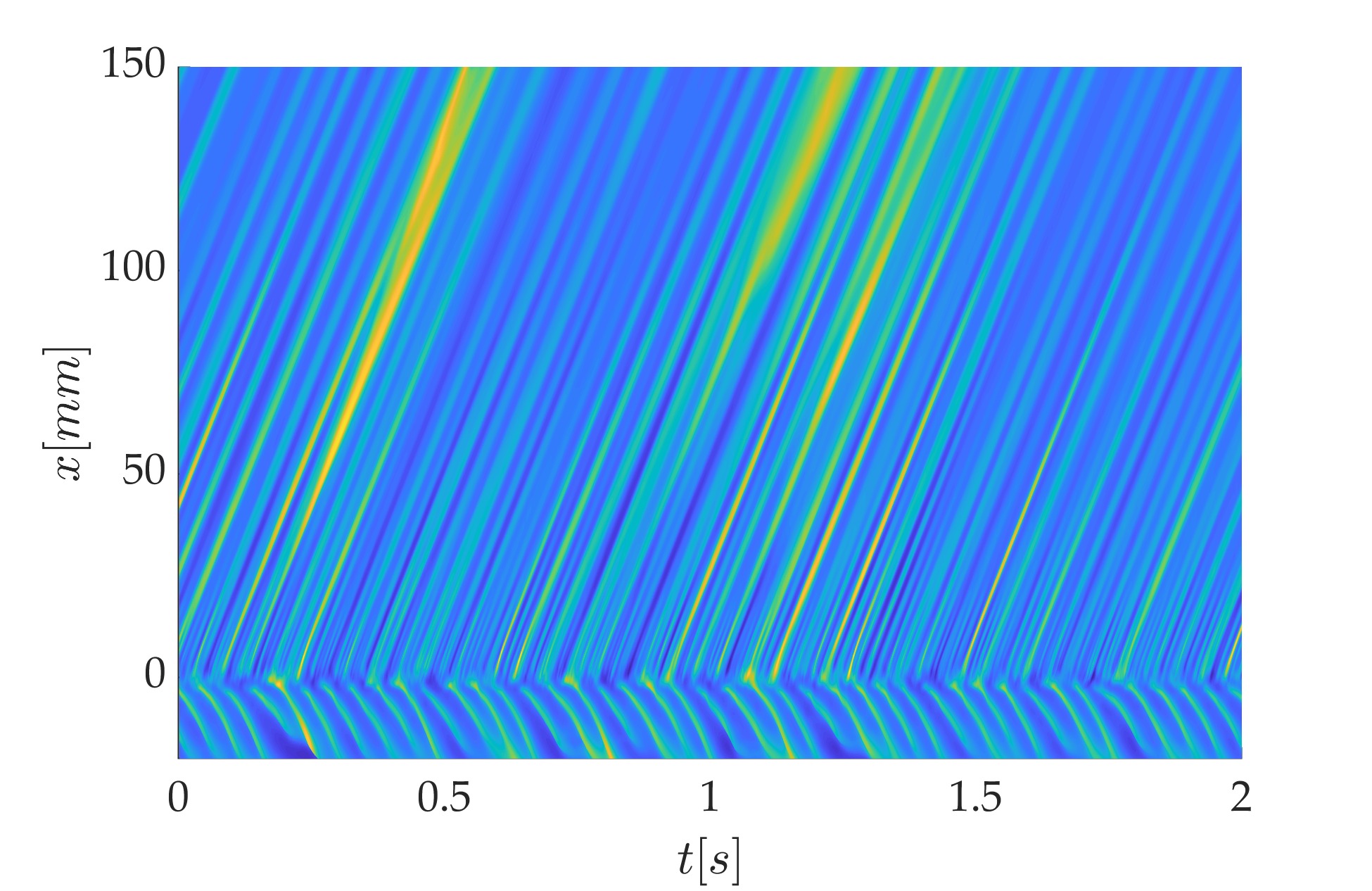}% Here is how to import EPS art
		\caption{\label{fig:thicknessMap_org} Characteristic map of the normalized liquid film thickness for Case 1. }
	\end{figure}

	An illustrative example of film thickness spatio-temporal evolution is shown in Fig.~\ref{fig:thicknessMap_org} for case 1. This contour shows the normalized thickness $\check{h}(x_i,t)$ in time and space at $z=15$ mm. The wave pattern in both the final coating film ($x>0$) and the run-back film ($x<0$) features wave merging and nonlinear interactions. These are analyzed in Section \ref{SecVIB}.
	
	\subsection{\label{SecIIIC}Correlation Analysis of the Gas Flow}
	
	Similarly to the data on the liquid film, the flow field from the gas jet is interpolated over a Cartesian grid prior to the modal decomposition. Velocity and pressure fields are the quantities of interest, re-sampled on several $z-constant$ planes $z=[5:15:25] \mbox{mm}$. These planes span a window $x \in [-20, 20]$ mm and $y \in [0, Z]$ mm with a spatial resolution of $\Delta x=\Delta y=200 \mu m$. The pressure resulting from the jet impact is sampled both at the wall, i.e. $p_w(x,z,t_k)=p_g(x,y=0,t_k)$ and at the gas-liquid interface, i.e. $p_i(x,z,t_k)=p_g(x,y=h(\mathbf{x}_i,t_k),t_k)$. 
	
	The modal analysis of the gas field is not considered in this work, which focuses on the correlation between the gas jet flow and the relevant patterns in the coating film. Accordingly, given $\psi_{\check{h},r}(t_k)$ the temporal structure of the $r-th$ mPOD mode of $\check{h}(\mathbf{x}_i,t_k)$, the most correlated gas structure is computed using the same projection as in Eq.~\ref{Proj}:
	
	\begin{equation}
	\label{Gas_STR}
	\phi_{g,r}(\mathbf{x})=\langle d_g(\mathbf{x},t),\psi_r(t)\rangle_T\,
	\end{equation} where the subscript $g$ denotes quantities from the gas jet flow (e.g. velocity, wall pressure or interface pressure). This correlation analysis can be seen as an extended mPOD analysis, similar to the extended POD presented by Bor{\'{e}}e \cite{Boree2003}.

	\section{A note on 2D Integral Models}\label{SecIV}
	
	We briefly recall here the fundamentals of 2D integral models of the jet wiping process, as the simplifying assumptions made in their derivation are tested in this work. For a more extensive treatment of these models, the reader is referred to Mendez \emph{et al.}\cite{Mendez2020}.
	
	Integral models are derived by integrating the Navier-Stokes (NS) equations across the film thickness, thus formulating the problem in terms of volumetric flow rate $q(x,t)=\int^{h(t)}_{0} u(x,y,t) dy$ and thickness $h(x,t)$. This integration is simplified by scaling the NS equation using a long-wave assumption, i.e. the reference length in the stream-wise direction $[x]$ is assumed to be much larger than the reference thickness $[h]$. The ratio between these two scales leads to a small parameter $\varepsilon=[h]/[x]$ which can be used to weight the contribution of each term of the equation. The long-wave formulation results in the assumption that the pressure gradient along the film thickness $\partial_y p$ is negligible, hence the velocity field within the liquid film is governed by a set of boundary layer like equations. The nondimensionalization and integration of these equations, considering the relevant boundary conditions at the wall and at the interface, yields the following system of PDEs:

	\begin{subequations}
		\label{BLEW_EQ}
		\begin{align}
		\partial_{\hat t} \hat h+\partial_{\hat x }\hat q&=0 \label{I1}\\
		\varepsilon Re \Bigl(\partial_{\hat t} \hat q+ \partial_{\hat x} {\mathcal{F}}\Bigr)&=\hat h\Bigl (1-\partial_{\hat x} \hat{p}_g+\partial_{\hat x\hat x\hat x}\hat h\Bigr) +\hat \tau_g-\hat \tau_w \label{I2}
		\end{align}
	\end{subequations}  where  $Re=[h]U_p/\nu=({{U}^3_p/g \nu})^{1/2}$ is the film Reynolds number, $\mathcal{F}=\int^{\hat{h}(\hat{t})}_0 u^2 dy$ is the integral advection term, $\hat{\tau}_g$ is the interface shear stress due to the gas and $\hat{\tau}_w=\partial_{\hat{y}} \hat{u} |_{\hat{y}=0}$ is the wall shear stress. The hats in Eq.~\ref{BLEW_EQ} denote dimensionless variables (e.g. $\hat{h}=h/[h]$), scaled with respect to the reference quantities in Table \ref{Scaling_Table}. This choice of reference scales leads to the definition of the film parameter as $\varepsilon=Ca^{1/3}$ with $Ca=U_p \mu_l/ \sigma$ the Capillary number.

	\begin{table}
		\caption{Reference quantities for the Shkadov scaling, for which $\varepsilon=Ca^{1/3}$, with $Ca=\mu_l\,U_p/\sigma$ the Capillary number.}
		\label{Scaling_Table}
		\small\addtolength{\tabcolsep}{1.5pt}
		\renewcommand{\arraystretch}{1.5}
		\centering
		\begin{tabular}{c|c|c}
			Reference Quantity & Definition & Expression \\
			$[h]$ & $(\nu_l\,[u]/g)^{1/2}$ & $(\nu_l\,U_p/g)^{1/2}$ \\
			$[x]$ & $[h]/\varepsilon$ & $[h]\,Ca^{-1/3}$ \\
			$[u]$ & $U_p$ & $U_p $ \\
			$[v]$ & $\varepsilon U_p$ & $U_p\,Ca^{1/3} $ \\
			$[p]$ & $\rho_l\,g\,[x]$ & $\rho_l \, g \, [h]\, Ca^{-1/3}$\\
			$[\tau]$ & $\mu_l\,[u]/[h]$ & $(\mu_l\,\rho_l\,g\,U_p)^{1/2}$\\
			$[t]$ & $[x]/[u]$ &  $(\nu_l/U_s\,g)^{1/2}\,Ca^{-1/3}$
		\end{tabular}
	\end{table}
	
	More details on the choice of the scaling quantities is given in previous works\cite{Gosset2019,Mendez2018}. The derivation and the closure of Eqs.~\ref{BLEW_EQ} build on three assumptions, herein briefly described: 1) the long-wavelength formulation, 2) the one-way coupling, 3) self-similarity.
	
	The long-wave assumption finds a theoretical justification only if $\varepsilon=Ca^{1/3}\ll1$. None of the test cases presented in this work satisfy this condition, as later discussed in Section \ref{SecVID}-- hence the interest in analyzing the consequences of significant departure from the `long-wave' modeling assumption.
	
	The one-way coupling assumption yields to the pressure gradient $\partial_{\hat{y}} \hat{p} (\hat{x},\hat{t})$ and the shear stress $\hat{\tau}_g (\hat{x},\hat{t})$ in Eqs.~\ref{BLEW_EQ} being independent from the liquid film dynamics ($\hat{h}$ and $\hat{q}$). The simplest formulation consists then in 
	modeling these terms via empirical correlations derived for the case of gas jet impinging on a dry wall.  
	
	Finally, the hypothesis of self-similarity is needed to close Eqs.~\ref{BLEW}, as both the advection and the wall shear stress term require some assumptions on the velocity field $\hat{u}(x,y,t)$. The simplest assumption postulates a parabolic dependency and self similarity, hence 
	
	\begin{equation}
	\label{Parabolic}
	\hat{u}(\hat{x},\hat{y},\hat{t})= C_2(\hat{x},\hat{t}) \frac{\hat{y}^2}{2}+ C_1(\hat{x},\hat{t}) + C_0
	\end{equation} where the coefficients can be found by imposing the boundary conditions ($u(\hat{x},0,\hat{t})=-1$ and $\partial_{\hat y} \hat{u} |_{\hat{y}=h}=\hat{\tau}_g$) and the flow rate definition. The resulting set is

\begin{subequations}
	\label{Vel_COEFF}
	\begin{equation}
	C_0=-1
	\end{equation}
	\begin{equation}
C_1=\frac{3\hat q}{\hat h}+\frac {3}{\hat h}- \frac{1}{2}\hat{\tau_g}
\end{equation}
	\begin{equation}
C_2=\frac{3\hat{\tau_g}}{2 \hat h}-\frac {3 \hat{q}}{\hat h}^3- \frac{3}{\hat{h}^2}
\end{equation}
\end{subequations}
	
Inserting these coefficients into the definition of the advection and the shear stress in \eqref{BLEW_EQ} terms gives
	
	\begin{subequations}
		\label{BLEW}
		\begin{equation}
		\mathcal{F}=\frac{ \hat h^3\, \hat\tau_g^2}{120}+\frac{ \hat h\, \hat q\, \hat \tau_g}{20}+\frac{ \hat h^2 \hat \tau_g}{20}+\frac{6\, \hat q^2}{5\,  \hat h}+\frac{2\, \hat q}{5}+\frac{ \hat h}{5}
		\end{equation}
		\begin{equation}
		\hat{\tau}_w=\frac{\hat{\tau}_g}{2}-\frac{3\hat{q}}{\hat{h}^2}-\frac{3}{\hat{h}}
		\end{equation}
	\end{subequations}
	
	The three simplifying assumptions introduced in this section are tested in Section \ref{SecVID}.

	\begin{figure}
		\begin{subfigure}{.48\textwidth}
			\centering
			% include first image
			\includegraphics[width=0.48\linewidth]{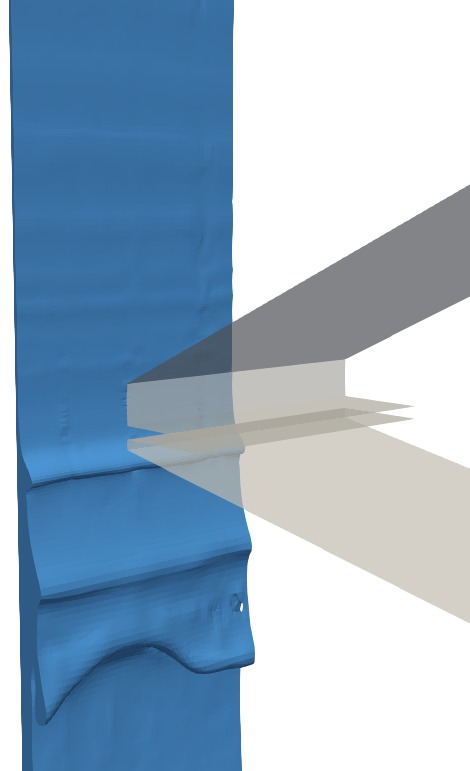}
			% include second image
			\includegraphics[width=0.48\linewidth]{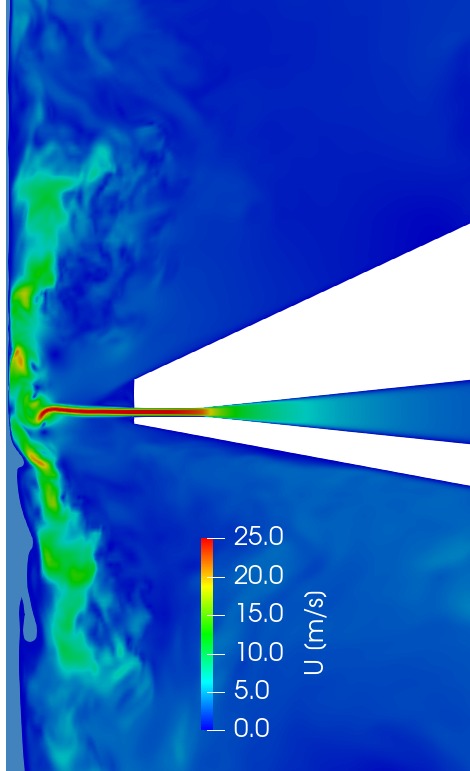}  
			\caption{Case 1: $\hat{Z}=14.2$ | $\Pi_g=0.16$}
			\label{fig:paraview_Z18_P425}
		\end{subfigure}
		
		\begin{subfigure}{.48\textwidth}
			\centering
			% include first image
			\includegraphics[width=0.48\linewidth]{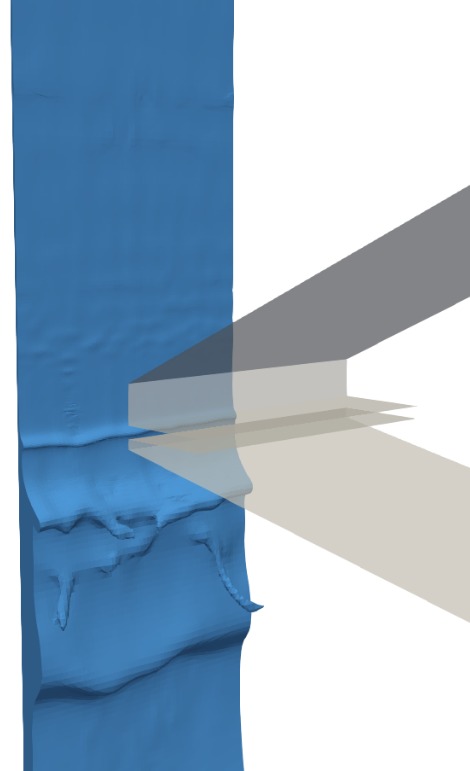}
			% include second image
			\includegraphics[width=0.48\linewidth]{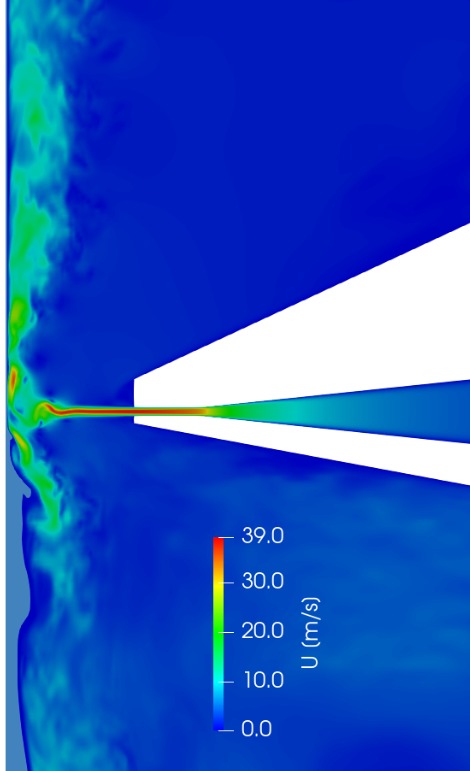}  
			\caption{Case 2: $\hat{Z}=14.2$ | $\Pi_g=0.33$}
			\label{fig:paraview_Z18_P875}
		\end{subfigure}
		
		\begin{subfigure}{.48\textwidth}
			\centering
			% include first image
			\includegraphics[width=0.48\linewidth]{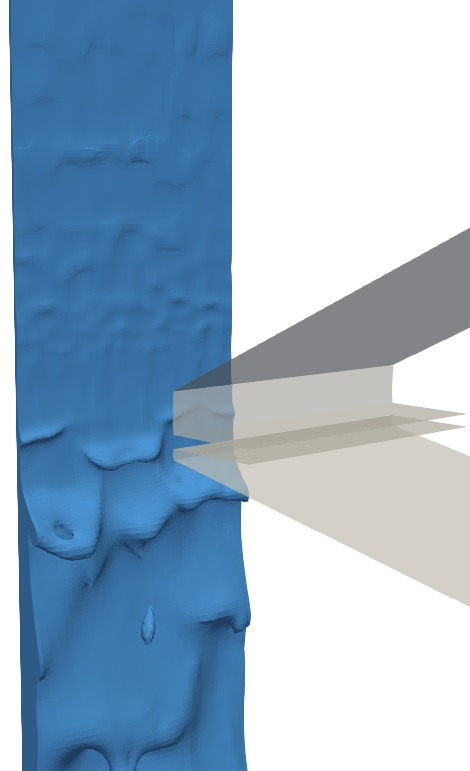}
			% include second image
			\includegraphics[width=0.48\linewidth]{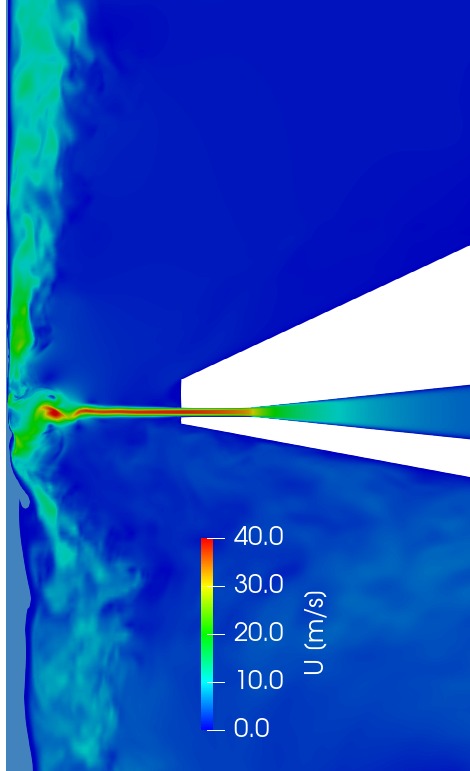}  
			\caption{Case 3: $\hat{Z}=19.4$ | $\Pi_g=0.18$}
			\label{fig:paraview_Z25_P875}
		\end{subfigure}
		\caption{\label{fig:paraview} Exemplary of the 3D reconstruction of the liquid film (left) and 2D velocity field (right) before data processing for Case 1 to 3.}
	\end{figure}

	\section{Results}\label{sec:results}
	
This section is divided into four parts. The first, in \ref{SecVIA}, opens with a general description of the investigated test cases and a comparison with experimental data for some of the relevant quantities of the process. The second part, in \ref{SecVIB}, presents the analysis of the undulation pattern in the liquid coat. The third part, in \ref{SecVIC}, analyzes the mechanism of undulation based on the correlated structures in the gas flow. Finally, this section closes in \ref{SecVID} with the assessment of the simplifying assumptions required in the derivation of integral models.

	\subsection{Case overview and Validation}\label{SecVIA}
	
	For the three test cases analyzed in this work (cf. Table \ref{tab:config} for the associated conditions), Fig.~\ref{fig:paraview} shows a snapshot of the 3D liquid film interface and a contourmap of the velocity magnitude in the gas flow in the middle plane $z=15mm$.
	
	The three cases produce very different undulation patterns. Case 1 ($\hat{Z}=14.2$, $\Pi_g=0.16$), is characterized by nearly bidimensional waves on the final coating film. Waves remains bidimensional also in the run-back flow for a distance of about 25 mm from the averaged impingement point. At larger distances, the steeping of the wave front is combined with the growth of 3D instabilities that eventually leads to the break-up of the liquid interface. Case 2, with the same stand-off distance but larger nozzle pressure ($\Pi_g=0.33$) produces a 3D wave pattern on both the final coating and the run-back flow regions. This configuration features splashing:  the breaking of the run-back flow occurs much closer to the wiping region, promoted by the high shear of the gas flow released from the impact. The momentum exchange between the two phases leads to the ejection of liquid droplets.

	Case 3 has a wiping number comparable to Case 1 but operates at a much larger stand-off distance. This case generates wave patterns that are similar to those of Case 2, but with more pronounced three-dimensionality and reduced splashing. The significant difference in the waves is thus mostly due to the different stand off distance and hence the gas flow field, analyzed in details in the following subsection.

	\begin{table}
		\caption{ Comparison of the spatio-temporal average of the final film thickness $\bar{h}_f$ (in microns) in the CFD calculations and experimental conditions\cite{Mendez2019}.}
		\begin{ruledtabular}
			\begin{tabular}{ccc}
				Case $\#$ & Numerical [$\mu$ m] & Experimental [$\mu$ m]  \\
				\hline
				\\
				1 &  $435$ & $449 \pm 5 \% $   \\
				2 &  $263$ & $311 \pm 5 \% $   \\
				3 &  $335$ & $399 \pm 5 \% $   \\
			\end{tabular}
		\end{ruledtabular}
	\label{tab:avg:thivkness}
	\end{table}

	A comparison of experimental and numerical results is shown in Table~\ref{tab:avg:thivkness} in terms of spatio-temporally averaged final coating thickness. This average is defined as 
	
	\begin{equation}
	\overline{h}_f=\frac{1}{x_2-x_1} \frac{1}{t_2-t_1}\int^{x_2}_{x_1}\int^{t_2}_{t_1} h(x',t') dt' dx'
	\label{MEAN}
	\end{equation} taking $[x_1,x_2]=[120,150]\mbox{mm}$ and $[t_1,t_2]=[0,1]\mbox{s}$. Here, the time $t=0$ is the one from which the data processing begins, once the flow is fully established.

	The CFD results predicts reasonably well the mean final thickness $h_f$, and captures correctly the trends due to varying process parameters: an increase of the nozzle pressure at constant $\hat{Z}$ decreases $h_f$, which is also the case when the nozzle standoff distance is increased at constant $\Delta P_N$. 
	
	Nevertheless, the average thickness $\overline{h}_f$ computed from the CFD appears systematically lower than what is observed in the experiments. This tendency was also documented in similar simulations by Myrillas\cite{Myrillas2011}. A possible explanation could be found in the slight differences between the numerical and the experimental settings reproduced in the Ondule laboratory at the von Karman Institute\cite{Gosset2007}. In this experimental facility, the moving substrate is simulated by a cylinder with a large diameter ($D=450$ mm) and the film thickness measurement were performed further downstream the wiping point (approximately 350mm after wiping) than in the numerical model. At such large distances, the change of substrate orientation due to the cylindrical configuration in the experiments might play a role in the final coating thickness.

	\begin{figure}
		\includegraphics[width=\linewidth]{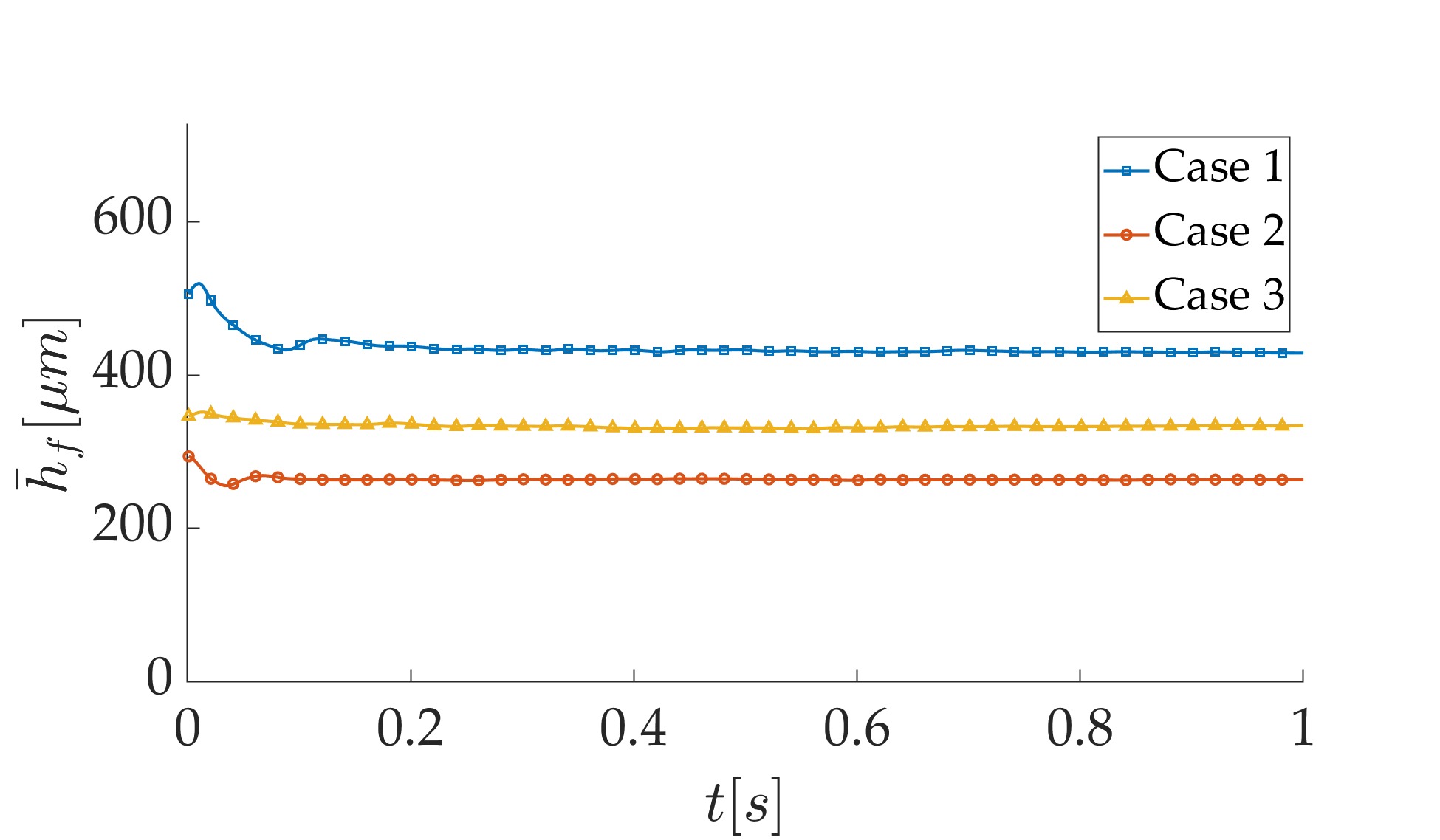}% Here is how to import EPS art
	    \\ (a)\\
		\includegraphics[width=\linewidth]{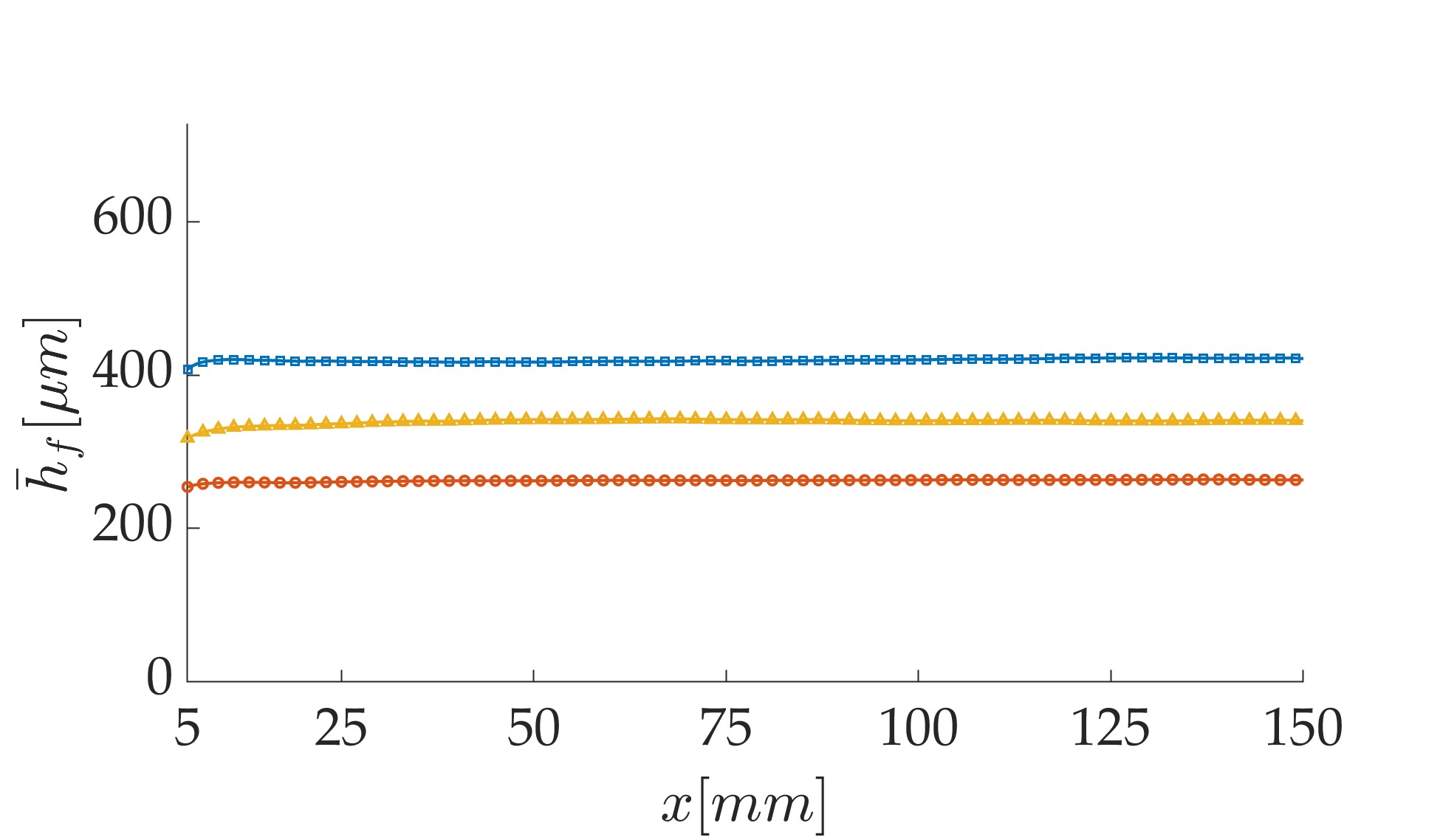}% Here is how to import EPS art
		\\ (b)
		\caption{ (a) time evolution of the spatial average and (b) spatial distribution of the temporal average of the final film thickness for the three analyzed test cases.}
		\label{fig:space_time_avg}
	\end{figure}

	\begin{figure*}
    \begin{subfigure}{.49\textwidth}
      \centering
      % include first image
      Liquid film
      \includegraphics[width=\linewidth]{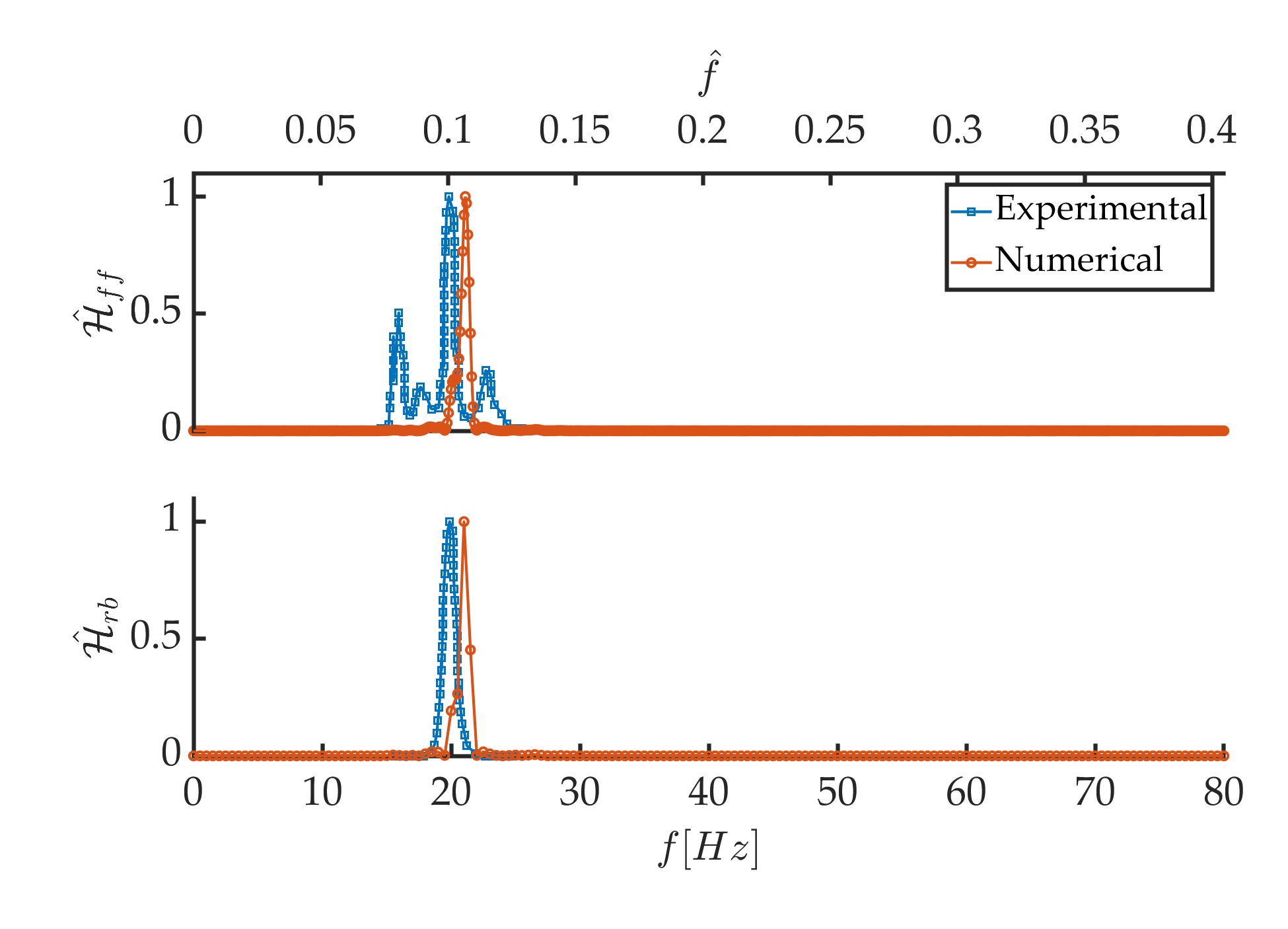}
      \caption{}
      \label{fig:spectra_liquid_Z18_P425}
    \end{subfigure}
    \begin{subfigure}{.49\textwidth}
      \centering
      % include first image
      Gas jet
      \includegraphics[width=\linewidth]{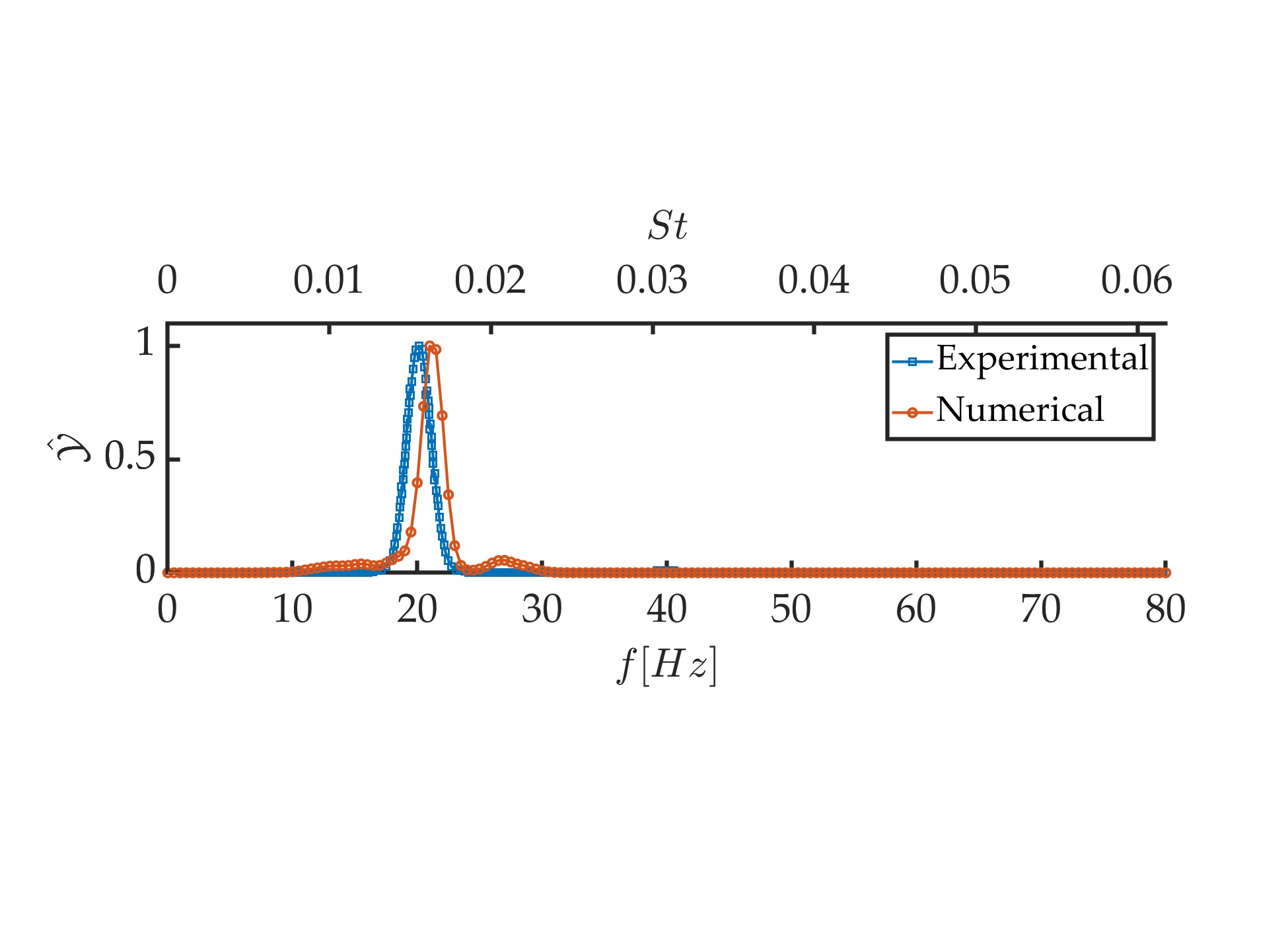}
      \caption{}
      \label{fig:spectra_gas_Z18_P425}
    \end{subfigure}\\
    Case 1: $\hat{Z}=14.2$ | $\Pi_g=0.16$ \\
    
    \begin{subfigure}{.49\textwidth}
      \centering
      % include first image
      \includegraphics[width=\linewidth]{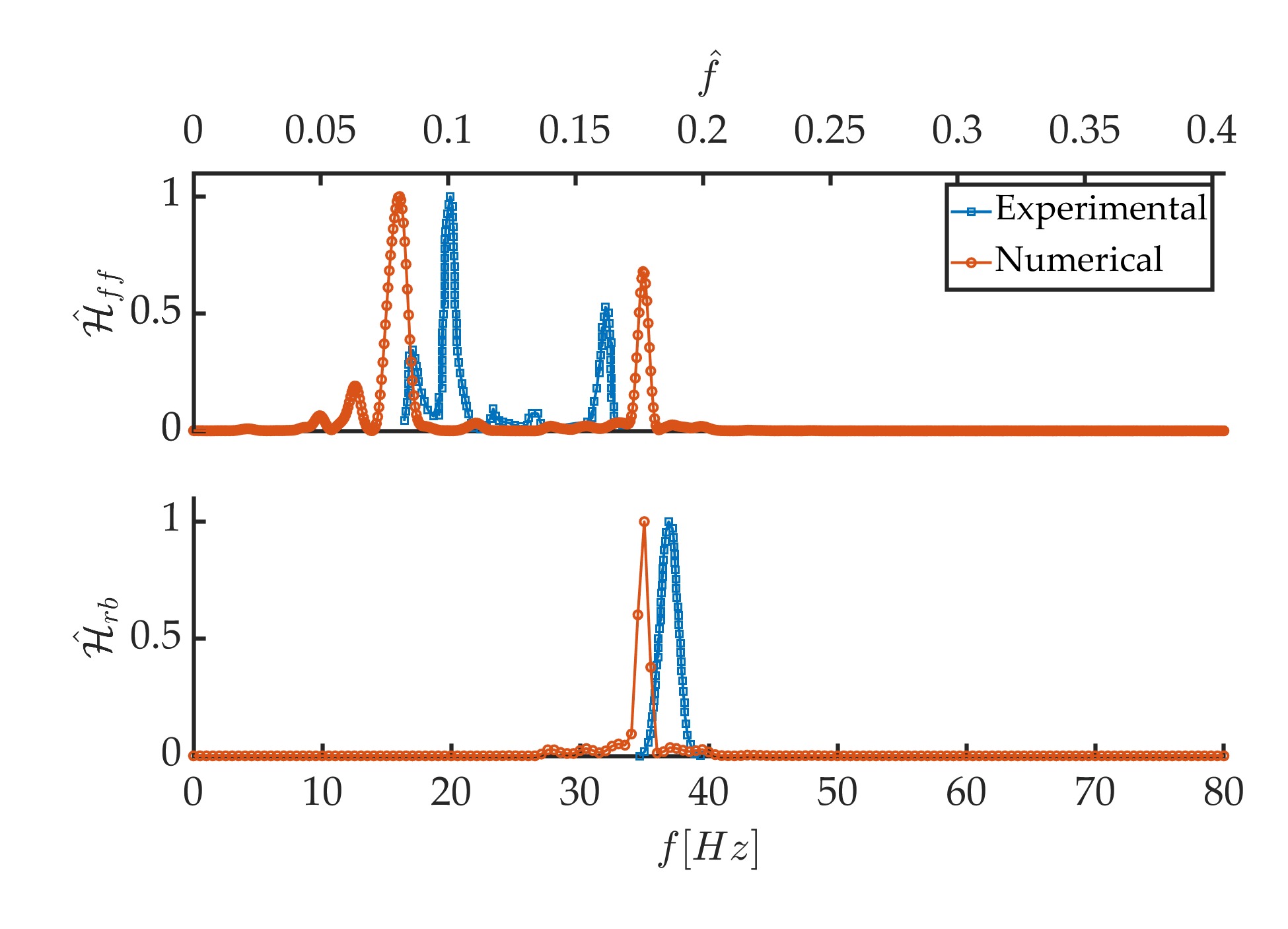}
      \caption{}
      \label{fig:spectra_liquid_Z18_P875}
    \end{subfigure}
    \begin{subfigure}{.49\textwidth}
      \centering
      % include first image
      \includegraphics[width=\linewidth]{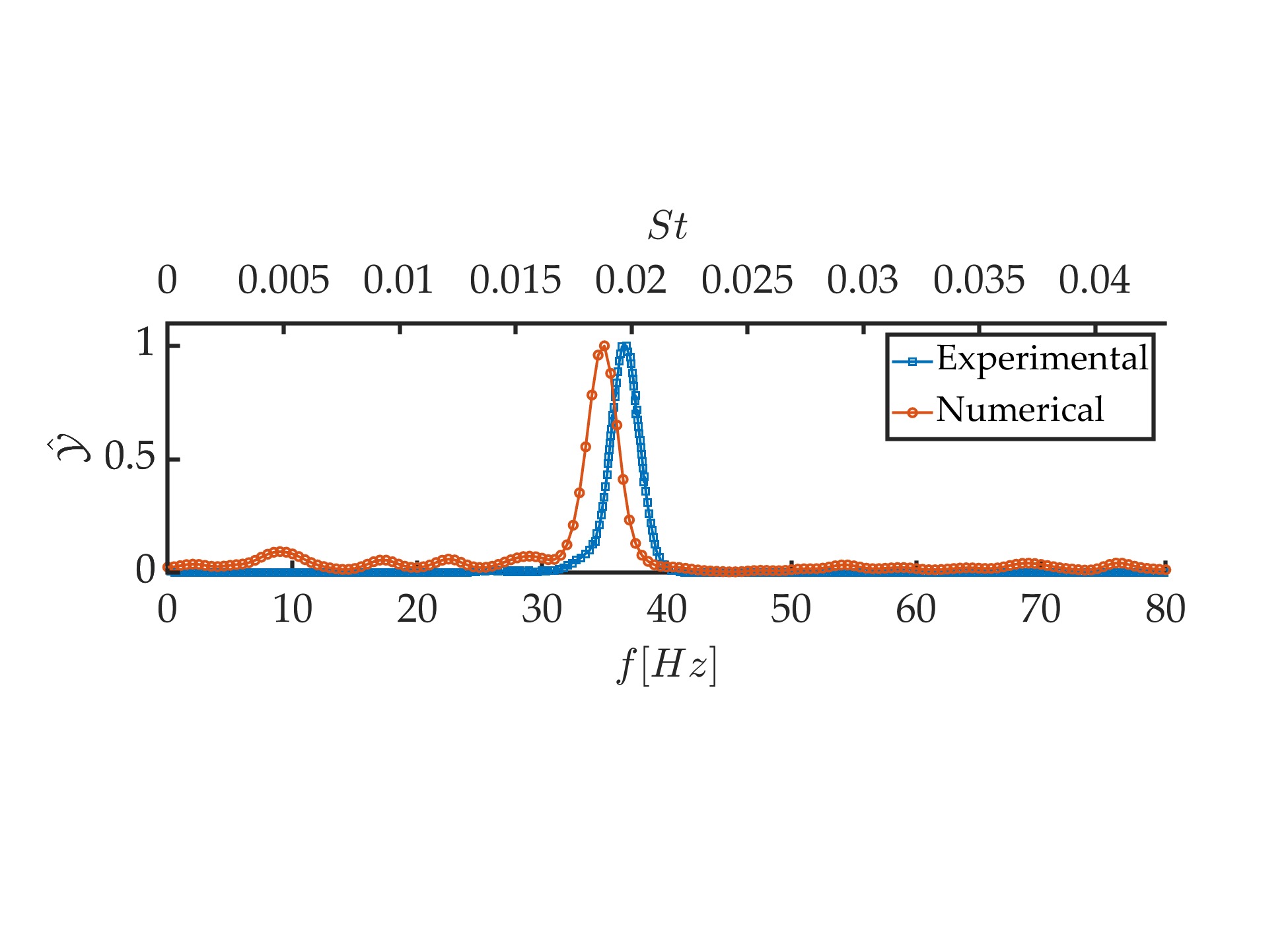}
      \caption{}
      \label{fig:spectra_gas_Z18_P875}
    \end{subfigure}\\
    Case 2: $\hat{Z}=14.2$ | $\Pi_g=0.33$\\
    
    \begin{subfigure}{.49\textwidth}
      \centering
      % include first image
      \includegraphics[width=\linewidth]{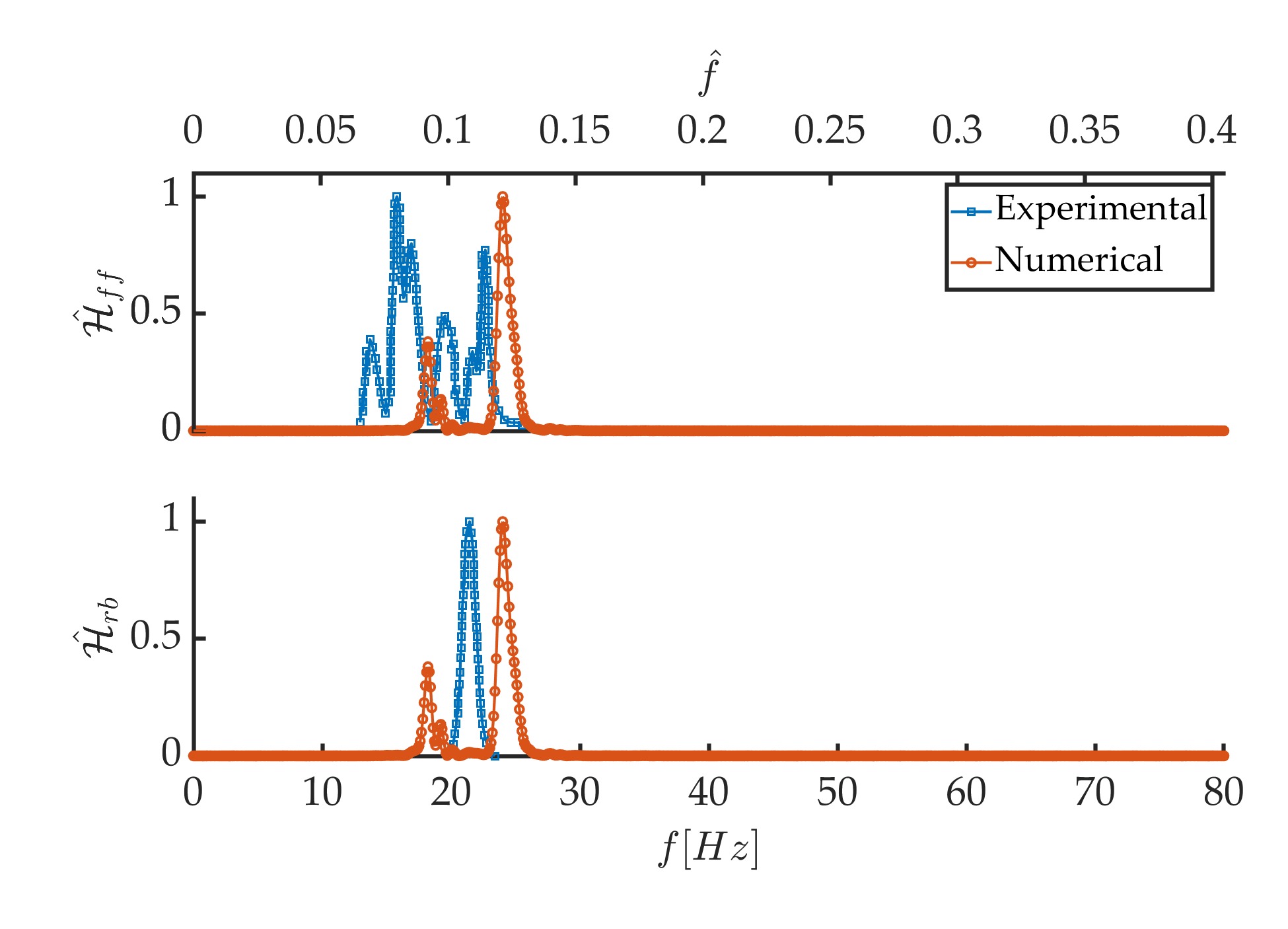}
      \caption{}
      \label{fig:spectra_liquid_Z25_P875}
    \end{subfigure}
    \begin{subfigure}{.49\textwidth}
      \centering
      % include first image
      \includegraphics[width=\linewidth]{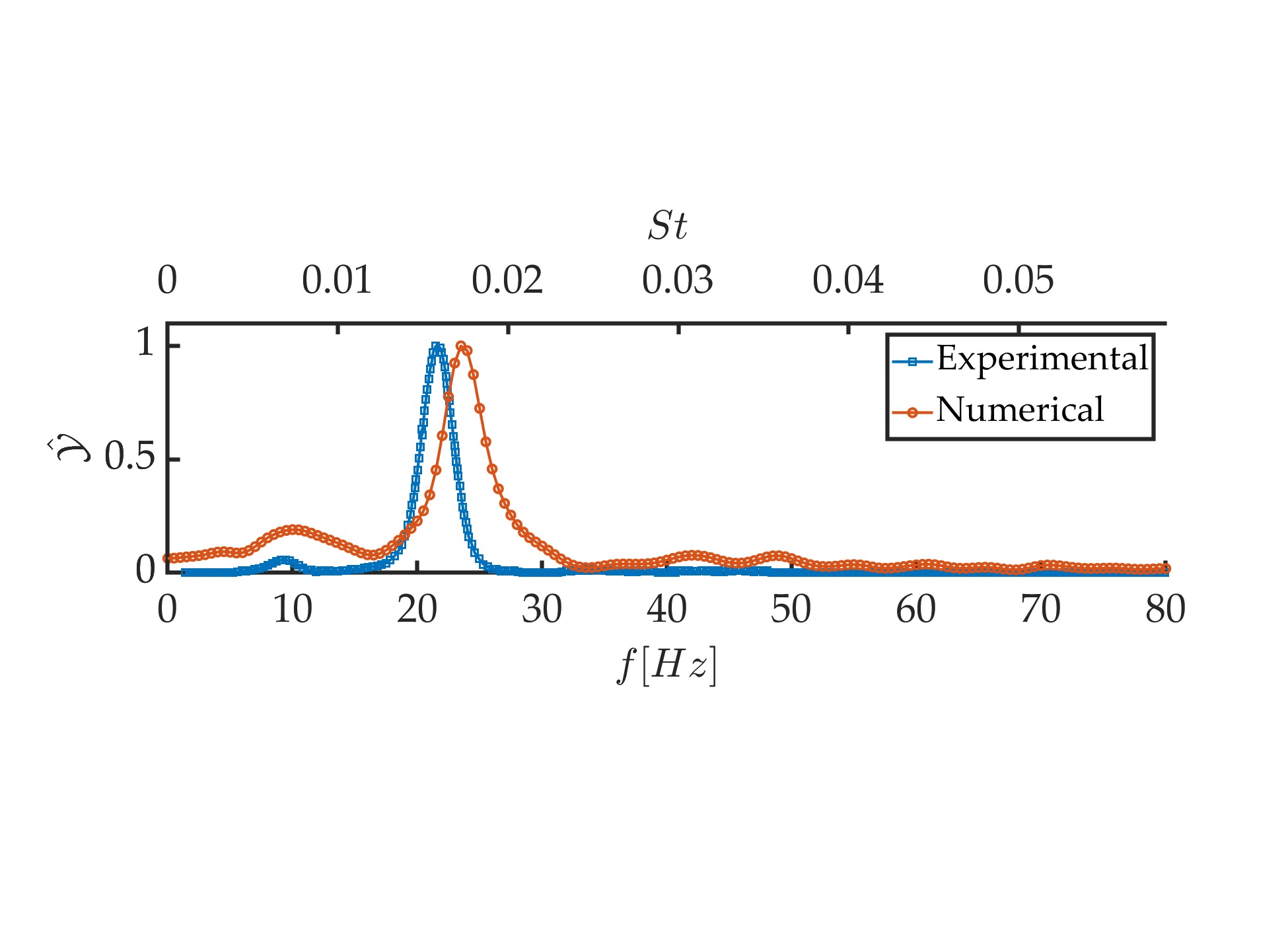}\\
      \caption{}
      \label{fig:spectra_gas_Z25_P875}
    \end{subfigure} \\
    	 Case 3: $\hat{Z}=19.4$ | $\Pi_g=0.18$ \\
    \caption{Comparison between the power spectral density in the leading mPOD modes for the experimental and the numerical data. The figures on the left correspond to the liquid film thickness data (LIF interface detection and light absorption measurement versus numerical results) while the figure on the right corresponds to the gas jet data (TR-PIV fields versus numerical results).}
    \label{fig:spectra}
    \end{figure*}

	The spatial and temporal convergence of the average thickness in Eq.~\ref{MEAN} is analyzed for the three cases in Fig.~\ref{fig:space_time_avg}. Figure a) shows $\overline{h}_f$ as a function of $t_2$, keeping $t_1=1 \mbox{s}$. Figure b) shows the same plot as a function of $x_2$, keeping $x_1=5 \mbox{mm}$. For all the investigated test cases, these figures show the excellent statistical convergence of the mean thickness prediction.

	To conclude this subsection, we proceed with the experimental validation in terms of frequency content in both the liquid film waves and in the gas jet flow. The comparison is carried out by analyzing the frequency spectra of the dominant mPOD modes in the thickness evolution and the gas velocity field. In the experimental data, the film thickness in the run-back flow was extracted from LIF-based interface tracking; the thickness contours in the final coating film were obtained via light absorption measurements, while the gas velocity field was measured via TR-PIV\cite{Mendez2019}.

	Fig.~\ref{fig:spectra} shows, on the left, the frequency content in the leading mPOD modes of the liquid film thickness data for the three test cases. In each sub-figure, the top plot refers to the final coating film and the bottom one to the run-back flow. The frequency axis is also shown in terms of dimensionless frequency $\hat{f}=f/[f]$ in the Skhadov-like scaling summarized in table \ref{Scaling_Table}. The frequency content is shown in terms of power spectral density, denoted as $\mathcal{H}$ and computed using Welch's method\cite{Welch1967}.
		
	The first test case shows a remarkable agreement between experimental and numerical data: both the final coating film and the run-back flow are characterized by the same dominant frequency ($\approx 20 Hz$) which is linked to the 2D wave pattern observed in both the final coating and the run-back flow, as further discussed in section \ref{SecVIB}. This frequency corresponds, in the dimensionless representation, to a range of $\hat{f}\approx 0.1-0.2$. This is well in the range of high sensitivity of the liquid film to finite gas jet perturbations, as found in previous theoretical works\cite{Mendez2020}.
	
	While in all the cases the run-back flow is characterized by a single dominant frequency, with fairly good agreement between experimental and numerical data, the same comparison for the cases 2 and 3 is of more difficult interpretation. These cases are characterized by three dimensional waves and are probably more influenced by the differences between the experimental and the numerical configurations. Interestingly, the dominant frequency in the final film, in case 2, is lower than the one observed in the run-back flow. This is observed in both numerical and experimental data although the position of the frequency peaks is different.

	Finally, Fig.~\ref{fig:spectra} shows, on the right, the power spectral densities of the leading mPOD modes in the gas jet for the three cases. The top axis represents the Strouhal number of the jet defined as $St=fZ/U_j$. The range of $St$ is close to the ones observed also in confinement-driven instability of impinging jet flows\cite{Mendez2018a}, and the matching between the spectra of the two flows is the main footprint of the liquid-gas coupling at the origin of the coating undulation\cite{Gosset2019,Mendez2019}.
	
	The agreement between numerical and experimental data is satisfactory and shows that the leading mechanisms driving the jet flow is a nearly harmonic behavior at the frequency of the run-back flow.
	Considering the slight difference in the flow configuration analyzed in the experimental and in the numerical works, and considering the differences in resolution, signal to noise ratio and tools employed for processing of experimental and the numerical data, it is reasonable to conclude that the numerical simulations presented in this work are able to capture the essential features of the flow dynamics. These are further investigated in the following subsections.

	\subsection{Modal Analysis of the Coating Waves}\label{SecVIB}

\begin{figure}
  \centering
  % include first image
  \includegraphics[width=\linewidth]{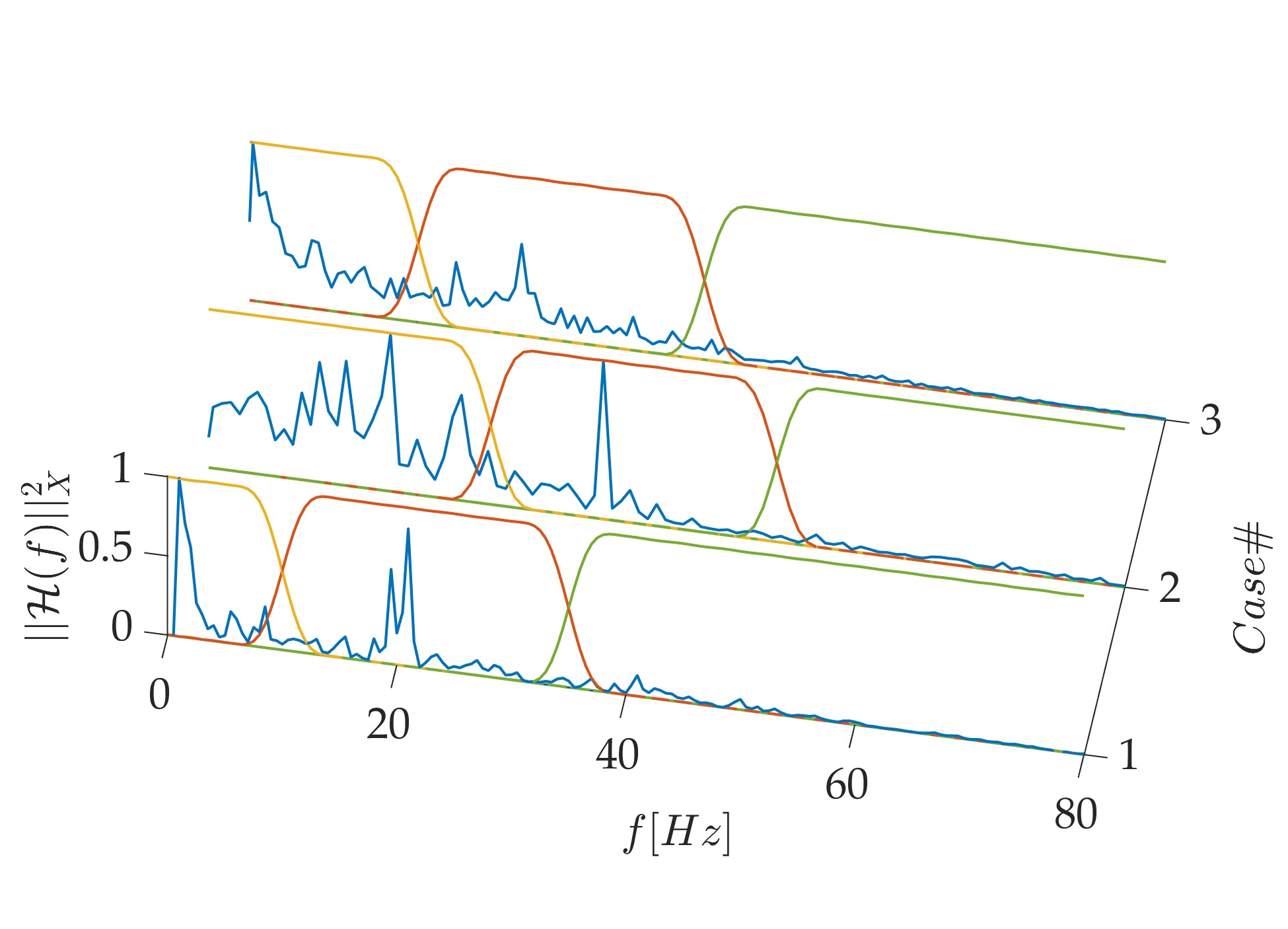}  
  \caption{Power Spectral Density (PSD) of the diagonal of the temporal correlation matrix together with the bands that define the scales of the decomposition.}
  \label{fig:PSD_K_Z18_P425}
\end{figure}

	We now analyze the wave pattern on the final coating film for the three test cases. Fig.~\ref{fig:PSD_K_Z18_P425} shows the spatially averaged power spectral density of the liquid film thickness, i.e:
	
	\begin{equation}
	    \label{PSD}
	    ||\mathcal{H}({f})||^2_{X}=\int_x \int_z |\mathcal{H}(x,z,f)|^2 dx dz\,
	\end{equation} where the subscript $X$ in the norm denotes the inner product in space, and
	
	\begin{equation}
	    \label{DFT}
	    \mathcal{H}(x,z,f)=\frac{1}{\sqrt{2\pi} }\int_{T} \check{h}(x,z,t) e^{-\mathrm{j} 2\pi t } dt
	\end{equation}is the Fourier transform in the time domain of the weighted thickness $\check{h}$, computed as in Eq.~\ref{eq:data_normalization}.
	
	Case 1 is characterized by a clear dominant frequency at about $\approx 20 \mbox{Hz}$, expected from the frequency content of the leading mPOD modes in Fig.~\ref{fig:spectra}. A similar peak around 20 Hz is observed in case 3, although much less pronounced. Case 2 is characterized by two ranges of frequencies: the first between 5 and 20 Hz, and the second, of larger intensity, around $35 Hz$.
	
	This figure also shows the modulus of the transfer function of the filters that were designed to isolate the three relevant scales of the data: one centered in the dominant frequency and the other two linked to the lowest and the highest portions of the spectra.

For the purpose of this work, the most relevant modes are those associated to traveling wave patterns. Because mPOD modes are real, such patterns must be described as a summation of two standing waves. For example, for a 1D harmonic traveling wave with wavenumber $k$ and frequency $f$, trigonometric identities dictates:

\begin{equation}
    \label{Trav}
    \begin{gathered}
    \sin (k x - \omega t)=\\= \sin(k x) \cos(\omega t)-\sin(k x+\frac{\pi}{2}) \cos(\omega t+\frac{\pi}{2}\bigl) 
    \end{gathered}.
\end{equation}

\begin{figure}
  \centering
  % include fourth image
  \includegraphics[width=\linewidth]{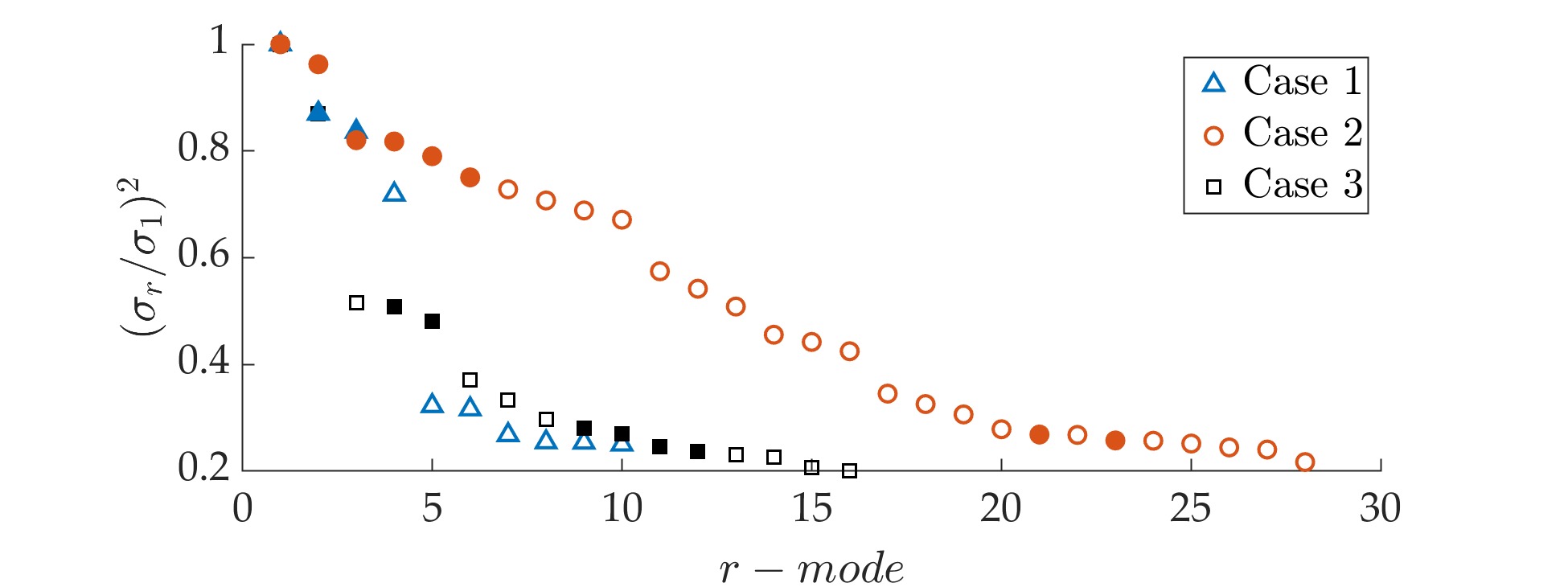}  
  \caption{Normalized energy content of the first 30 mPOD modes on the film thickness contourmaps.}
  \label{fig:sigma_mPOD}
\end{figure}

\begin{figure*}
\begin{subfigure}{.33\textwidth}
  \centering
  $\check{h}(x,z)$
  % include first image
  \includegraphics[width=\linewidth]{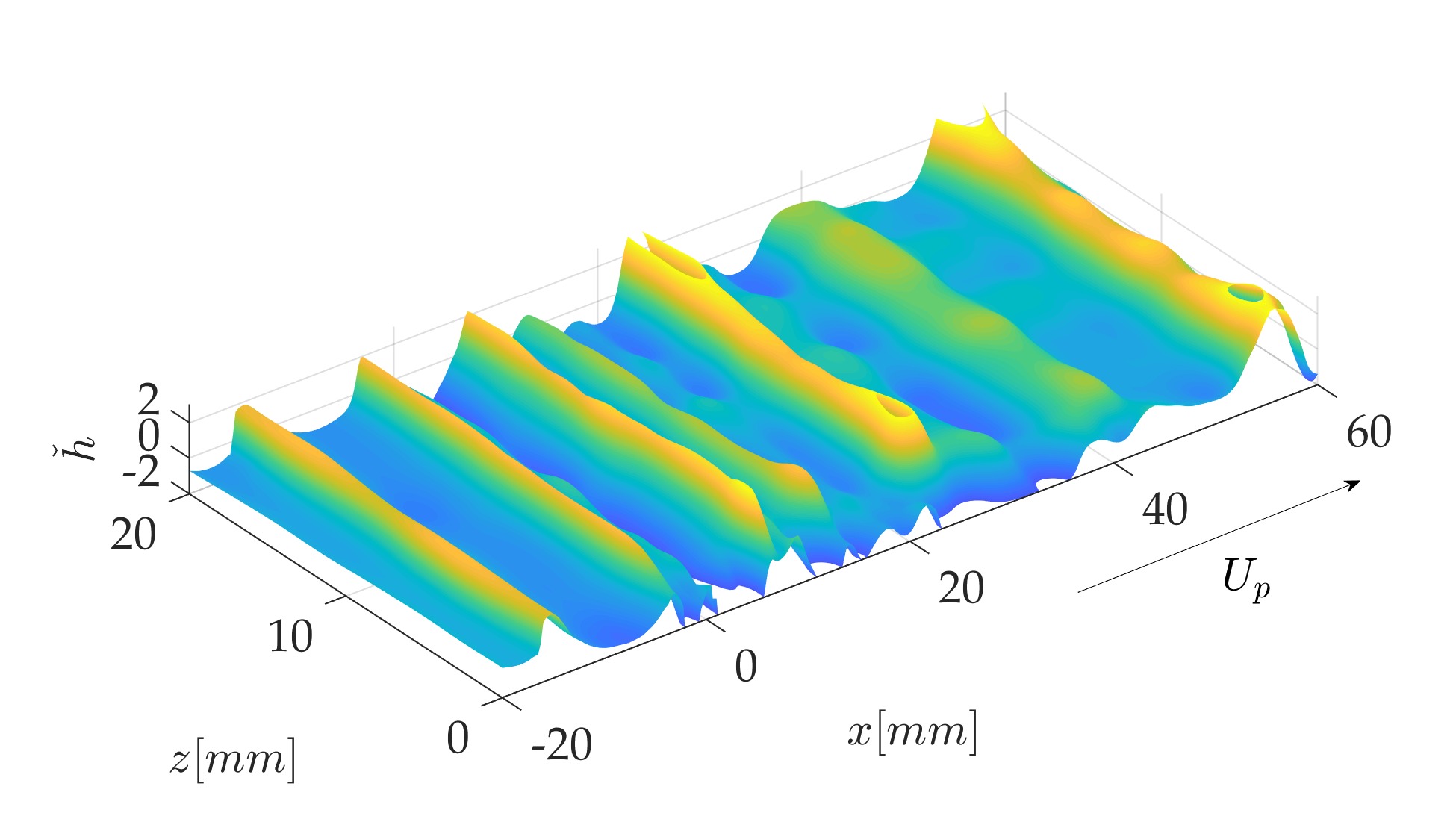}
\end{subfigure}
\begin{subfigure}{.33\textwidth}
  \centering
  $\phi_{2}(x,z)$
  % include first image
  \includegraphics[width=\linewidth]{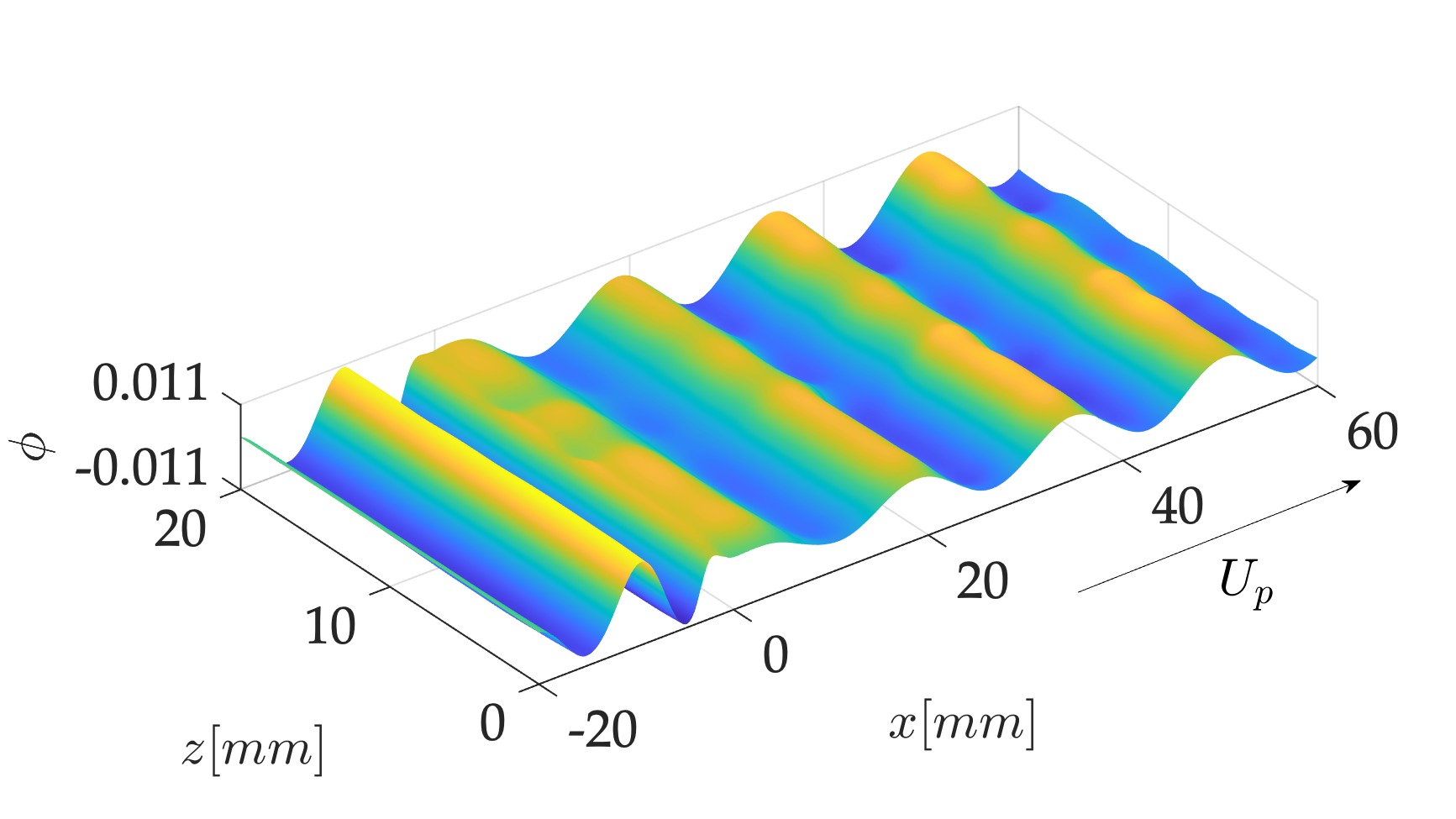}
\end{subfigure}
\begin{subfigure}{.33\textwidth}
  $|\hat{\psi}_{2,3}|(x,z)$
  \centering
  % include second image
  \includegraphics[width=\linewidth]{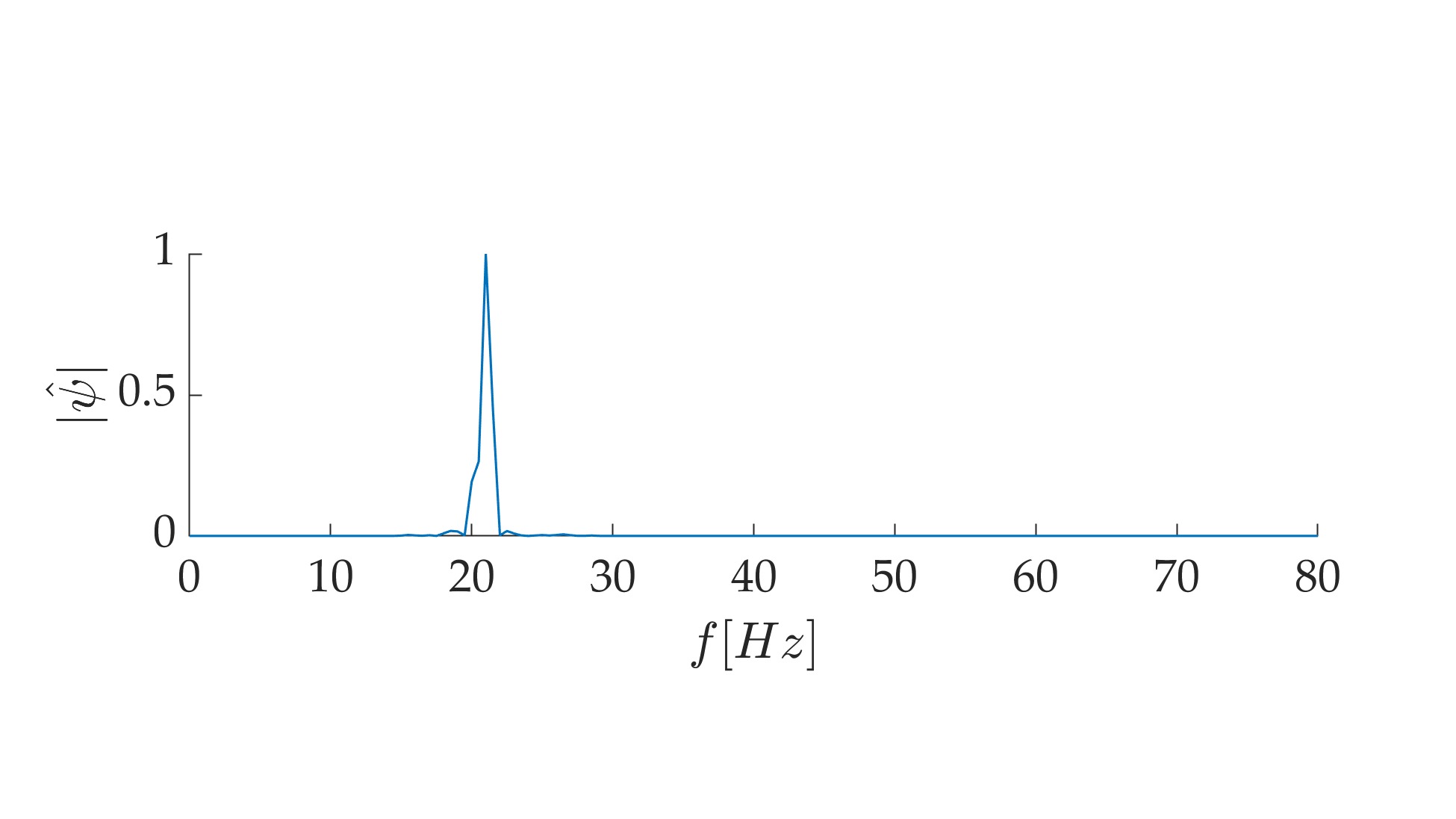}  
\end{subfigure}
\caption{Detected traveling wave pattern in test case 1. A snapshot of the normalized thickness $\check{h}$ is shown on the left. The figure in the middle shows the most dominant traveling wave structure. The associated frequency spectrum is shown on the right.}
\label{fig:mPOD_liquidfilm_Case1}
\end{figure*}

Similarly, in the decomposition of the thickness maps, mPOD modes linked to traveling waves are expected to be paired in couples having comparable amplitude, wavenumber and frequency spectra, and a $\pi/2$ phase delay in both space and time. These constraints were used to automatically detect travelling wave patterns using a simple minimization problem, described as follows. Given 

\begin{equation}
    \widehat{\psi}_j(\omega)=\frac{1}{\sqrt{2 \pi}}\int\psi_j(t) e^{-\omega t} dt\,,\\
\end{equation} the frequency spectra of the j-th mPOD mode and given 

\begin{equation}
\widehat{\phi}_j(k_x,z)=\frac{1}{\sqrt{2 \pi}}\int\phi_j(x,z) e^{-k_x x} dx
\end{equation} its stream-wise wavenumber spectra, a wave pattern traveling along the $x$ direction is identified by the pair $i,j$ such that all the following quantities are minimized:

\begin{equation}
\label{MIN}
\begin{gathered}
    \bigl | |\widehat{\psi}_j|-|\widehat{\psi}_i|\bigr|_O=E_1 \\
    \bigl | |\widehat{\phi}_j|-|\widehat{\phi}_i|\bigr|_K=E_2 \\
    \bigl | |\mbox{arg}\{\widehat{\psi_j}\}|-|\mbox{arg}\{\widehat{\psi_i}\}|-\pi/2\bigr|_O=E_3\\
     \bigl | |\mbox{arg}\{\widehat{\phi_j}\}|-|\mbox{arg}\{\widehat{\phi_i}\}|-\pi/2\bigr|_K=E_4,
    \end{gathered}
\end{equation} where the subscript $O$ and $K$ denotes norms along the angular frequencies $\omega$ and the wavenumber domain $k_x$. 

The normalized energy content of the mPOD modes $(\sigma_r/\sigma_1)^2$ in the three cases is shown in Fig.~\ref{fig:sigma_mPOD}. Only modes with $\sigma^2_r/\sigma^2_1>1/5$ are shown and those linked to traveling waves according to the minimization of Eq.~\ref{MIN} are indicated with a full marker. The remaining ones can be seen as `standing wave' corrections to these patterns and are of no interest for the analysis that follows. 
	
A qualitative picture of the different levels of complexity in the observed wave patterns is given in Fig.~\ref{fig:sigma_mPOD}. In case 2, the energy decay with respect to the mode number is considerably more gentle than in the rest of the cases. This implies that a larger number of modes is required to approximate the film dynamics for a given accuracy. This test case is the one in which the interaction between the gas flow and the liquid film is the strongest, having the smallest $\hat{Z}$ and the largest $\Pi_g$.

In case 1, only one pair of modes linked to traveling waves is identified (modes 2 and 3). The associated spatial structure is shown in Fig.~\ref{fig:mPOD_liquidfilm_Case1} ($\phi_{2}$, in the middle), together with its frequency spectrum ($|\widehat{\psi}_{2,3}|$, on the right) and a snapshot of the original scaled thickness map ($\check{h}$, on the left). Comparing the snapshot with the detected structure gives a qualitative indication of their relative importance and hence how well these modes approximate the data. 

For this test case, featuring almost bi-dimensional waves, the identified traveling pattern accounts for a significant portion of the undulation amplitude. It is worth noticing that in this case the liquid film and the gas jet are locked at the same frequency (see Fig. \ref{fig:spectra}) and the detected pattern is present in the \emph{entire domain}, i.e. both the final film and the run-back flow. In these two regions, waves have significantly different shapes, travel in opposite direction with different speed and evolve over a film with largely different thickness. It is thus remarkable that, despite these differences, waves in both regions are produced at the same frequency, which is also strongly present in the gas flow field.

\begin{figure*}
  \centering
  \begin{minipage}{.33\linewidth}
      $\check{h}(x,z)$ 
      {\includegraphics[width=\linewidth]{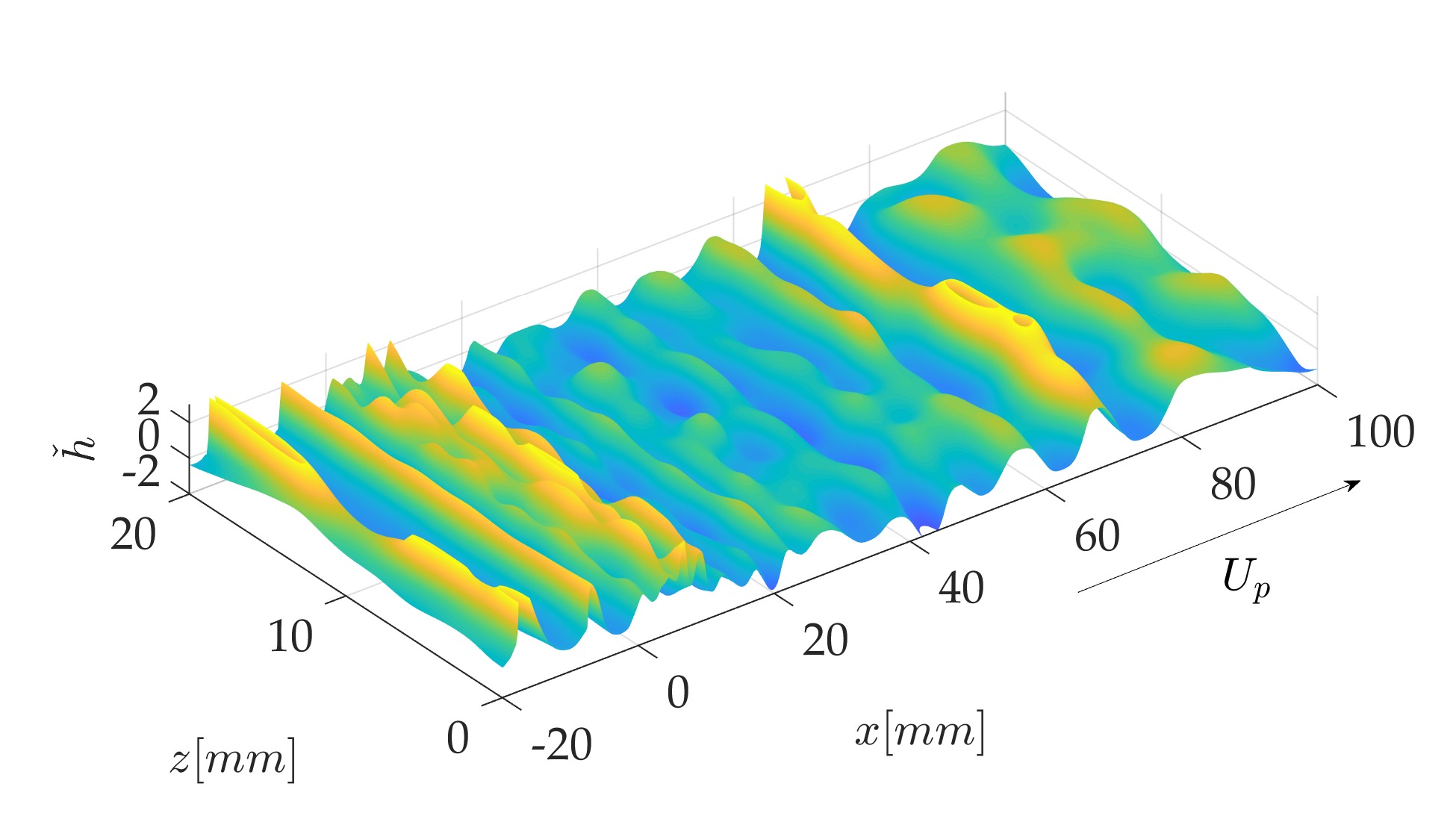}}%
  \end{minipage}%
  \hfill
  \begin{minipage}{.33\linewidth}
      $\phi_{1}(x,z)$\\
      {\includegraphics[width=\linewidth]{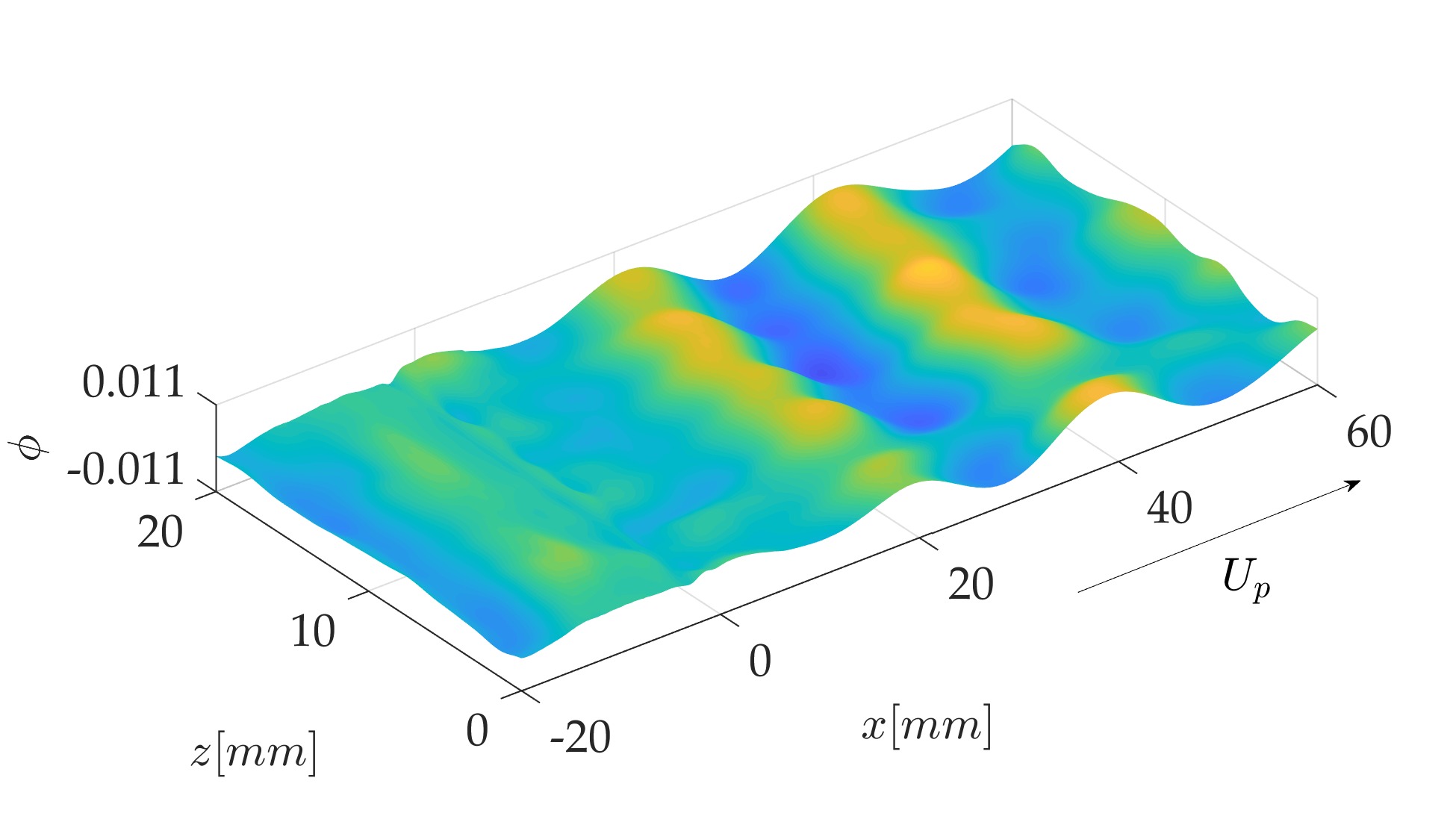}}
      $\phi_{3}(x,z)$\\
      {\includegraphics[width=\linewidth]{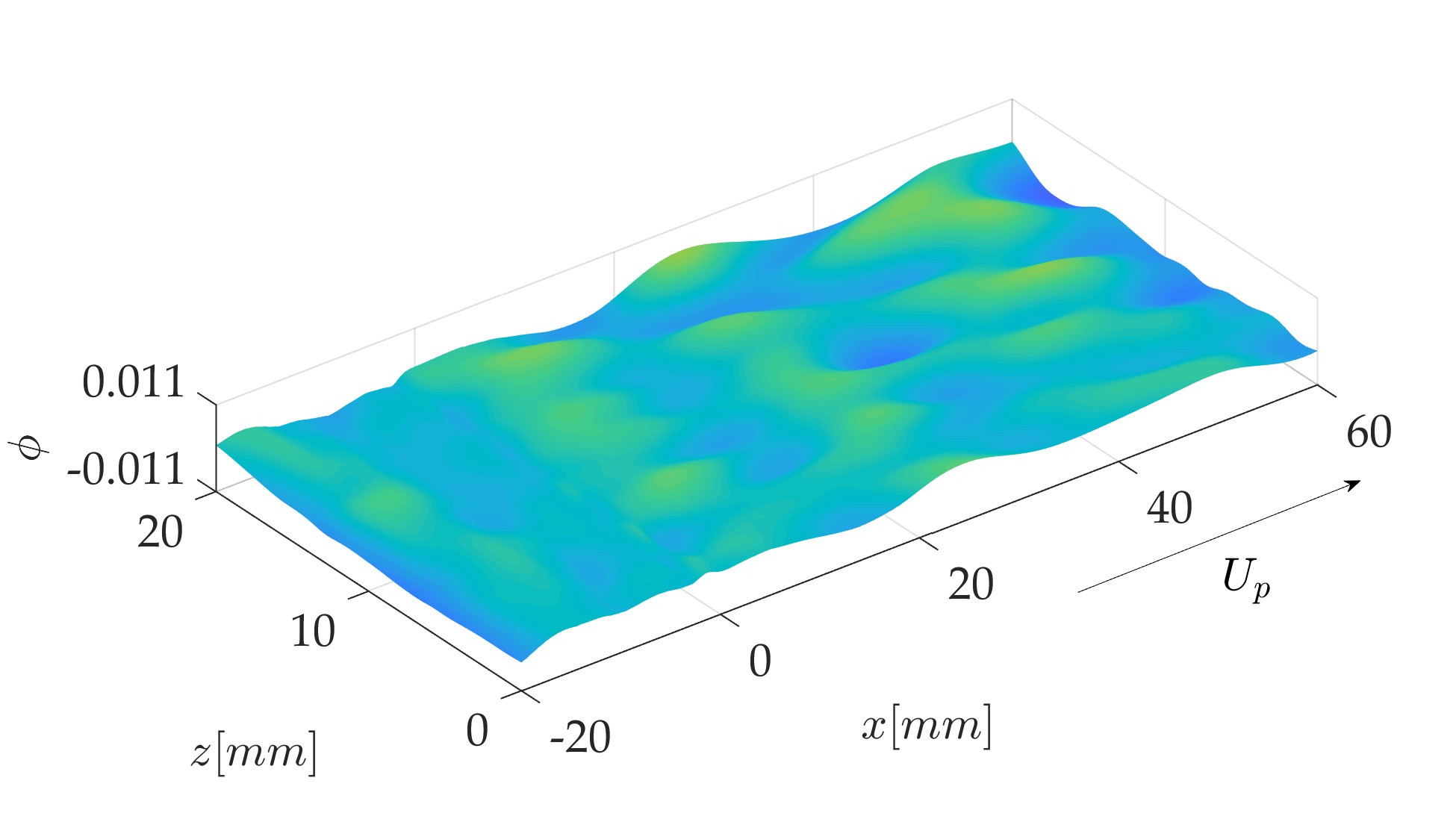}}
      $\phi_{5}(x,z)$\\
      {\includegraphics[width=\linewidth]{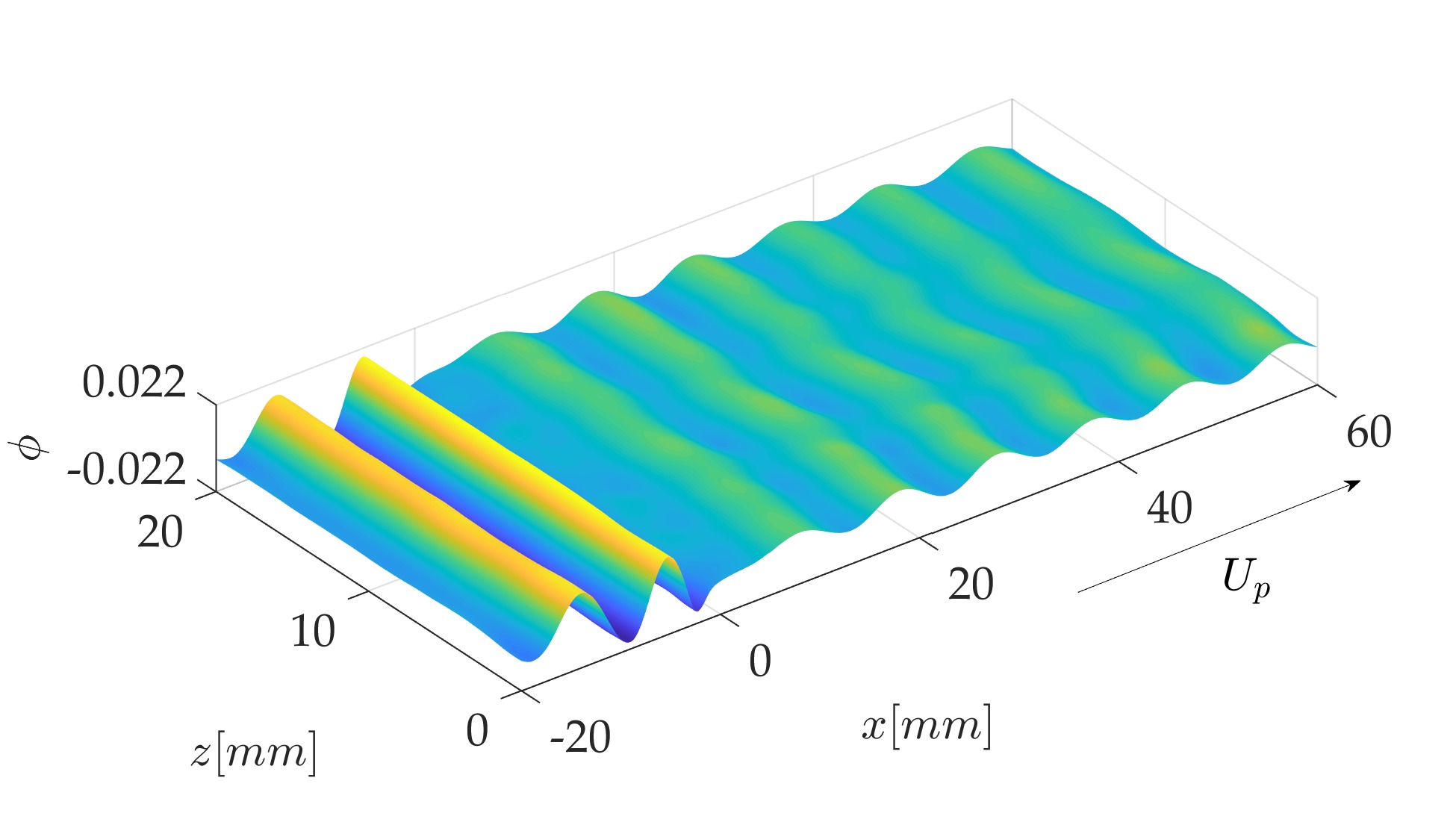}}
      $\phi_{21}(x,z)$\\
      {\includegraphics[width=\linewidth]{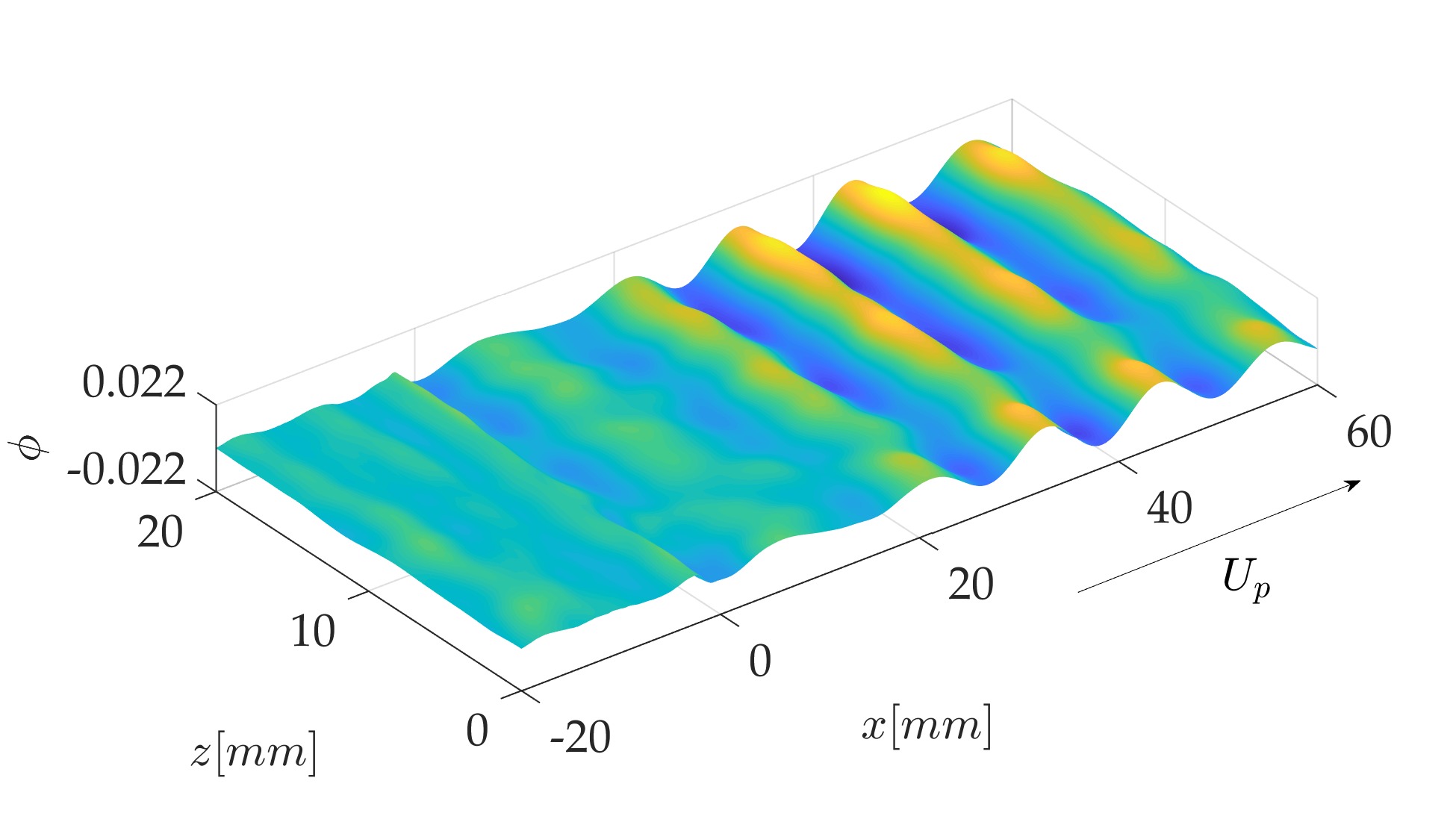}}%
  \end{minipage}%
    \hfill
    \begin{minipage}{.33\linewidth}
      $|\hat{\psi}_{1,2}|(x,z)$\\
      {\includegraphics[width=\linewidth]{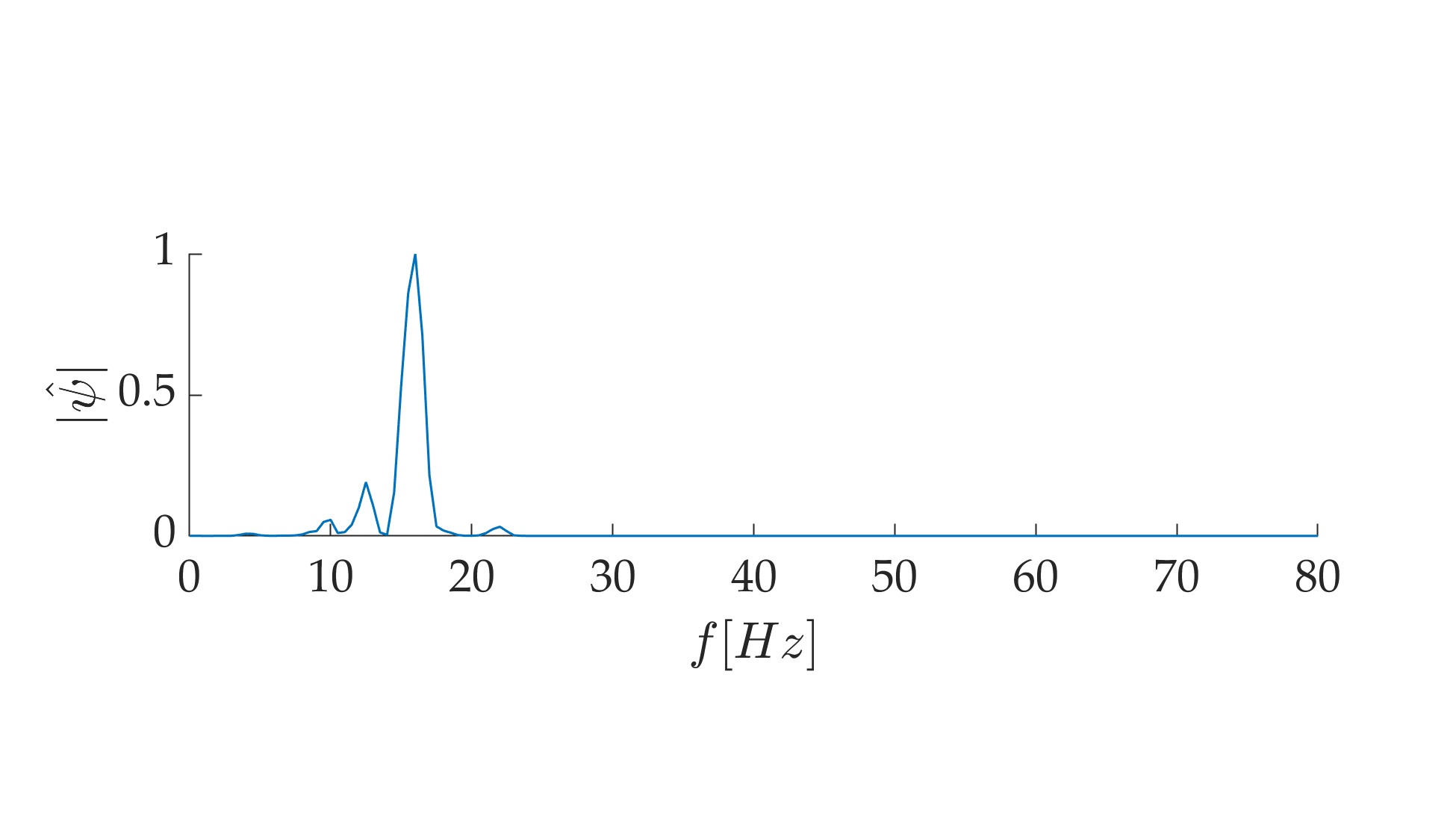}}
      $|\hat{\psi}_{3,4}|(x,z)$ \\
      {\includegraphics[width=\linewidth]{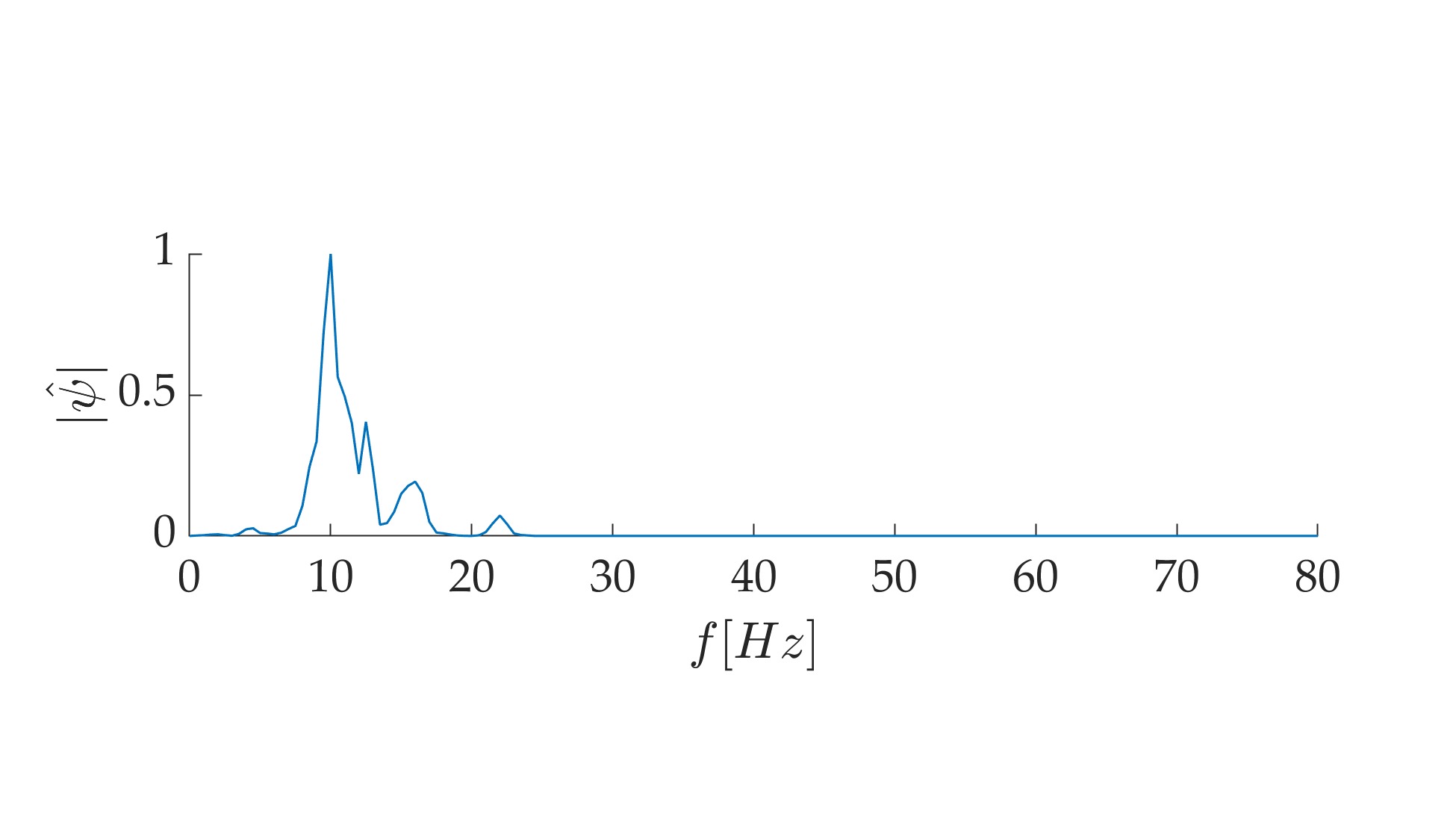}}
      $|\hat{\psi}_{5,6}|(x,z)$\\
      {\includegraphics[width=\linewidth]{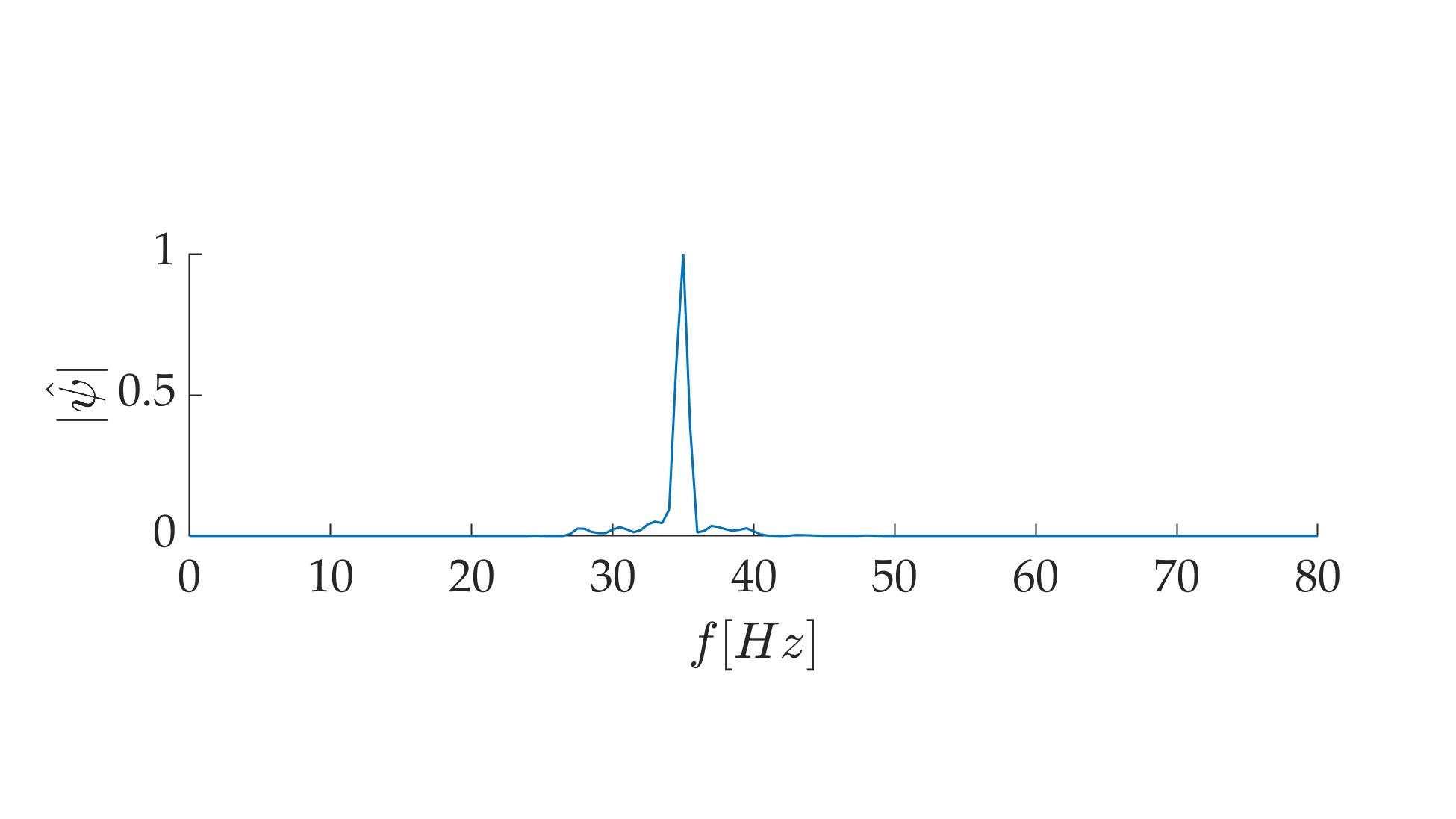}}
      $|\hat{\psi}_{21,23}|(x,z)$\\
      {\includegraphics[width=\linewidth]{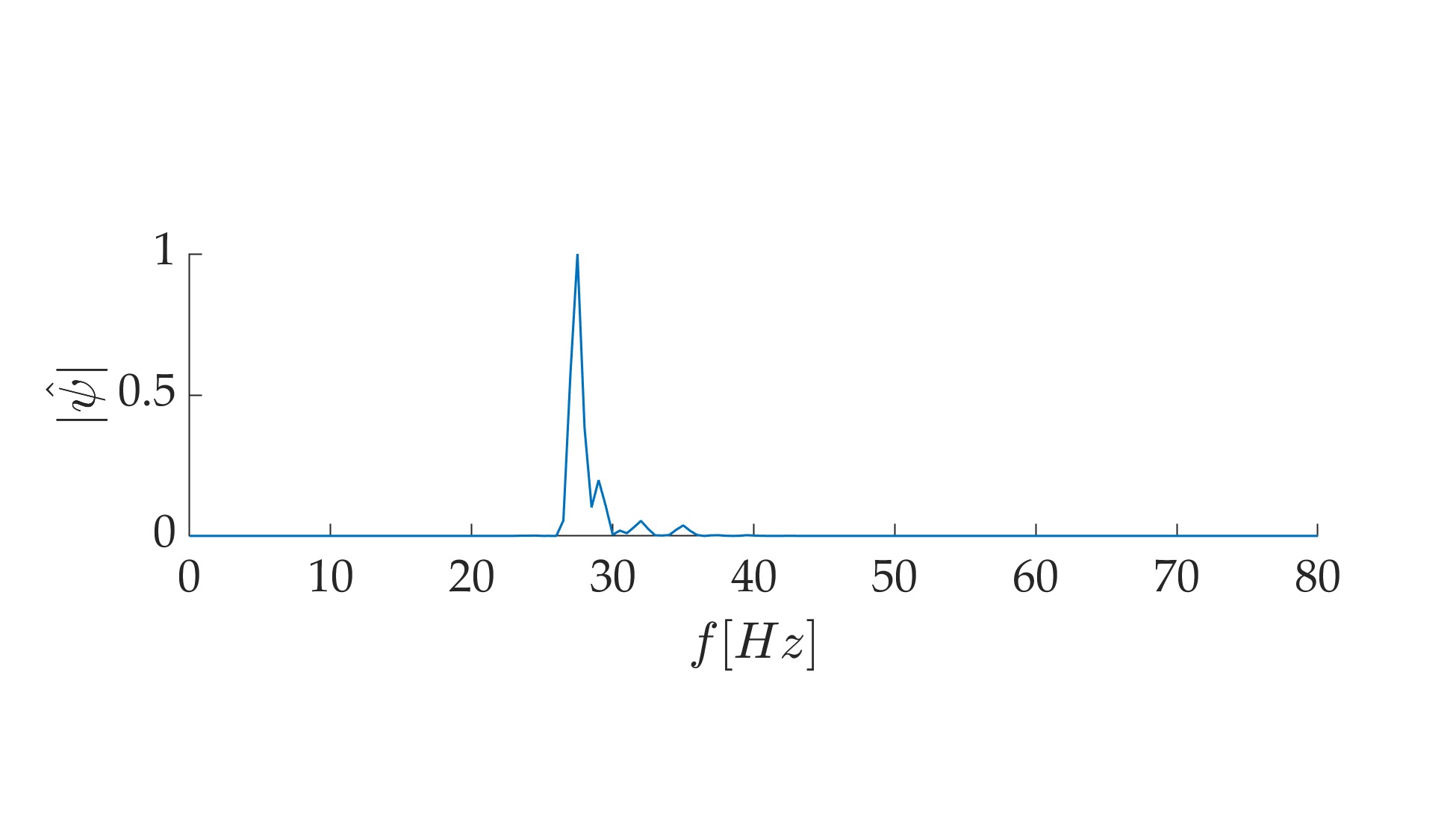}}%
  \end{minipage}%
  \caption{Same as in Fig.~\ref{fig:mPOD_liquidfilm_Case1} for Case 2.}
\label{fig:mPOD_liquidfilm_Case2}
  \end{figure*}

Fig.~\ref{fig:mPOD_liquidfilm_Case2} shows the same plots for test case 2, in which four traveling wave patterns are detected (modes 1 and 2, 3 and 4, 5 and 6, 21 and 23). The preliminary analysis in Fig.~\ref{fig:spectra_liquid_Z18_P875} shows that, in both experimental and numerical data, the frequency content in the final film and the run-back flow are different. The identified traveling pattern suggests possible causes for such a difference. The pattern in $\phi_1$, having the highest amplitude, is characterized by a wavelength of about $20mm$ and a frequency of $\approx 18 Hz$. Interestingly, this mode is nearly absent in the run-back flow and in the impingement region and seems thus to be associated to the downstream evolution of the waves. The same is true for the traveling pattern in modes 3 and 4 ($\phi_3$): these modes account for the three-dimensional evolution of the waves. The frequency spectra of this pattern is shifted towards lower ranges, with a peak at about $9 Hz$.

The remaining pairs are both bidimensional and have very different locations in space. The waves in modes 5 and 6 ($\phi_5$) span the entire domain, similarly to the pattern observed in Figure \ref{fig:mPOD_liquidfilm_Case1} for case 1. The minor difference with respect to the equivalent mode in case 1 is its stronger presence in the run-back flow and weaker in the final coating film. On the other hand, similarly to the pattern in case 1, the frequency content of these modes ($\approx 35$ Hz) is closely matching the one of the jet flow (cf. Fig.~\ref{fig:spectra_gas_Z18_P875}). Finally, the waves corresponding to modes 21 and 23 are solely present in the final coating, far downstream the impingement region, and are thus associated to the wave evolution. 

\begin{figure*}
   \centering
  \begin{minipage}{.33\linewidth}
      $\check{h}(x,z)$
      {\includegraphics[width=\linewidth]{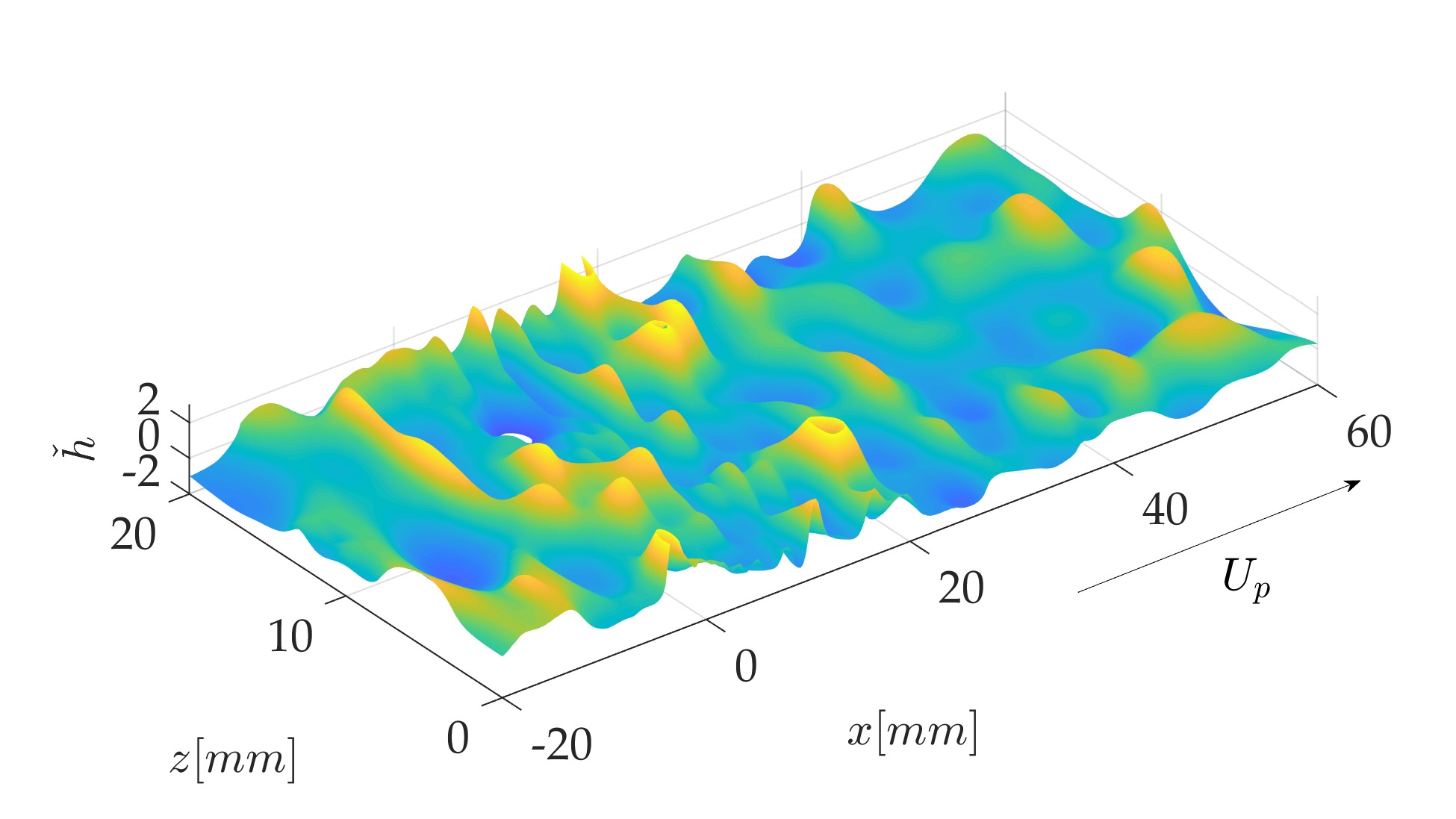}}%
  \end{minipage}%
  \hfill
  \begin{minipage}{.33\linewidth}
      $\phi_{4}(x,z)$\\
      {\includegraphics[width=\linewidth]{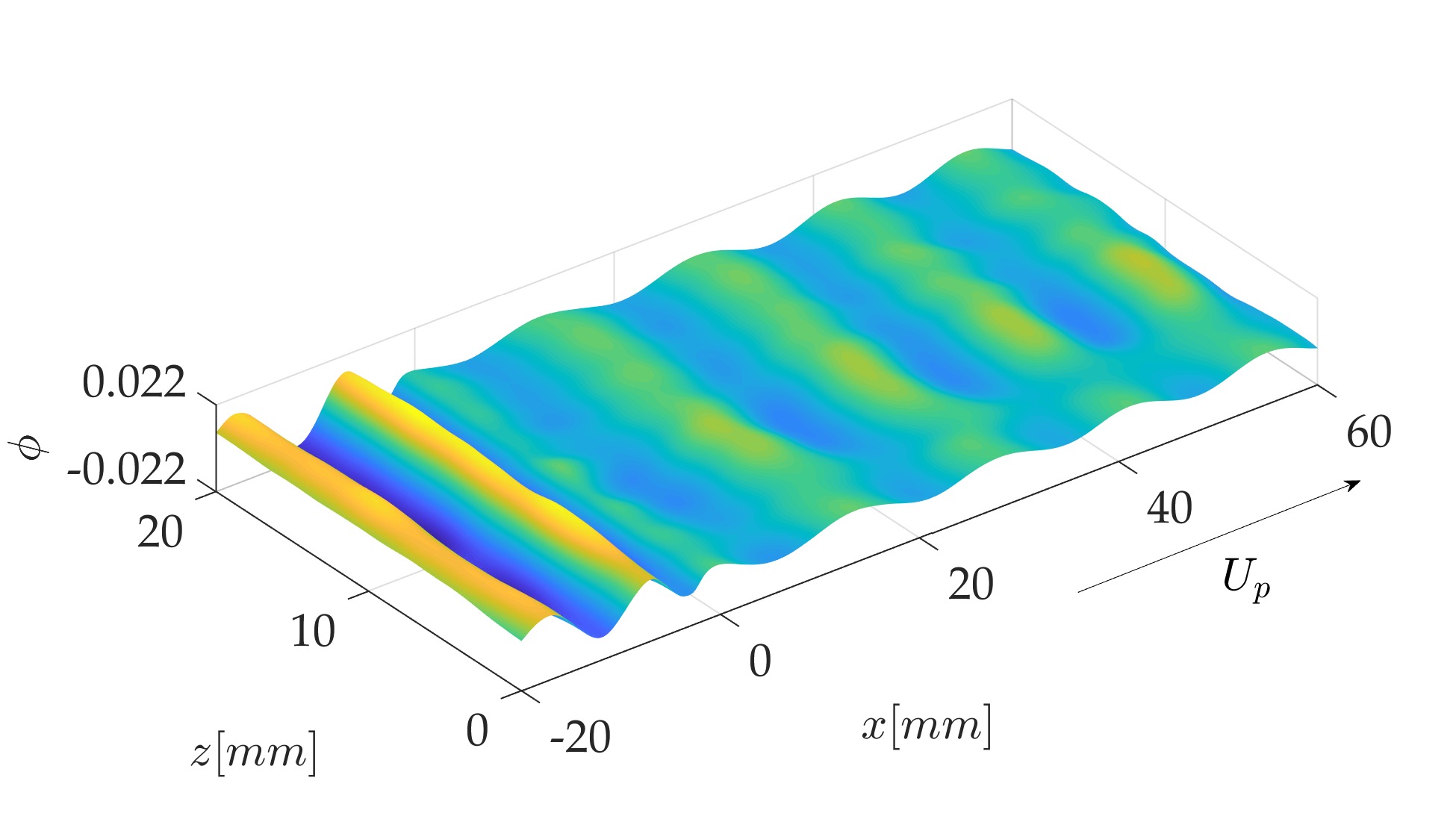}}
      $\phi_{9}(x,z)$\\
      {\includegraphics[width=\linewidth]{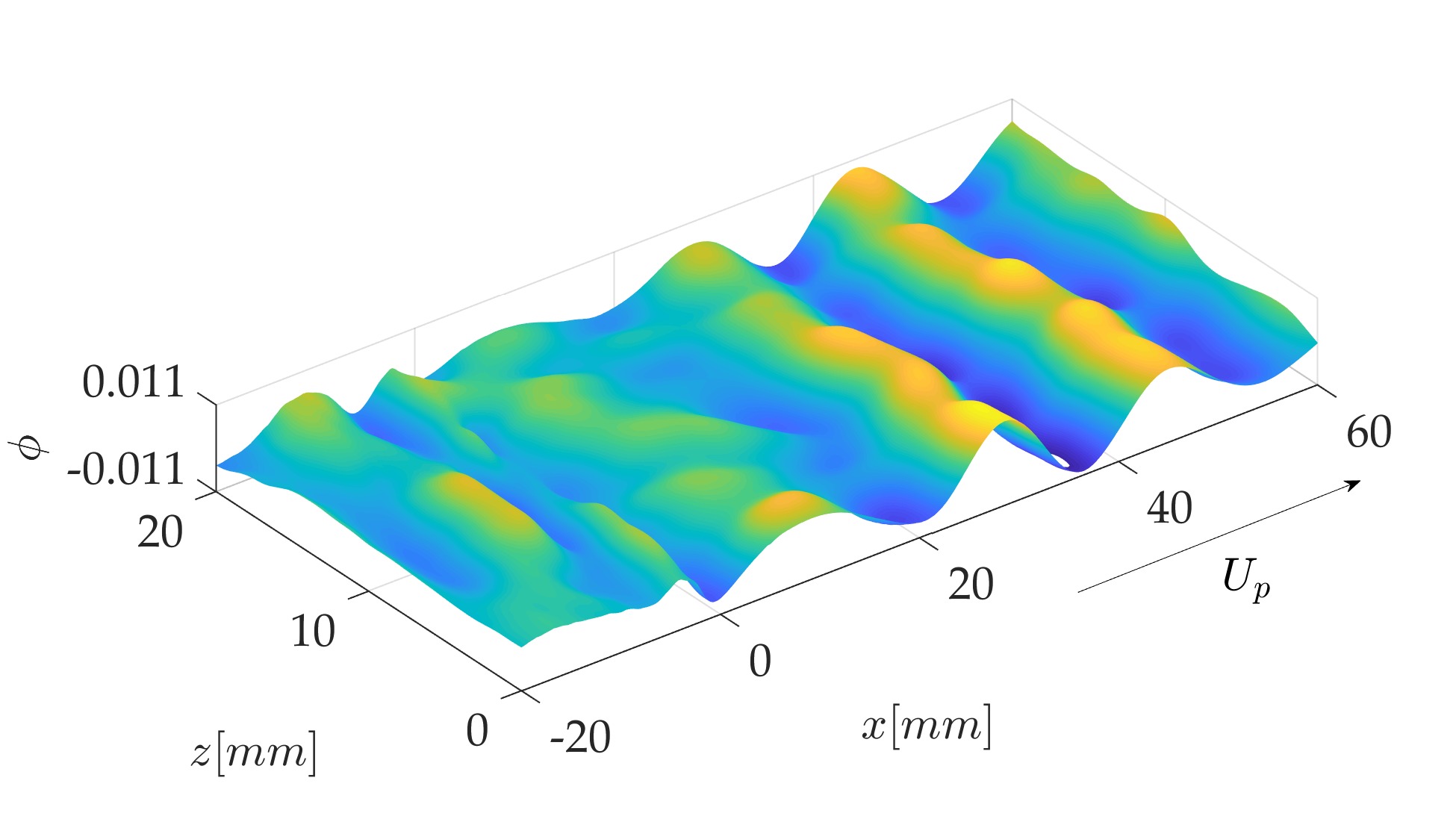}}
      $\phi_{11}(x,z)$\\
      {\includegraphics[width=\linewidth]{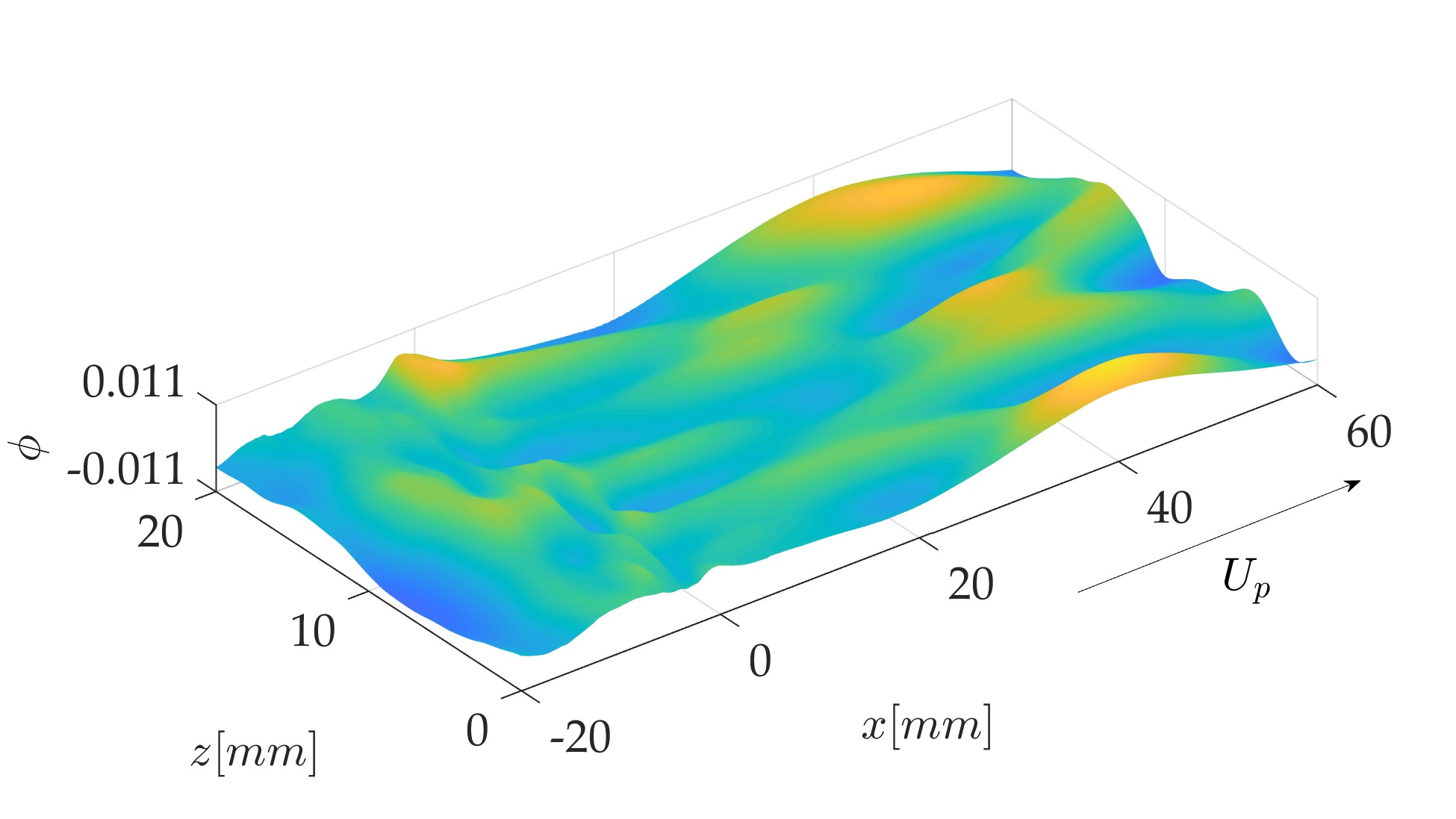}}%
  \end{minipage}%
  \hfill
    \begin{minipage}{.33\linewidth}
      $|\hat{\psi}_{4,5}|(x,z)$
      {\includegraphics[width=\linewidth]{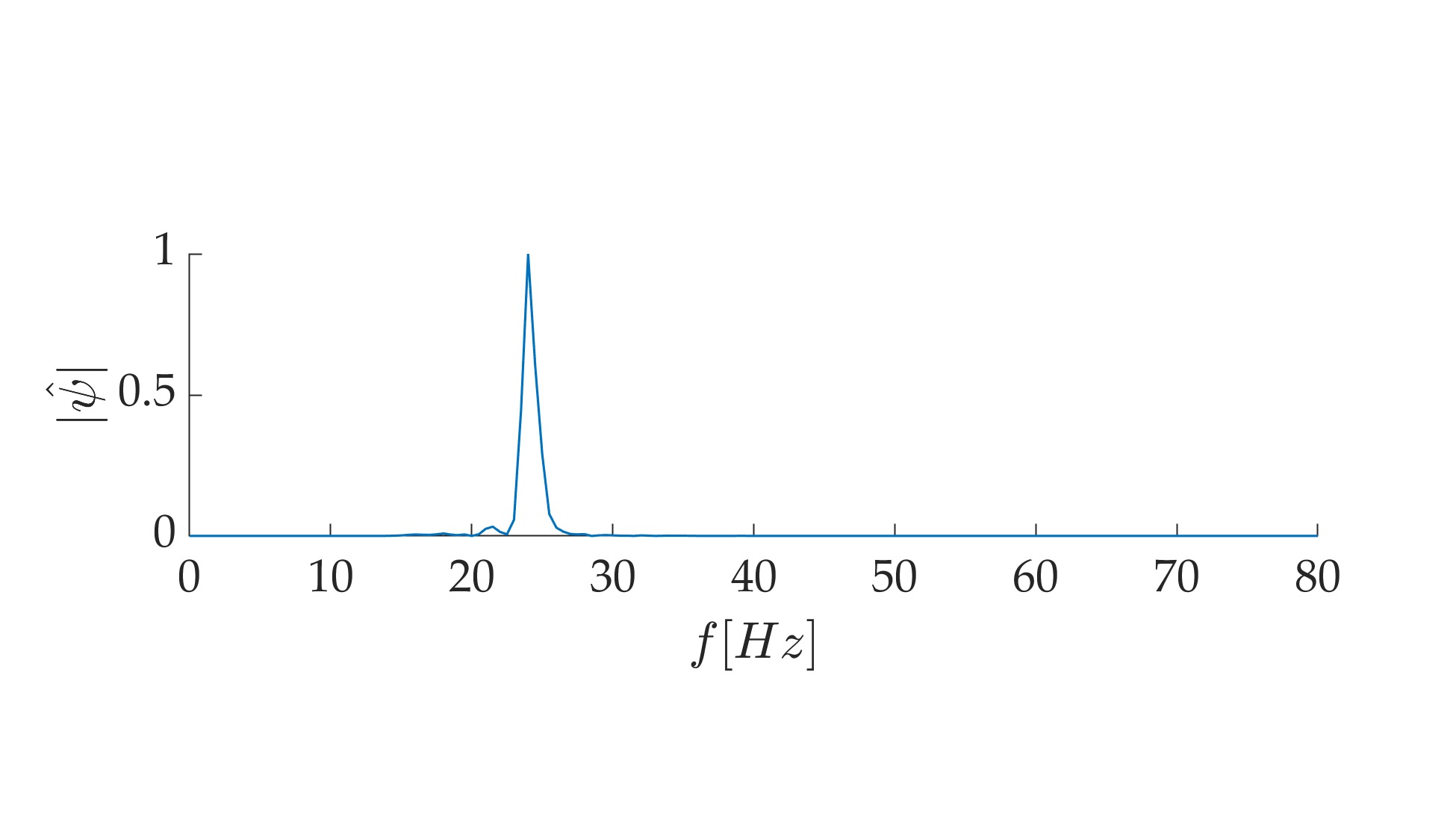}}
      $|\hat{\psi}_{9,10}|(x,z)$ 
      {\includegraphics[width=\linewidth]{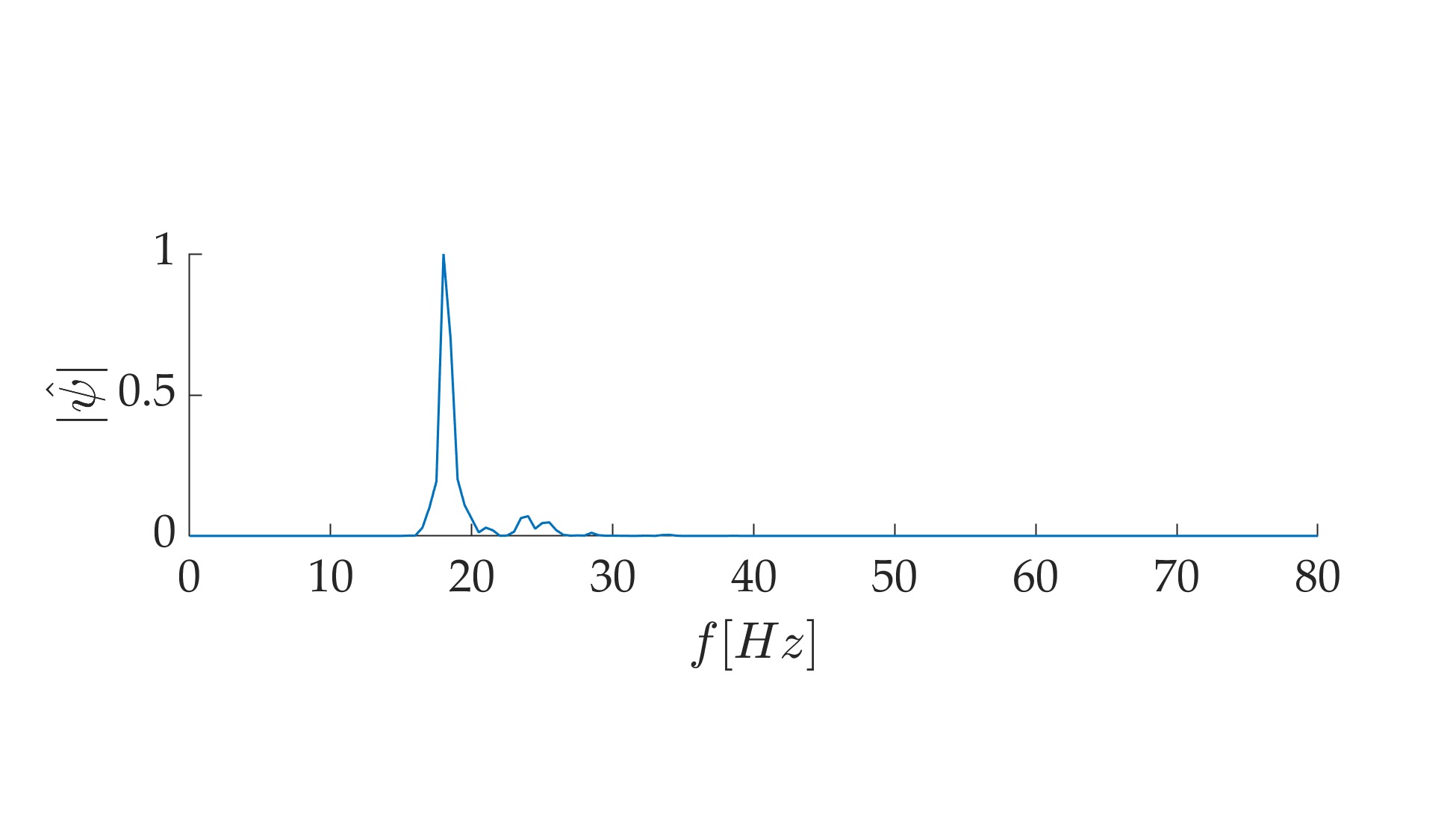}}
      $|\hat{\psi}_{11,12}|(x,z)$
      {\includegraphics[width=\linewidth]{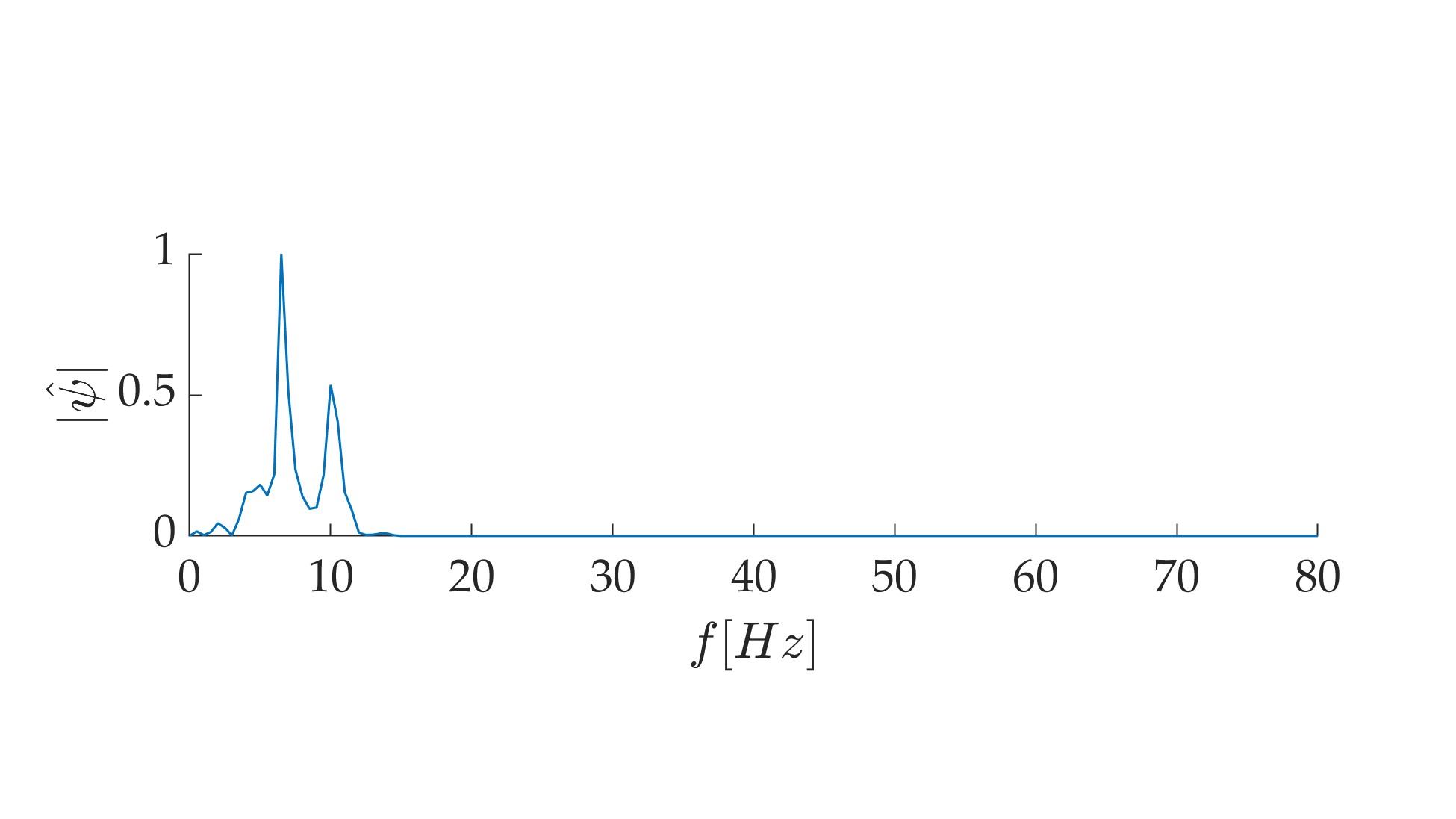}}%
  \end{minipage}%
\caption{Same as in Fig.~\ref{fig:mPOD_liquidfilm_Case1} for Case 3.}
\label{fig:mPOD_liquidfilm_Case3}
\end{figure*}

The same analysis is presented in Fig.~\ref{fig:mPOD_liquidfilm_Case3} for the last test case, in which three traveling wave patterns have been identified (modes 4 and 5, 9 and 10, 11 and 12). Of these, the first pattern is similar to the ones already observed in the other two cases, characterized by 2D waves which span the entire domain. The remaining two patterns account for 3D distortions and the downstream evolution of the coating waves.

The analysis of the waves reveals an undulation pattern present in \emph{all} the investigated test cases albeit with different intensity: it consists in the formation of a bi-dimensional wave whose spectral content coincides with the one of the gas jet (modes 2,3 for case 1, modes 5,6 for case 2 and modes 4,5 for case 3). This component is hereinafter referred to as \emph{2D coupled undulation}. The fact that the waves originating from this mechanism share the same frequency in both the final coating and the run-back flow region, despite the largely different thicknesses and traveling speeds, further corroborates the hypothesis of their common origin and their link to the jet flow unsteadiness. The coupling between phases is investigated further in Section \ref{SecVIC}. 

In general, the energy content of the modes is weak in the run-back flow except for the 2D coupled undulation. The other modes are mostly active in the final film, and in this sense, this region seems to be more capable to sustain a wider portion of the spectra of gas jet perturbations. The fact that the frequencies of the gas-liquid interaction are several orders of magnitude lower than the natural frequency of the jet impinging on a dry flat plate supports the idea of a two-way coupling between phases.

The frequency of the 2D coupled mode in case 3 is remarkably close to the one in case 1 ($\approx 22$ Hz). This was expected as the undulation frequency is known to scale well with the wiping parameter $\Pi_g$ \cite{Gosset2019}, almost identical in both cases (table \ref{tab:config}). It demonstrates that the spectral content of the undulation patterns far downstream wiping (as it is the case in experiments \cite{Gosset2019}) is mostly linked to this coupled 2D mode, even if the waves undergo transverse modulation and levelling due to surface tension and viscosity \cite{orchard_surface_1963,eres_three-dimensional_1999}. 

It is now interesting to trace back the location at which these 2D waves originate and analyze their phase domain. To this end, we consider the approximation $\tilde{h}$ of the film thickness evolution in the mid-plane, i.e. $h(x,z=15mm,t)$, constructed by summing only the traveling wave patterns $j=2,3$ in case 1, $j=5,6$ in case 2 and $j=4,5$ in case 3.

\begin{figure*}
    \begin{subfigure}[t]{.33\textwidth}
      \centering
      % include first image
      Case 1 \\
      $\hat{Z}=14.2$ | $\Pi_g=0.16$
      \includegraphics[width=\linewidth]{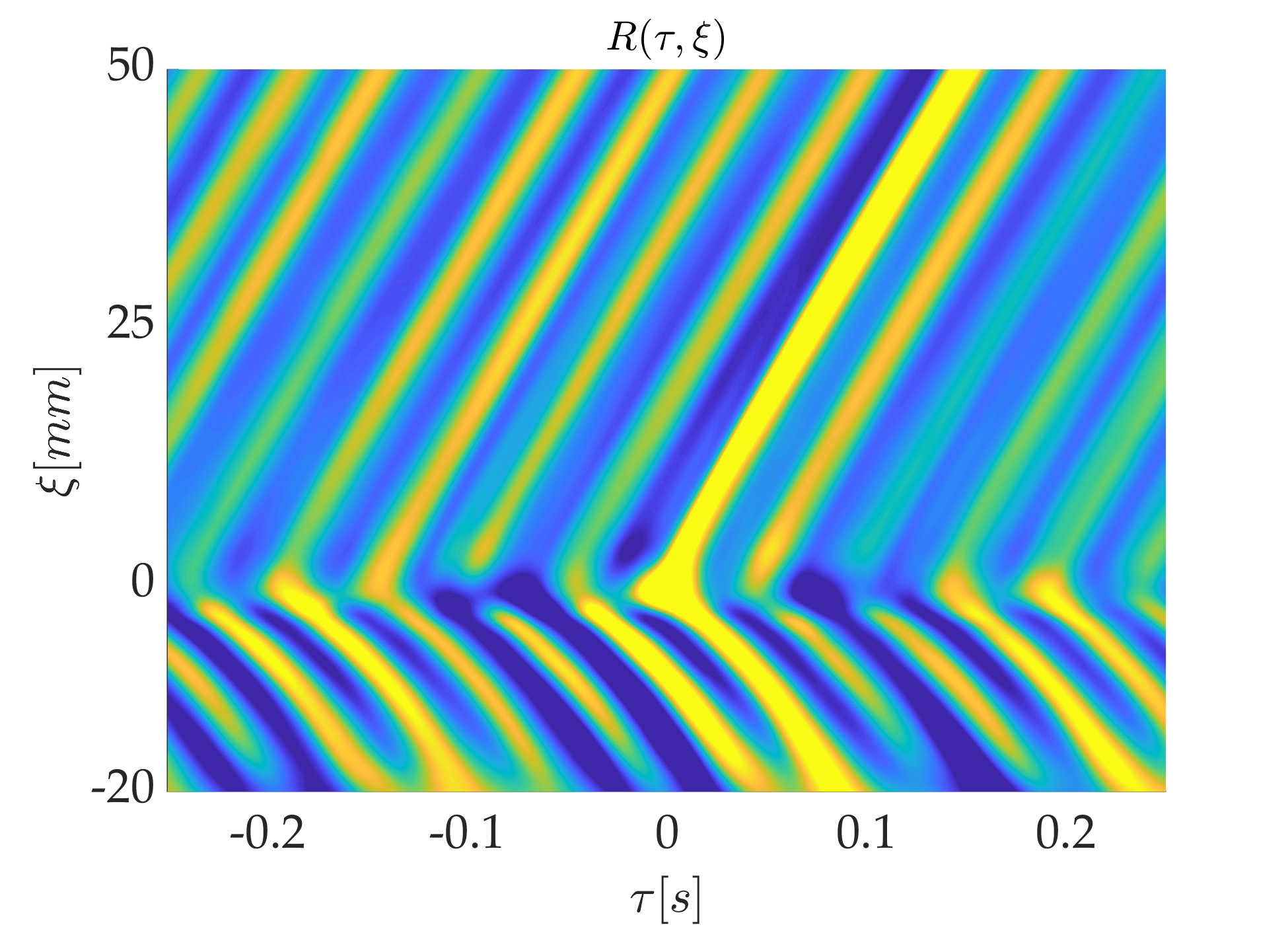}
    \end{subfigure}
    \begin{subfigure}[t]{.33\textwidth}
      \centering
      % include first image
      Case 2 \\
      $\hat{Z}=14.2$ | $\Pi_g=0.33$
      \includegraphics[width=\linewidth]{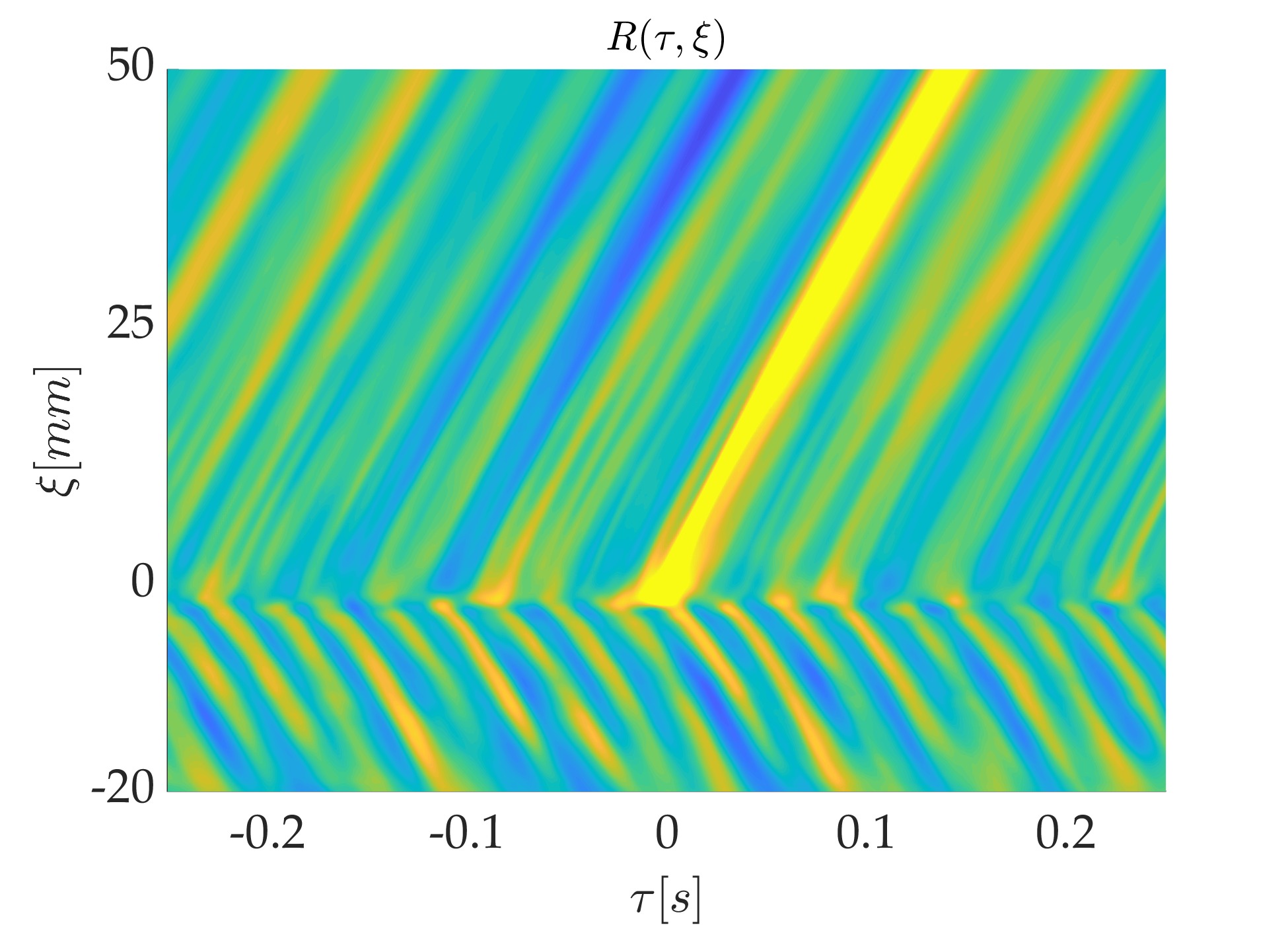}
      \caption{Original}
      \label{fig:xcorr_org}
    \end{subfigure}
    \begin{subfigure}[t]{.33\textwidth}
      \centering
      Case 3 \\
      $\hat{Z}=19.4$ | $\Pi_g=0.18$
      % include second image
      \includegraphics[width=\linewidth]{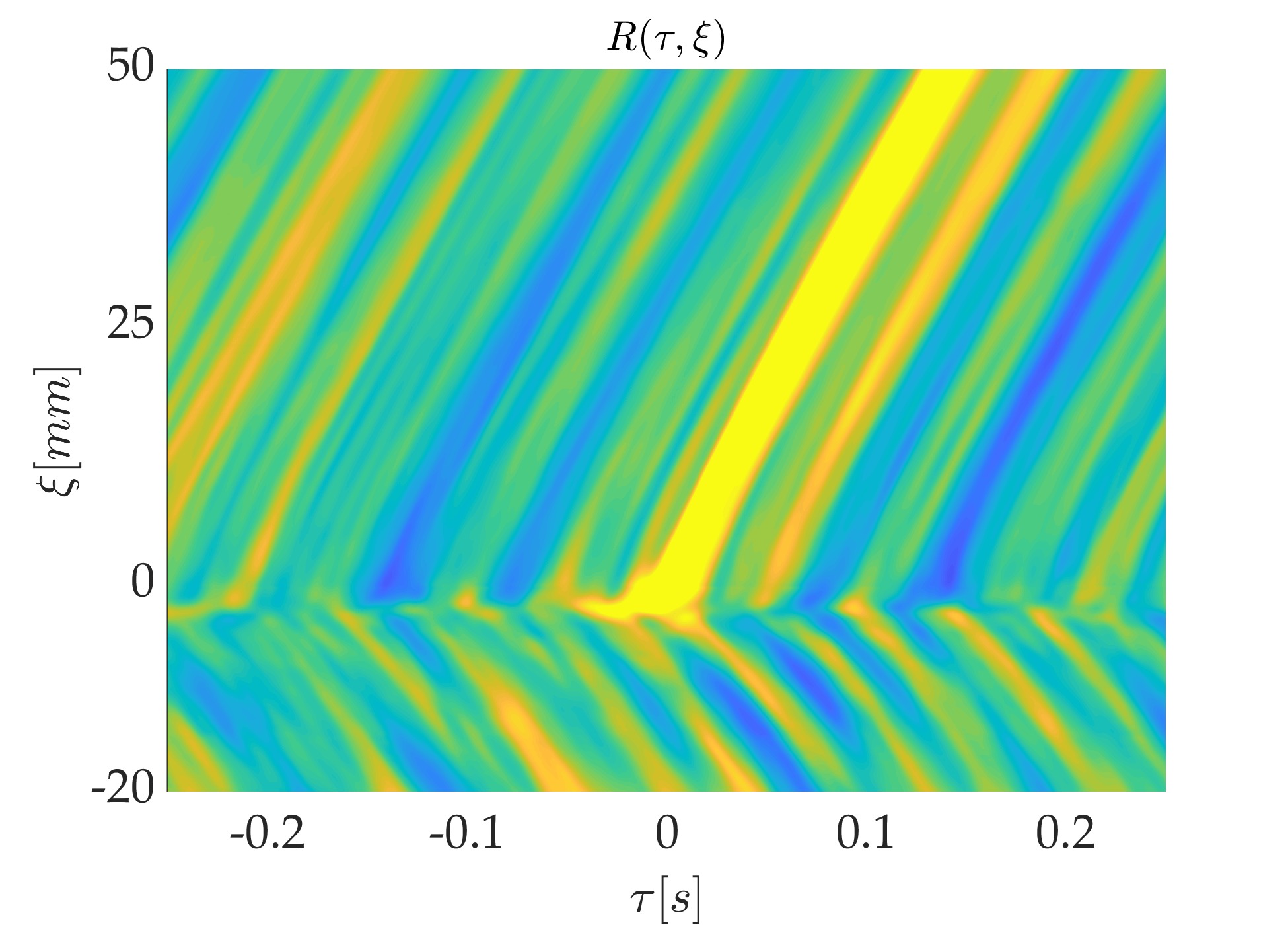}  
    \end{subfigure}

    \begin{subfigure}[t]{.33\textwidth}
      \centering
      % include first image
      \includegraphics[width=\linewidth]{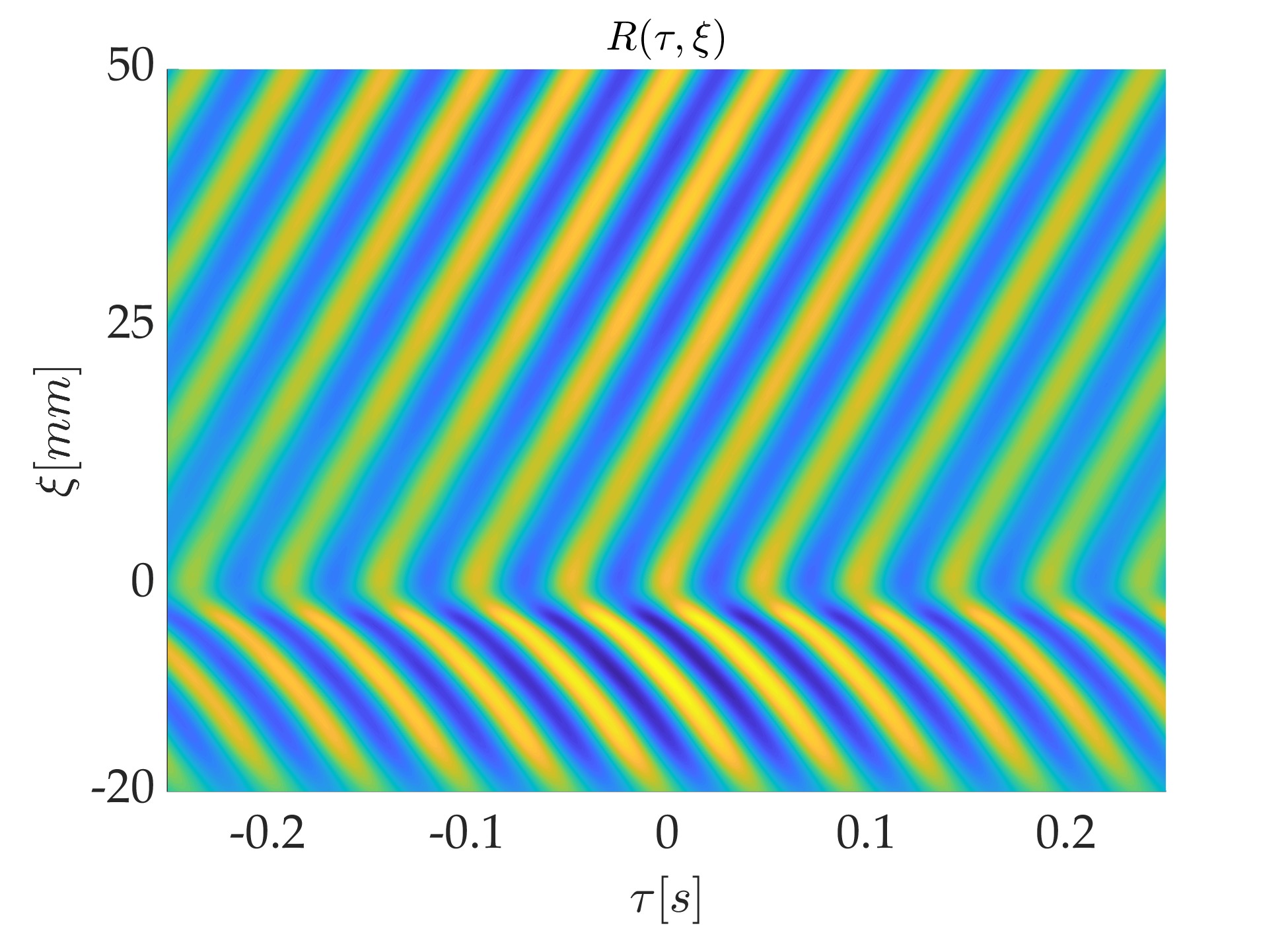}
    \end{subfigure}
    \begin{subfigure}[t]{.33\textwidth}
      \centering
      % include first image
      \includegraphics[width=\linewidth]{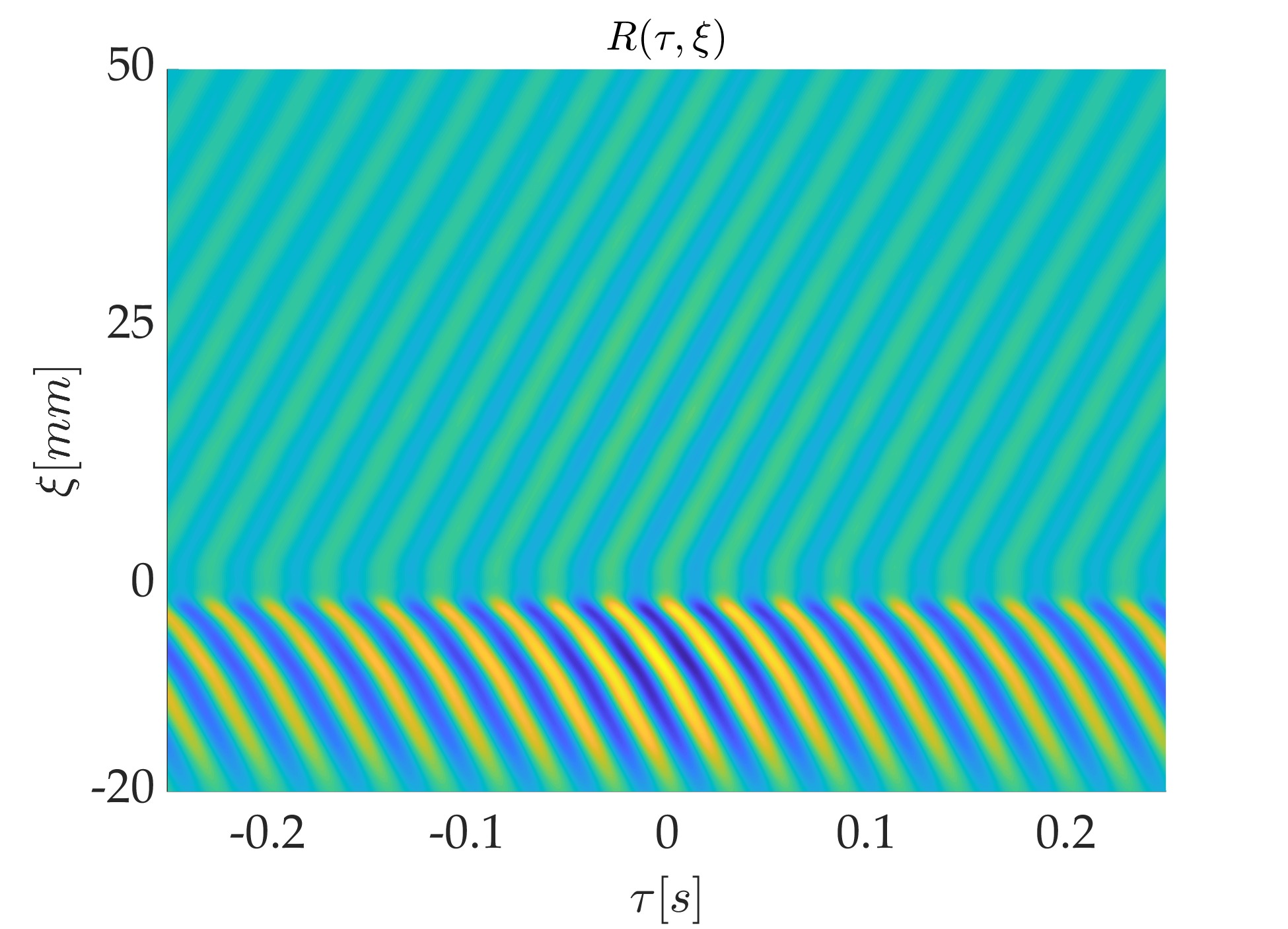}
      \caption{mPOD filtered}
      \label{fig:xcorr_mPOD}
    \end{subfigure}
    \begin{subfigure}[t]{.33\textwidth}
      \centering
      % include second image
      \includegraphics[width=\linewidth]{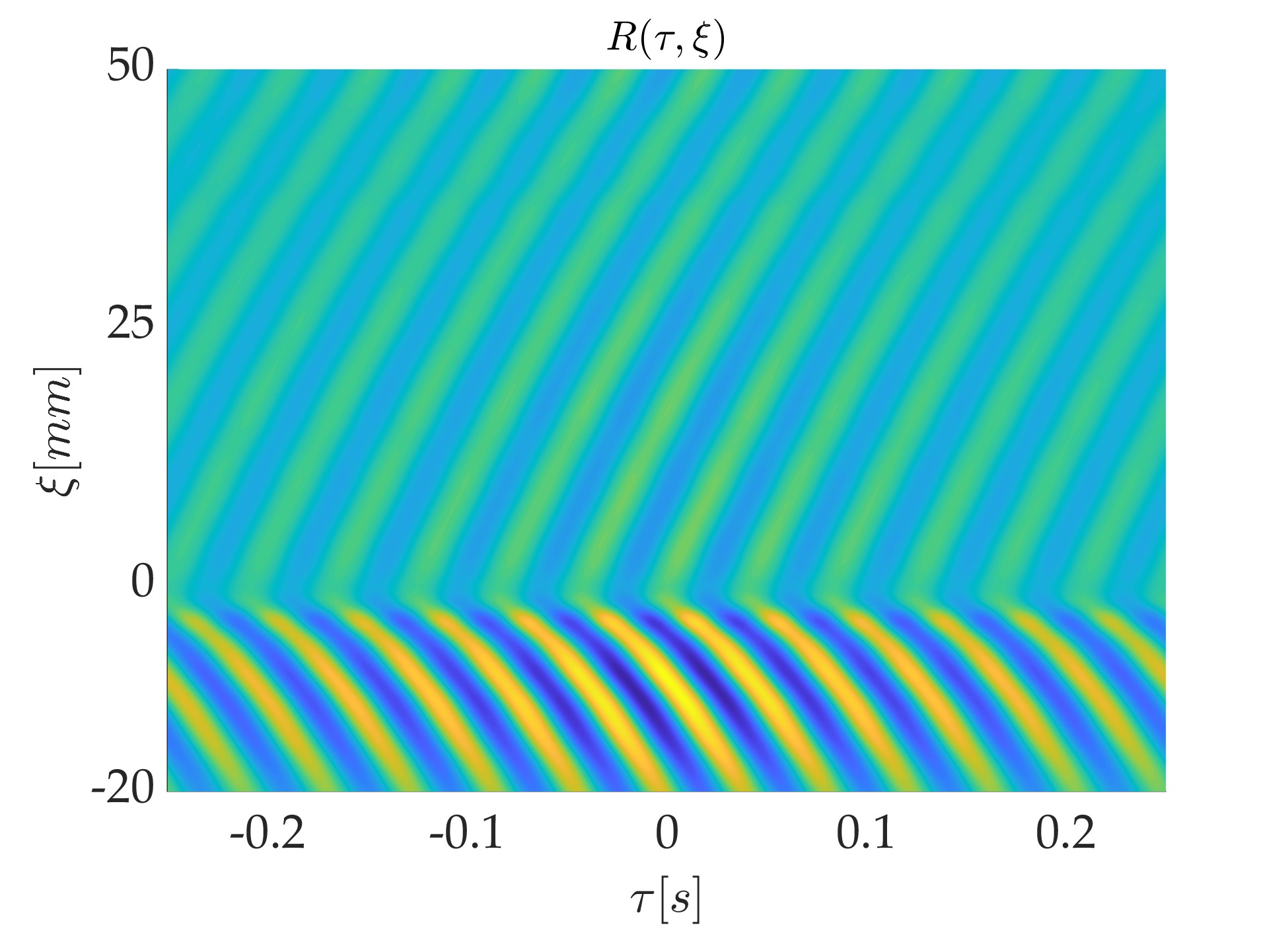}  
    \end{subfigure}
    \caption{Comparison between the original (a) and mPOD filtered (b) thickness cross correlation colormaps for the three cases of study organized in columns.}
    \label{fig:xcorr}
\end{figure*}

The spatio-temporal autocorrelation function of the thickness map $\tilde{h}(x,t)$ is defined as:

\begin{eqnarray}
\label{eq:auto-correlation}
 R(\tau,\xi)= \int \int {\tilde{h}(x,t)   \tilde{h}(x-\xi,x-\tau)} dx \,dt
\end{eqnarray} 
The contourmap of $R(\tau,\xi)$ is shown in Fig.~\ref{fig:xcorr} considering both the original thickness map (top row) and the single traveling wave approximation (bottom row). A comparison of the two allows assessing how well the detected traveling wave pattern approximates the statistics of the liquid interface. As expected from the amplitude decay of the modes, the approximation is more accurate for test case 1, while in cases 2 and 3, its accuracy is mostly limited to the run-back flow region. 

The velocity of the waves is computed from the slope of the characteristic lines in the auto-correlation maps. This also gives qualitative information about the force balance in the liquid film. Shear stress and gravity dominate viscosity in the run-back flow while the reverse is true in the final coating film. A constant value of 0.34 m/s, equal to the substrate speed, is observed in the final film for all the cases. The shearing effect of the gas seems to be weak here since the characteristic lines are straight. 

On the other hand, the run-back waves are slightly accelerated as they travel towards the coating bath. The velocities range from 0.15 m/s near the wiping region, to 0.23 m/s approximately 20 mm below. In case 2, featuring the larger wiping number $\Pi_g$, this effect is less pronounced due to the higher prevalence of the shearing effect of the gas.

Most interestingly, these maps show that the waves in both the run-back flow and final coating originate \emph{at the same location}, slightly below $x=0$, which is the expected impact point of the jet. This occurs in all the three cases.
	
Noteworthy, the location at which the waves originate does not significantly vary in time and the formation of the waves in both directions are in phase. Therefore, from previous theoretical work\cite{Mendez2020}, we conclude that in the investigated test cases these patterns are produced mostly by an \emph{intensity modulation} of the wiping actuators (pressure gradient and shear stress) and not their \emph{displacement}. %The specific perturbation of the wiping actuators that triggers these patterns on the film is investigated further in the following section.
It is thus of interest now to analyze the gas flow structures (and the corresponding wiping actuators) correlated with the modes found in the film. It is a necessary step to elucidate the liquid-gas coupling that leads to wave formation on the coating film. 

	\subsection{The main mechanism of undulation}\label{SecVIC}
	
\begin{figure*}
  \begin{minipage}{.33\linewidth}
    $P'_w$
  \end{minipage}
   \begin{minipage}{.33\linewidth}
    $\tau_{wx}$
  \end{minipage}
   \begin{minipage}{.33\linewidth}
    $\tau_{wz}$
  \end{minipage}
  \noindent\makebox[\linewidth]{\rule{\linewidth}{0.4pt}}  
  \\
  \begin{minipage}{.33\linewidth}
      $\phi_{1}(x,z)$\\
      {\includegraphics[width=\linewidth]{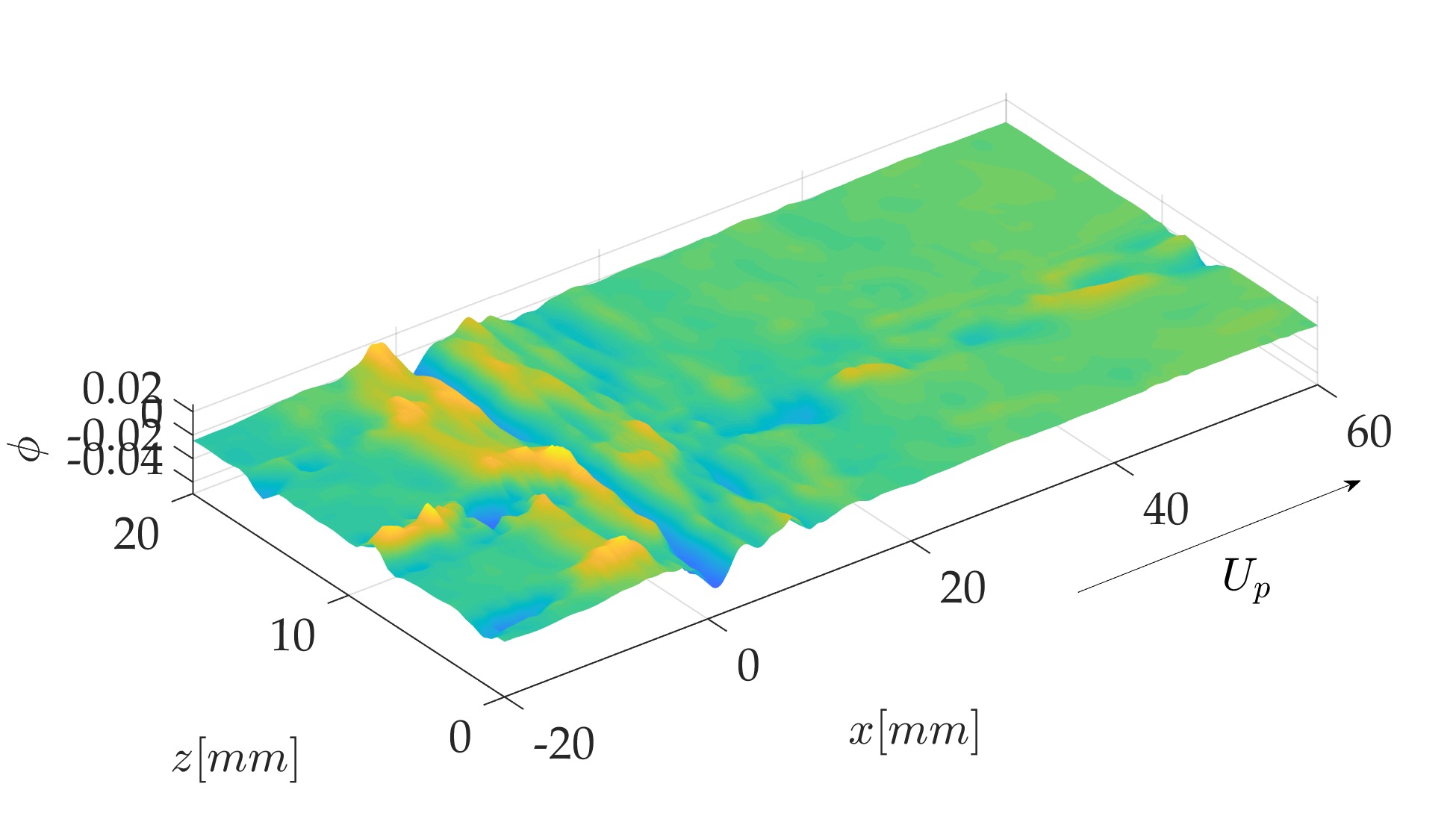}}
      $\phi_{3}(x,z)$\\
      {\includegraphics[width=\linewidth]{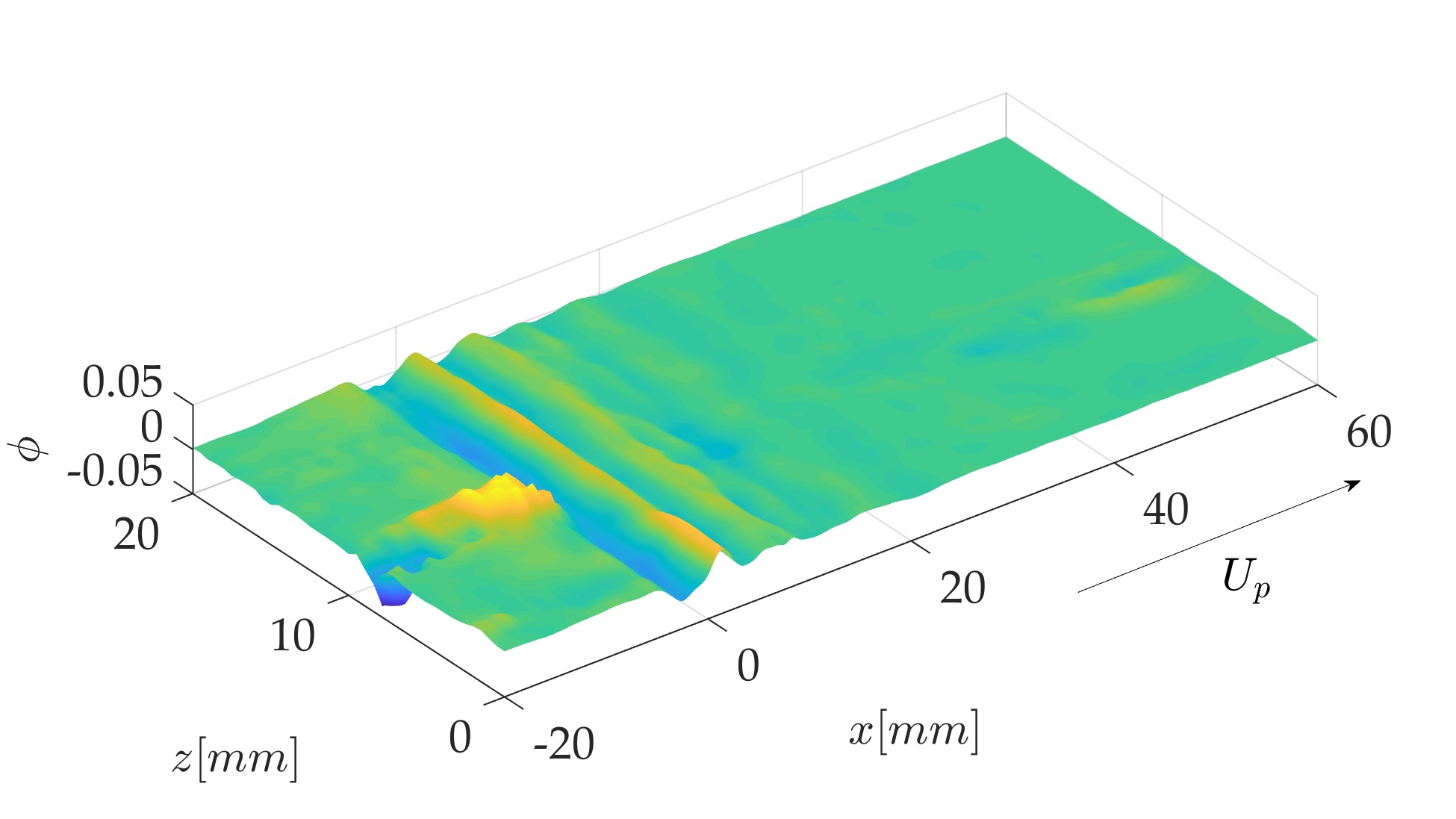}}
      $\phi_{5}(x,z)$\\
      {\includegraphics[width=\linewidth]{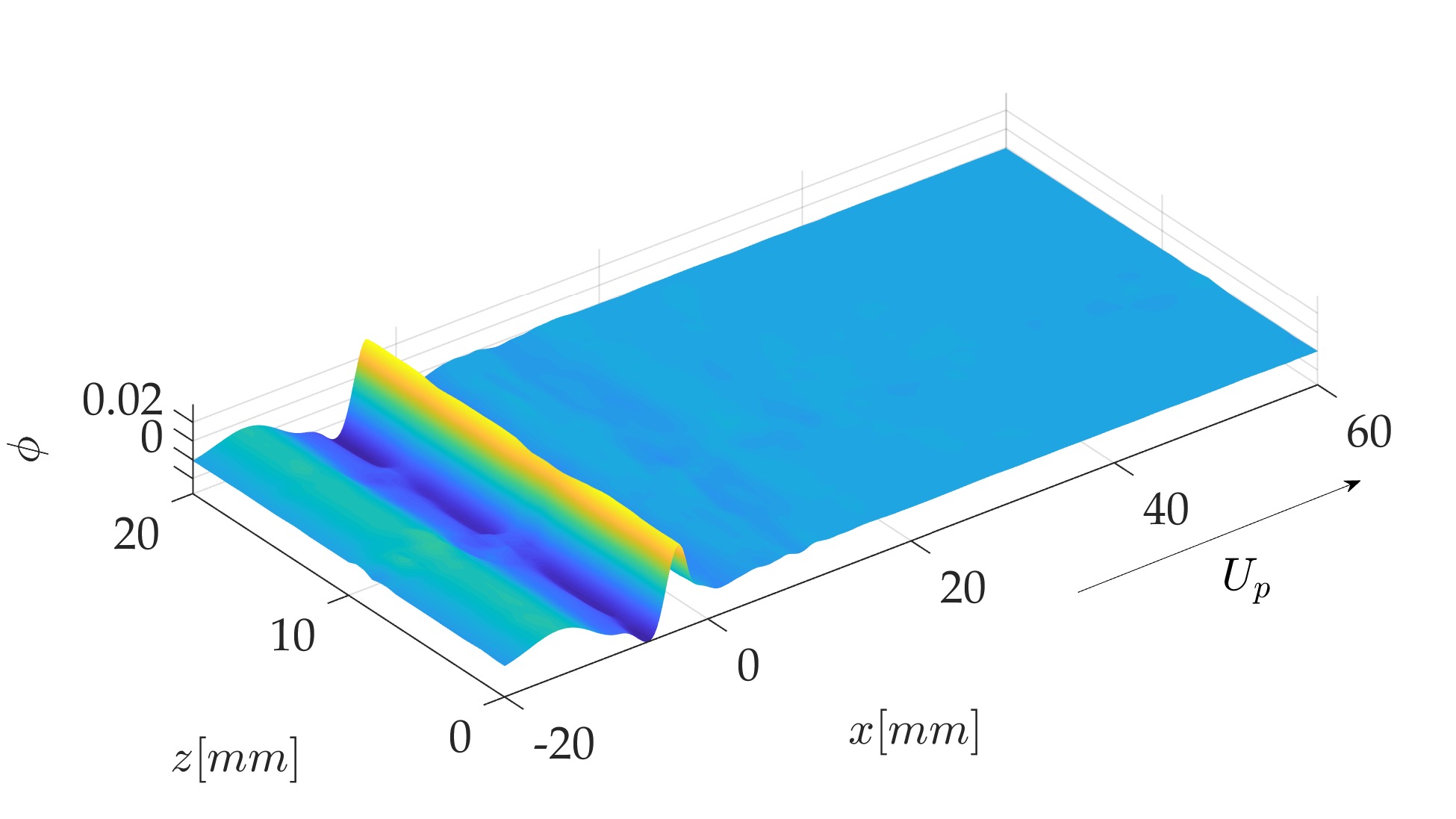}}
      $\phi_{21}(x,z)$\\
      {\includegraphics[width=\linewidth]{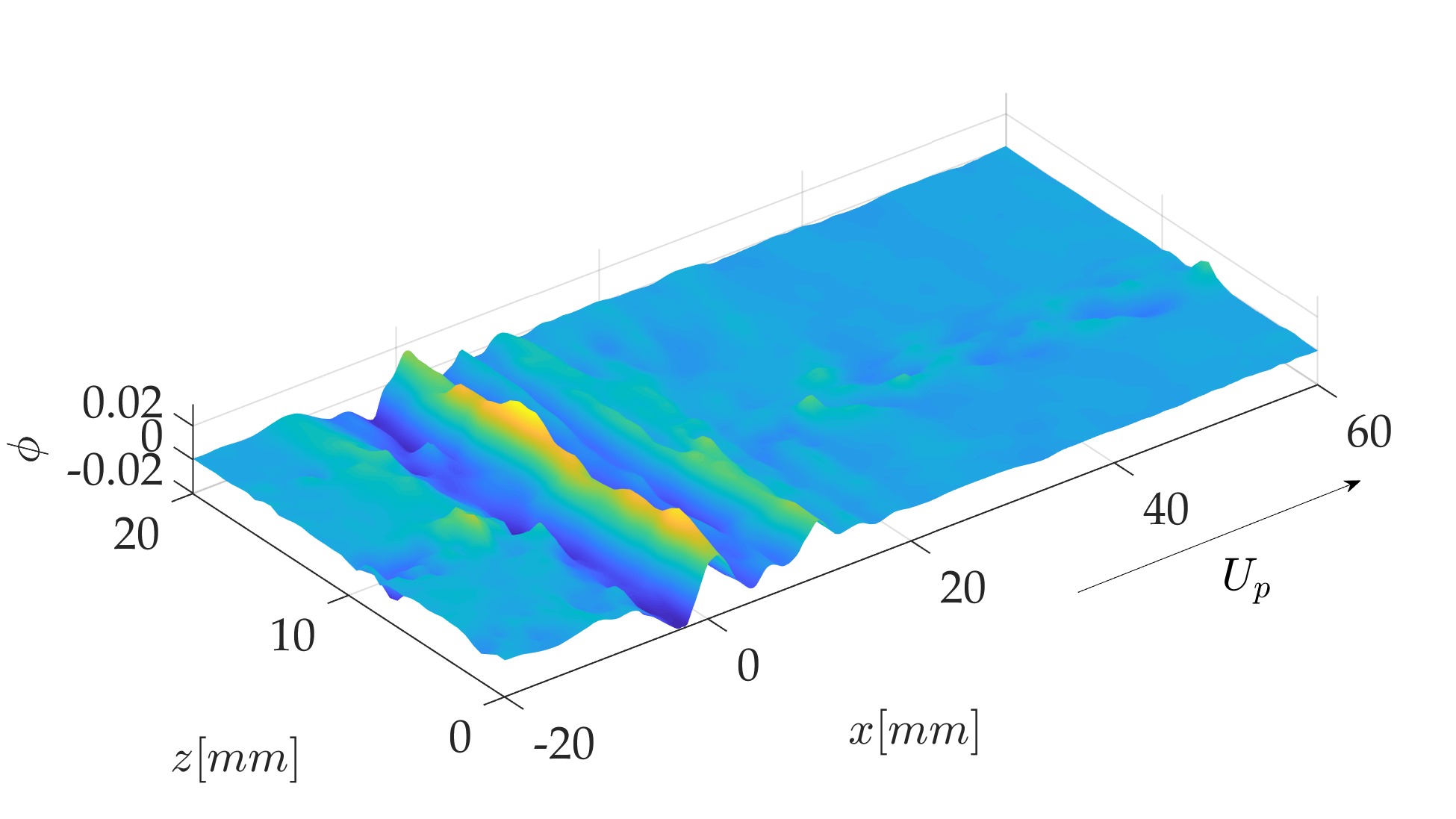}}%
  \end{minipage}%
    \hfill
    \begin{minipage}{.33\linewidth}
      $\phi_{1}(x,z)$ \\
      {\includegraphics[width=\linewidth]{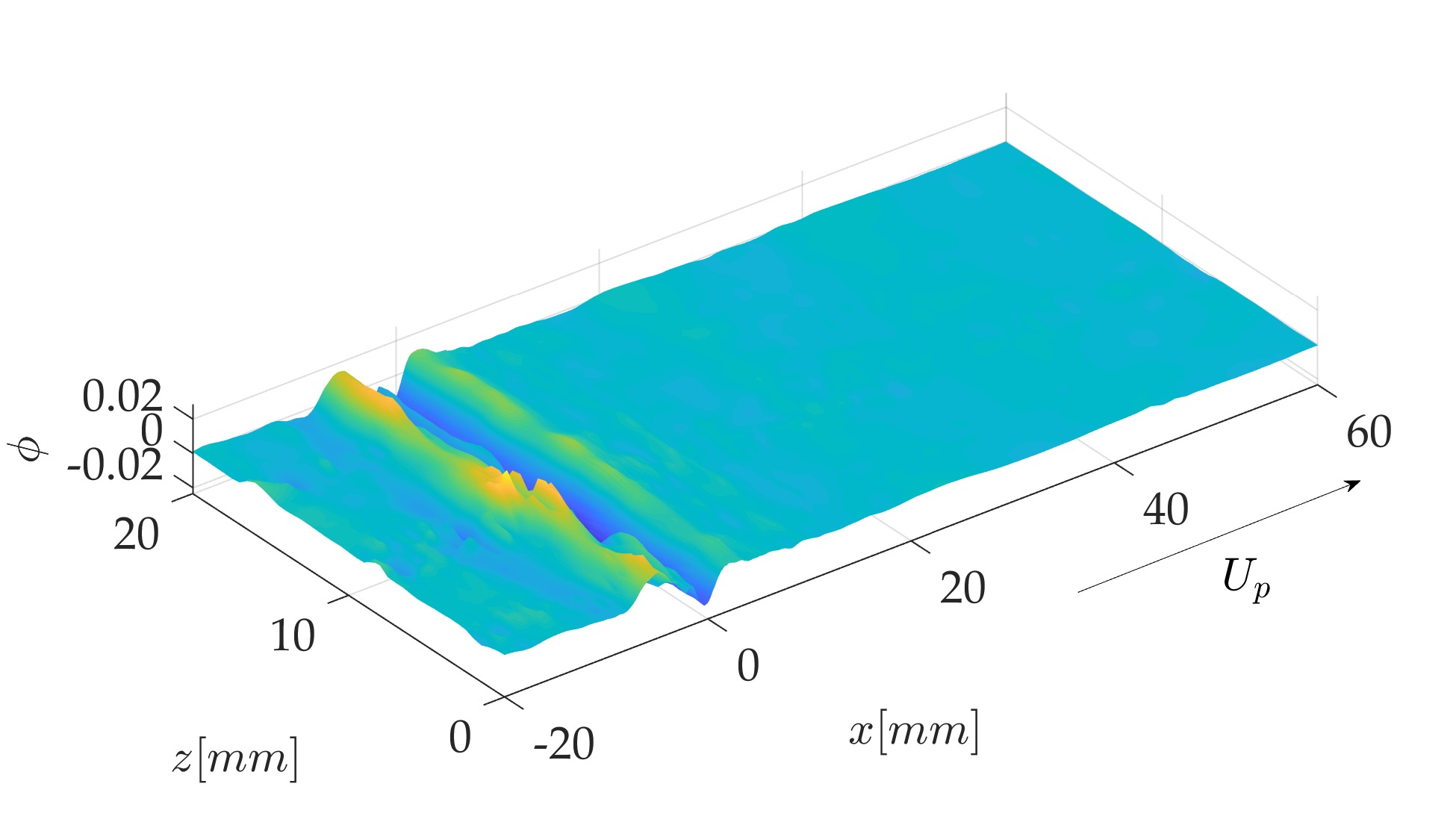}}
      $\phi_{3}(x,z)$ \\
      {\includegraphics[width=\linewidth]{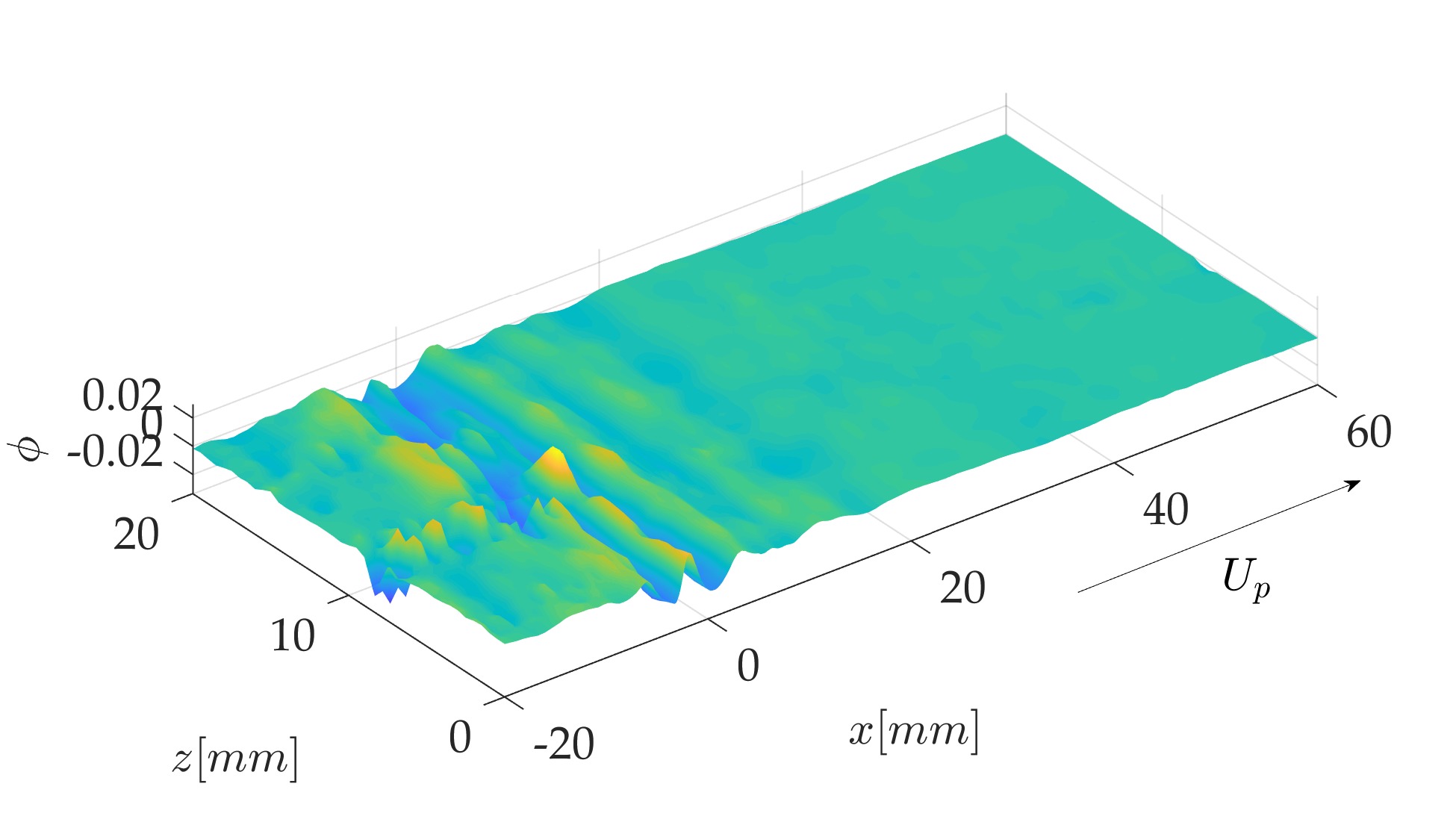}}
      $\phi_{5}(x,z)$ \\
      {\includegraphics[width=\linewidth]{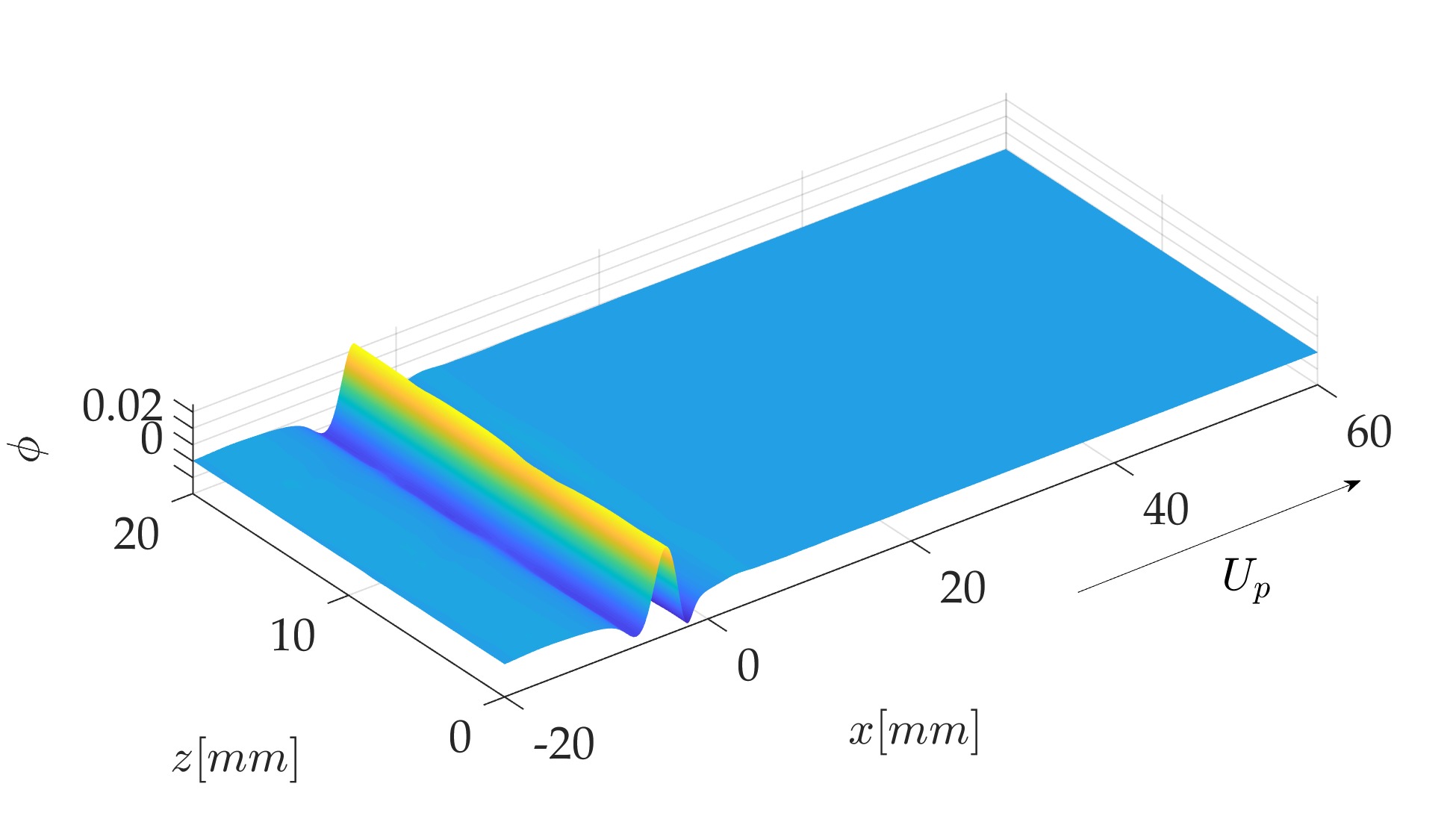}}
      $\phi_{21}(x,z)$\\
      {\includegraphics[width=\linewidth]{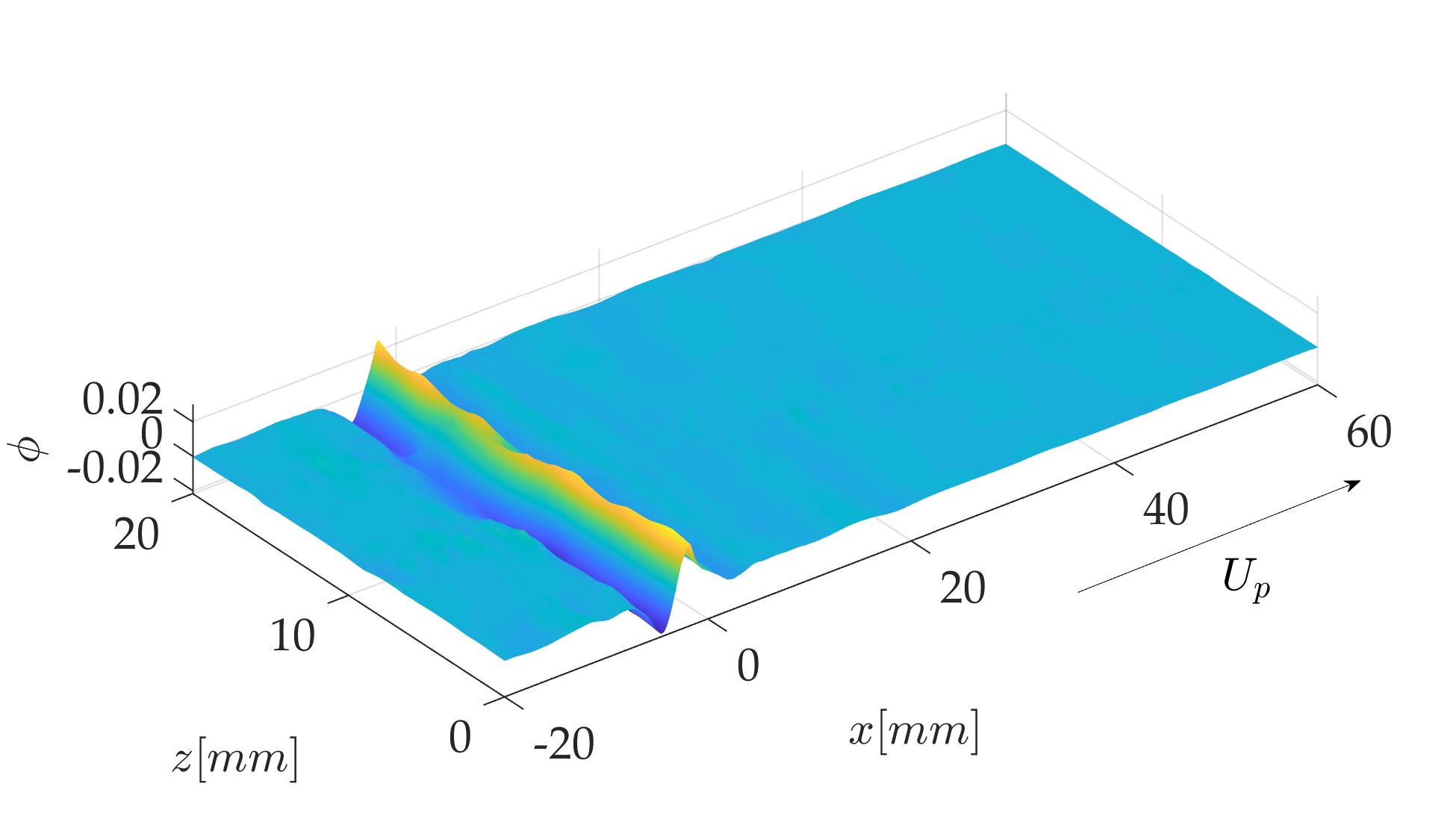}}%
  \end{minipage}%
    \hfill
    \begin{minipage}{.33\linewidth}
      $\phi_{1}(x,z)$ \\
      {\includegraphics[width=\linewidth]{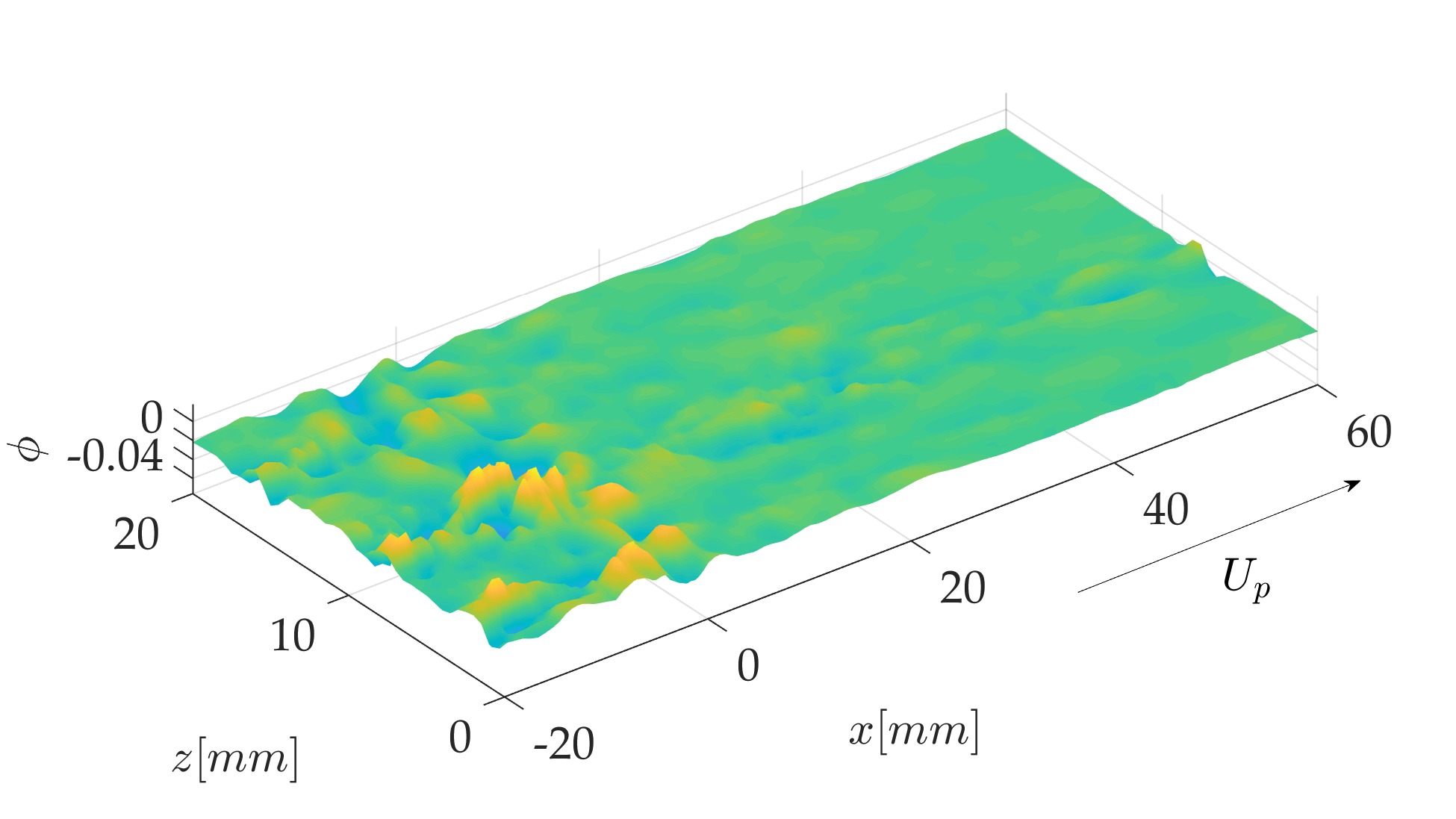}}
      $\phi_{3}(x,z)$ \\
      {\includegraphics[width=\linewidth]{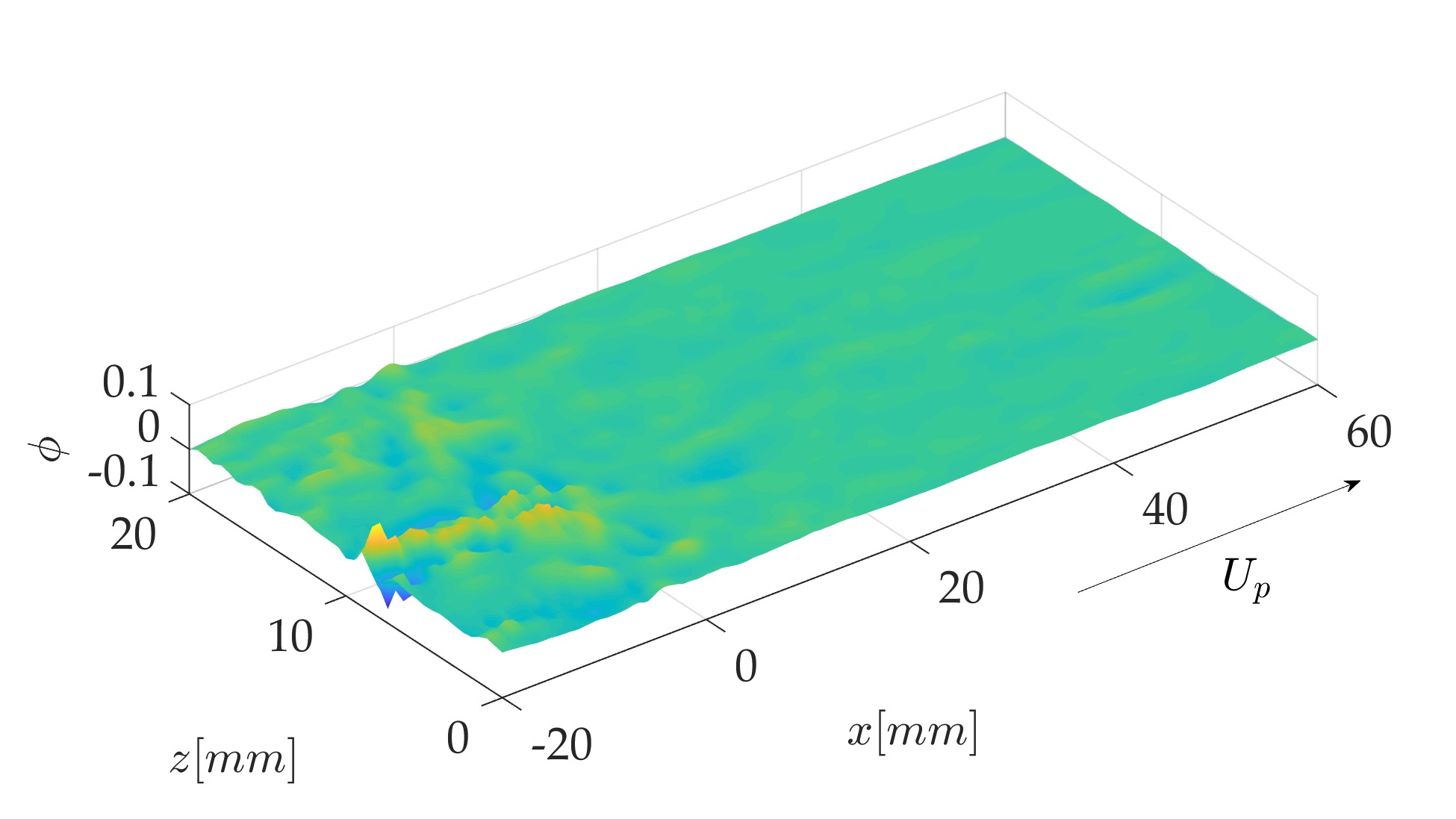}}
      $\phi_{5}(x,z)$\\
      {\includegraphics[width=\linewidth]{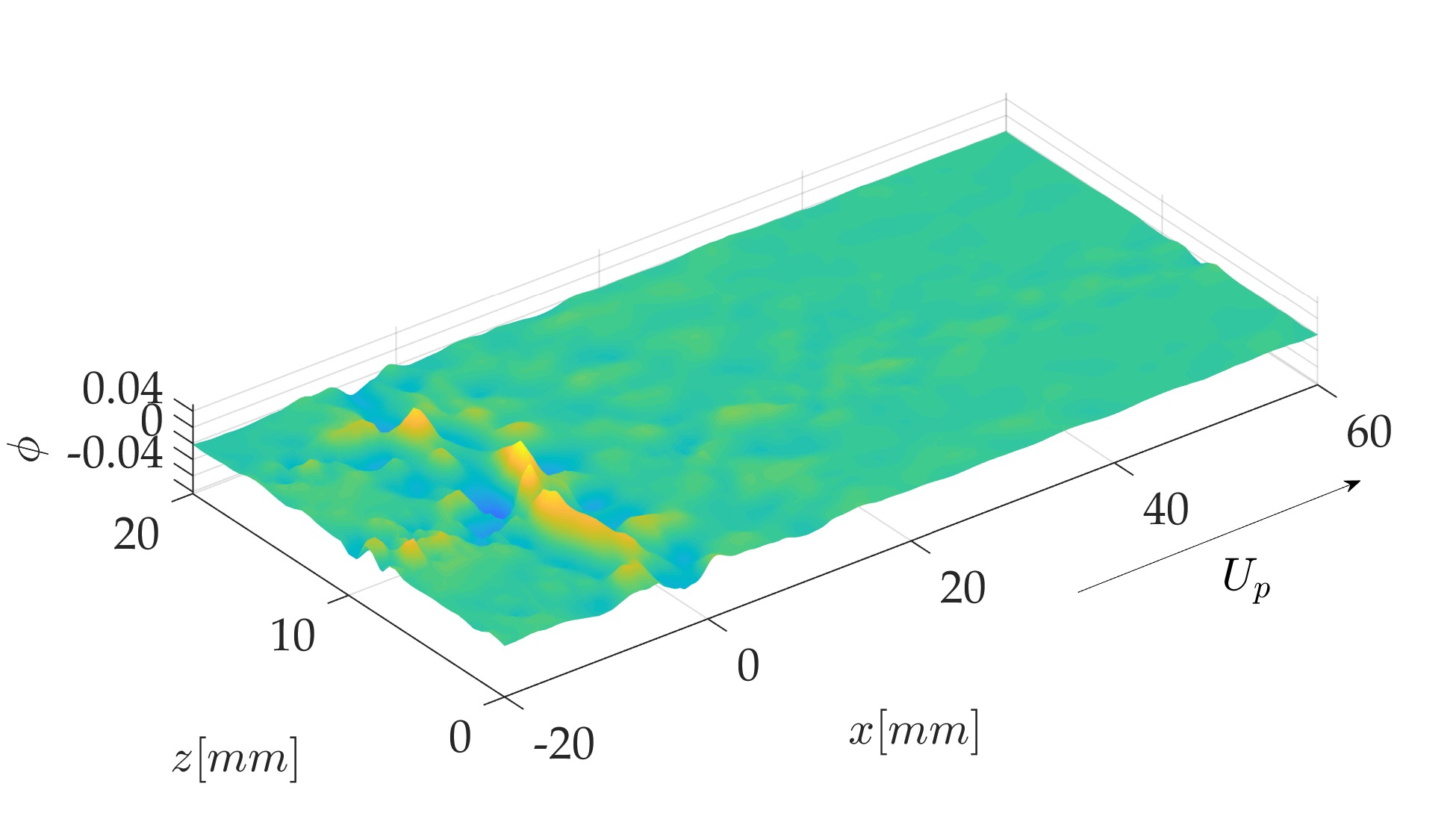}}
      $\phi_{21}(x,z)$ \\
      {\includegraphics[width=\linewidth]{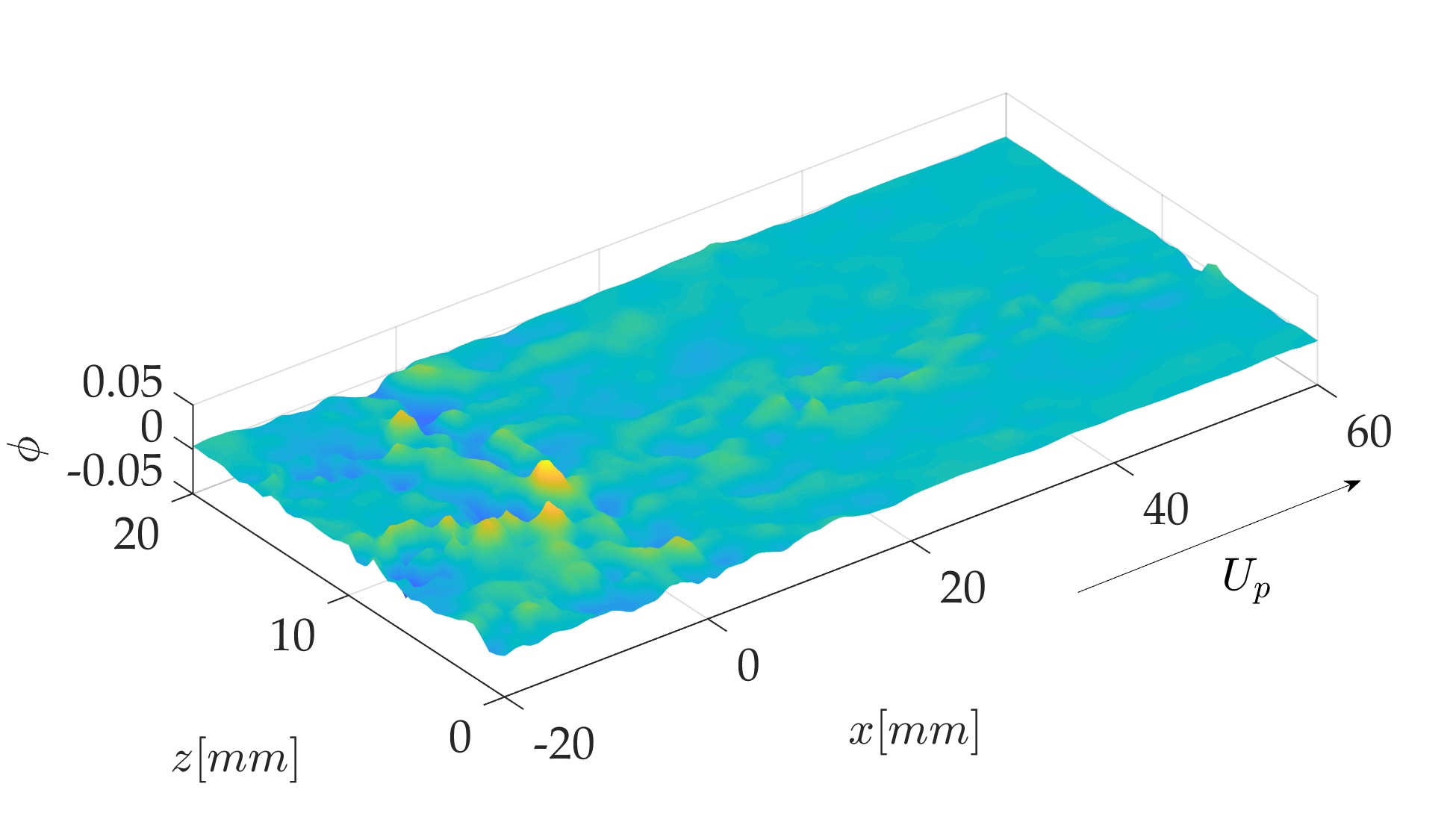}}%
  \end{minipage}%  
  \caption{Spatial structure of several wall quantities resulting from the projection into the temporal evolution of the modes in Case 2: pressure at the wall $P'_w$ (left), stream-wise $\tau_{wx}$ (center) and transverse $\tau_{wz}$ (right) wall shear stress.}
\label{fig:emPOD_gas_Case2}
\end{figure*}

An extended mPOD analysis was performed to correlate, using Eq.~\ref{Gas_STR}, the 2D coupled undulation with specific flow structures in the gas jet. 

Firstly, the correlation between the three dimensionality in the liquid film and the gas jet is examined. For that purpose, case 2 is considered because the energy decay of the modes is more gentle than in the other cases, and in this sense it is the most challenging: more modes are needed to represent the dataset for a given level of accuracy. The analysis is first carried out for three wall quantities linked to the wiping action of the gas jet: the wall pressure and the two wall shear stress components. It is worth noticing that in the boundary layer approximation framework (see Sec. \ref{SecIV} and \ref{SecVID}), the pressure field is independent of the cross-stream axis $\hat{y}$. Moreover, the shear stress at the wall is related to the shear stress at the interface; the link between the two depends on the local thickness and the shape of the velocity profile (see Sec. \ref{SecVID}).

Fig.~\ref{fig:emPOD_gas_Case2} shows the detected structures for the corresponding fields: pressure on the left, stream-wise shear stress in the center and transverse shear stress on the right. These plots should be analyzed together with Fig.~\ref{fig:mPOD_liquidfilm_Case2}. While the basic mechanism in coupled undulation is strongly bi-dimensional in all the cases (shown in Sec. \ref{SecVIB}), some of the detected modes in the wall quantities display some three-dimensionality, especially in the transverse shear stress, within the region $z=9\pm3$ mm (see modes 1, 3 and 21) along the substrate width. These are linked to the entrainment of bubbles in the liquid film and are of no interest to the present analysis.

On the other hand, it is remarkable that the modes linked to the \emph{2D coupled undulation} (5 and 6, cf. Fig. \ref{fig:mPOD_liquidfilm_Case2}) are perfectly bi-dimensional in both pressure and stream-wise shear stress at the wall. The downstream evolution of the waves, which eventually results in three-dimensional patterns for this condition (see Fig. \ref{fig:mPOD_liquidfilm_Case2}), does not have any relevant correlation with spatial structures in the gas. This result highlights a comparatively little correlation between the three-dimensionality in the liquid film wave patterns and the three-dimensionality in the gas flow field. The same observation holds for cases 1 and 3, which are thus not further discussed. This analysis corroborates the hypothesis that the mechanism responsible for the undulation in the final coating is, in essence, bidimensional. Therefore, the following representations of the gas flow fields are limited to a z-normal plane taken in the middle of the domain depth ($z=15mm$).

We now investigate the 2D flow structures involved in the coupling between phases and, most importantly, their link with the wiping actuators. Following the correlation analysis in Sec \ref{SecVIB}, the gas flow fields are projected onto the temporal basis of the liquid film flow (Eq.\ref{Gas_STR}). 

\begin{figure*}
    \begin{minipage}{.24\linewidth}
       %$\tilde{U'}$
       A
      {\includegraphics[width=\linewidth]{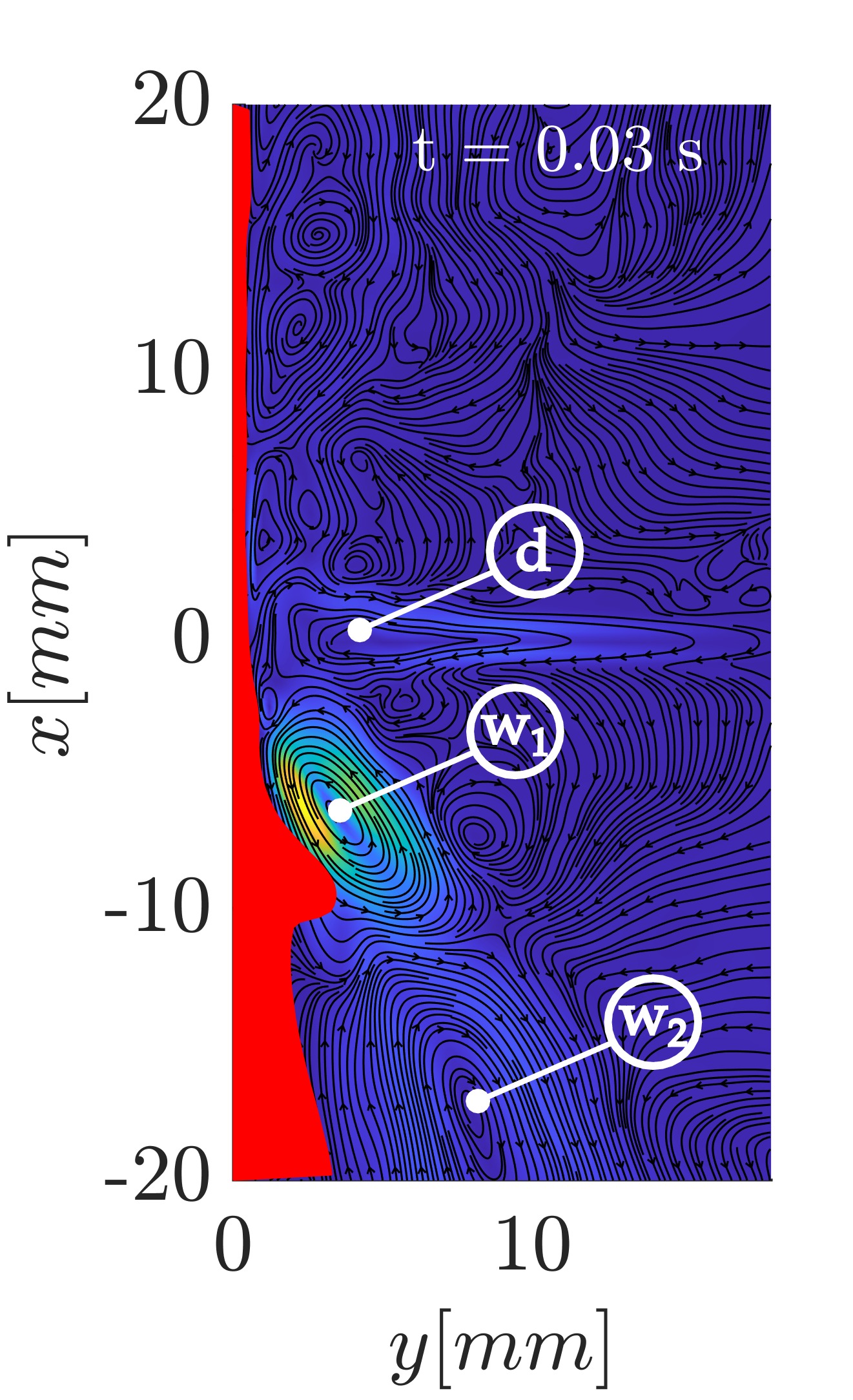}}
      {\includegraphics[width=\linewidth]{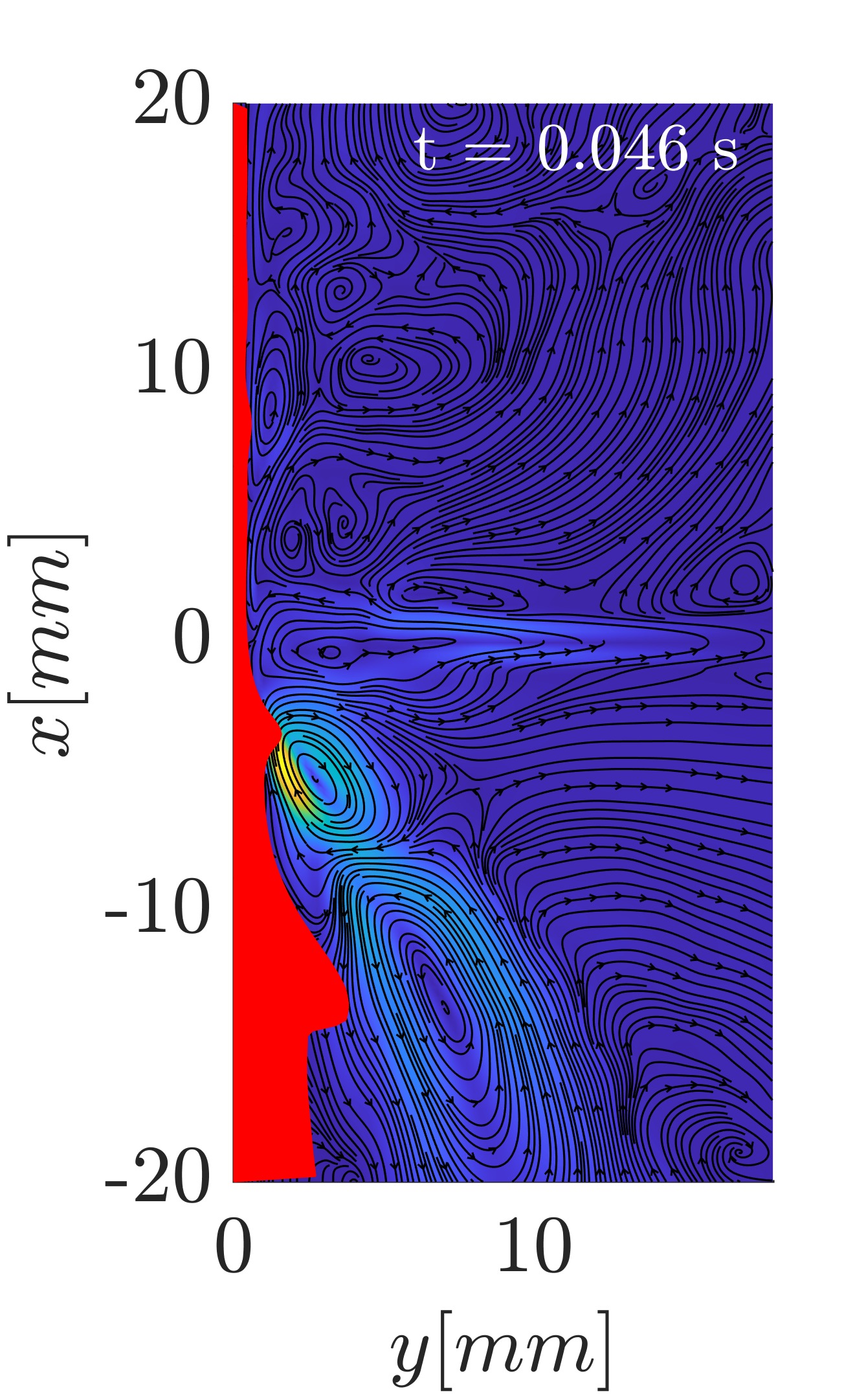}}
      {\includegraphics[width=\linewidth]{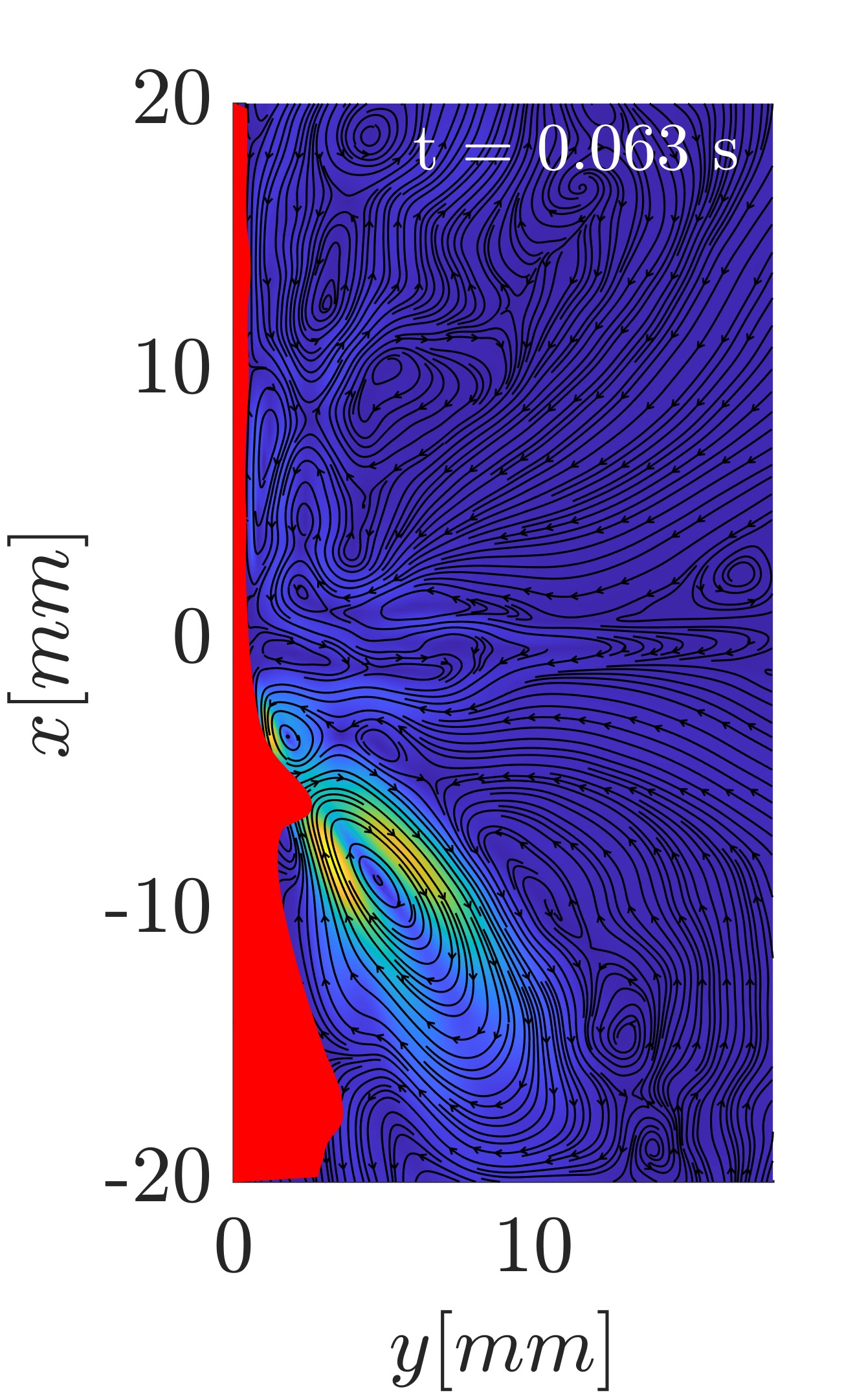}}%
    \end{minipage}%
  \hfill
    \begin{minipage}{.24\linewidth}
      %$U$
      B
      {\includegraphics[width=\linewidth]{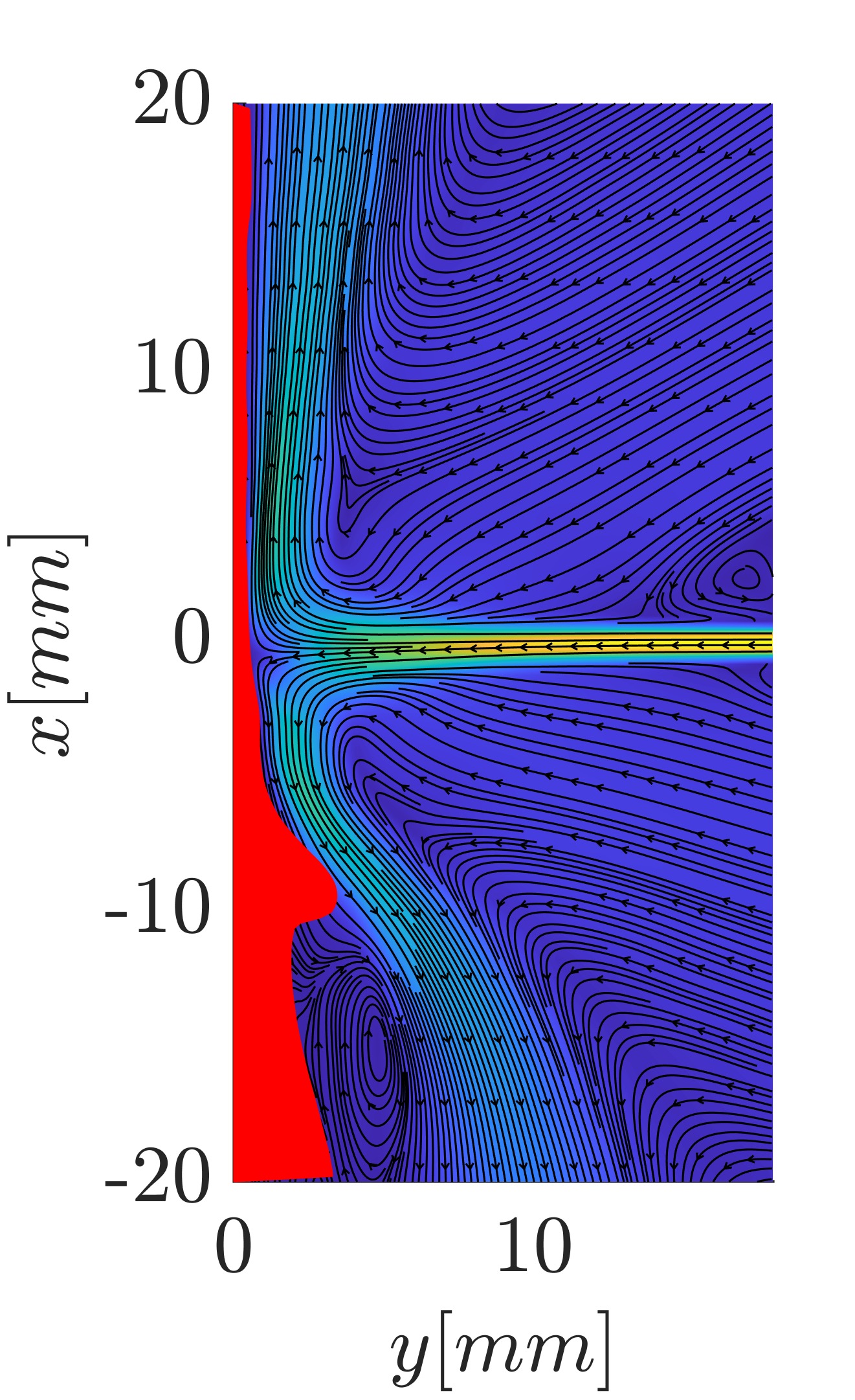}}
      {\includegraphics[width=\linewidth]{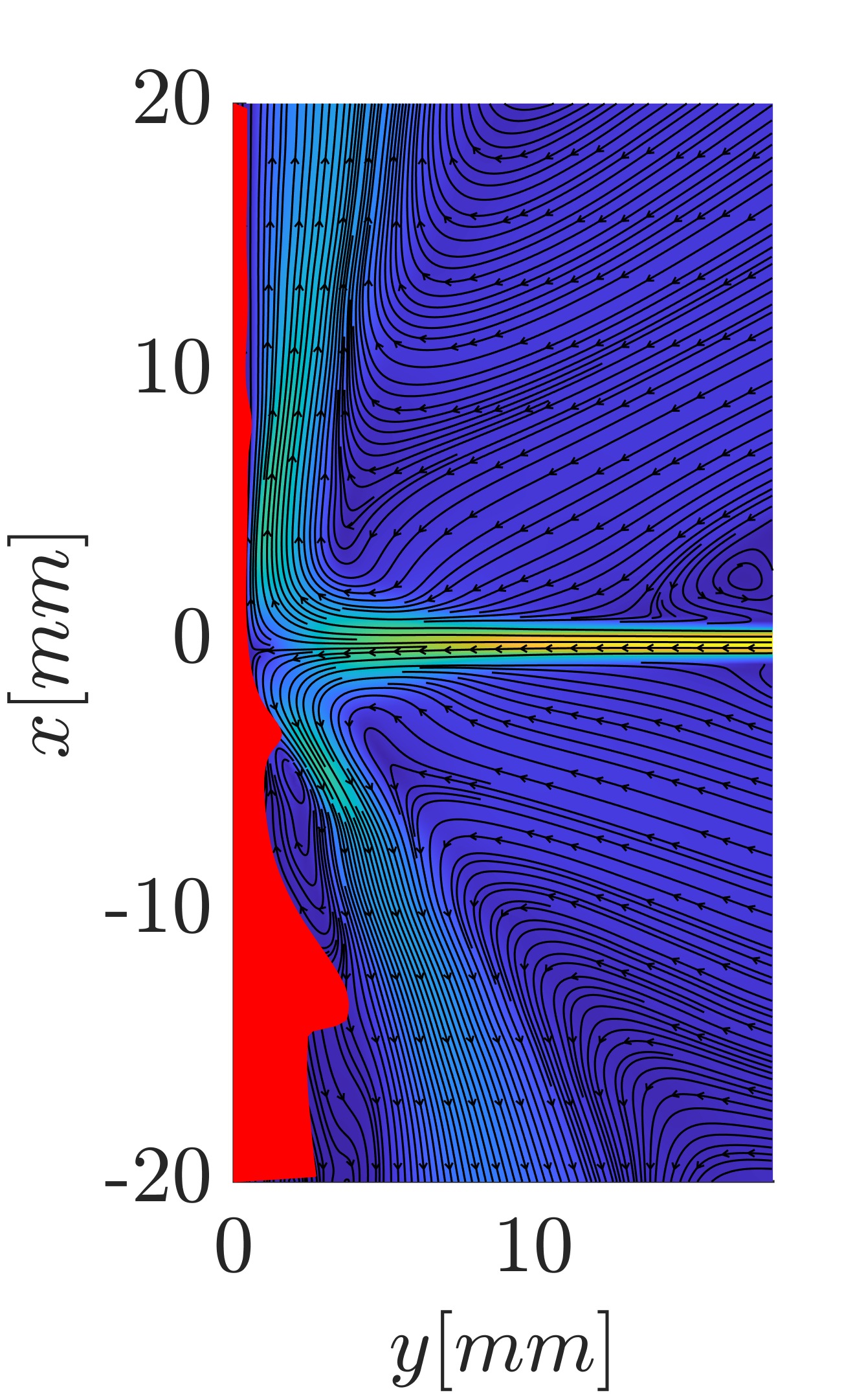}}
      {\includegraphics[width=\linewidth]{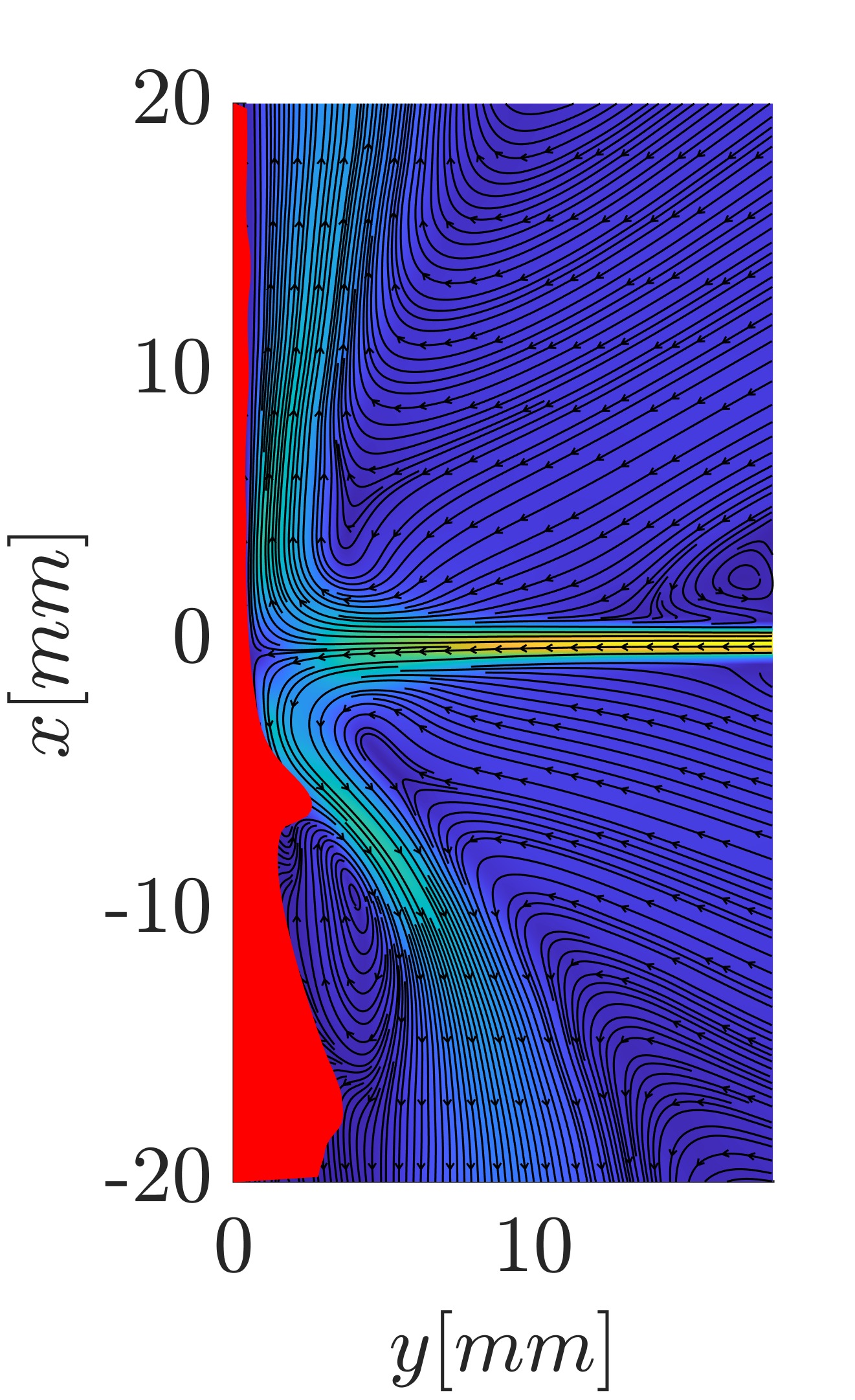}}%
    \end{minipage}%
  \hfill
    \begin{minipage}{.24\linewidth}
      %$U$
      C
      {\includegraphics[width=\linewidth]{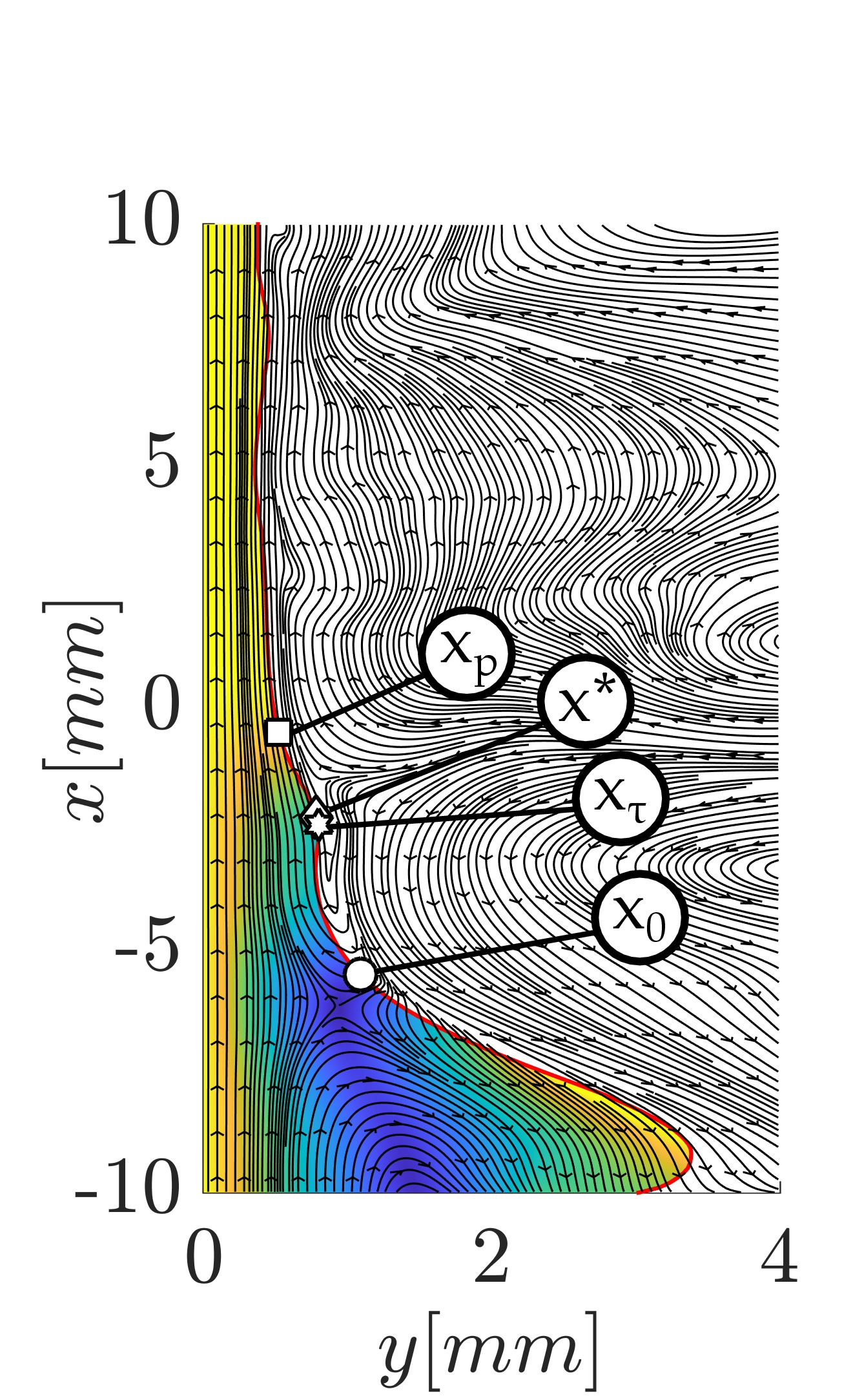}}
      {\includegraphics[width=\linewidth]{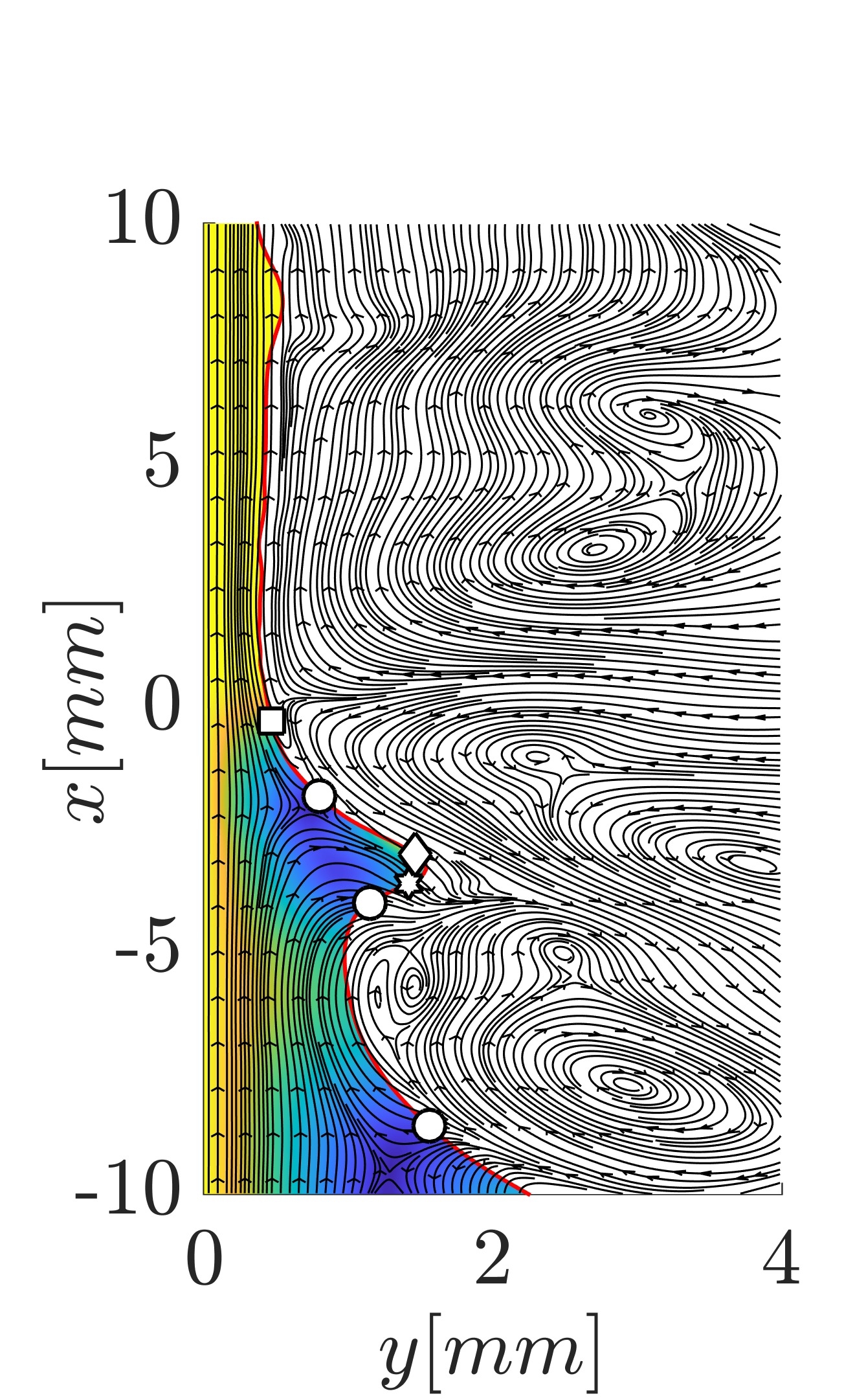}}
      {\includegraphics[width=\linewidth]{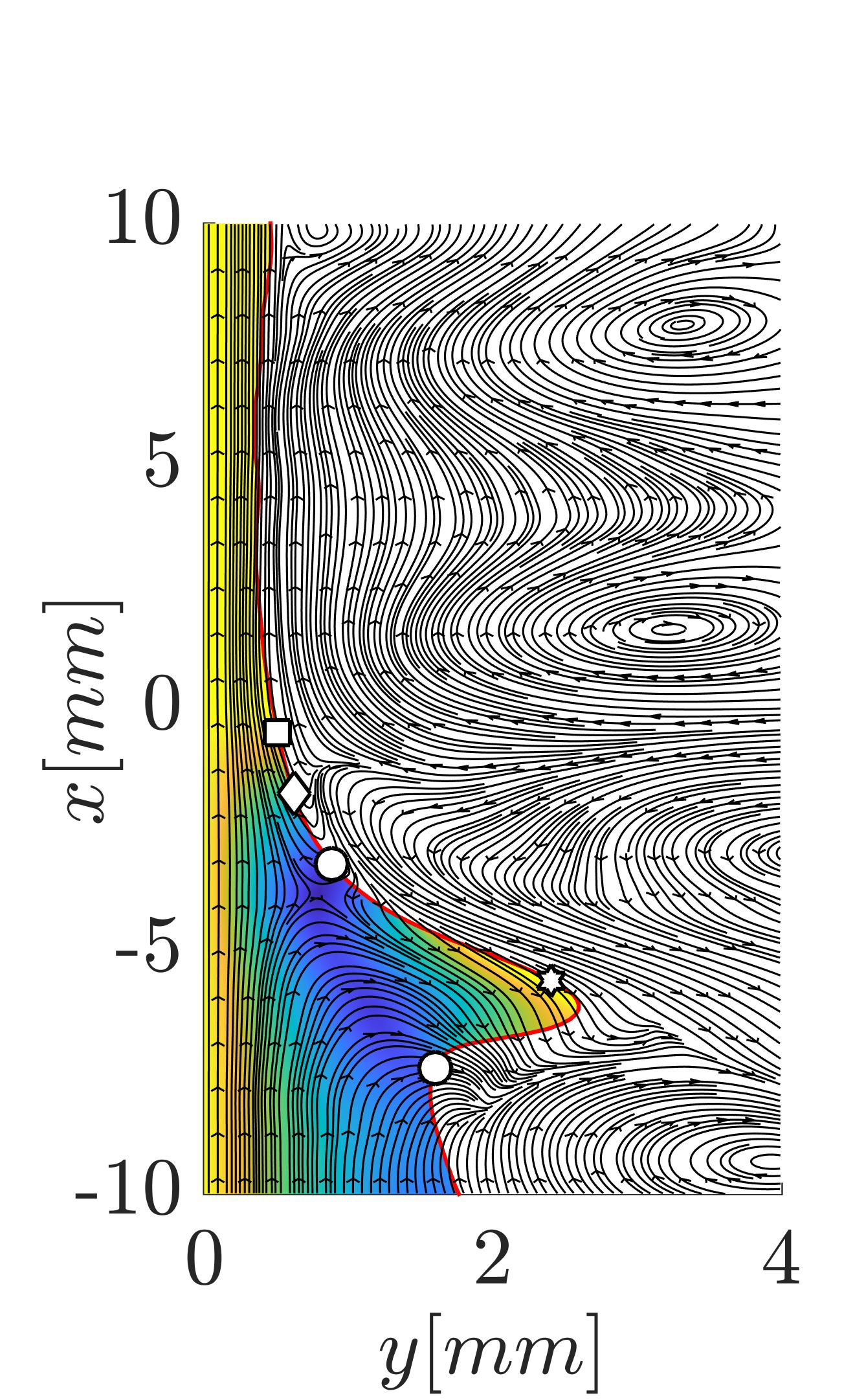}}%
    \end{minipage}%
  \hfill
    \begin{minipage}{.24\linewidth}
      %$\tau_x | \partial_x P$
      D
      {\includegraphics[width=\linewidth]{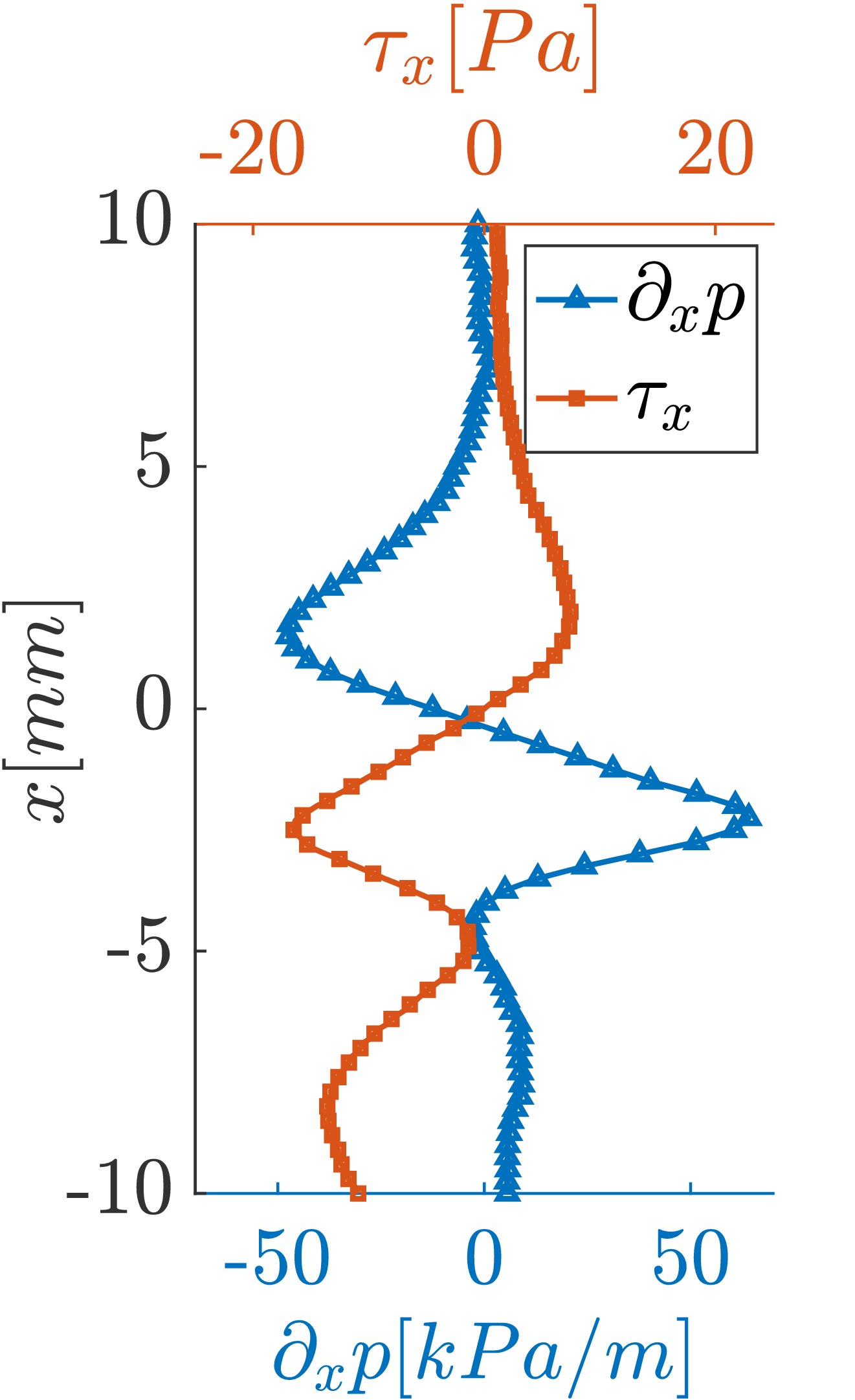}}
      {\includegraphics[width=\linewidth]{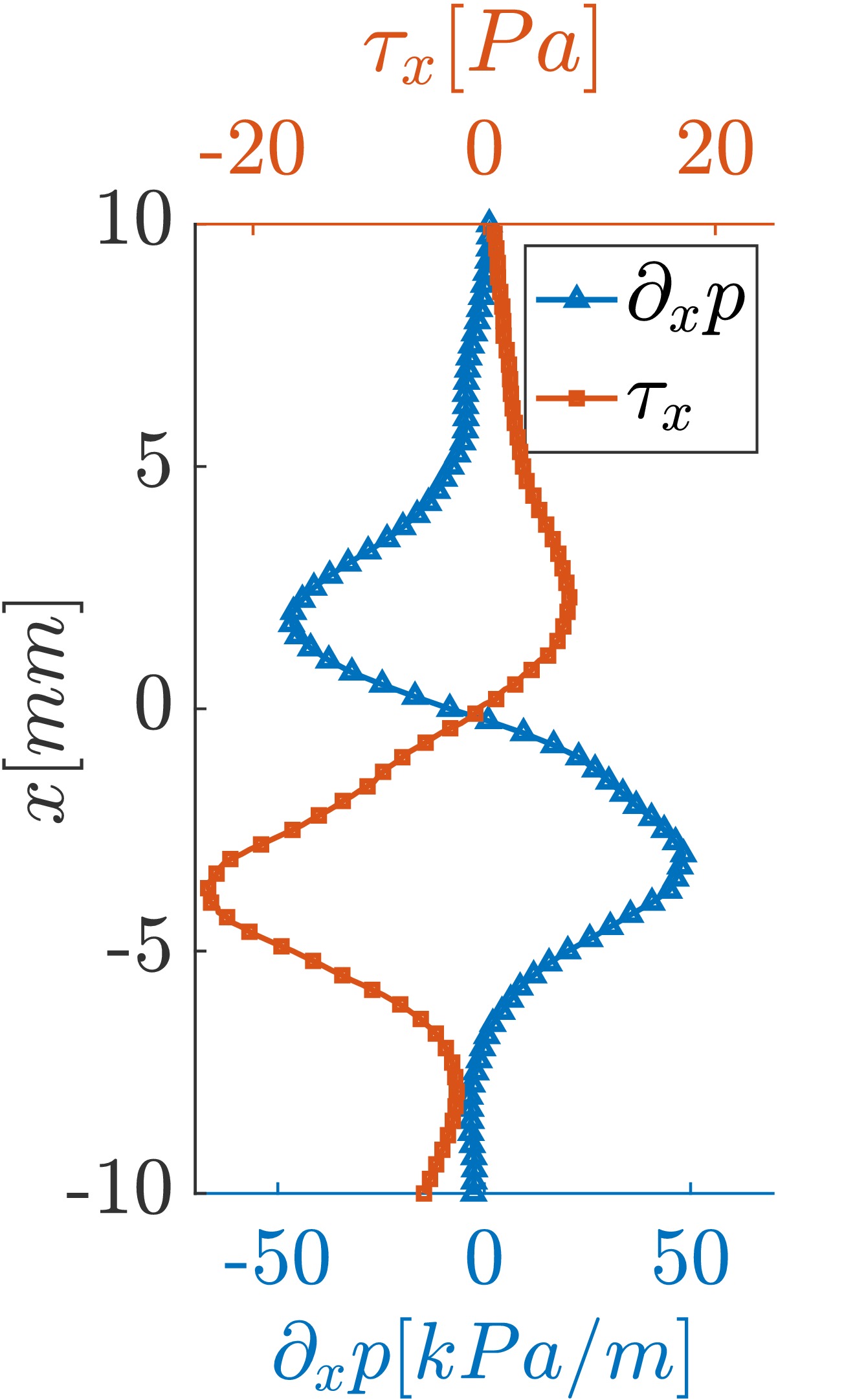}}
      {\includegraphics[width=\linewidth]{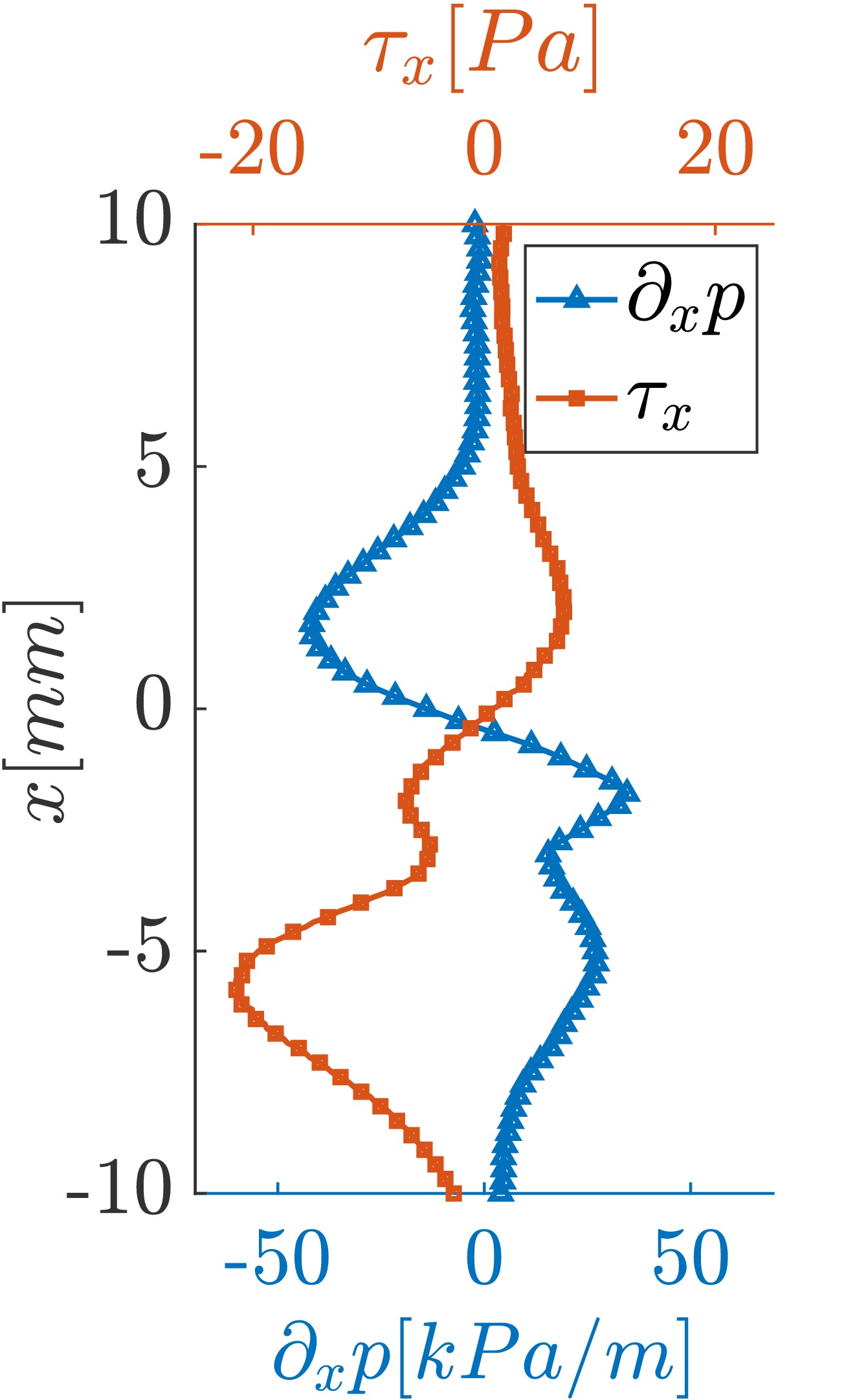}}%
    \end{minipage}%
 \caption{Results from the flow gas flow field analysis and interface dynamics for the test case 1 (Multimedia view).  The first column (A) collects three snapshots of the gas flow field constructed considering only the 2D coupled undulation modes. In the second column (B), the time - averaged velocity field is added. The third column (C) shows a close up of the impingement region, with a contour of the velocity magnitude within the liquid film. The fourth column (D) shows the distributions of pressure gradient and streamwise shear stress at the film interface. The points in column C represent the impact point ($x_p$), the wiping point ($x^*$), the shear point ($x_{\tau}$) and the interface stagnation points ($x_0$). These are defined in the text.}
\label{fig:gas_emPOD_Case1}
\end{figure*}

\begin{figure*}
    \centering
    \begin{minipage}{.24\linewidth}
       A
      {\includegraphics[width=\linewidth]{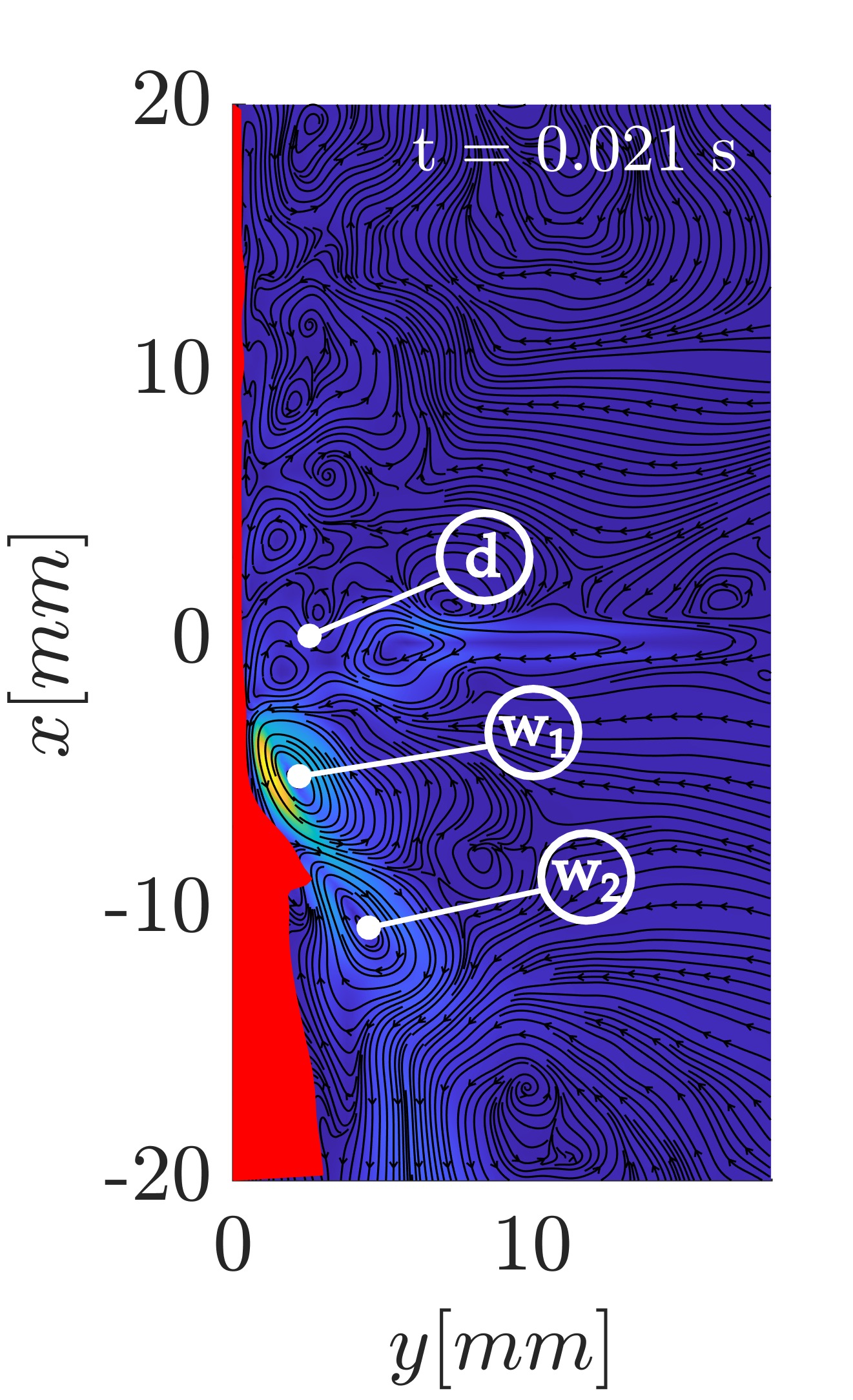}}
      {\includegraphics[width=\linewidth]{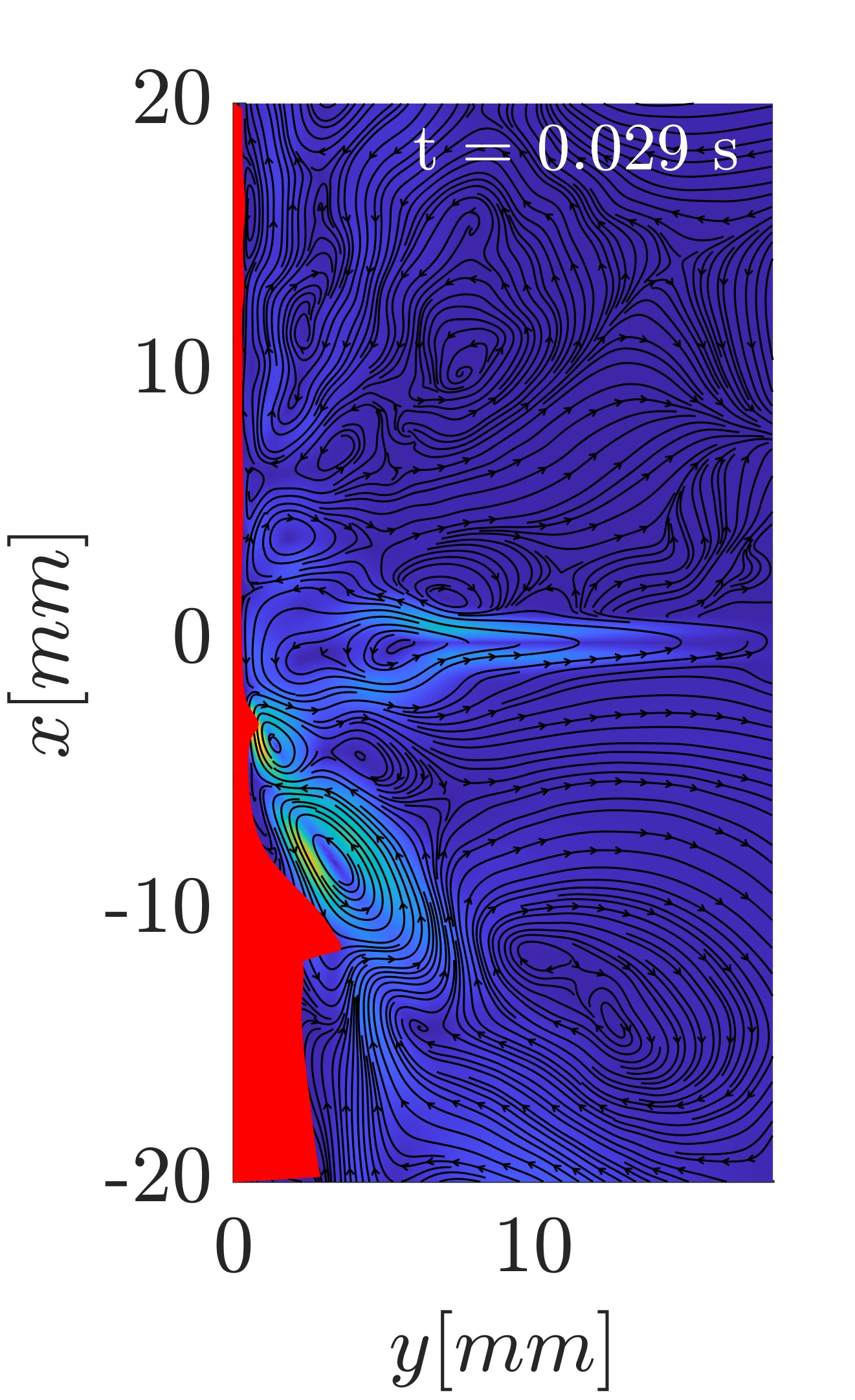}}
      {\includegraphics[width=\linewidth]{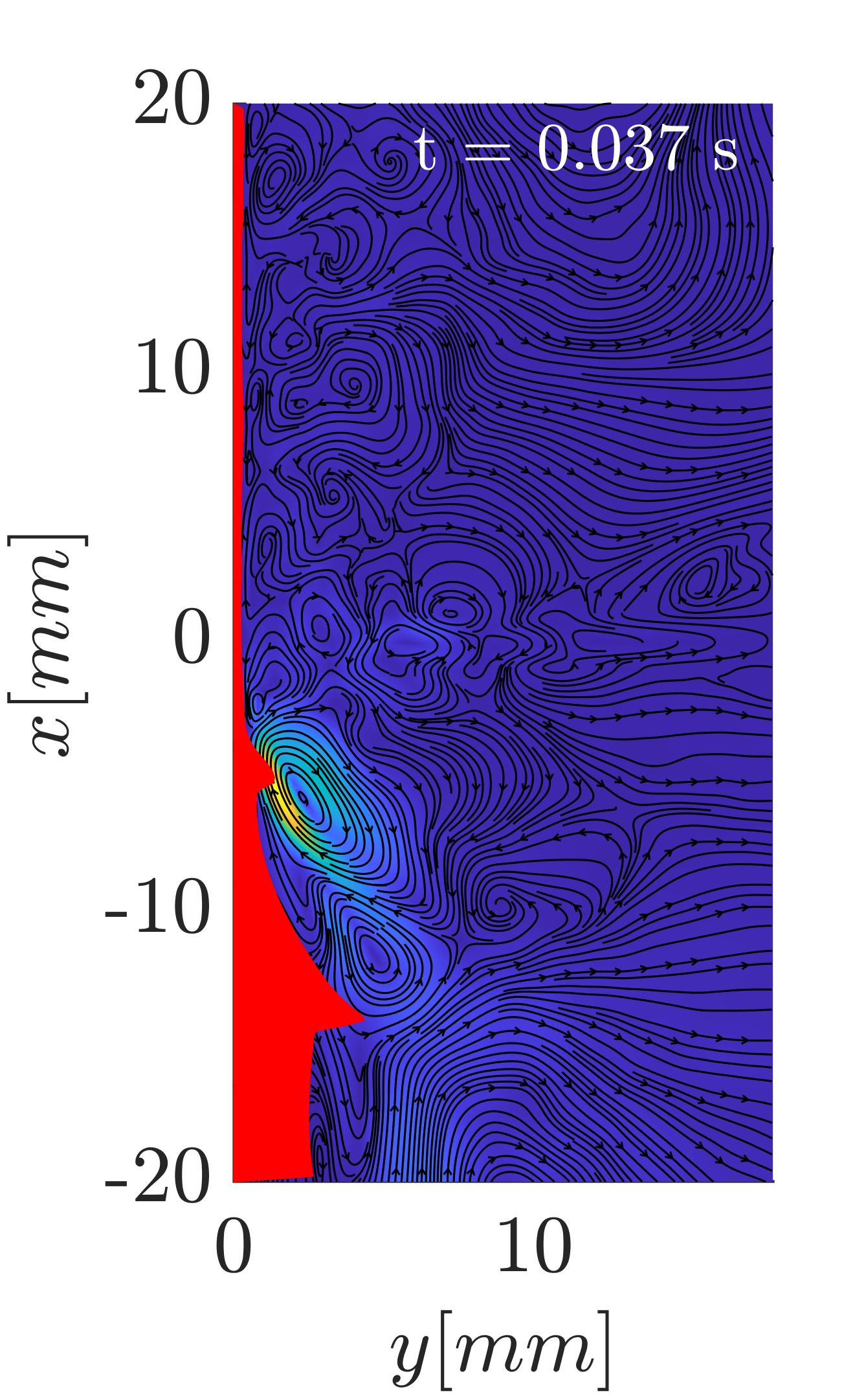}}%
    \end{minipage}%
  \hfill
    \begin{minipage}{.24\linewidth}
      B
      {\includegraphics[width=\linewidth]{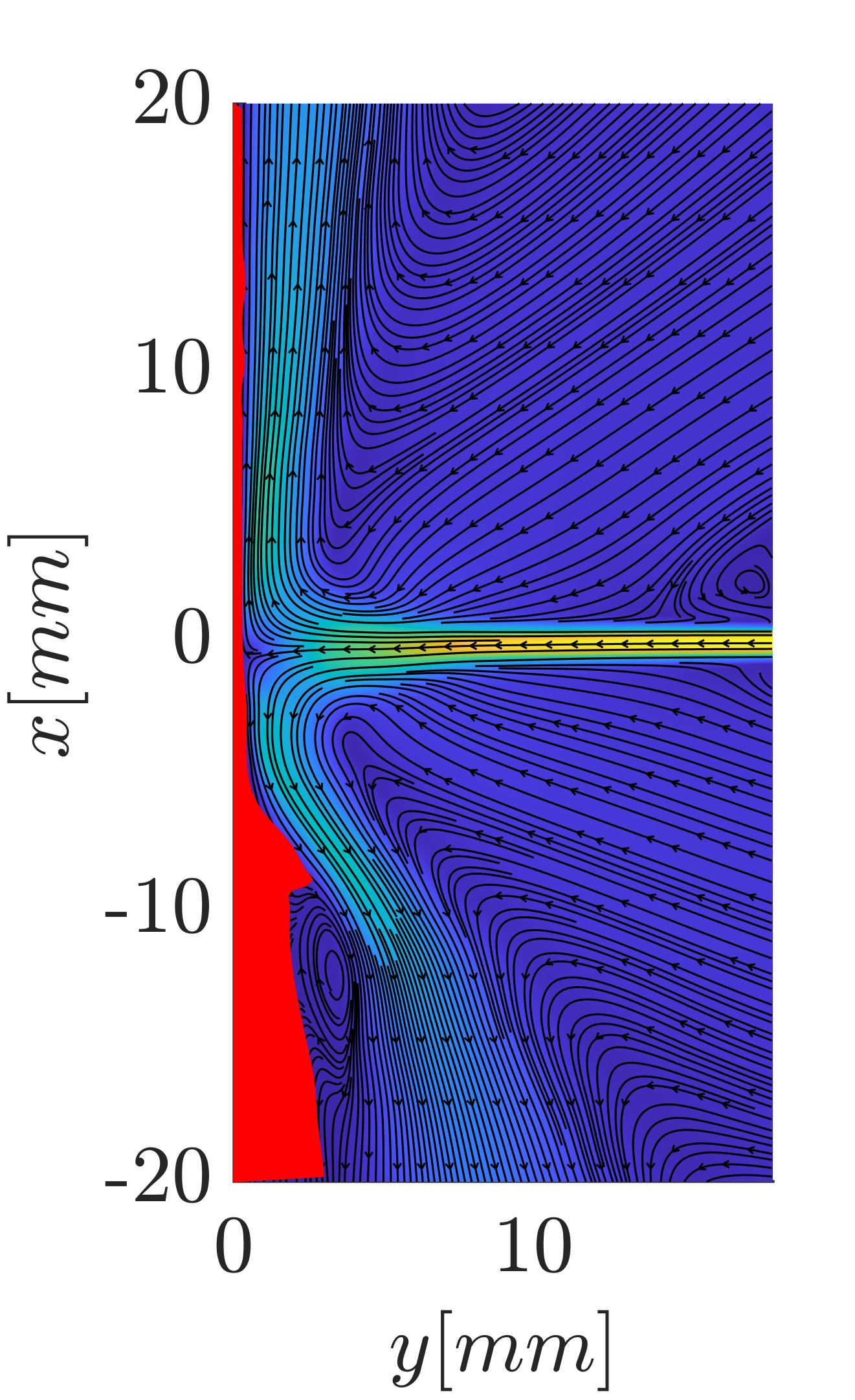}}
      {\includegraphics[width=\linewidth]{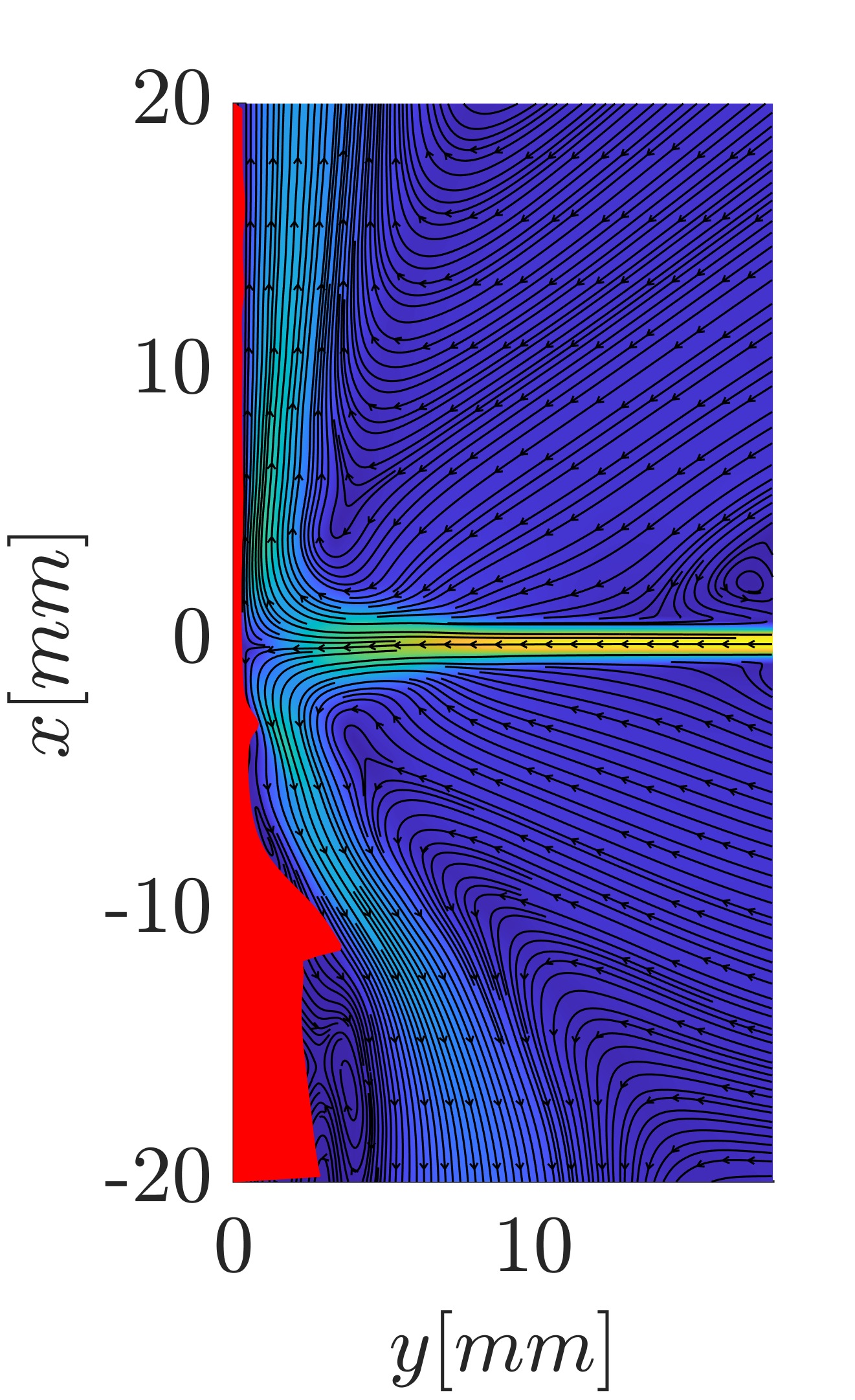}}
      {\includegraphics[width=\linewidth]{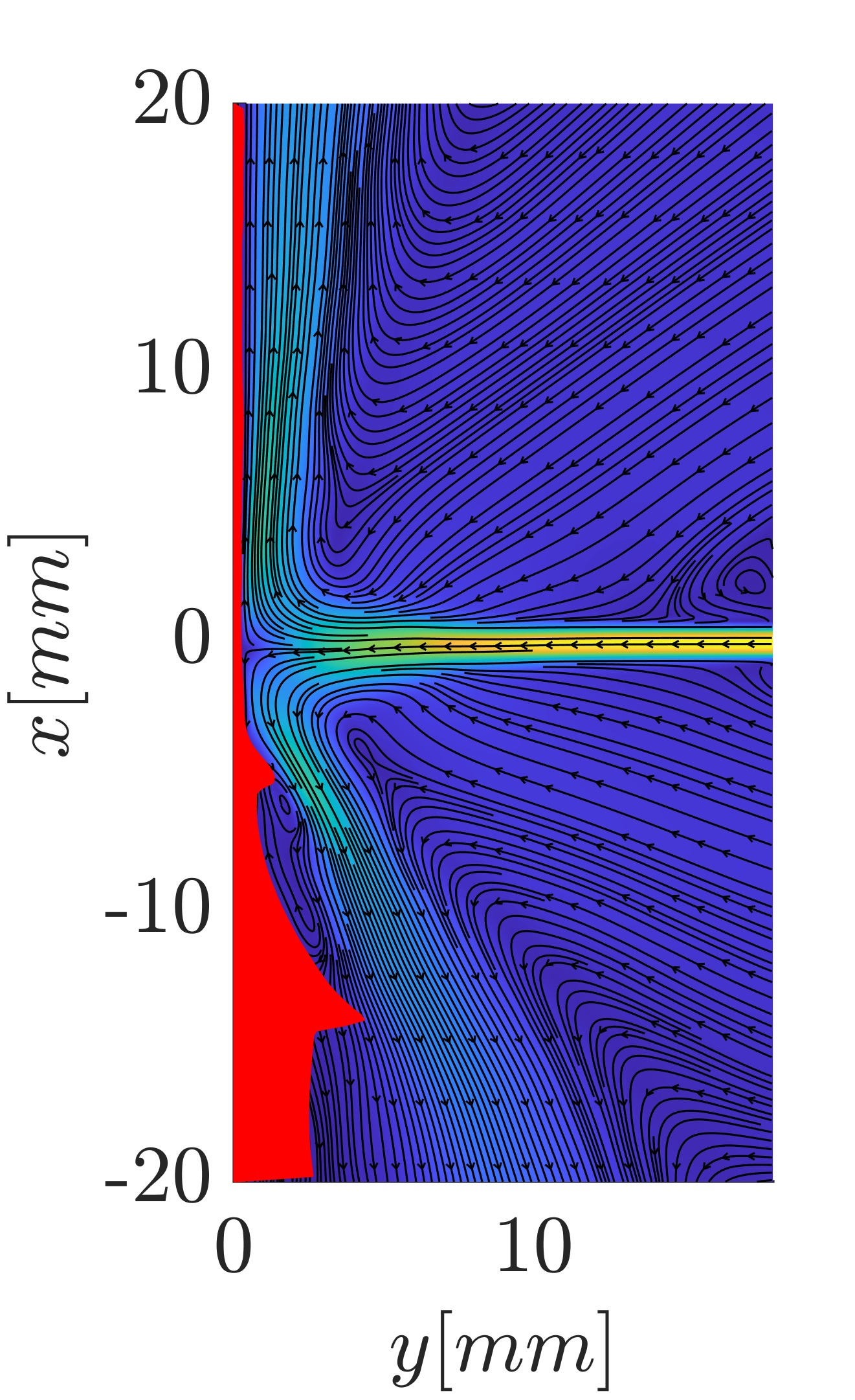}}%
    \end{minipage}%
  \hfill
    \begin{minipage}{.24\linewidth}
      C
      {\includegraphics[width=\linewidth]{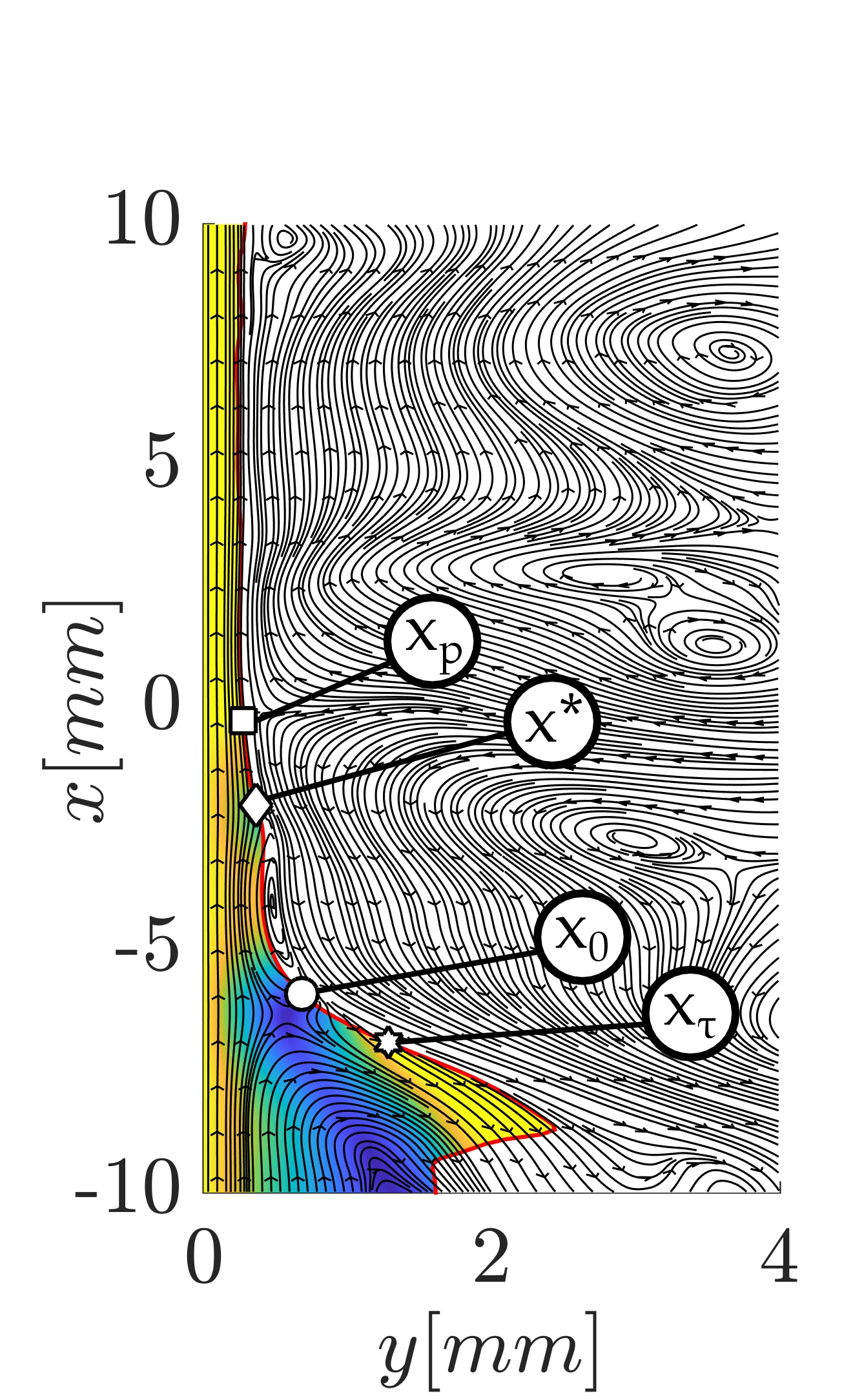}}
      {\includegraphics[width=\linewidth]{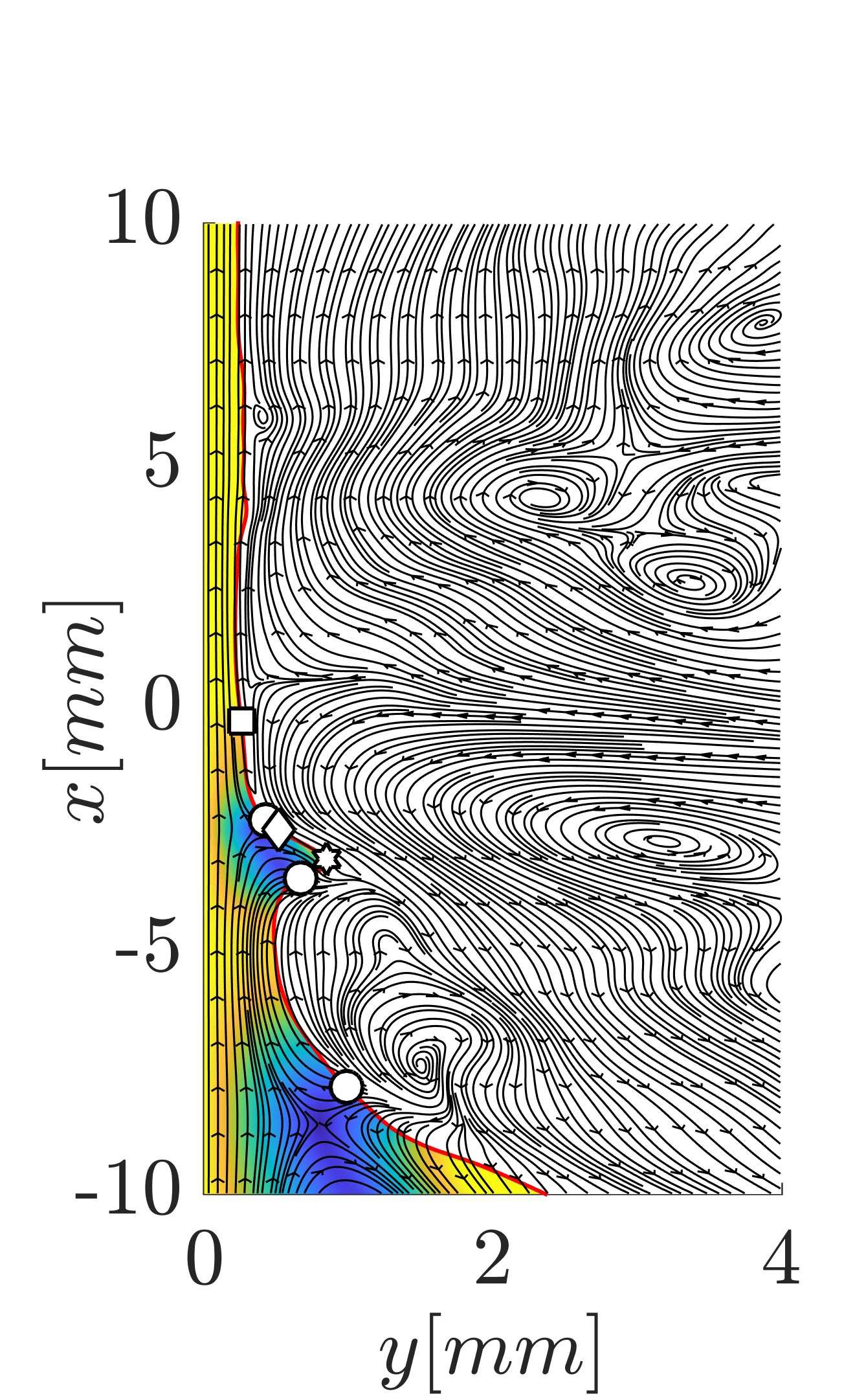}}
      {\includegraphics[width=\linewidth]{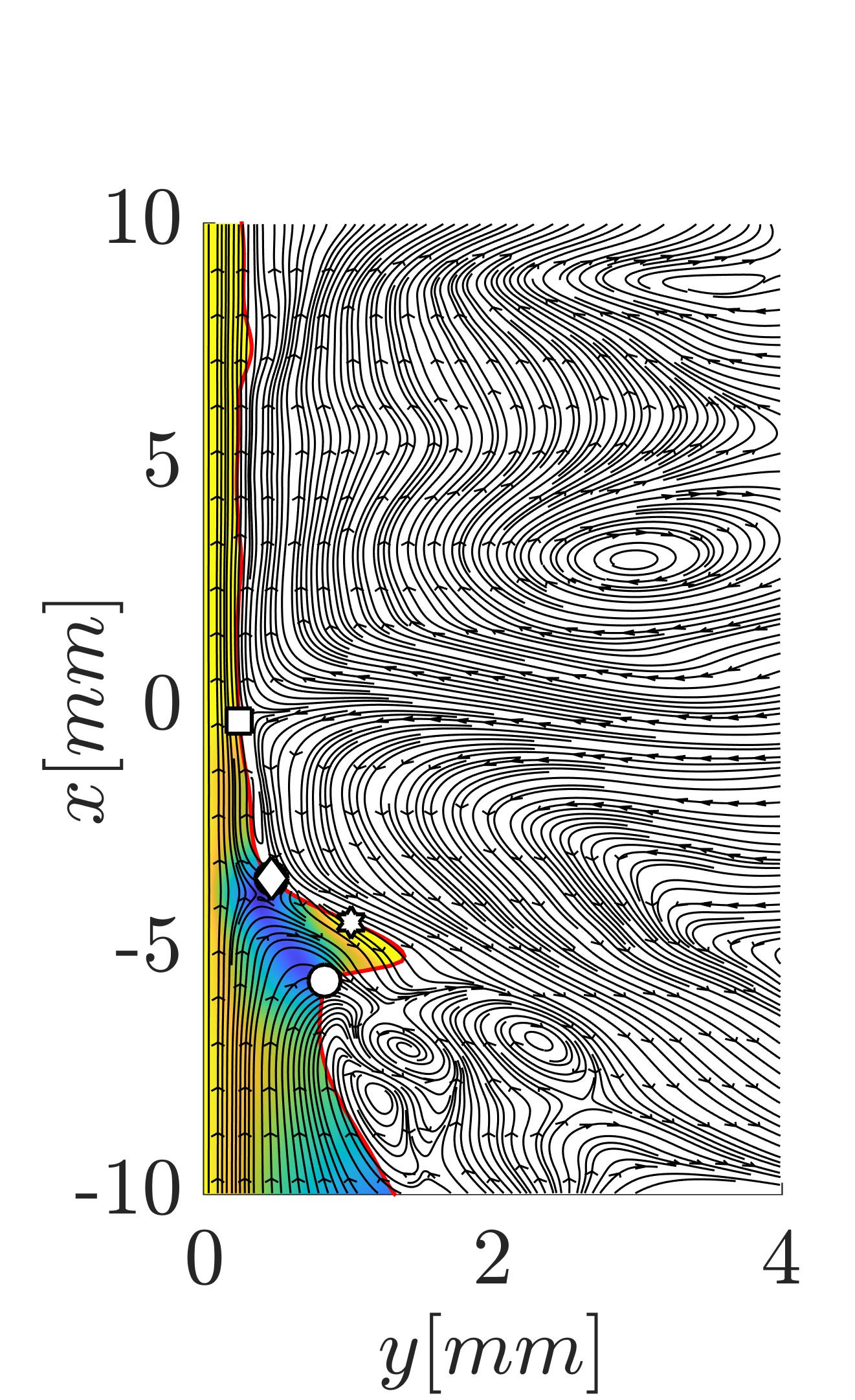}}%
    \end{minipage}%
  \hfill
    \begin{minipage}{.24\linewidth}
      D
      {\includegraphics[width=\linewidth]{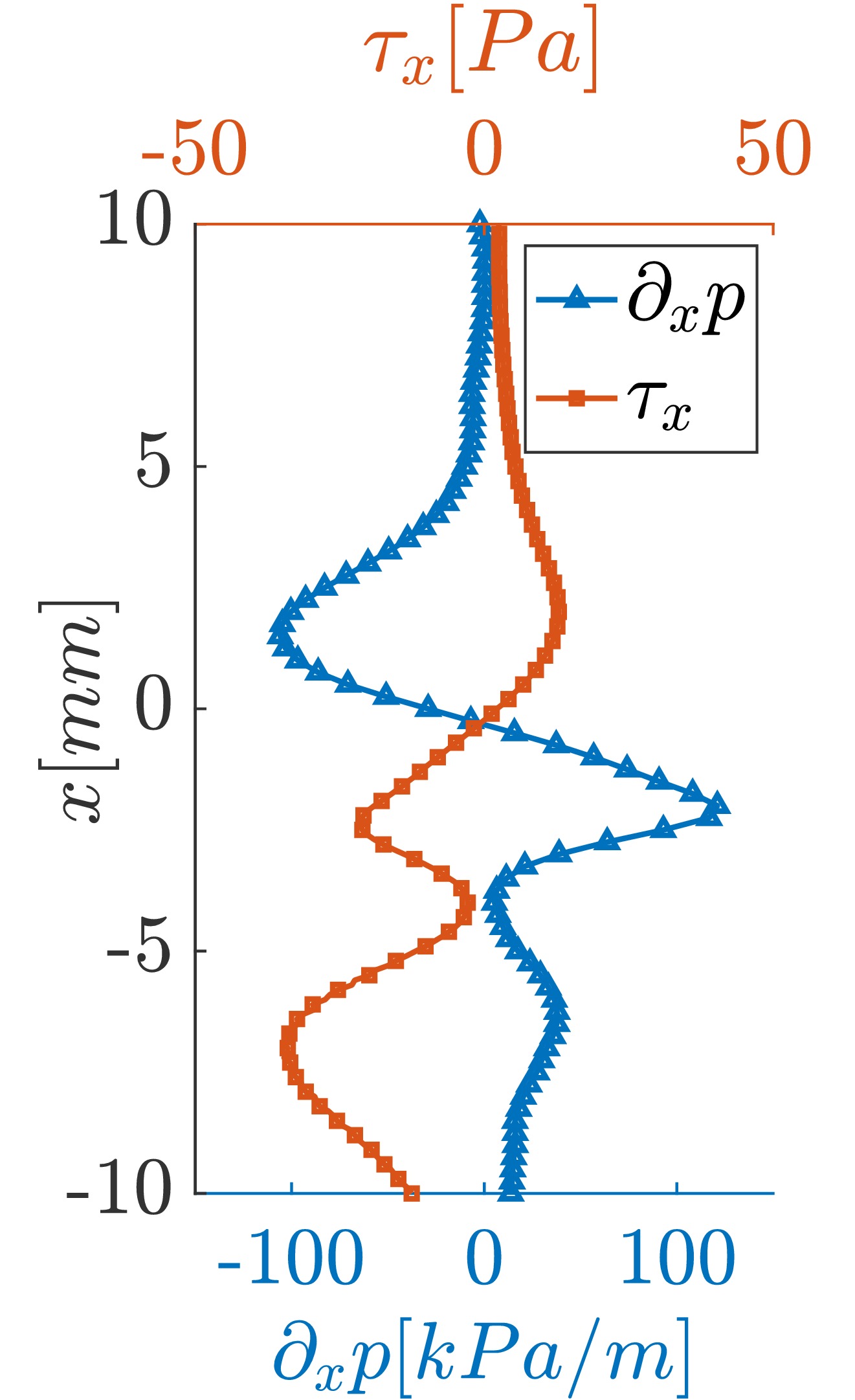}}
      {\includegraphics[width=\linewidth]{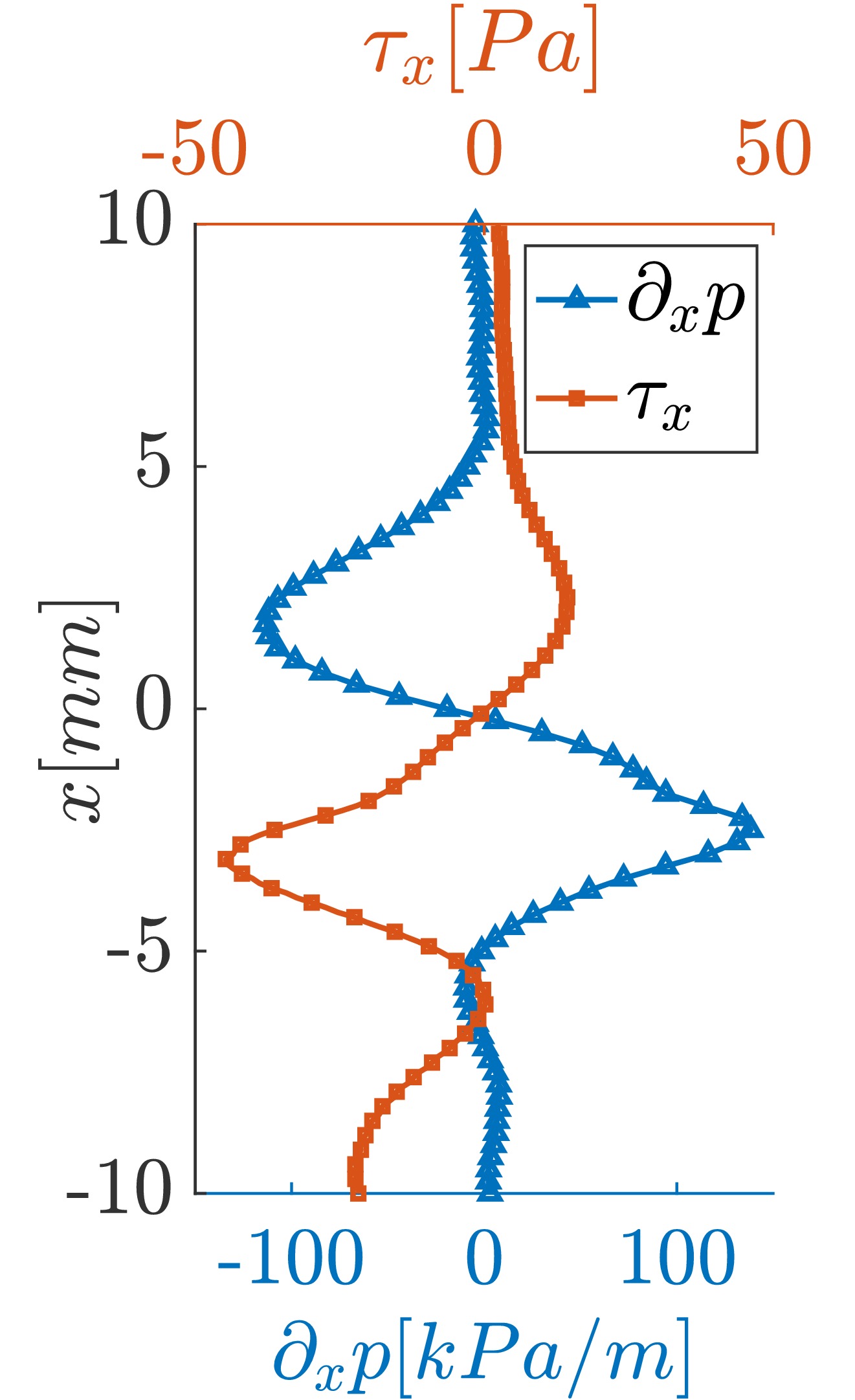}}
      {\includegraphics[width=\linewidth]{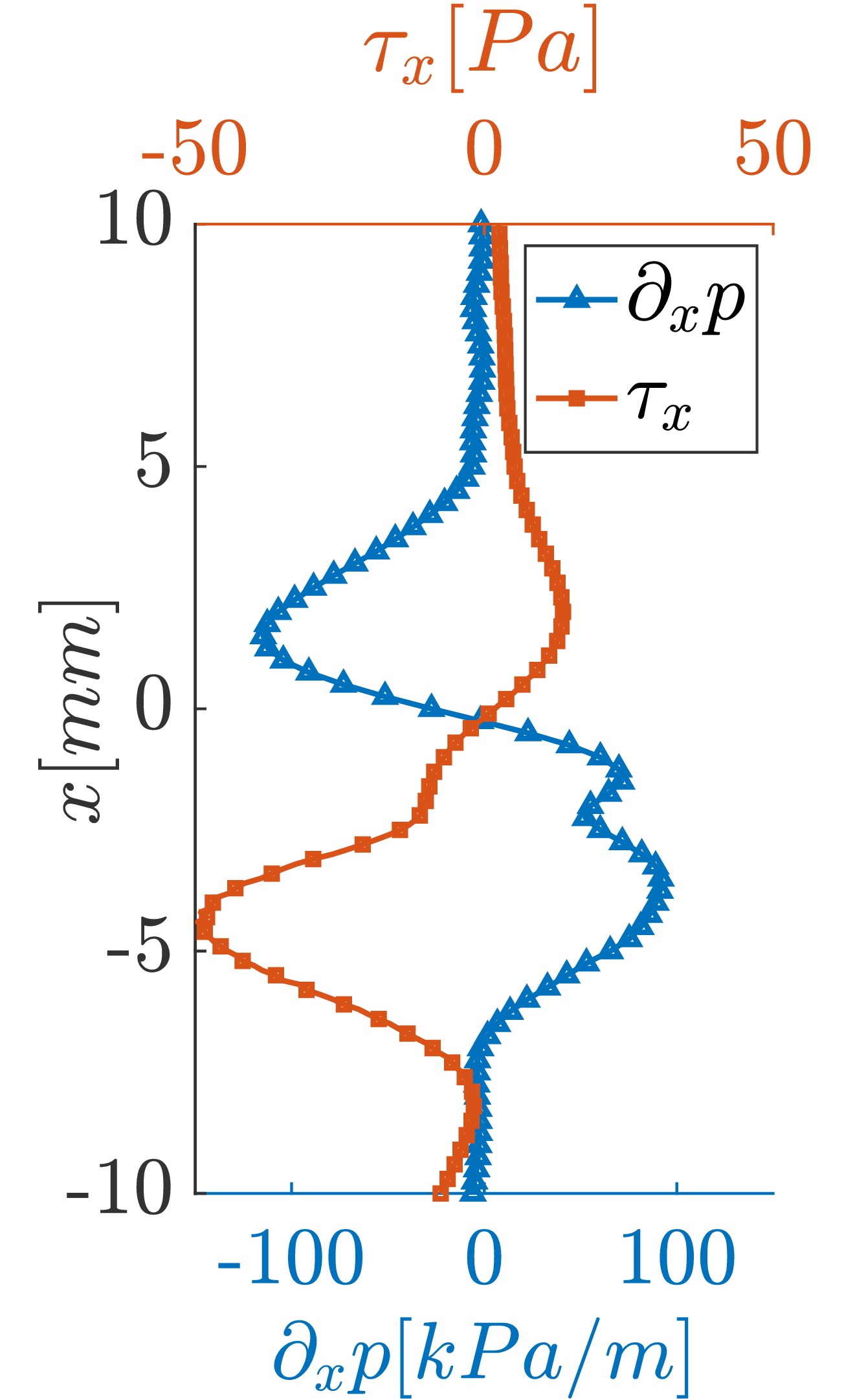}}%
    \end{minipage}%
 \caption{Same as in Fig.~\ref{fig:gas_emPOD_Case1} for Case 2 (Multimedia view).}
\label{fig:gas_emPOD_Case2}
\end{figure*}

\begin{figure*}
    \centering
    \begin{minipage}{.24\linewidth}
       A
      {\includegraphics[width=\linewidth]{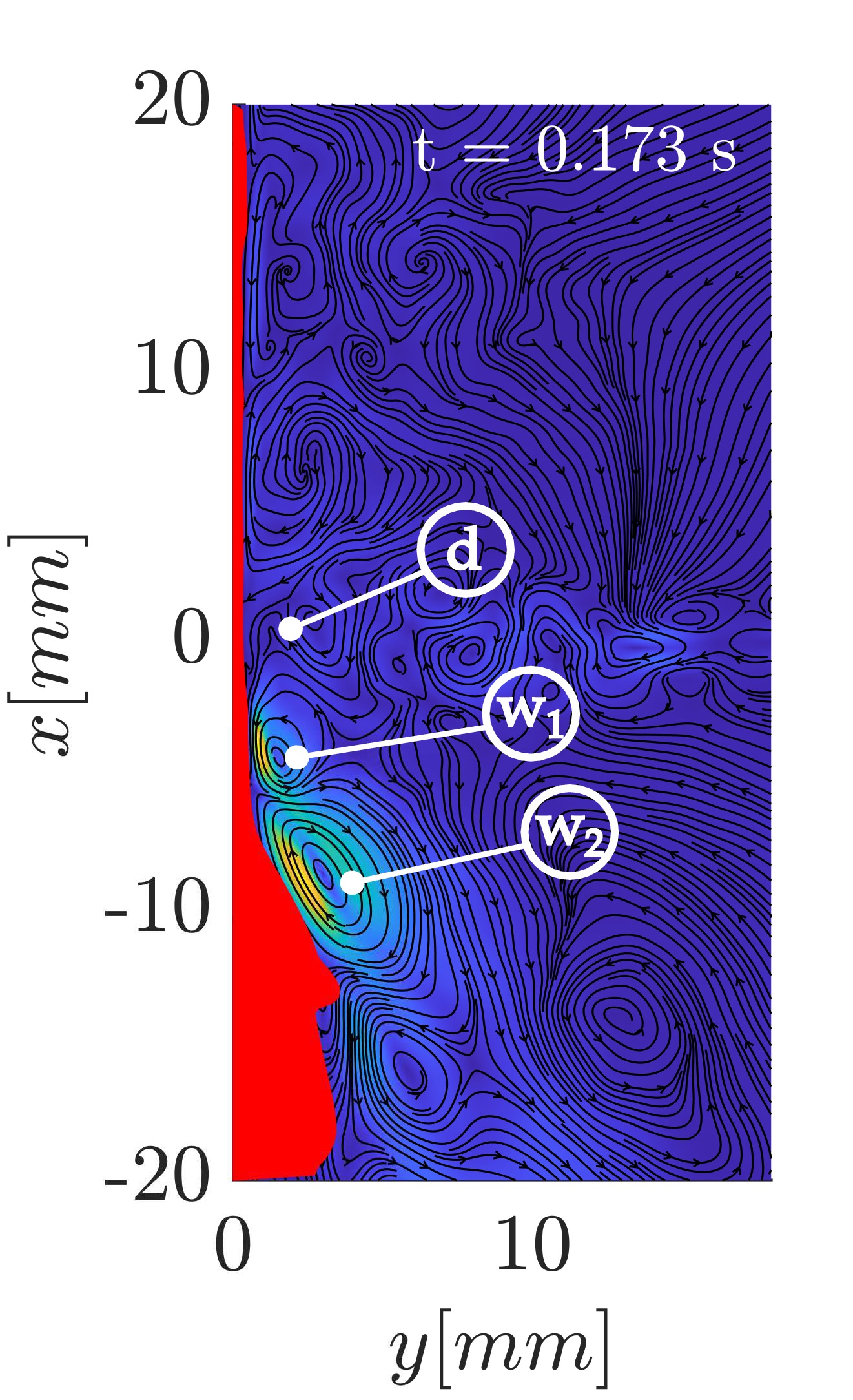}}
      {\includegraphics[width=\linewidth]{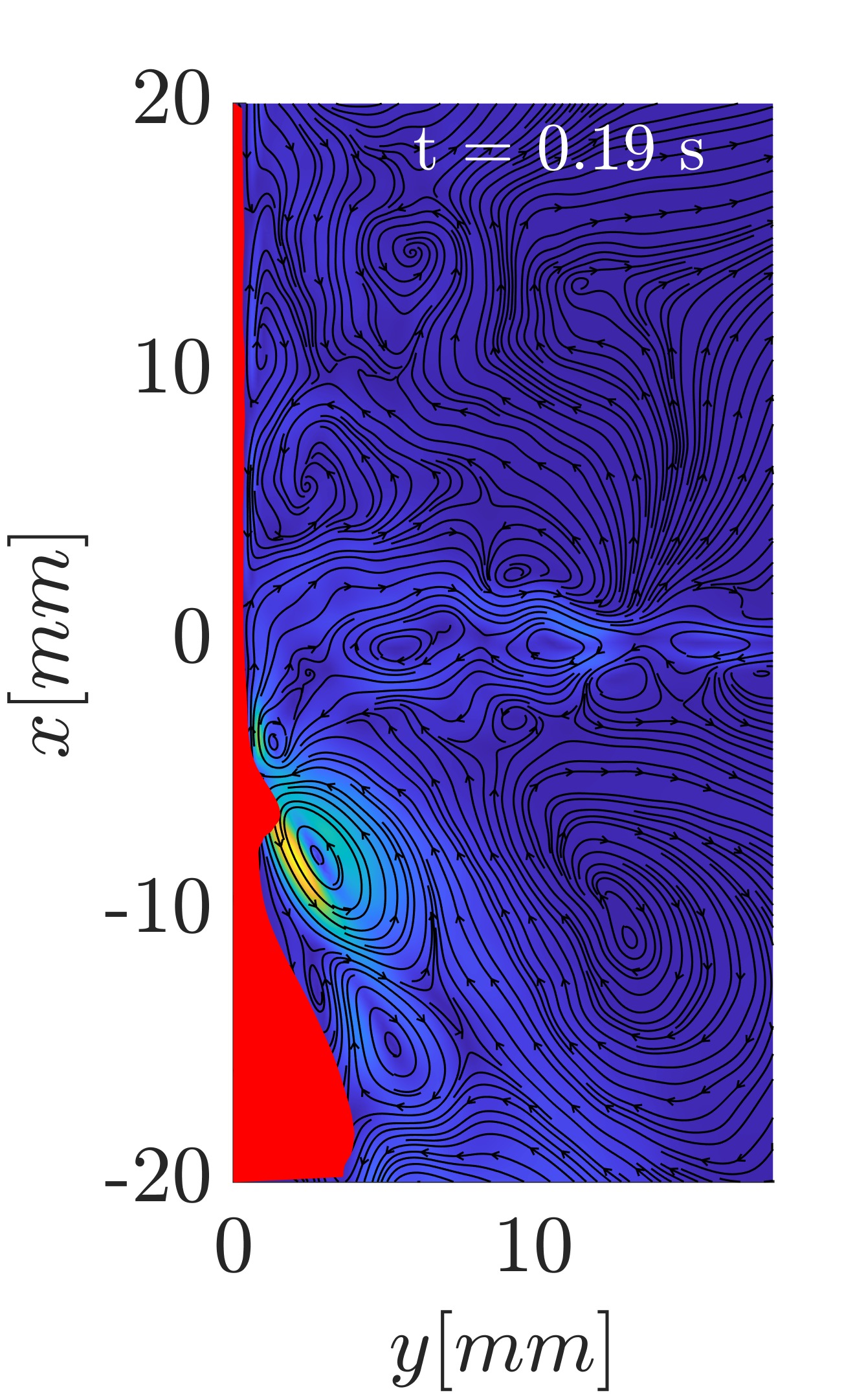}}
      {\includegraphics[width=\linewidth]{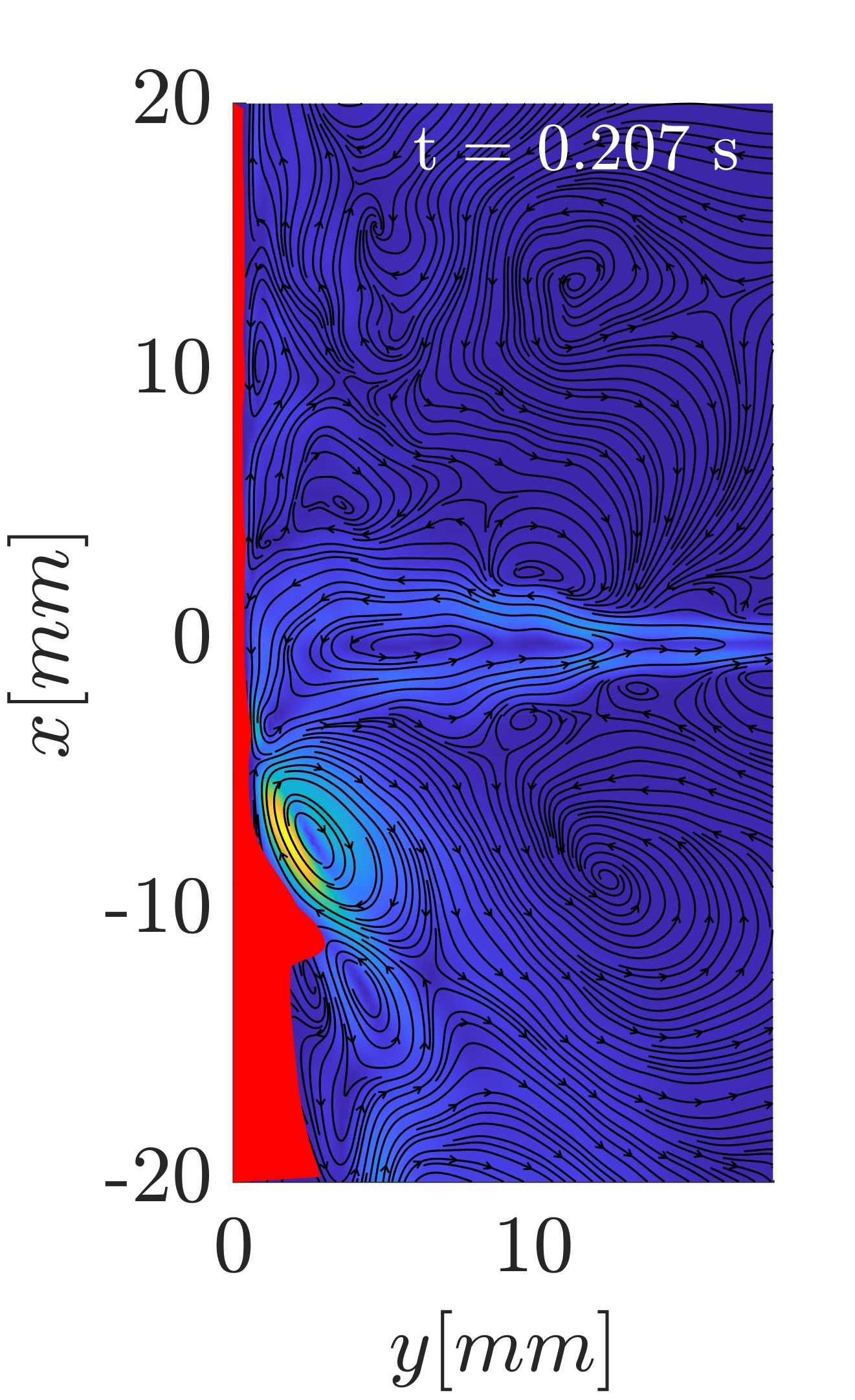}}%
    \end{minipage}%
  \hfill
    \begin{minipage}{.24\linewidth}
      B
      {\includegraphics[width=\linewidth]{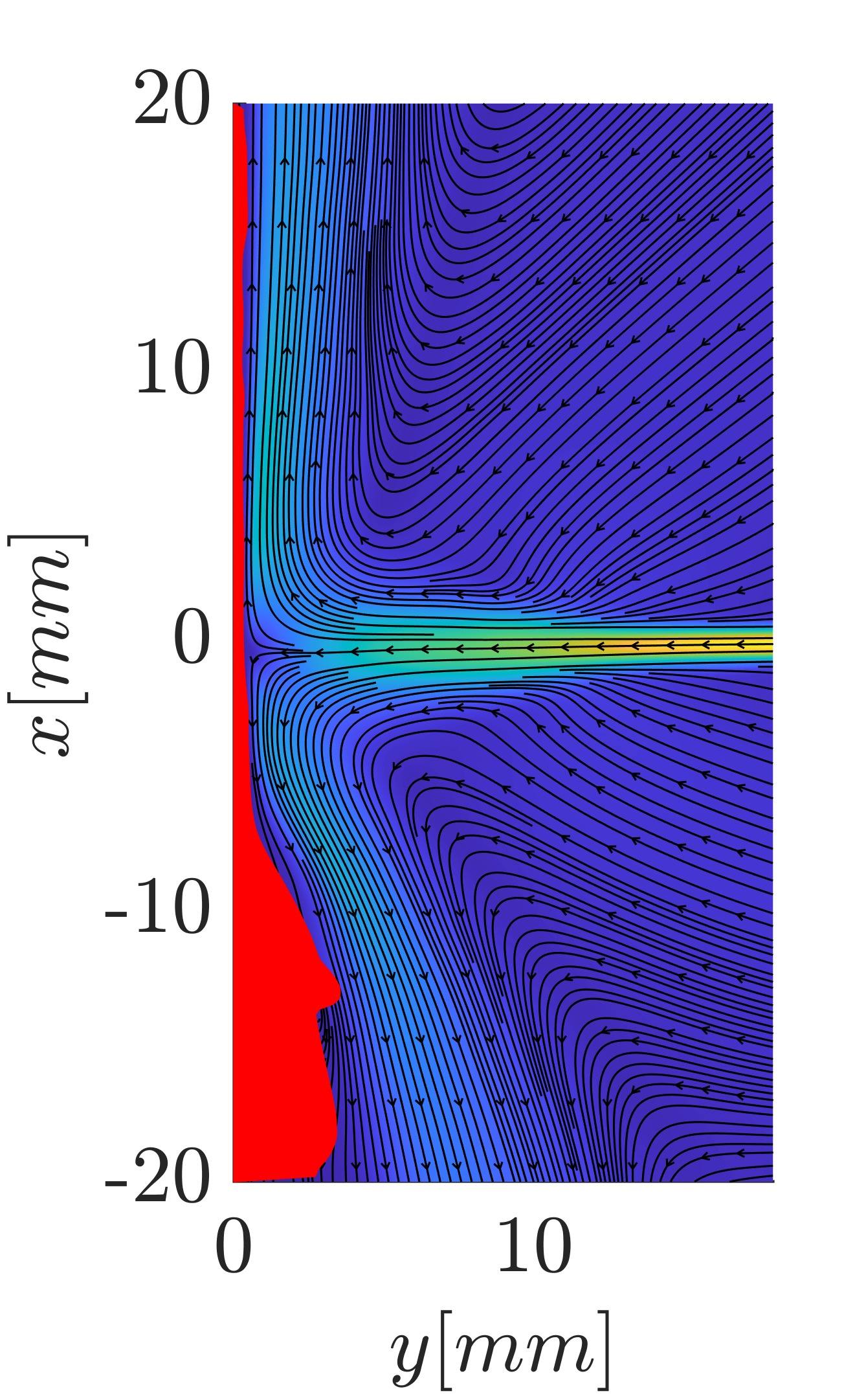}}
      {\includegraphics[width=\linewidth]{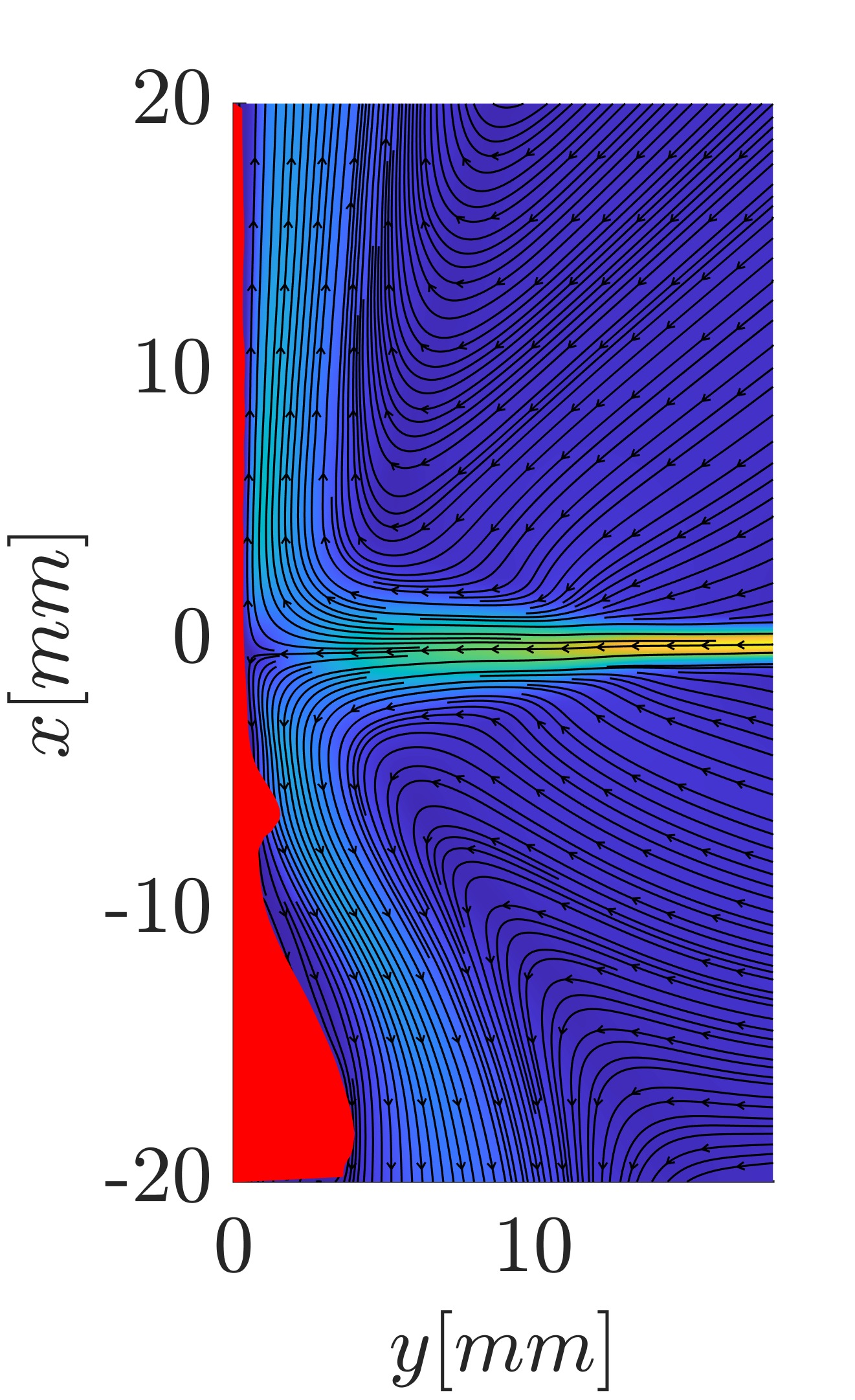}}
      {\includegraphics[width=\linewidth]{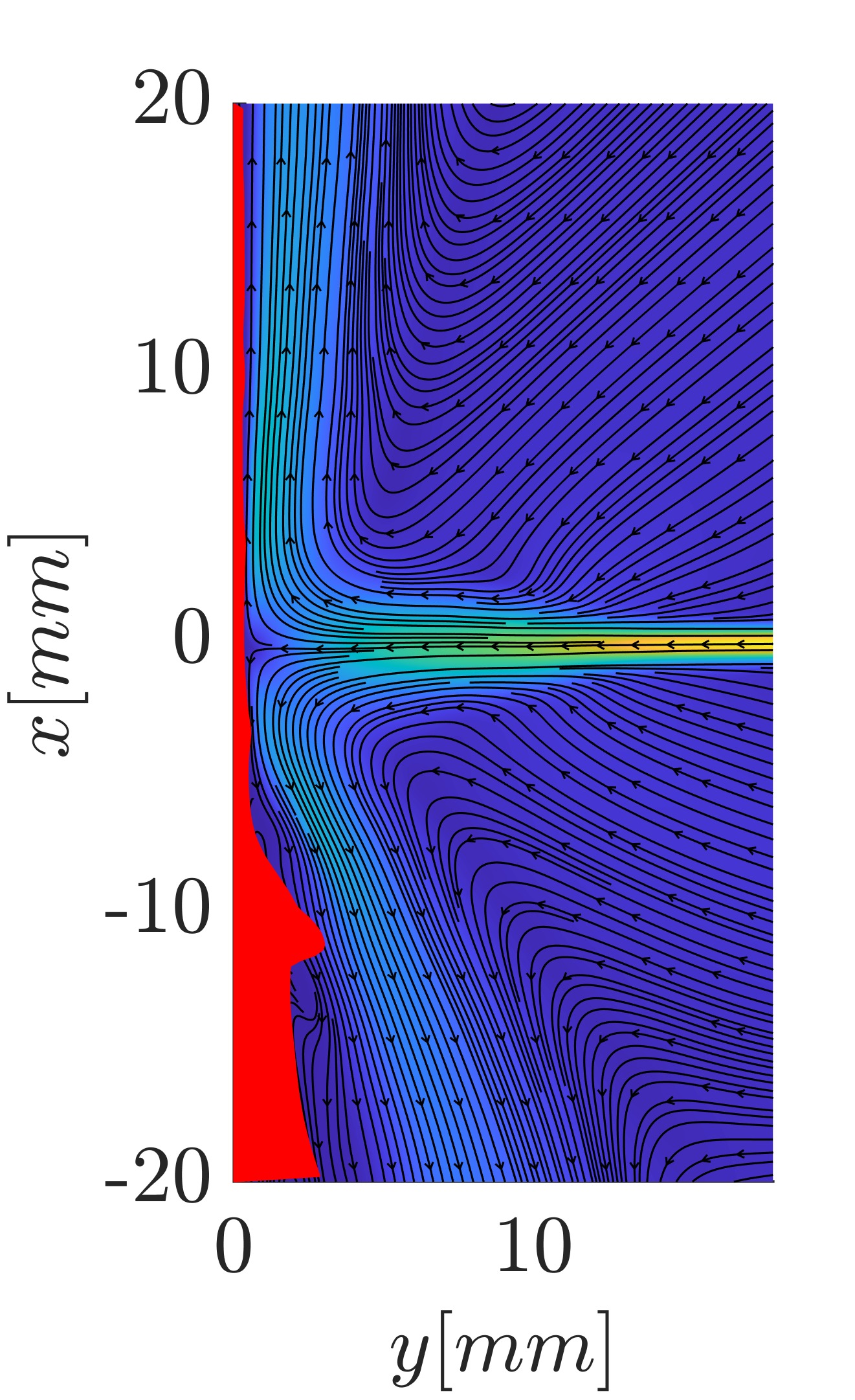}}%
    \end{minipage}%
  \hfill
    \begin{minipage}{.24\linewidth}
      C
      {\includegraphics[width=\linewidth]{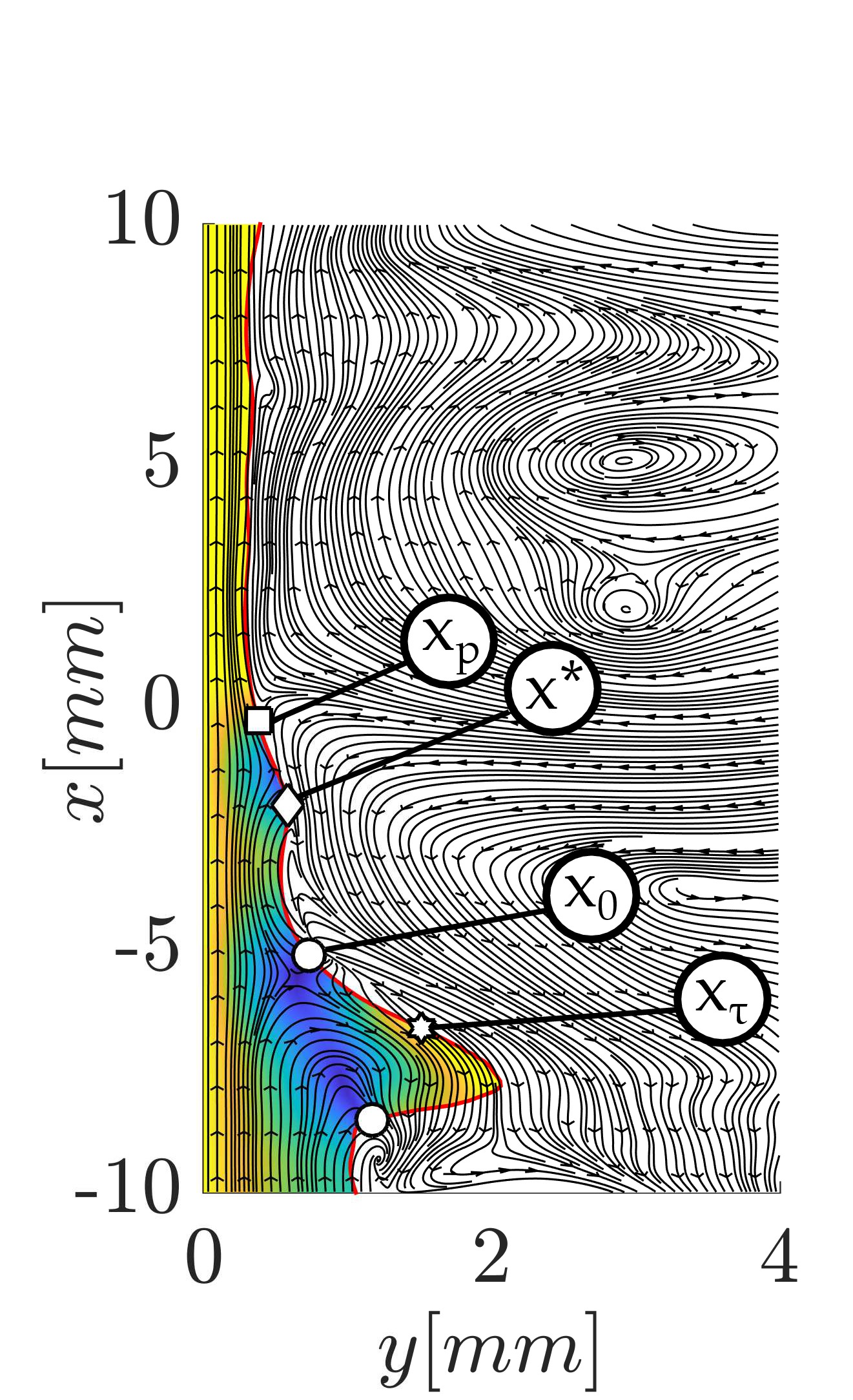}}
      {\includegraphics[width=\linewidth]{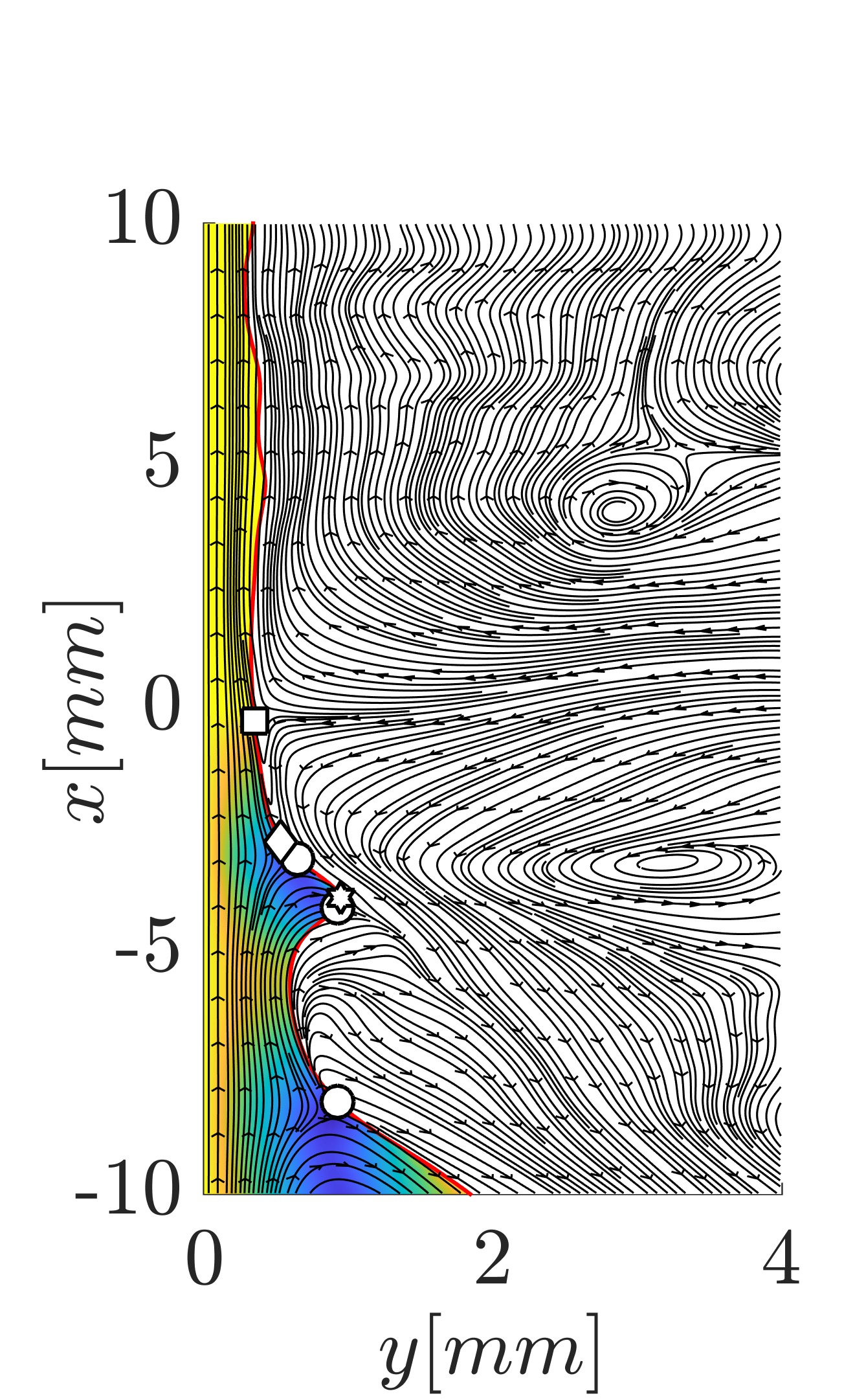}}
      {\includegraphics[width=\linewidth]{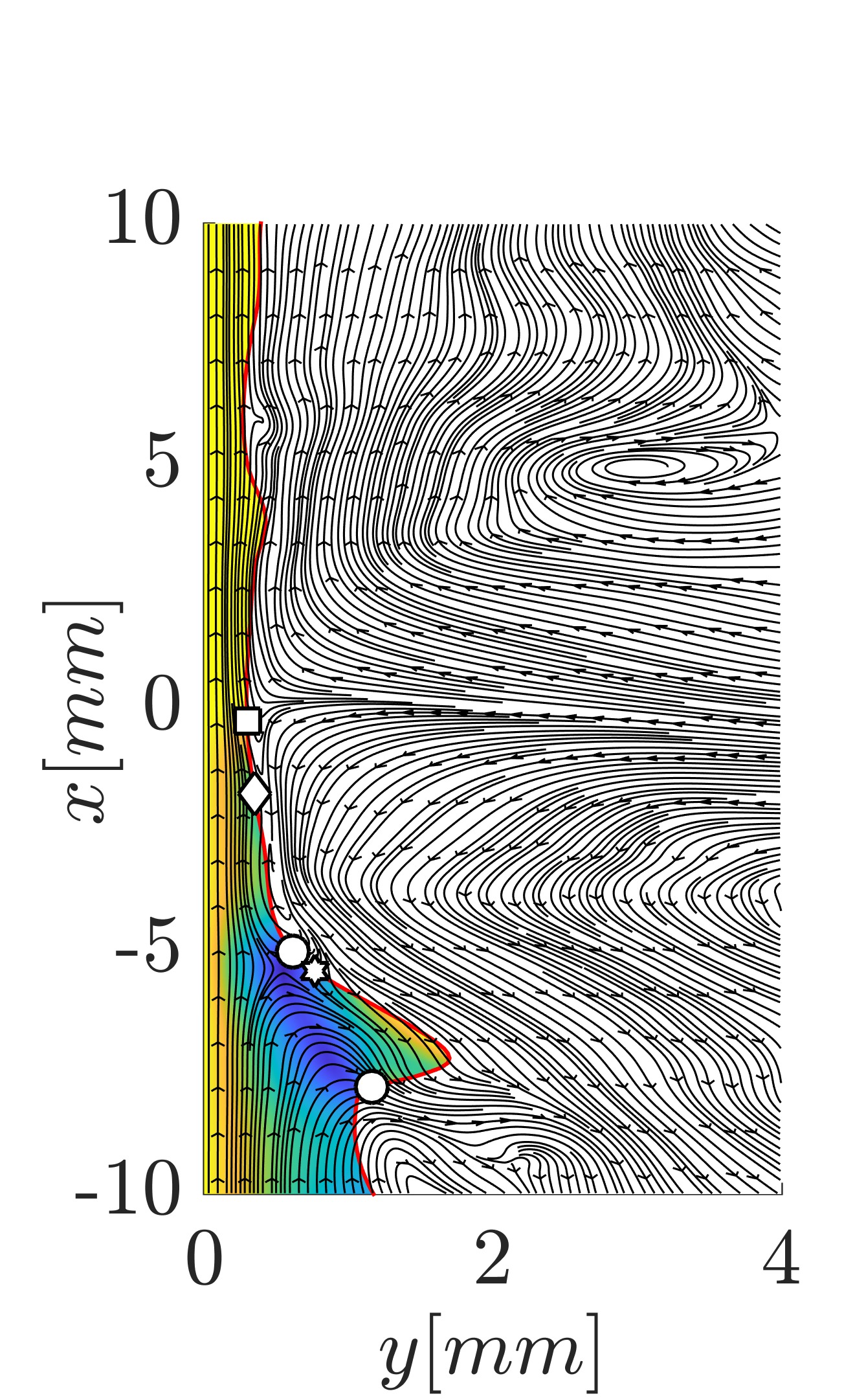}}%
    \end{minipage}%
  \hfill
    \begin{minipage}{.24\linewidth}
      D
      {\includegraphics[width=\linewidth]{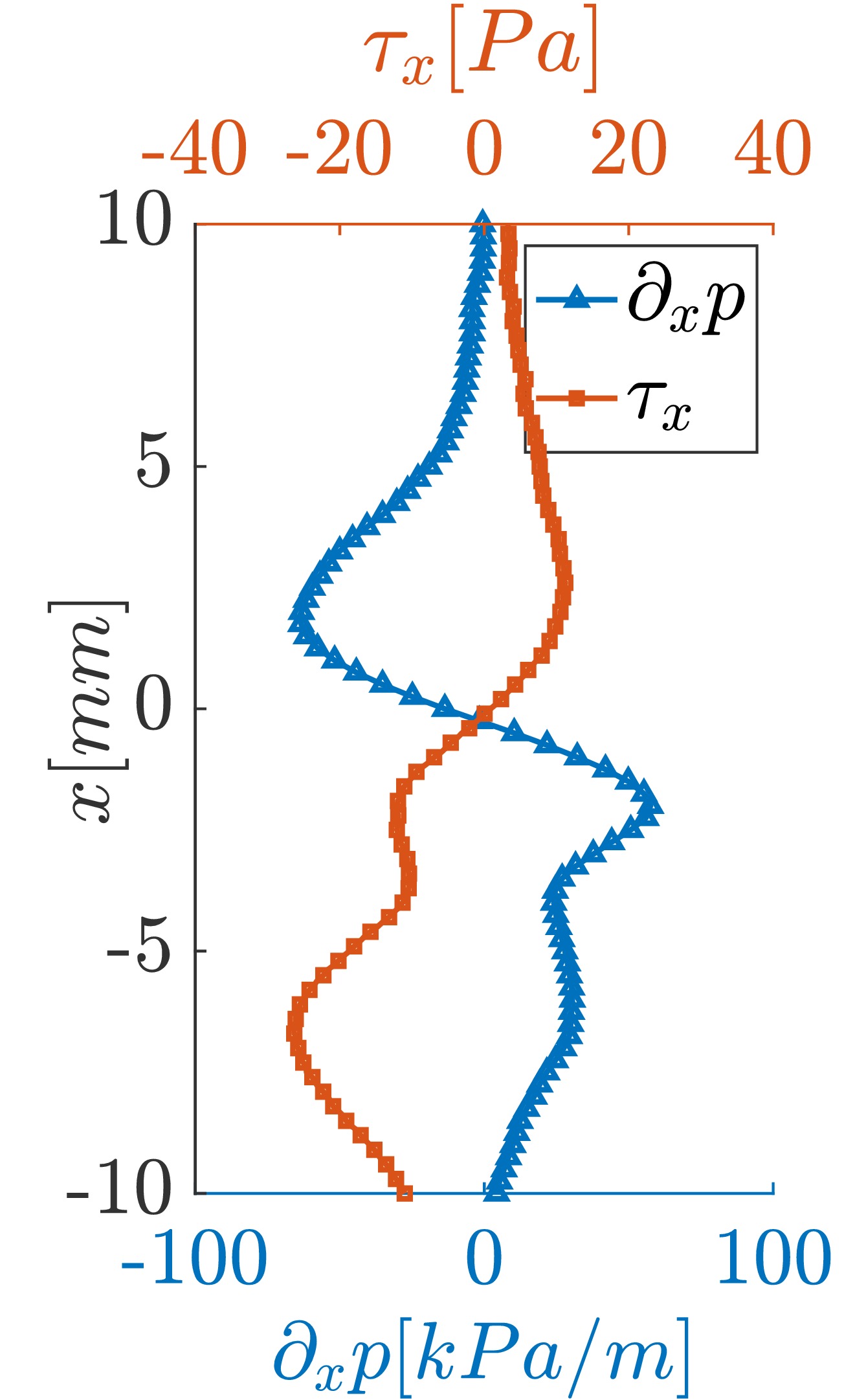}}
      {\includegraphics[width=\linewidth]{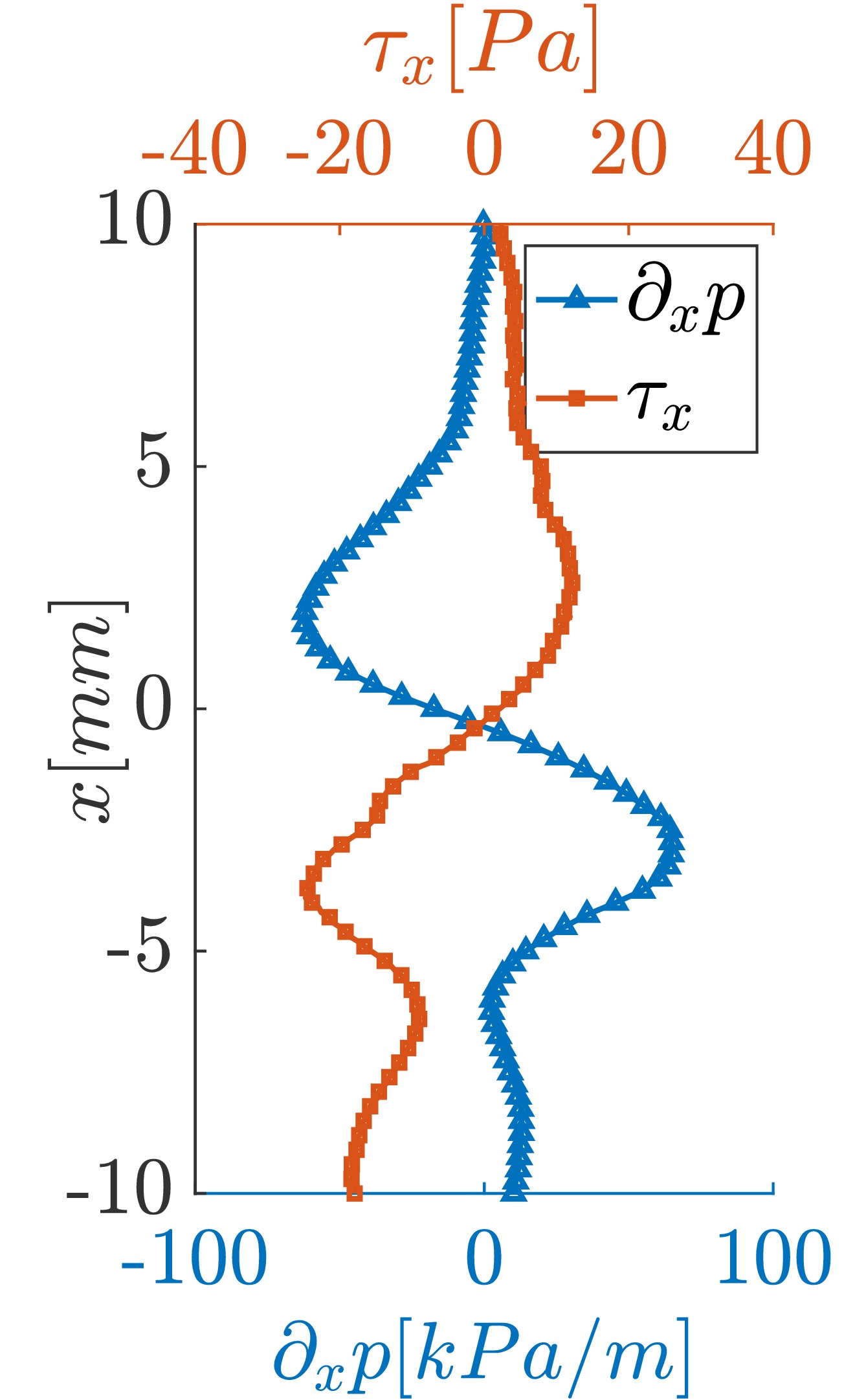}}
      {\includegraphics[width=\linewidth]{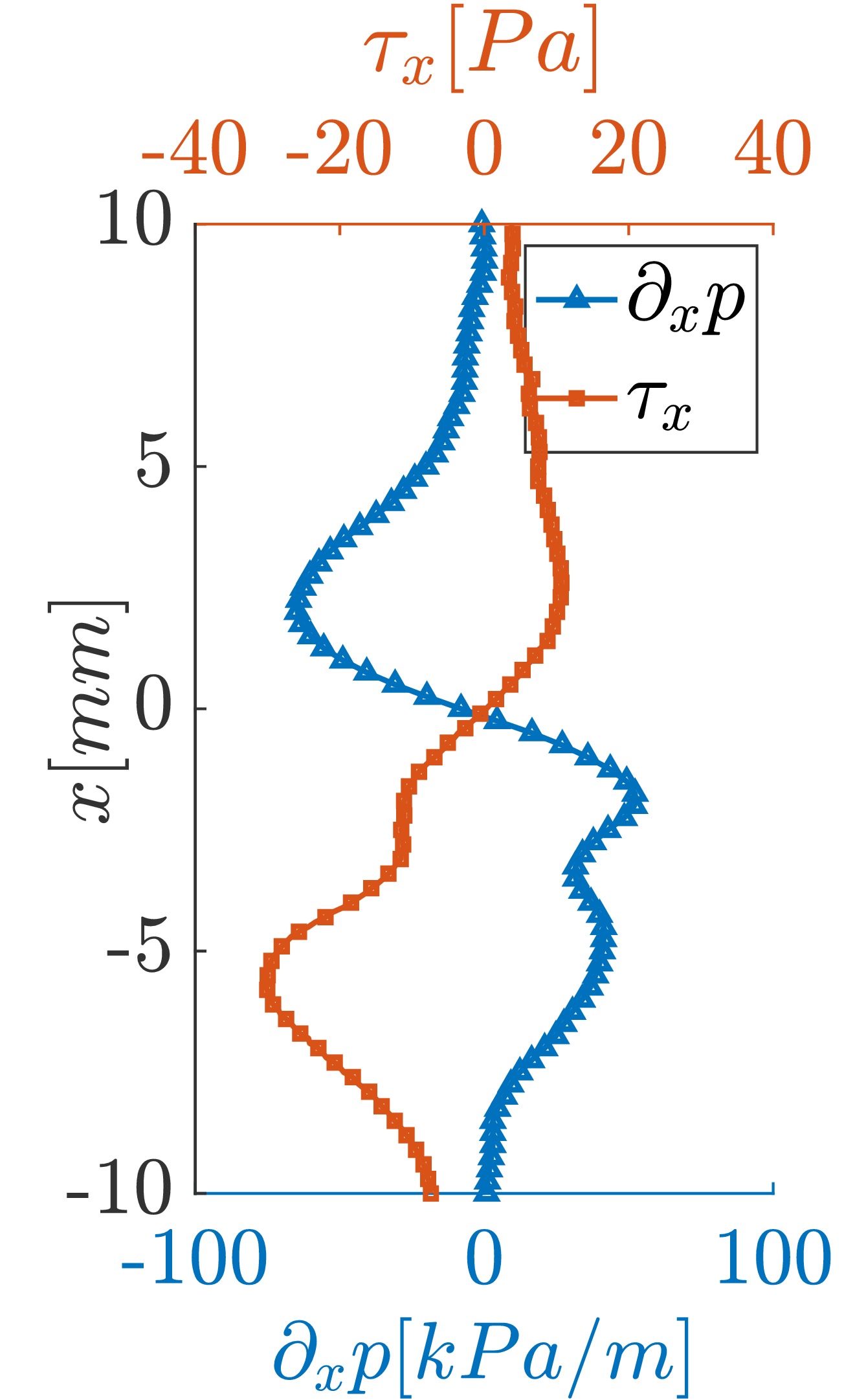}}%
    \end{minipage}%
 \caption{Same as in Fig.~\ref{fig:gas_emPOD_Case1} for Case 3 (Multimedia view).}
\label{fig:gas_emPOD_Case3}
\end{figure*}

The results are shown in Fig.~\ref{fig:gas_emPOD_Case1} - \ref{fig:gas_emPOD_Case3} for the three cases (one figure for each case). All figures are organized in three rows, one for each different time instant within a wave period, and four columns.
The first column (A) collects three snapshots of the gas flow field constructed considering only the 2D coupled undulation modes previously discussed. In the second column (B), the time - averaged velocity field is added, to show the interaction of these modes with the impinging gas jet flow. The third column (C) shows a close up of the impingement region, with a contour of the velocity magnitude within the liquid film. In these plots, the original data is considered (i.e. including all the modes). The film interface is highlighted by a red line. The fourth column (D) shows the distributions of pressure gradient and streamwise shear stress at the film interface considering only the 2D undulation modes. The computation of the shear stress at the interface is detailed in Section\ref{SecVID}. An animation of these images is provided as complementary material to the article.

Despite the different operating conditions, the three cases display similar features. The velocity fields in the modes linked to the undulation (column A) display a periodic shedding of counter-rotating vortices on the run-back flow side, labelled as `$w_1$' and `$w_2$'. The shedding of these vortices is perfectly synchronized with the waves in the liquid film on both final coating and run-back flow side. The interaction of these coherent structures with the jet flow is better revealed by summing their evolution to the time averaged flow (column B).

The structure $w_1$ lifts the gas flow's boundary layer to accommodate the formation and the growth of waves (see snapshots $t_1$, column B) in the run-back flow. This phase is critical in the momentum exchange between the two phases, because the falling waves cannot be accelerated up to the gas velocity and are large enough to interact with the impinging flow. When the wave in the run-back flow grows and travels downward, the gas flow remains attached to its tail and deflects. The structure $w_2$ originates mostly as a consequence of the subsequent step: when the wave is sufficiently large and steep, the gas flow separates at its crest, which acts as a sort of backward facing step (see time step 3 in all cases).

In the kinematic decomposition observed in these extended mPOD modes, the interplay of the structures $w_1$ and $w_2$ describes the dynamic interactions between the two phases that are mostly correlated with the formation of the undulation pattern in the final coating. Interestingly, no appreciable structures are visible in the gas flow on the side of the final coating film ($x>0$), revealing that the dynamics of the flow in these region does not correlate with the undulation pattern. Moreover, it is worth recalling that the structures $w_1$ and $w_2$ were also detected in the TR-PIV characterization of the gas flow (therein referred to as `$w$' in Mendez \emph{et al.}\cite{Mendez2019}), but were not fully identified due to the light reflections on the liquid interface. Although in that experimental work it was understood that these constituted the footprint of the run-back wave passage, their mechanism of formation and their interaction with the liquid flow was not clear and it is now uncovered with the present CFD data.

The mechanism previously described results in a slight deflection of the impinging jet labelled as `$d$' (as in the experimental work\cite{Mendez2019}) due to the region of low pressure induced by the wave-induced vortices. The process is reminiscent of a \emph{Coanda effect}\cite{Mendez2019,Mendez2018a}. On the other hand, the oscillation is limited to the region next to the interface, close to impingement, and its amplitude is much smaller than what was experimentally observed. The impact that such oscillations can have on the wiping actuators is discussed in the remaining of this section with the help of columns C and D from Fig.~\ref{fig:gas_emPOD_Case1}-\ref{fig:gas_emPOD_Case3}.

The closed-up views in columns C are further enriched by marking and labelling several relevant points. Besides the previously defined wiping point $x^*$, labeled with a diamond label, the figures also identify the impact point $x_{p}$ (at which $p(x_p)=max(p(x))$), with a square marker, the shear point $x_{\tau}$ (at which $\tau(x_{\tau})=max(\tau)$), with a star marker and the interface stagnation points (at which $u(x_0,y=h(x_0))=0$), with a circle marker. These points' location should be analyzed together with the figures in column D, plotting the streamwise pressure gradient and shear stress distribution.

The first snapshot of each sequence, in all cases, captures an instant in which the last wave in the run-back flow is far from the impingement point, and the gas-liquid interface is gently sloped in the impingement region. The gas flow's overall confinement is low, and the pressure gradient (column D) is relatively symmetric about $x_p$. On the other hand, the shear stress is consistently higher in the run-back flow region because of the higher relative velocity between the two phases. The jet's wiping action is focused in the vicinity of the impingement point, i.e. $x_p$ and $x^*$ are close.

In these first snapshots, the liquid velocity field is entirely directed upwards, and the stagnation point at the interface $x_0$ is located about $6$ to $8$ $\mbox{mm}$ upstream, depending on the test case. This point sits on the tail of a wave, above a saddle point in the liquid's velocity field, where the bifurcation of the flow due to the wiping originates. As the wave grows (second snapshot in each figure), the gas flow field is significantly influenced. The extent to which the evolution of the pressure gradient and the shear stress profile at the wall is due to the wave passage itself (and thus the liquid film dynamics) or by the wave-induced deflection of the gas flow (and thus a varying wiping capability) is unclear. On the other hand, the results show that the wave passage is always combined with a pulsation in the pressure gradient intensity and a shift of the point $x_{\tau}$ further down $x_0$: this is the region where the liquid flow experiences the largest acceleration and multiple local maxima of shear stress are observed in the first snapshot (see column D). 

The interplay of pressure gradient and shear stress on the wave dynamics is partially revealed by the figures and better captured by the animation in the supplemented material. In the investigated wiping regimes, the shear stress does not appreciably contribute to the film thickness reduction\cite{Gosset2019,Mendez2020}, but influences the shape and the traveling velocity of the waves in the liquid film. One might thus expect that this interplay governs the spectral content of the undulation.

Referring to the classification between \emph{coupled} oscillations and \emph{fluidic} oscillations made in previous experiments\cite{Mendez2018a}, it is clear that the oscillations observed in this work belongs to the first category. However, the amplitude of these oscillations is surprisingly low, as revealed by the negligible displacement of the impact point $x_p$ observed in Fig.~\ref{fig:gas_emPOD_Case1}-\ref{fig:gas_emPOD_Case3}. The modulation in the pressure gradient evolution is nevertheless well correlated with the dynamics of the film thickness.

\begin{figure}
    \begin{subfigure}{.35\textwidth}
      \centering
      % include first image
      \includegraphics[width=\linewidth]{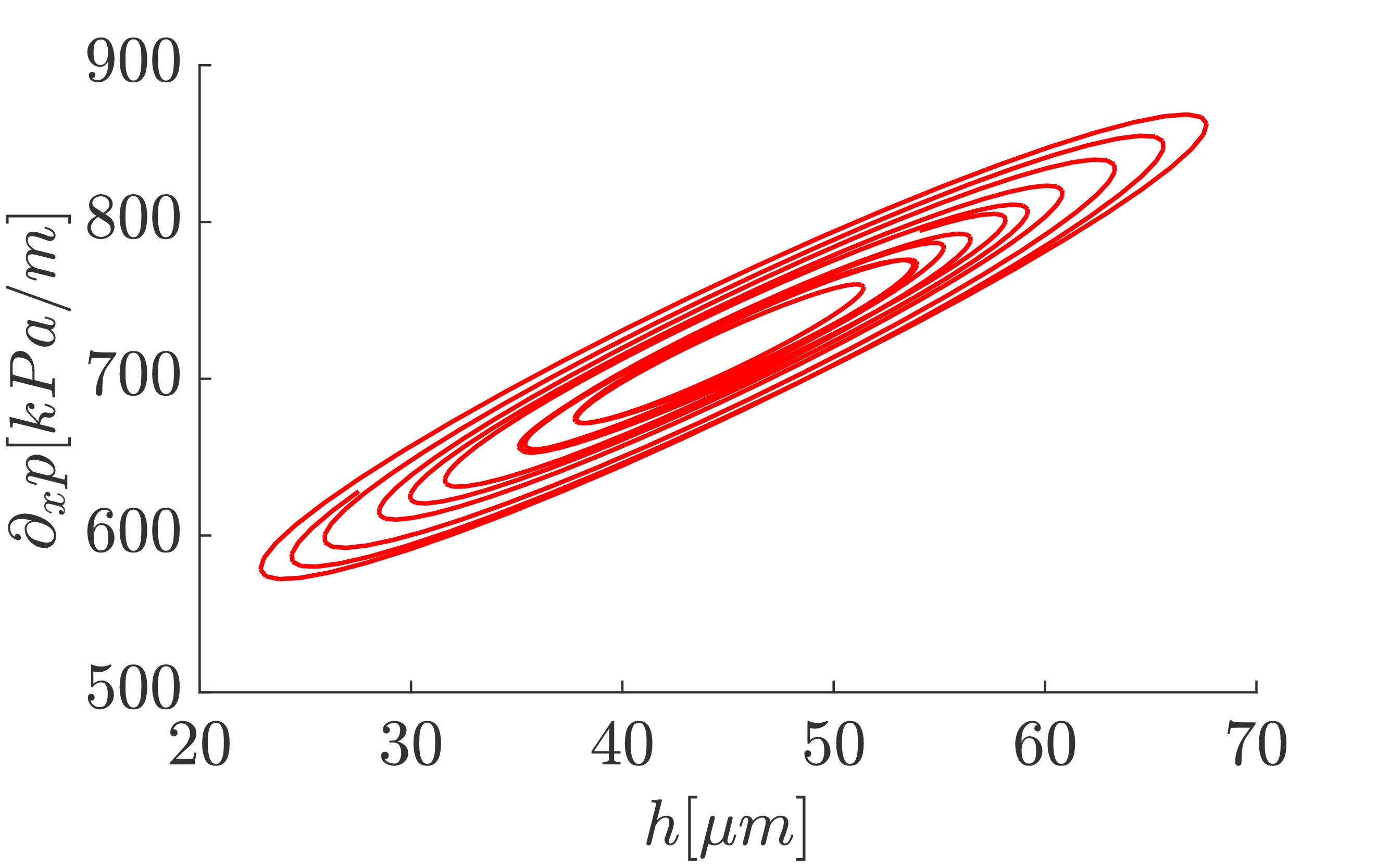}
      %\caption{$\hat{Z}=14.2$ | $\Pi=0.16$}
      \label{fig:orbit_Z18_P425}
    \end{subfigure}
    \begin{subfigure}{.35\textwidth}
      \centering
      % include first image
      \includegraphics[width=\linewidth]{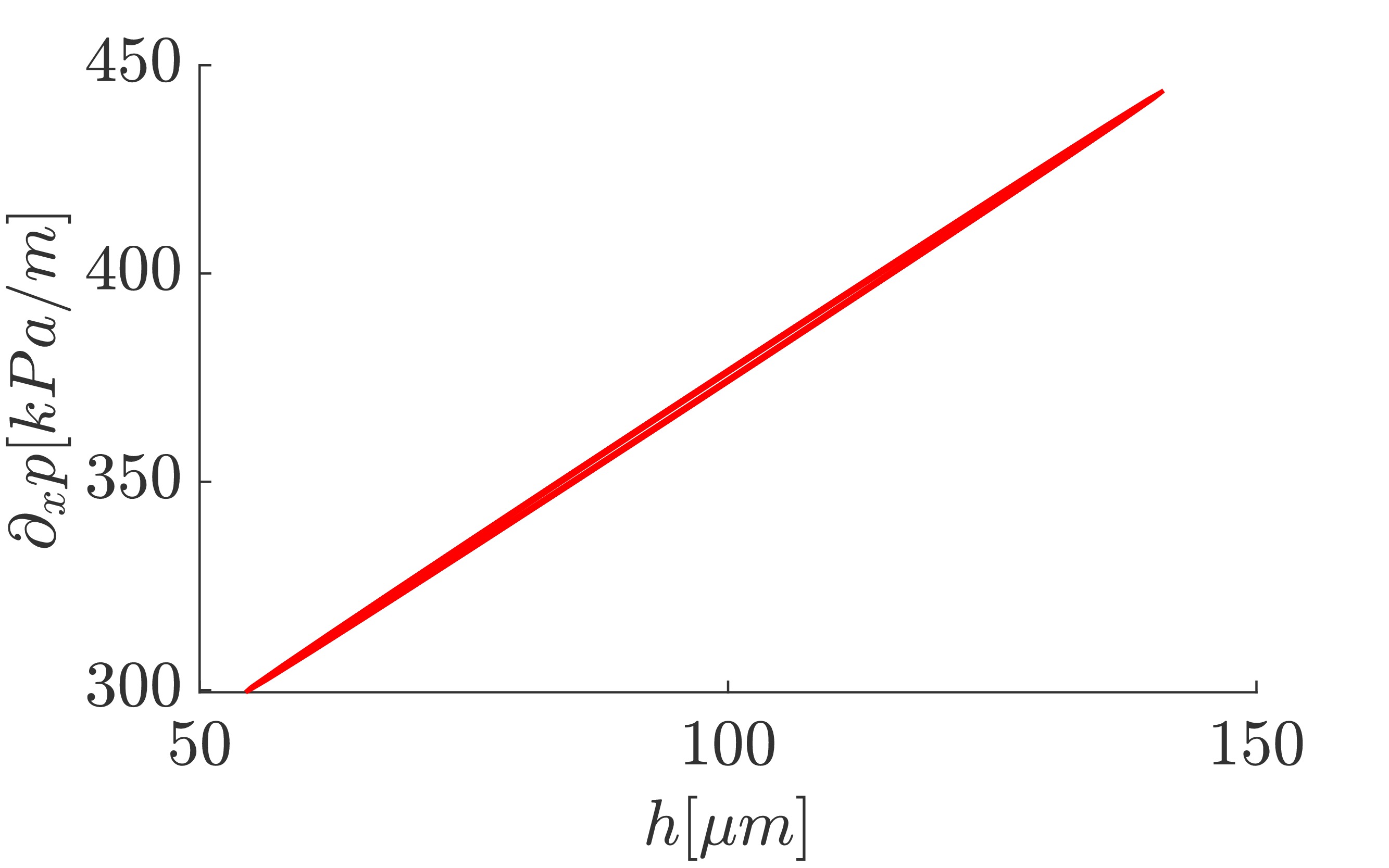}
      %\caption{Case 3: $\hat{Z}=19.4$ | $\Pi=0.18$}
      \label{fig:orbit_Z18_P875}
    \end{subfigure}
    \begin{subfigure}{.35\textwidth}
      \centering
      % include second image
      \includegraphics[width=\linewidth]{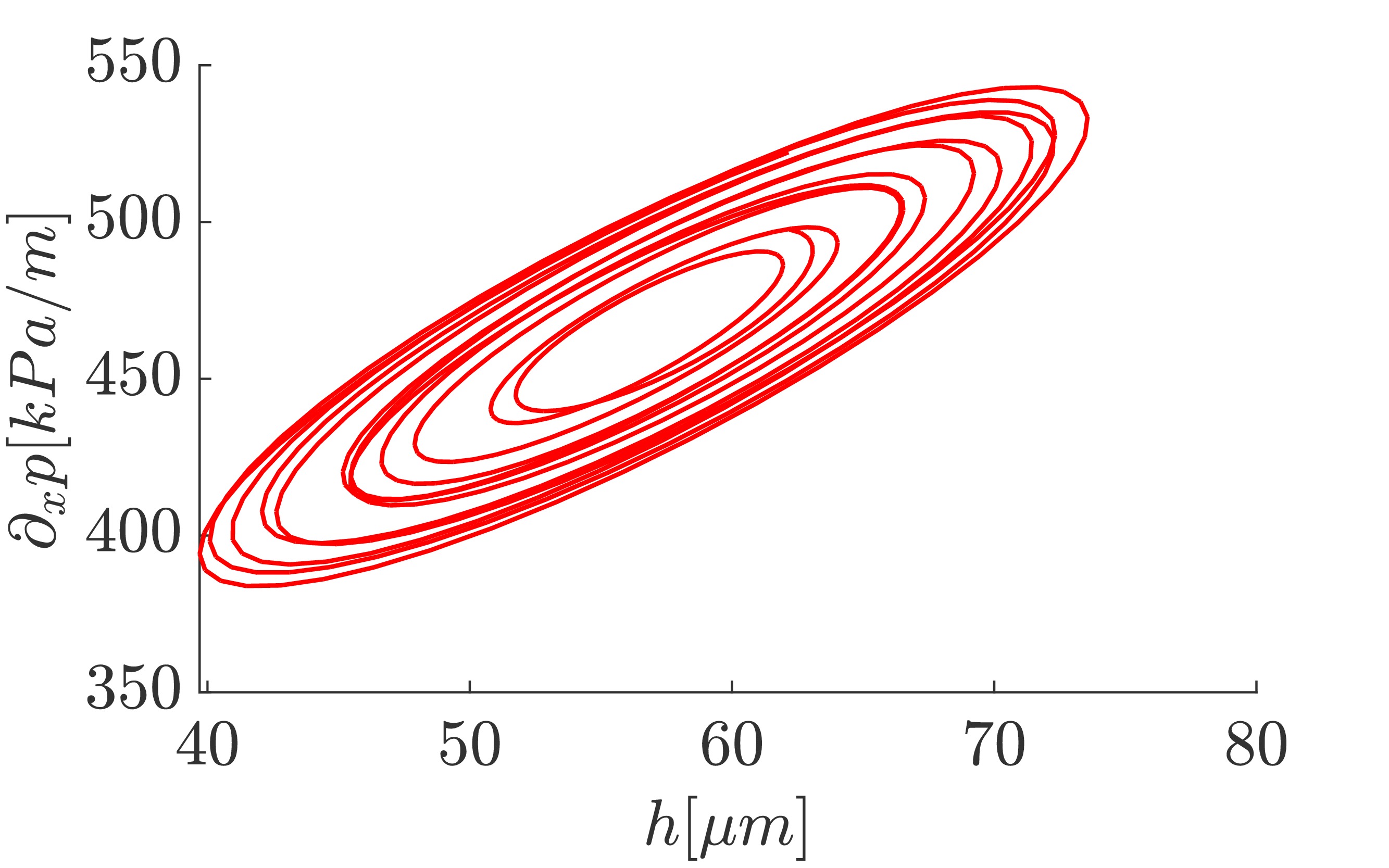}  
      %\caption{$\hat{Z}=14.2$ | $\Pi=0.33$}
      \label{fig:orbit_Z215_P875}
    \end{subfigure}
    \caption{Lissajous curves for the pressure gradient $\partial_x p$ and the thickness $h$ at the position of the maximum of the temporal averaged pressure gradient.}
    \label{fig:orbits}
    \end{figure}

We conclude this section by describing the phase portrait of the film thickness evolution versus the pressure gradient evolution. This is illustrated in Fig.~\ref{fig:orbits}, showing the Lissajous curves built by extracting these quantities from the 2D coupled undulation modes for the three cases. Both signals are taken at the location of the maximum time averaged pressure gradient, $\bar{x^*}$. This location coincides with the one of the largest instantaneous maximum gradient, produced when the wiping meniscus is rather flat, and the wave formation is at its earliest stage. The perfect frequency match results in ellipsoidal orbits, whose inclination and width capture the effects of inertia in the liquid film. The thickness signal has a phase delay of about $\pi/4$ with respect to the pressure gradient in cases one and three, which have comparable wiping strength and thus film thickness, while these are perfectly in phase in case 2 where the wiping is stronger. In all cases, however, the largest values of film thickness are observed in instants in which the pressure gradient is largest, hence defying any intuitive description of the flow interaction based on steady-state modeling of the flow.

   \subsection{Validation of Integral Formulations}\label{SecVID}
   
   In this final section we analyse the validity of the assumptions supporting the integral modelling formulation\cite{Mendez2020}, recalled in \ref{SecIV}. We focus on test case 1 and we begin by assessing the long-wave assumption. This test case is characterized by $\epsilon=0.93$ in the Skhadov-like scaling (cf. Table\ref{Scaling_Table}), meaning that the long wave assumption is not expected to hold.
   
\begin{figure*}
    \centering
    \begin{minipage}{.2\linewidth}
      {\includegraphics[width=\linewidth]{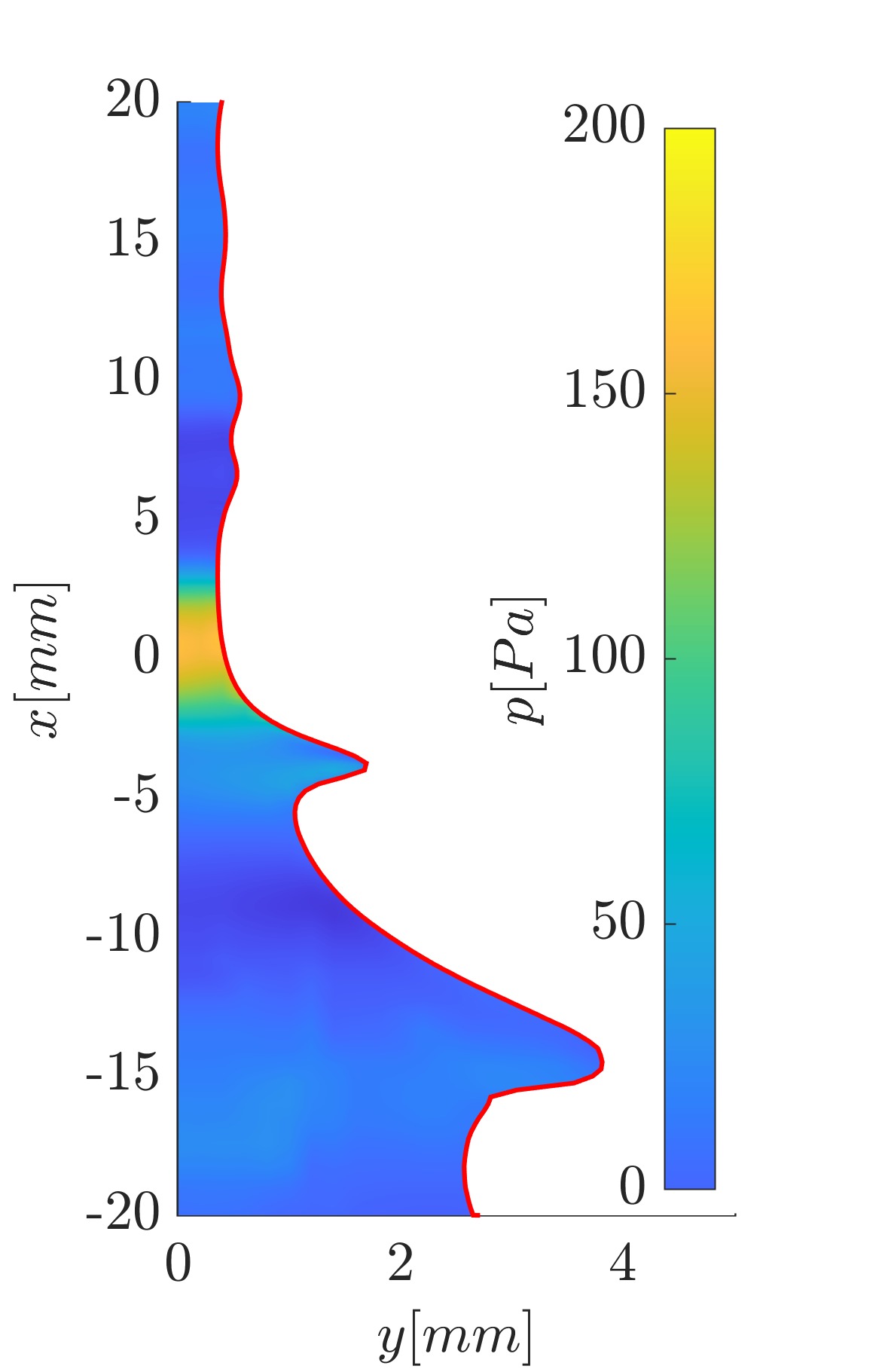}}
    \end{minipage}%
  \hfill
    \begin{minipage}{.2\linewidth}
      {\includegraphics[width=\linewidth]{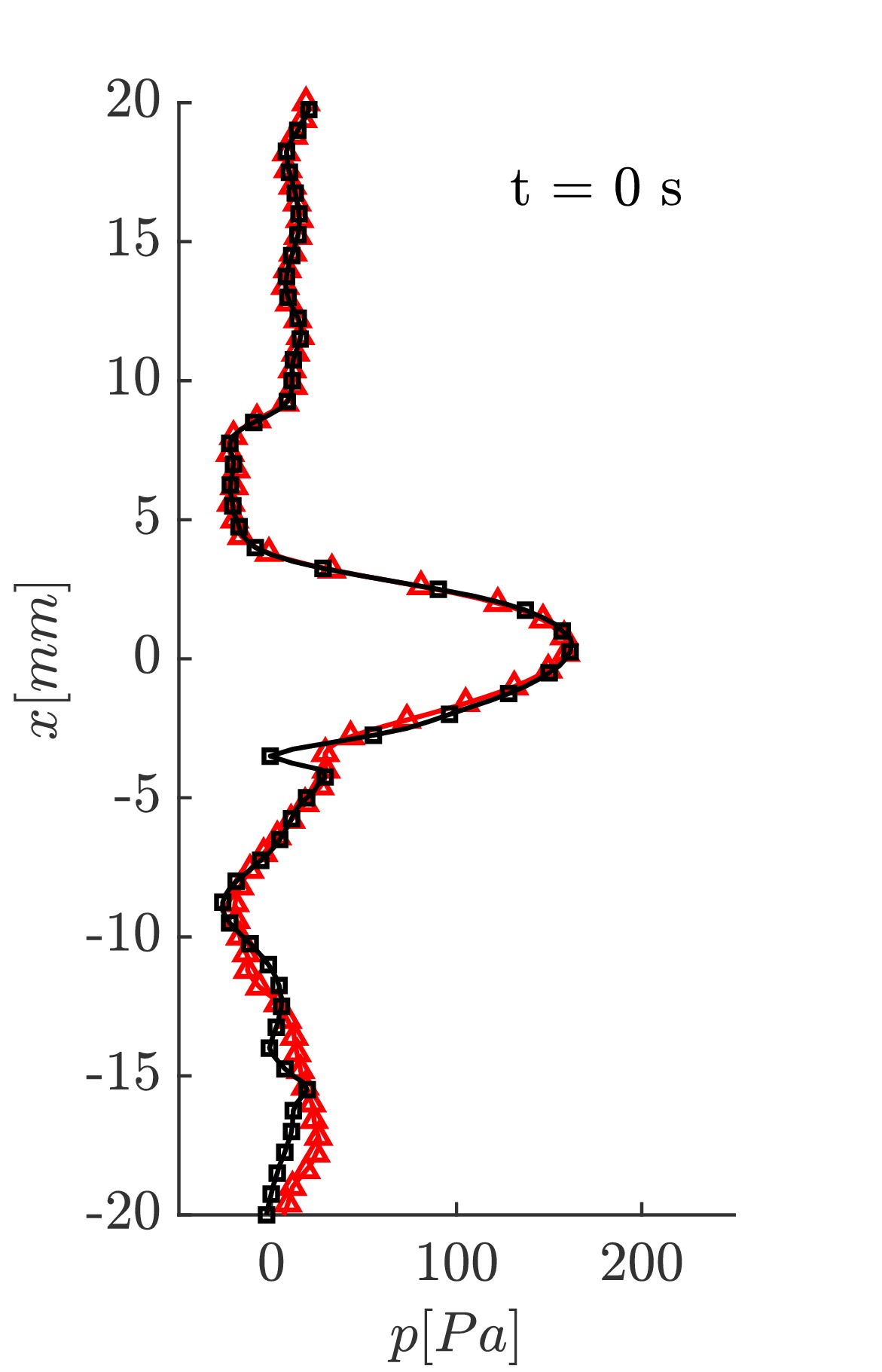}}
    \end{minipage}%
  \hfill
    \begin{minipage}{.2\linewidth}
      {\includegraphics[width=\linewidth]{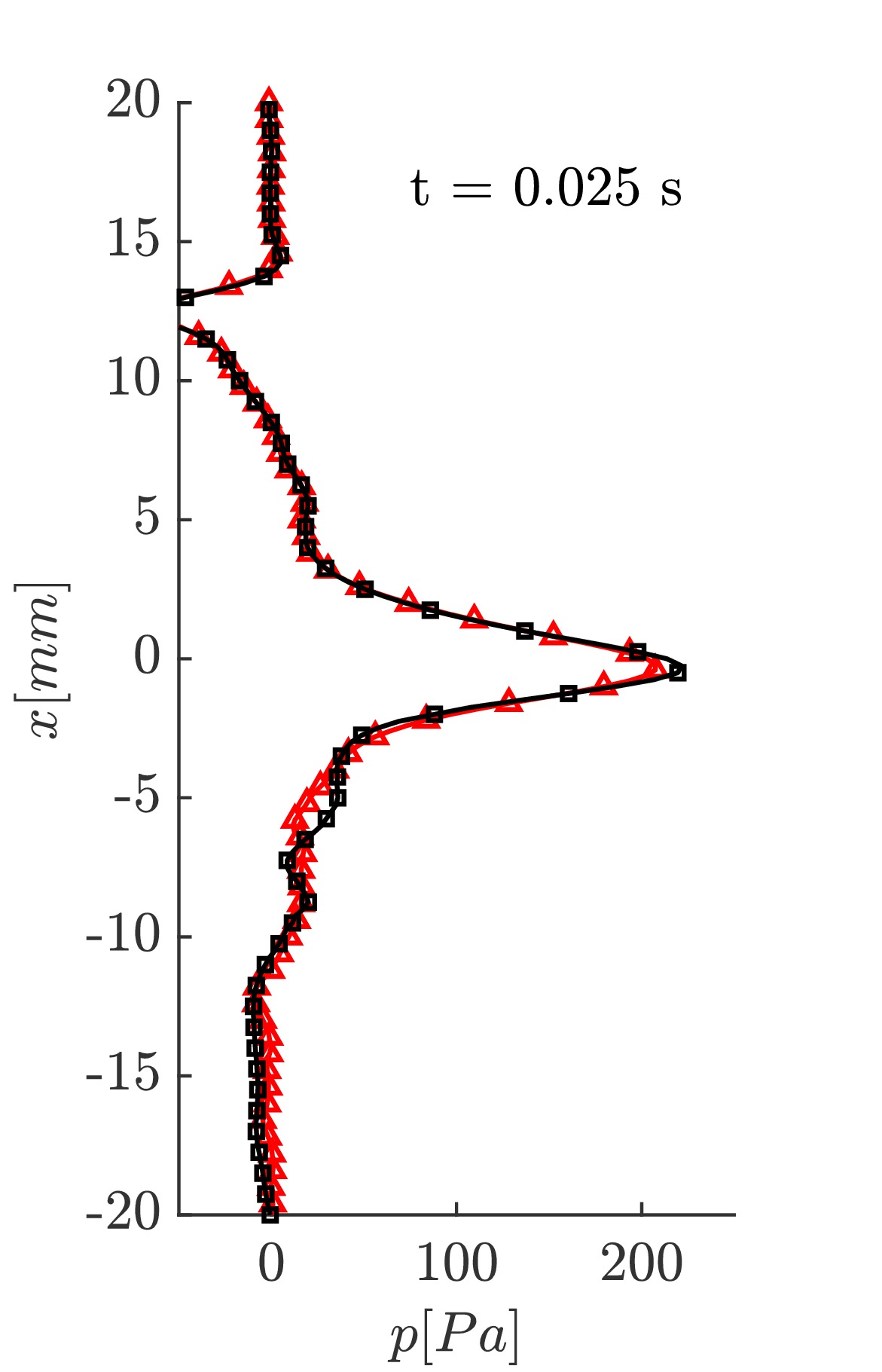}}
    \end{minipage}%
  \hfill
    \begin{minipage}{.2\linewidth}
      {\includegraphics[width=\linewidth]{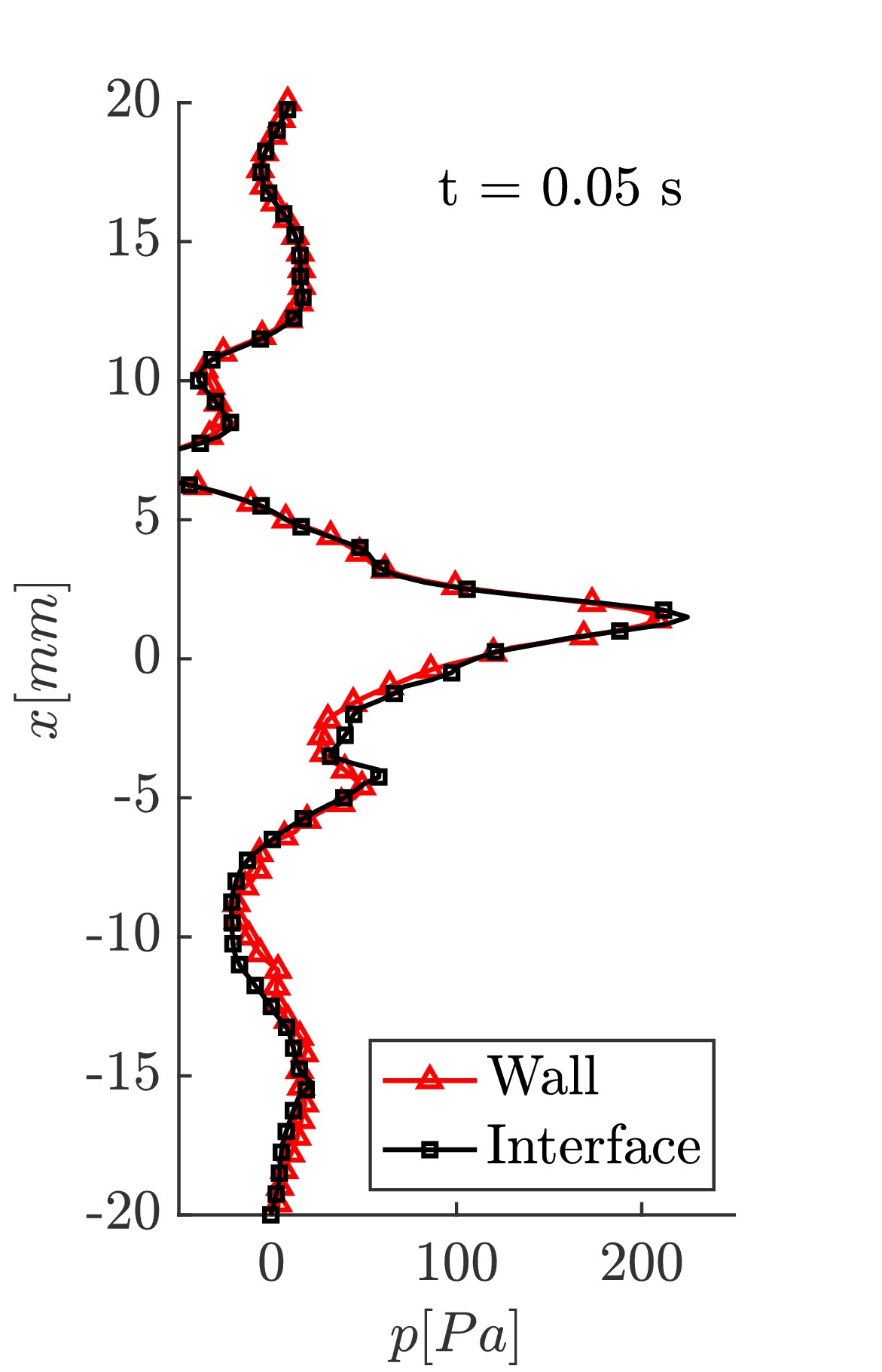}}
    \end{minipage}%
     \caption{Instantaneous pressure field within the liquid film and corresponding pressure distributions at the wall and at the interface.}
    \label{fig:P_liquid}
    \end{figure*}

   Fig.~\ref{fig:P_liquid} shows, on the left, an instantaneous snapshot of the liquid film interface, with a contourplot of the pressure field within the liquid film. The aspect ratio of the axes should be carefully examined. The snapshot shows that the wavelength in the run-back flow's wave is of the order of $\lambda\approx 10\mbox{mm}$ while the average thickness is of the order of $\overline{h}\approx 2\mbox{mm}$. The factor $\lambda/\overline{h}\approx 5$ shows that the flow cannot be considered fully 'long-wave', but yet the Skhadov-like scaling does not do justice to the actual scales of the flow.
   
   Nevertheless, the contour map shows that the pressure is approximately uniform along the cross-stream direction (i.e. $\partial_y p\approx0$) and hence the boundary-layer approximation remains valid. This is further illustrated in the three snapshots in Fig.~\ref{fig:P_liquid}, which compare the pressure distributions at the wall with the pressure distributions at the interface. No appreciable differences are found, expect for some minor differences in the run-back flow region.
   
\begin{figure}
	    \includegraphics[width=\linewidth]{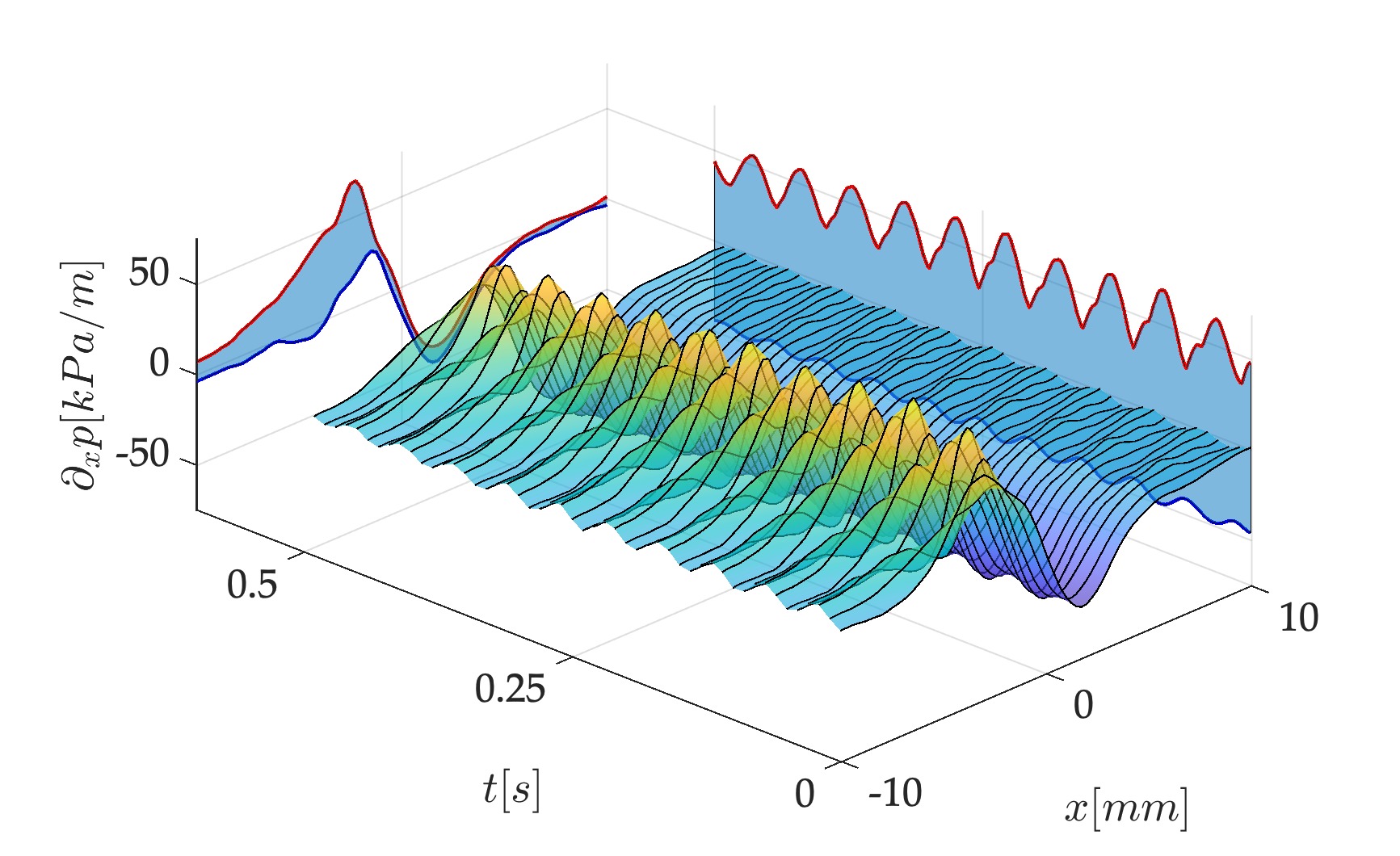}% Here is how to import EPS art
	    \caption{\label{fig:gradP_envelope} Envelope of the pressure gradient in time correlated with the 2D coupled undulations for Case 1.}
    \end{figure}

   We then analyze in Fig.~\ref{fig:gradP_envelope} the spatio-temporal evolution of the pressure gradient distribution considering only the coupled modes in the gas jet. It is worth recalling that in the investigated wiping conditions the shear stress distribution plays a negligible role in wiping. This figure shows the envelopes in the time and space domain, jointly visualized by means of a 3D plot with projections. The red and the blue lines are used to highlight the loci of maxima and minima in both the space (plane $\partial_x p-x$) and in the time domain (plane $\partial_x p-t$). The pressure gradient envelopes in time clearly highlight the pulsation at the origin of the coating undulation, while the envelopes in space highlight the region where these pulsations are stronger. That is the region $x<0$, i.e. on the side of the run-back flow. As expected, this is the region mostly influenced by the passage of waves in the liquid film. In the analyzed test cases, this result shows that the evolution of the pressure gradient is closer to the 'pulsating' case than the 'oscillating' case described in the previous theoretical work \cite{Mendez2020}. In addition, it is found in this last reference that the relative wave amplitude on the run-back is substantially larger than the one on the final film when the integral film model is fed with a pulsation of the wiping actuators, which is coherent with the present observations. 
   
\begin{figure}
		\includegraphics[width=0.46\linewidth]{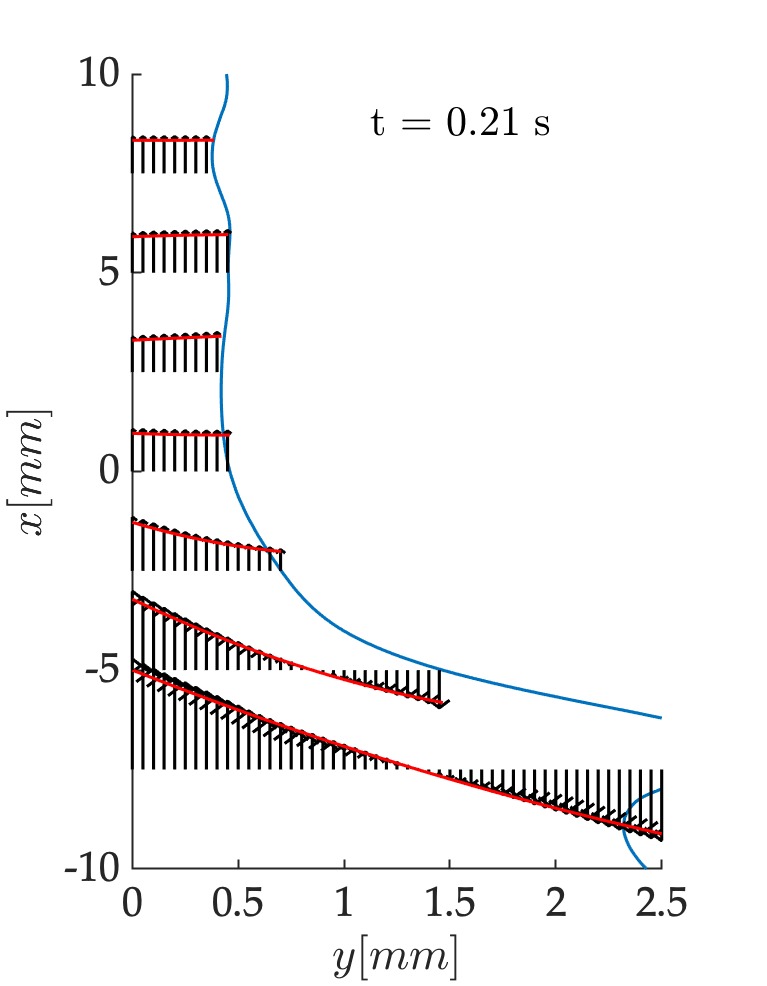}
		\includegraphics[width=0.46\linewidth]{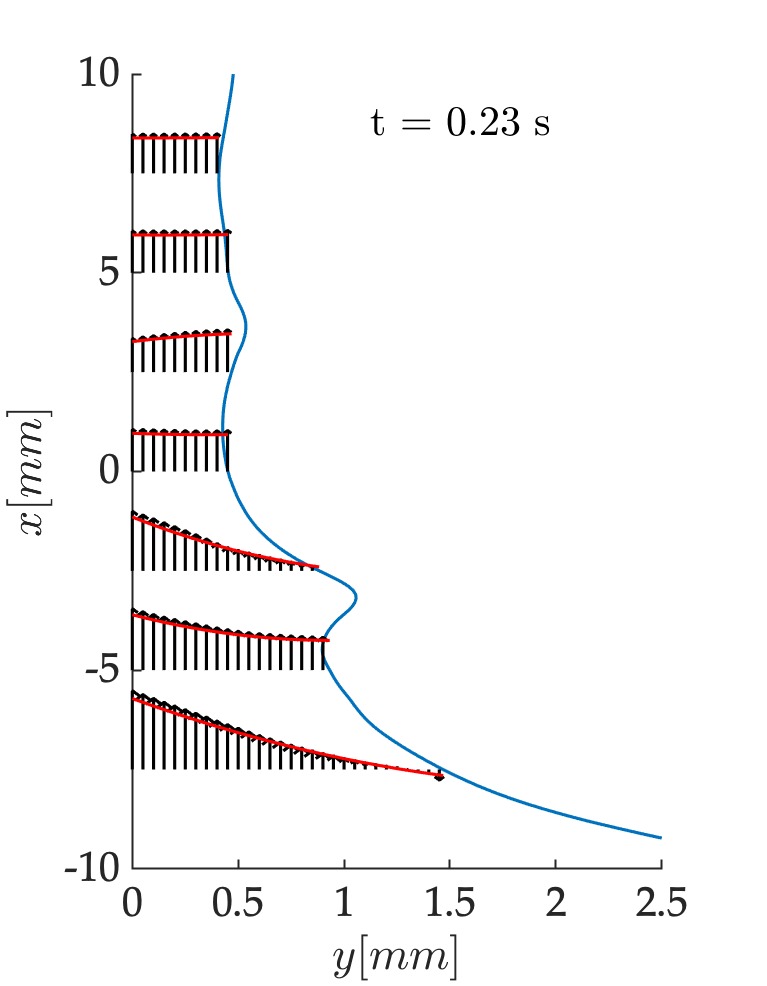}
		\includegraphics[width=0.46\linewidth]{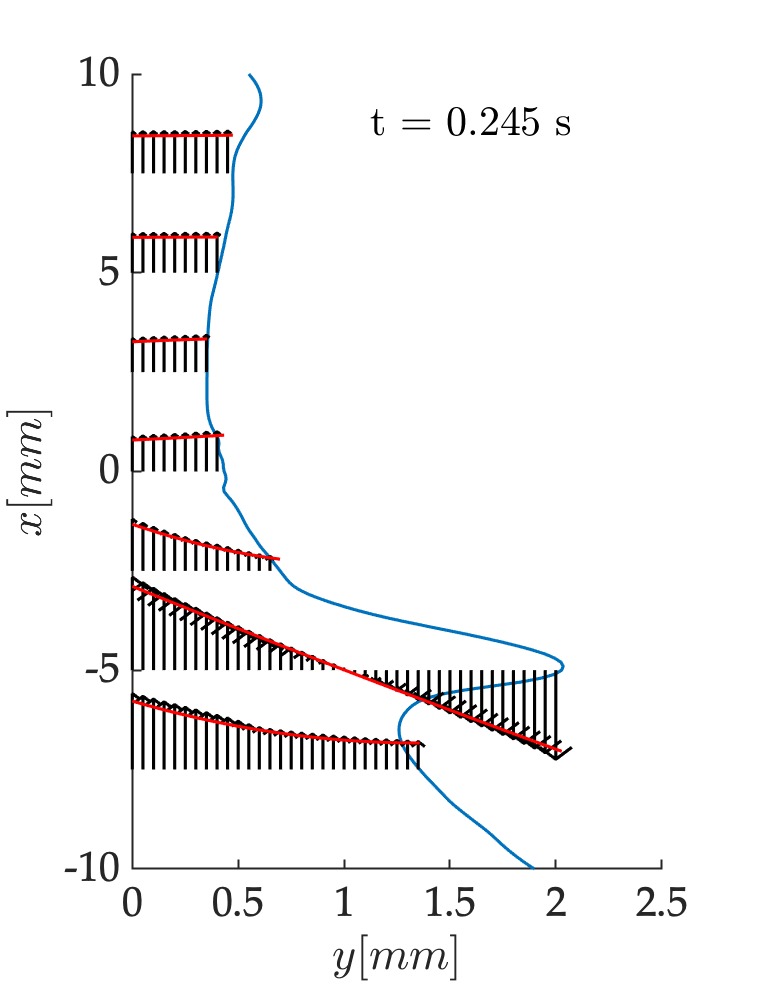}
		\includegraphics[width=0.46\linewidth]{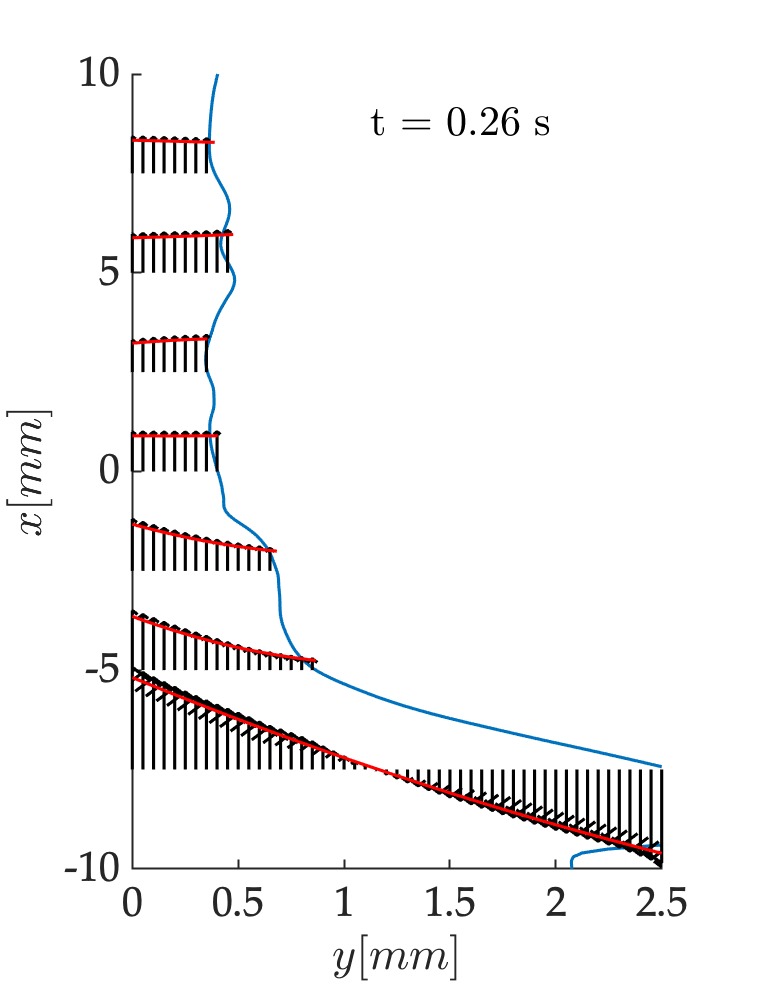}
	\caption{\label{fig:U_liquid} Instantaneous x-velocity profiles for four snapshots corresponding to the same undulation cycle. The CFD data is represented with black arrows, while the red lines correspond to the theoretical profiles computed from Eq.~\ref{Vel_COEFF}.}
    \end{figure}

   Finally, building on the validation of the boundary layer formulation, we conclude by analyzing the velocity profiles within the liquid film at different locations and different instants. These are shown in  Fig.~\ref{fig:U_liquid}. 
   All profiles are remarkably close to parabolic, as assumed in the integral formulation. Therefore, Eq.~\ref{Vel_COEFF} can be used to extract important quantities such as flow rate per unit width ($q$) and interface shear stress $\tau_g$ from a simple polynomial regression of the velocity profiles at each location. This is how the shear stress distributions in Fig.~\ref{fig:gas_emPOD_Case1}-\ref{fig:gas_emPOD_Case3} were computed.
   %% I propose an additional sentence because finishing the whole paper on this sentence sounds a little strange.
   
   All these elements suggest that the integral model might an interesting alternative to CFD, provided that the time-dependent wiping actuators are correctly modelled.

	\section{\label{sec:conclusions}Conclusions}
	
The dynamics of the gas-liquid interaction in the jet wiping process has been analyzed via CFD simulations and modal analysis. The numerical model showed a good accuracy in the prediction of the characteristics of the undulation observed in previous experiments \cite{Mendez2019}, in spite of slight differences between both configurations. 

The Multiscale Proper Orthogonal Decomposition (mPOD) has been extended to correlate coherent patterns in the liquid coating and coherent structures in the jet gas flow. In particular, traveling wave patterns have been identified in the liquid film and their temporal evolution correlated with the gas flow field. Remarkably, the dominant wave patterns were found to be highly bidimensional in all cases, and present in both the final coating region and the run-back flow region. The spatial structures, frequency content and contribution to the coating film thickness fluctuations are discussed. 

The extended mPOD revealed the gas structures coupled and correlated with the coating undulation. The dynamics of those structures was analyzed along with their impact on the pressure gradient and shear stress distributions at the  film interface. Jet oscillations, correlated with coating waves have been revealed. Nevertheless, the amplitude of these oscillations turned out to be significantly lower than what was observed experimentally.  Whether this result can be generalized to other wiping conditions remains to be investigated. In any case, the present work shows that the mechanism for wave formation in the coating film is, for the conditions analyzed, linked to a pulsation of the wiping actuators, rather than an oscillation of the later.

Finally, some key aspects of the integral model formulation have been analyzed using the CFD data. The numerical results were found to meet the main assumptions of the model: the cross-stream pressure gradient is negligible, confirming the validity of the boundary layer approximation, and the velocity profiles are parabolic, in agreement with the self-similarity hypothesis. The spatio-temporal evolution of the wiping actuators revealed a pulsation of the pressure gradient mostly concentrated in the run-back flow side.

The next step in this investigation will be the full validation of the integral model for wiping conditions at lower $\epsilon=Ca^{1/3}$, in which the Shkadov scaling is expected to hold better.

	\begin{acknowledgments}
		D.Barreiro-Villaverde is financially supported by Xunta de Galicia with the pre-doctoral grant "Programa de axudas á etapa predoutoral" (ED481A-2020/018) and the research project is founded by Arcelor-Mittal. The authors also wish to thank the “Red Española de Supercomputación” for the attribution of special computational resources at FinisTerrae II (CESGA) and Tirant (UV) (FI-2018-3-0040, FI-2019-1-0044). 
	\end{acknowledgments}
	
	\section*{Data availability}
         The data that support the findings of this study are available on request from the corresponding author. The data are not publicly available due to privacy restrictions.

	\section*{References}
	%\nocite{*}
	\bibliography{biblio} % Produces the bibliography via BibTeX.
	
\end{document}